\pdfoutput=1
\documentclass[%
acmsmall,%
screen,%
]{acmart}

\usepackage{array}
\usepackage[fatsemi,only,lbag,rbag,llbracket,rrbracket,llparenthesis,rrparenthesis,arrownot,Arrownot]{stmaryrd}
\usepackage{mathpartir}
\usepackage{mathtools}
\usepackage{xspace}
\usepackage{enumitem}
\usepackage{cleveref}   %
\usepackage{thmtools}
\usepackage{bbold}      %

\definecolor{behav}{RGB}{128,0,32}
\definecolor{lav}{RGB}{96,0,148}
\definecolor{snow}{RGB}{246, 248, 248}
\definecolor{fgreen}{RGB}{34, 139, 34}
\definecolor{anncol}{RGB}{130,0,115}
\definecolor{intcol}{RGB}{200,0,160}

\newif\ifio
\iotrue

\newif\ifunsub
\unsubtrue

\newif\ifminted
\mintedtrue
\InputIfFileExists{use_listings_please.tex}{}{}

\newif\ifcomments
\commentsfalse
\commentsfalse          %

\newif\ifbehavior
\behaviortrue
\behaviorfalse          %

\newcommand{\marginnote}[2]{%
  \ifcomments%
    $^{\color{magenta}\mathclap\star}$%
    \marginpar[
        \flushright\tiny\sf\textbf{#1}: #2
    ]{
        \flushleft\tiny\sf\textbf{#1}: #2
    }%
  \fi%
}
\ifbehavior
\newcommand{\behavior}[1]{{\color{behav}#1}}
\else
\newcommand{\behavior}[1]{}
\fi 

\ifcomments
\newcommand{\lucam}[1]{\marginnote{Luca}{\color{blue}#1}}
\newcommand{\luca}[1]{\noindent\textbf{[Luca: }{\color{blue}#1}\textbf{]}}
\newcommand{\beppem}[1]{\marginnote{Beppe}{\color{magenta}#1}}
\newcommand{\beppe}[1]{\textbf{[Beppe: }{\color{magenta}#1}\textbf{]}}
\newcommand{\amine}[1]{\textbf{[Amine: }{\color{lav}#1}\textbf{]}}
\newcommand{\danielem}[1]{\marginnote{Daniele}{\color{fgreen}#1}}
\newcommand{\daniele}[1]{\noindent\textbf{[Daniele: }{\color{fgreen}#1}\textbf{]}}
\else
\newcommand{\lucam}[1]{}
\newcommand{\luca}[1]{}
\newcommand{\beppem}[1]{}
\newcommand{\beppe}[1]{}
\newcommand{\amine}[1]{}
\newcommand{\danielem}[1]{}
\newcommand{\daniele}[1]{}
\fi

\newlength{\elixircodevspace}
\ifminted
  \usepackage[frozencache=true,cachedir=minted-cache]{minted}
  \newcommand{\elixirfontsize}{\fontsize{8.5}{10.2}\selectfont}
  \setminted[elixir]{%
    frame=lines,
    bgcolor=snow,
    fontsize=\elixirfontsize,
    firstnumber=last,
    linenos=true,
    numbersep=.5em,%
    xleftmargin=0em,%
    escapeinside=!!}
  \setmintedinline[elixir]{bgcolor={},fontsize=\elixirfontsize}
  \newmintinline[eelixir]{elixir}{}   %
  \newcommand{\felixirfontsize}{\fontsize{7}{8.4}\selectfont}
  \newmintinline[felixir]{elixir}{fontsize=\felixirfontsize}
  \newminted[elixircode]{elixir}{firstnumber=last}    %
  \makeatletter
  \@ifundefined{minted@colorbg}{}{%
  \renewenvironment{minted@colorbg}[1]{%
    \setlength{\OuterFrameSep}{0pt}%
    \let\minted@tmp\FV@NumberSep
    \edef\FV@NumberSep{%
      \the\numexpr\dimexpr\minted@tmp+\number\fboxsep\relax sp\relax}%
    \vskip\elixircodevspace\relax
    \begin{minted@snugshade*}{#1}}%
   {\end{minted@snugshade*}%
    \vskip\elixircodevspace\relax\noindent}%
  }
  \makeatother
\else
  \usepackage{listings}
  \lstdefinelanguage{elixir}{%
    morekeywords={def,defp,defmodule,do,end,fn,receive,case,cond,when,after,%
      if,else,unless,spawn,send,self,use,import,alias,require,type,or},%
    sensitive=true,%
    morecomment=[l]{\#},%
    morestring=[b]",%
    morestring=[b]'%
  }
  \def\eelixir{\lstinline}
  
  \lstnewenvironment{elixircode}{}{}  %
\fi

\providecommand{\elixirfontsize}{}
\DeclareRobustCommand{\kwinterface}{{\ttfamily\elixirfontsize\color{intcol}\$interface}}
\DeclareRobustCommand{\kwselfinterface}{{\ttfamily\elixirfontsize\color{intcol}\$self\_interface}}

\newcommand {\enc}[1]{\langle\!\langle#1\rangle\!\rangle}
\newcommand {\dec}[1]{\mathsf{dec}(#1)}   %
\newcommand {\EncTypes} {\mathcal{K}}      %
\newcommand {\sem}[1]{\llbracket#1\rrbracket}
\newcommand {\semD}[1]{\sem{#1}_{\Domain}}   %
\newcommand {\accept} [1] {{\lbag #1 \rbag}}  
\newcommand {\testof} [1] {{\lceil #1 \rceil}}   %
\newcommand{\eqdef}{\stackrel{\smash{\textsf{\upshape\tiny def}}}=}

\newcommand\angles[1]{\langle#1\rangle}
\newcommand\braces[1]{\{#1\}}
\newcommand\bracks[1]{[#1]}

\newcommand\kwcolor{\color{blue}}
\newcommand\constcolor{\color{purple}}
\newcommand\atomcolor{\color{teal}}
\newcommand\anncolor{\color{anncol}}

\newcommand\mkkeyword[1]{\mathsf{\kwcolor#1}}
\newcommand\mkconstant[1]{\mathsf{\constcolor#1}}

\DeclareRobustCommand\ann[1]{{%
  \let\kwcolor\anncolor
  \let\constcolor\anncolor
  \let\atomcolor\anncolor
  \anncolor#1}}

\newcommand\seqof[1]{\overline{#1}}
\newcommand\subst[2]{\braces{#1/#2}}
\newcommand\match[2]{#1/#2}

\newcommand{\rulename}[1]{\textnormal{\textsc{\small[#1]}}\xspace}

\newcommand\VarSet{\mathsf{Vars}}
\newcommand\PIDSet{\mathsf{PIDs}}
\newcommand\AtomSet{\mathsf{Atoms}}

\newcommand\Wch{\mkconstant{Pid}}
\newcommand\Wfun{\mkconstant{Fun}}
\newcommand\Cnil{\mkconstant{nil}}
\newcommand\CPoint{\mkconstant{noval}}

\newcommand\Tany{\mathbb{1}}
\newcommand\Tempty{\mathbb{0}}

\newcommand\TFun{\mkkeyword{fun}} %
\newcommand\TAtom{\mkkeyword{atom}}
\newcommand\TBool{\mkkeyword{bool}}
\newcommand\TInt{\mkkeyword{int}}
\newcommand\TTuple{\mkkeyword{tuple}}

\newcommand\TDyn{\mkkeyword{dynamic}}
\newcommand\TChan[2][]{\mkkeyword{pid}\ifblank{#1}{}{^{#1}}\ifblank{#2}{}{(#2)}}

\makeatletter
\newcommand\annmv[1]{\@ifnextchar_{\annmv@sub{#1}}{\ann{\textsf{#1}}}}
\def\annmv@sub#1_#2{\ann{\textsf{#1}_{\textsf{#2}}}}
\makeatother
\newcommand\ta{\annmv{t}}
\newcommand\sa{\annmv{s}}
\newcommand\decorate[2][]{\textsf{decorate}_{#1}(#2)}
\newcommand\erase[1]{\textsf{erase}(#1)}
\newcommand\typeerase[1]{\textsf{typerase}(#1)}

\newcommand {\domega} {\partial}
\newcommand {\dagemo} {\delta}
\newcommand {\iffdef}  {\hbox{\;\;$\iff$\hspace*{-5.94mm}\raise5pt\hbox{\rm\scriptsize def}\hspace*{4mm}}}
\newcommand {\E}{\mathbb{E}}
\newcommand {\extint}[1]{\E(#1)}
\newcommand {\Types}{\mathcal{T}}
\newcommand {\Pd} {\mathcal{P}}
\newcommand {\Pf} {\mathcal{P}_{\!\textup{fin}}}
\newcommand {\Domain} {\mathcal{D}}
\newcommand {\DomainFCB} {\mathcal{D}_{\textsf{FCB}}}
\newcommand {\jzero} {\jmath_0}
\newcommand {\Constants} {\mathcal{C}}
\newcommand {\ConstantsInBasicType} {\mathbb{B}}

\newcommand {\Set}[1] {\{#1\}}
\newcommand*\oline[1]{%
   \vbox{%
     \hrule height 0.6pt%
     \kern0.25ex%
     \hbox{%
       \kern-0em%
       \ifmmode#1\else\ensuremath{#1}\fi%
       \kern-0em%
     }%
   }%
}
\newcommand\x{x}
\newcommand\y{y}

\newcommand\Atom[1][m]{\mathsf{\atomcolor#1}}

\newcommand\Const{\mkconstant{c}}
\newcommand\PID[1][a]{\mathsf{#1}}

\newcommand\Fun{\lambda}
\newcommand\Tuple[1]{(#1)}

\newcommand\Send[2]{\mkkeyword{send}(#1,#2)}
\newcommand\Recv[1]{\mkkeyword{receive}\,\braces{#1}}

\newcommand\Case[2]{\mkkeyword{case}\;#1\,\braces{#2}}
\newcommand\Spawn[2]{\mkkeyword{spawn}(#1,#2)}
\newcommand\Self{\mkkeyword{self}}

\newcommand\Eval{\mathcal{E}}
\newcommand\Hole{\bracks\,}

\newcommand\Branch[2]{#1\mathstrut\Rightarrow #2}

\newcommand\parop{\mathrel\|}
\newcommand\EmptyMailbox{\varepsilon}
\newcommand\Process[4]{\angles{#1^{#2}, #3, #4}}

\newcommand\red{\rightarrow}

\newcommand\EmptyContext\emptyset
\newcommand\Context{\Gamma}
\newcommand\ContextD{\Delta}

\newcommand\ptenv[2]{{#1}/{#2}}
\newcommand\ptrel[4]{{#1}/\!\!/{#2} \rightsquigarrow (#3\,;\,#4)}

\NewDocumentCommand{\wte}{ O{i} O{o} m m m }{
  #3 \vdash_{#2}^{#1} #4 : #5
}
\NewDocumentCommand{\wtec}{O{???}O{???}O{{\color{red}???}}mmmm}{
  #4 \vdash_{#3}^{#1 \vartriangleright #2} #5 : #6 \vartriangleright #7
}
\newcommand\wtc[1]{\vdash #1}
\NewDocumentCommand{\wtee}{ O{i} O{o} m m m }{
  #3 \Vdash_{#2}^{#1} #4 : #5
}
\newcommand\wtce[1]{\Vdash #1}
\newcommand\gen{\mathsf{gen}}

\newcommand{\sboxc}[1]{%
  \hbox to 6.5pt{\hss$\scriptstyle #1$\hss}%
}
\newcommand{\sboxa}[1]{%
  \hbox to 5.5pt{\hss$\scriptscriptstyle #1$\hss}%
}
\newcommand {\deco}[2]{#1[#2]}

\NewDocumentCommand{\wbe}{ O{i} O{o} m m m O{\varepsilon} O{\varepsilon}}{
  #3 \vdash_{\deco{#2}{#7\hfill}}^{\deco{#1}{#6\hfill}} #4 : #5
}

\NewDocumentCommand{\tinf}{ O{i} O{o} m m m O{\varnothing}}{
  #3 \vdash_{#2}^{#1} #4 : #5 \;\triangleright\; #6
}
\NewDocumentCommand{\telab}{ O{i} O{o} m m m O{\varnothing} m }{
  #3 \vdash_{#2}^{#1} #4 : #5 \;\triangleright\; #6 \;\rightsquigarrow\; #7
}

\newcommand\subt{\leq}
\newcommand\issubt{\mathrel{\dot{\leq}}}

\newcommand\dom[1]{\mathsf{dom}(#1)}

\newcommand\cov[1]{\mathsf{cov}(#1)}
\newcommand\rcv[1]{\mathsf{rcv}(#1)}
\newcommand\dem[1]{\mathsf{dem}(#1)}

\AtEndPreamble{%
  \theoremstyle{acmdefinition}%
  \theoremstyle{remark}%
  \newtheorem{remark}[theorem]{Remark}%
}

\makeatletter
\def\@parfont{\bfseries\itshape}
\@mparswitchfalse
\makeatother

\acmJournal{TOPLAS}
\acmVolume{1}
\acmNumber{1}
\acmArticle{1}
\acmYear{2026}
\acmMonth{1}

\setcopyright{none}
\renewcommand\footnotetextcopyrightpermission[1]{}
\AtBeginDocument{%
  \csappto{ps@standardpagestyle}{\fancyfoot{}}%
  \csappto{ps@firstpagestyle}{\fancyfoot{}}%
  \pagestyle{standardpagestyle}%
}

\begin{document}

\title[You Have Mail: Set-Theoretic Types for Actors]{You Have Mail: Set-Theoretic Types for Actors}

\author{Giuseppe Castagna}
\affiliation{%
  \institution{IRIF, CNRS, Université Paris Cité}
  \city{Paris}
  \country{France}}

\author{Luca Padovani}
\affiliation{%
  \institution{DISI, Università degli Studi di Bologna}         
  \country{Italy}}

\author{Daniele Varacca}
\affiliation{%
  \institution{LACL, Université Paris-Est Créteil}         
  \country{France}}

\author{Amine Nasri}
\affiliation{%
  \institution{LACL, Université Paris-Est Créteil}        
  \country{France}}

\begin{abstract}
We present a type system for the processes of Elixir, a dynamic language
based on the actor model whose 
static type system covers only its
functional core. Each process is typed by \texttt{pid($t$)}, where the
``interface'' $t$ is the type of the messages the process expects in its
mailbox: sends are checked against the recipient's interface, and receives
rely on the interface of the process that performs them.

One of the difficulties of typing Elixir is that mailboxes are implicit:
a function shipped in a message may evaluate \texttt{self()} and execute
receive instructions, and both refer to whichever process eventually runs
the function---\texttt{self()} returns that process's pid, receive reads
that process's mailbox---so the mailbox they act on is not known where the
function is written. We solve this problem by inferring, for every
function, \emph{bounds} on the type of the mailboxes it can be used with;
as a result, the system types higher-order functions that travel in
messages even when they contain reference to \texttt{self()} and receive instructions.

The only new annotation the programmer writes is the interface of each spawned
process; everything else is inferred by reusing the existing subtyping and
constraint-resolution machinery of Elixir, and the system is proved sound.

For legacy code, the interface can be reconstructed from the receive clauses: what the
branches \emph{do} with their messages bounds it from above, and the most
permissive solution is sharp enough to recover received pids and closures.

\behavior{Finally, we
explore extending interfaces with a limited form of mailbox types \emph{à la}
de'Liguoro and Padovani, checking that every send is matched somewhere by a receive clause
that handles it and vice versa---a balance that rules out entire classes of
protocol errors, though short of liveness properties such as deadlock
freedom.}

\end{abstract}

\maketitle

\section{Introduction}
\label{sec:intro}
In the actor model~\cite{Hewitt1973,Hewitt1977,Agha1986}, concurrency is
achieved without shared memory: each actor (or \emph{process}) owns a private
mailbox into which other processes (or the process itself) deposit messages
asynchronously. A process scans its own mailbox and selects a message to handle,
usually by pattern matching. The absence of shared memory makes concurrent
programs easier to write and to reason about, which is a large part of why the
actor model is so widely used in practice.

A major 
implementation of the actor model is the BEAM~\cite{beam}, the virtual machine originally designed
for Erlang. The BEAM is also the runtime of Elixir, a dynamic language recently
endowed with a static gradual type system based on set-theoretic types and semantic
subtyping~\cite{CDV24,CD26,duboc-phd}. The Elixir type system, which has been
steadily integrated into Elixir since release 1.17,
covers the functional core of Elixir, but still lacks types for two of its
important features: first-class modules, generically typed by
\eelixir{module()}, and processes, generically typed by \eelixir{pid()} (for
\emph{\texttt{p}rocess \texttt{id}entifier}). 

This work fills the second gap, introducing a type system for Elixir processes
in which each process is typed by \texttt{pid($t$)}, where $t$ is the process
``interface'', that is, the type of the messages the process expects to find in
its mailbox. This information serves two purposes: to statically check that
every message sent to a process is compatible with the process's interface, and
to statically infer the type of the messages a process receives. For instance,
if a process \eelixir{p} has interface \eelixir{integer()}
then any
message sent to \eelixir{p} must be guaranteed to be an integer and, conversely,
\eelixir{p} can rely on the assumption that every message in its mailbox is an
integer. Our extension is conservative by design: interfaces, and the typing of
shipped code, compile into the set-theoretic types Elixir already uses, so the
existing subtyping and inference machinery, formal and implemented, applies
unchanged; type annotations are erased at compilation, and typed programs run without the need of modifying the BEAM.

Typing the actor model is not new. It has been studied in academic settings
(mailbox types that constrain the contents of a mailbox~\cite{DP18,FAD23},
session types that discipline the conversations a process engages
in~\cite{MV11,NY17}, type-parameterized actor references in the style of
TAkka~\cite{TAkka}, and processes typed by the very set of messages they
accept~\cite{CPS97,FLW17}), and these ideas have reached industry, most
prominently in Akka Typed~\cite{AkkaTyped}, the industrial reference, which
descends from the TAkka line of work.  As far as we know, however, all of these
approaches share one simplifying assumption, namely that the identity of the
process running a piece of code is known when that code is typed. This
assumption is enforced in one of two ways. In some cases, functions are not
first-class: there is a global pool of function definitions, so nothing with a
captured environment can travel at all. In other cases, code can travel, but it
is \emph{sealed at its definition site}: every process identity and send
capability it uses must be captured or passed explicitly, and is typed where the
closure is written, not where it runs. 

Such simplifying assumption is not satisfied by Erlang and Elixir. In these
languages, functions are first-class, meaning not only that it is possible to
write higher-order functions, but also that functions can travel in messages. In
addition, the sealing discipline is not possible because of the nature of the
primitive operations (\eelixir{receive} and \eelixir{self()}) that concern a
process's \emph{own} mailbox: these operations \emph{implicitly} refer to the
mailbox of whichever process runs them, so there is nothing for a closure to
capture or pass explicitly.
The operation
\eelixir{receive} scans the mailbox of the process that \emph{runs} it
(there is no way to scan and consume messages of another process's mailbox).
Thus, a \eelixir{receive} captured inside a closure reads a different mailbox,
and finds differently-typed messages in it, depending on which process
eventually runs the closure. The operation \eelixir{self()} returns the pid of
the process that \emph{runs} it. Its binding is thus late: the same
occurrence of \eelixir{self()}, captured inside a closure, denotes a different
process, and must be typed differently, depending on which process eventually
runs that closure. This is not peculiar to the BEAM: since Hewitt's seminal
work~\cite{Hewitt1977}, any self-reference inside the behavior of an
actor is not resolved where the behavior is defined (languages differ in when
they resolve it, see Section~\ref{sec:related}). The late binding of
\eelixir{self()} is thus a feature of the actor model at large, not just of
Elixir and Erlang.

In order to cope with first-class functions that may contain implicit references
to the process's mailbox through the \eelixir{receive} and \eelixir{self()}
operations, we decorate the type of functions with an interval of types called
\emph{bracket}. The bracket allows us to identify the range of interfaces of the
processes that can safely apply the function.
To illustrate the role and meaning of brackets more concretely, consider the
following Elixir function definition encapsulating a remote procedure call:
\begin{elixircode}
def remote_procedure_call(server) do
  send(server, {:request, self()})
  receive do {:response, y} -> y + 1 end
end
\end{elixircode}
The function accepts the pid of a \eelixir{server}, to which it sends a message
\eelixir'{:request, self()}', that is, the tuple made of the \emph{atom} \eelixir{:request} (a constant with a symbolic name) along with the
pid of the process that runs the function.
It then waits for a \eelixir{:response}-tagged message
(supposedly the one sent by the server through the pid returned by
\eelixir{self()}) whose payload \eelixir{y} is assumed to be an integer. The
function terminates returning \eelixir{y + 1}.

What is the interface $t$ of a process that can safely call the function
\eelixir'remote_procedure_call'? The key observation is that there is not a
unique answer.
On the one hand, the process's mailbox must \emph{at least} accept \eelixir{:response}-tagged messages with integer payloads. Thus the interface $t$ should be \emph{bounded from below} by the type
\begin{center}
  \hfill\eelixir'{:response, integer()}' \hfill (\refstepcounter{equation}\theequation)\label{eq:rpc-lower-bound}
\end{center}
because that is the type of the message(s) that \eelixir{server} sends back to
the process through the pid returned by \eelixir{self()}. Any process whose
interface $t$ does not include this bound risks finding in its mailbox an
unexpected \eelixir|:response|-tagged message.

On the other hand, the process's mailbox can safely contain other types of message as long as its interface ensures that any \eelixir{:response}-tagged message has integer payload. Thus, the interface $t$ should be \emph{bounded from above} by the type
\begin{center}
  \hfill\eelixir'{:response, integer()} or not({:response, not(integer())})'. \hfill (\refstepcounter{equation}\theequation)\label{eq:rpc-upper-bound}
\end{center}
where we use the connectives \eelixir{or} and \eelixir{not} to respectively
express unions and negations, in the spirit of set-theoretic types. This upper bound excludes an interface such as \eelixir'{:response, integer() or atom()}', that cannot ensure that the \eelixir{y} occurring in \eelixir{y + 1} is an integer.

The types in~\eqref{eq:rpc-lower-bound} and~\eqref{eq:rpc-upper-bound} are the lower and upper bounds of the bracket that decorates the type of \eelixir'remote_procedure_call'.
Any process whose interface is not included in these bounds may have messages in
its mailbox that match the pattern \eelixir'{:response, y}' in the body of
\eelixir'remote_procedure_call', but where \eelixir{y} is not guaranteed to be
an integer.
It is fine if the mailbox contains other messages (those described by the type
\eelixir'not({:response, not(integer())})'), because the function executes a
\emph{selective receive} that ignores any message not matching the pattern.

Note that distinguishing the two bounds is important, in the sense that
collapsing the bracket one way or another (or even to a type in the middle) would be a unnecessary
restriction: we would rule out processes that can safely run this function.
For instance if we decrease the upper bound so that
it matches the lower bound, then the function is only callable by those
processes that do not allow anything other than messages of type
\eelixir'{:response, integer()}' in their mailbox.

In the rest of the paper, we
devise a type system
guaranteeing that no message with unexpected type is ever sent or selectively
received in the presence of first-class functions. For instance, in the example above, the function \eelixir'remote_procedure_call' can be given type
\begin{center}
   \hfill \eelixir|pid({:request, pid({:response, integer()})}) -> integer()|\hfill (\refstepcounter{equation}\theequation)\label{eq:rpc-type}
\end{center}
since it expects as argument the pid of a server accepting messages tagged with \eelixir{:request} and with the runner's pid as payload. This arrow type will be decorated with the types in \eqref{eq:rpc-lower-bound} and \eqref{eq:rpc-upper-bound}.

Ideally, for our approach to be applicable to \emph{existing} Elixir code, brackets and interfaces should be inferrable. We show that brackets can indeed be inferred, and their inference is complete and decidable (Section~\ref{sec:ann-inference}). We also show that interfaces can be reconstructed from the code of the spawned process, but only with a weaker guarantee: a reconstructed interface that can be written as an annotation is one the programmer could have declared, whereas a programmer-written declaration can say more than reconstruction can derive (Section~\ref{sec:reconstruction}).

\subsection*{Summary of Contributions}

\begin{itemize}[leftmargin=*,itemsep=2pt,topsep=2pt]
    \item \emph{Set-theoretic interfaces for processes.} Processes are typed by
    the set of messages they accept: \eelixir{pid(t)} is contravariant in
    \eelixir{t} and comes with unions, intersections, negations, and genuine
    subtyping, where existing systems rely on sealed nominal protocols.
    \item \emph{Typed higher-order messages.} Functions can be passed to other
    functions, shipped as messages, and called by a process different from the one that defined them: the system checks what a function may read from and deposit into the mailbox of
    whichever process \emph{calls} it at the definition site. %
    No existing
    actor type system %
    types shipped code
    at its runner, or without fixing in advance the exact interface of that
    runner: in our work the same function runs in every process
    whose interface satisfies the minimal requirements of the function.
    \item \emph{Soundness.} The system is proved sound
    (Section~\ref{sec:properties}): every message sent is guaranteed to belong to the
    recipient's interface, and every \eelixir{receive} can rely on its own
    process's interface: messages neither go astray on the way in nor arrive
    mistyped on the way out, even when they carry code.
    \item \emph{Bracket inference.} The only new annotation required is the interface of
    a freshly spawned process; the brackets in arrow types are
    inferred, resolving to a \emph{single} call to constraint resolution
    (Section~\ref{sec:ann-inference}). %
    The interface of the running process is kept
    \emph{symbolic}: a function that reads the mailbox is typed \emph{once},
    which yields a \emph{requirement} on the interface of whatever process runs
    it, and each \eelixir{spawn} of a process that runs it only checks that the
    new interface satisfies this requirement. A function can thus be shared by
    processes with different interfaces, typically with no loss of precision
    with respect to typing its body separately for each process that runs
    it. Annotations have no runtime effect, so typed
    programs run on the stock BEAM. 
    \item \emph{Interface reconstruction.} For legacy code that carries no
    annotation at all, interfaces are reconstructed from the \eelixir{receive}
    clauses, with a soundness guarantee
    (Section~\ref{sec:reconstruction}). What the branches of a \eelixir{receive} \emph{do} with a
    message is a requirement on the interface, and the reconstruction returns the
    \emph{greatest} interface satisfying all of them (or a canonical maximal one, when a
    received closure makes the requirement mention the interface).
 \behavior{
    \item \emph{A behavioral layer.} Mailbox types tracking multiplicities
    and alternatives, built on the same foundations
    (Section~\ref{sec:behavioral-types}) and checking a two-way balance between sends and receives, rule out entire classes of protocol errors.\lucam{Puff, dal nulla compare il livello comportamentale. Non avendo spiegato le limitazioni dei controlli ``statici'', non è chiaro quale sia il problema che risolve questo layer comportamentale e introdurlo così sembra molto forzato. Forse c'è troppa roba.}}
\end{itemize}

The rest of the paper is organized as follows. Section~\ref{sec:example}
illustrates the system on a more complex running example, previewing all its
main ingredients. Section~\ref{sec:syntax} defines the syntax of a core calculus
for Elixir processes, and Section~\ref{sec:types} its types, their set-theoretic
interpretation, and the encoding that reduces their subtyping to the existing
algorithms. Section~\ref{sec:opsem} gives the dynamic semantics (presented after
subtyping because the runtime pattern matching semantics depends on it), and
Section~\ref{sec:typing} the static semantics; the properties of the system,
notably type soundness, are stated in Section~\ref{sec:properties}.
Section~\ref{sec:ann-inference} closes the gap between the calculus and Elixir:
programmers annotate each spawn with an interface written in current Elixir type
syntax (enriched with \eelixir+pid(t)+ types), and everything else is inferred with the existing subtyping and constraint-resolution
machinery.
Section~\ref{sec:reconstruction} reconstructs the interfaces that are not
annotated,
\behavior{ Section~\ref{sec:behavioral-types}
presents the optional behavioral layer,}and Section~\ref{sec:elixir} discusses
the integration of our system with Elixir. Related work is discussed in Section~\ref{sec:related} and
Section~\ref{sec:conclusion} concludes. An online REPL prototype that types a suitable
    subset of Elixir is available at \url{https://url.irif.fr/elixirproc}. It is written in Elixir and it
    implements the inference of Section~\ref{sec:ann-inference} and the reconstruction of
    Section~\ref{sec:reconstruction} by directly calling the
    API of the Elixir compiler (the development version) for the type
    representation, subtyping checking, and  constraint solving.\lucam{Nell'intro parliamo poschissimo di ``actors'' e moltissimo di ``Elixir''. Nel titolo c'è ``actor'' e non c'è ``Elixir''. Continuo a pensare che il titolo non sia del tutto rappresentativo.}
    \beppem{Wadler ha intitolato il suo lavoro "Typecasting Actors: from Akka to TAkka." anche se il lavoro concerne solo Scala a tal punto che è stato pubblicato nel workshop di Scala. Lasciamolo così!}

\section{A Running Example}
\label{sec:example}

To
illustrate all the aspects outlined in the introduction, we present a classic example of a server that
accepts clients and spawns one \emph{counter} worker per client; the worker then
serves that client directly, maintaining a counter that the client can
increment, increase by a given integer, or query.

\begin{elixircode}
defmodule Server do
  !\color{anncol}\$type! worker = :bump or {:add, integer()} or {:get_total, pid({:total, integer()})}  !\phantomsection\label{workerdef}!

  !\color{anncol}\$! (-> none())
  def dispatcher do                                     # CODE EXECUTED BY THE SERVER
    receive do
      {:connect, client} ->                             # wait for a client to connect
        !\kwinterface! worker
        spawn(fn -> serve(client) end)                  # spawn a worker for that client !\phantomsection\label{spawnclosure}!
        dispatcher()
    end
  end

  !\color{anncol}\$! pid({:ready, pid(worker)}) -> none()                               !\phantomsection\label{serveann}!
  defp serve(client) do                                 # CODE EXECUTED BY THE WORKER 
    send(client, {:ready, self()})                      # self() is the worker (late-bound)         !\phantomsection\label{self1}!
    send(self(), :bump)                                 # counts the connection as one bump  !\phantomsection\label{self2}!
    loop(0)                                                                               !\phantomsection\label{loopzero}!
  end

  !\color{anncol}\$! integer() -> none()                                                  !\phantomsection\label{loopann}!
  defp loop(n) do                                                                         !\phantomsection\label{loopdef}!
    receive do
      :bump              -> loop(n + 1)                      # increment the counter!\phantomsection\label{bumpbranch}!
      {:add, k}          -> loop(n + k)                      # add k to the counter       !\phantomsection\label{addbranch}!
      {:get_total, from} -> send(from, {:total, n}); loop(n) # return the counter's value  !\phantomsection\label{total}!
    end
  end
end
\end{elixircode}

The code above is valid Elixir code except for the \kwinterface{}
annotation, which declares the type of the messages the spawned process expects
in its mailbox.\footnote{The $\color{anncol}\texttt{\$}$-prefixed annotations
are not part of Elixir yet, but their introduction is planned for a future
release~\cite{CDV24,Elixir-1.20}.} 

The server is implemented by a zero-argument function \eelixir{dispatcher} that loops forever, waiting for clients to connect; its type, as specified by the annotation preceding its definition, is ``\eelixir{(-> none())}'', meaning that it takes no argument and returns a value in the empty type \eelixir{none()}, i.e. it never returns. 
The dispatcher waits for a client to connect,
and when it does, it spawns a worker to serve that client. 
The worker's
interface is declared as \eelixir{worker}, which is a union of three message
types: \eelixir{:bump} (i.e., the singleton type of the atom \eelixir|:bump|) to increment the worker's internal counter,
the tuple type \eelixir'{:add, integer()}' to add an integer to the counter, and
the tuple type \eelixir'{:get_total, pid({:total, integer()})}' to ask for the
total count, which is returned to the client.\lucam{Avrei scelto inc, add, get}\beppem{Si può cambiare... via Claude}
The \eelixir{or} separating them denotes a union type.

Note that, in this example and in the rest of the paper, we follow the convention of the Elixir documentation and write the parentheses
on the predefined types (\eelixir{integer()}, \eelixir{none()},
\eelixir{pid(t)}, \dots) and omit them on the type names defined by the
programmer (\eelixir{worker}, \eelixir{job}, \dots), so that the two are told
apart at a glance.

\paragraph{The interface as a contract.}
Declaring \kwinterface{} turns the mailbox into a \emph{contract}, checked
at both ends. On the way in, every message aimed at the worker is verified
against the \eelixir{worker} interface: a client holding the worker's pid
\eelixir{w} that sends either of
\begin{elixircode}
send(w, {:aadd, 12})        # typo in the tag: :aadd                !\phantomsection\label{typo}!
send(w, {:get_total, 42})   # wrong payload: not a pid  !\phantomsection\label{total42}!
\end{elixircode}
is \emph{rejected} statically. The first message (line~\ref{typo}) has a tag,
\eelixir{:aadd}, that no message of the interface carries. The second
(line~\ref{total42}) has the right tag but the wrong payload: under \eelixir{:get_total} the
interface expects a reply address, of type \eelixir|pid({:total, integer()})|, not an
integer. Elixir today accepts both sends: at run time the first message is never
matched and stays in the worker's mailbox forever, while the second crashes the
worker when it executes \eelixir|send(from, {:total, n})| with \eelixir{from} bound to
42, far from the client where the error is. On the way out, the interface \emph{types} what the worker
receives: in the \eelixir|{:add, k}| branch of \eelixir{loop} (line~\ref{addbranch}) the contract
guarantees \eelixir{k : integer()}, so \eelixir|loop(n + k)| typechecks as
it stands: \eelixir|n + k| is an integer, as the annotation of \eelixir{loop}
requires, because the interface promised one. \emph{Messages caught on the way in, payloads typed on the way out: this is what
interface annotations are for}.

\paragraph{The dispatcher's interface.}
The interface of the dispatcher is specified when spawning a new
dispatcher:
\begin{elixircode}
!\kwinterface! {:connect, pid({:ready, pid(Server.worker)})}           # dispatcher's interface !\phantomsection\label{dispiface}!
spawn(fn -> Server.dispatcher() end)
\end{elixircode}
The interface, where \eelixir{Server.worker} is the type \eelixir{worker} defined
in the module \eelixir{Server}, states that a dispatcher expects a \eelixir{:connect} message
containing the pid of a client, that is, a process that accepts \eelixir{:ready}
messages with a worker's pid payload. A client, whose code we do not show, connects
by executing \eelixir|send(d, {:connect, self()})|, where \eelixir{d} is the pid
returned by the \eelixir{spawn} above. When the client is the main process---not one spawned by another process---it
declares its own interface, the worker's pid and then the total, by a \kwselfinterface{} top-level
annotation:\lucam{main interface?}
\begin{elixircode}
!\kwselfinterface! {:ready, pid(Server.worker)} or {:total, integer()}    # client's interface !\phantomsection\label{clientiface}!
\end{elixircode}
Note how sharply the dispatcher's interface (line~\ref{dispiface}) describes
the client: not just as a pid, but as a pid of type
\eelixir|pid({:ready, pid(Server.worker)})|. Why is that so, given that
\eelixir{dispatcher} itself never sends anything to \eelixir{client}? Because it
passes \eelixir{client} to \eelixir{serve} (line~\ref{spawnclosure}), whose
annotation (line~\ref{serveann}) requires exactly this type, and a type can only
be \emph{weakened} downstream by subsumption, never sharpened. So the interface of
a process must anticipate not just the sends its own code performs but everything
reachable through the calls that code makes: here, whatever
\eelixir{serve(client)} requires of \eelixir{client}. Had line~\ref{dispiface}
declared, say, \eelixir+{:connect, pid(none())}+ (``some pid, assume nothing''),
the call \eelixir{serve(client)} would not typecheck.

Conversely, notice what the dispatcher's interface does \emph{not} say. The
client's interface (line~\ref{clientiface}) also contains
\eelixir+{:total, integer()}+, the reply to the total requests the client sends to
the worker, yet line~\ref{dispiface} never mentions it. First, subtyping hides
that part of the client's interface when the client connects: a pid that accepts
the messages of line~\ref{clientiface} accepts, in particular, those of type
\eelixir+{:ready, pid(Server.worker)}+, so the \eelixir{self()} that the client
sends to the dispatcher fits the type line~\ref{dispiface} expects. Second, the
pid to which the worker sends the total does not come from the dispatcher: it
arrives with the request itself, as the payload \eelixir{from} of a
\eelixir+{:get_total, from}+ message (line~\ref{total}), whose type the
\eelixir{worker} interface (line~\ref{workerdef}) fixes. So clients that never ask
for the total owe nothing: the obligation falls only on those who ask, exactly
when they ask.

\paragraph{Late-bound \eelixir{self()}.} Checking sends against a declared
interface is what any typed-actor system does; what matters here is
\emph{whose} interface a given \eelixir{self()} carries. The two occurrences of
\eelixir{self()} in \eelixir{serve} (lines~\ref{self1} and \ref{self2}) denote
the \emph{worker}, even though the \eelixir{spawn} that creates the worker runs
in the dispatcher: \eelixir{self()} denotes the process that \emph{evaluates}
it, not the one that wrote it. Its interface is therefore the worker's, and
the client duly receives \eelixir+{:ready, pid(worker)}+.

Late binding also allows a process to send messages to \emph{itself}, and
\eelixir{serve} does so to prime the counter, with \eelixir{send(self(), :bump)}
(line~\ref{self2}), counting the connection as the first bump. A self-sent
message lands in the very mailbox its sender reads from, so it is checked
against the sender's \emph{own} interface rather than, as for an ordinary send, against the recipient's. Here \eelixir{:bump} is declared in the worker's interface and
all is well; had \eelixir{serve} contained a stray self-send such as
\begin{elixircode}
send(self(), :reset)  # rejected                                     !\phantomsection\label{reset}!
\end{elixircode}
(a typo, or a leftover from a half-removed feature), the \eelixir{spawn} creating the worker would be rejected statically,
for \eelixir{:reset} is not in the interface and this time there is no other
process to blame.

\paragraph{Shipping code, not just data.}
So far the only function that travels is the one handed to \eelixir{spawn}. The
distinctive strength of the system is that a function can also be shipped through
an ordinary \emph{message} and run by whoever receives it. Functions are
thus first-class messages, and they remain typed in transit. Let the worker
accept jobs by adding a new message type to its interface:
\begin{elixircode}
!\color{anncol}\$type! job    = (-> term())
!\color{anncol}\$type! worker = :bump or {:add, integer()} or {:get_total, pid({:total, integer()})} !\phantomsection\label{workerinterface}!
                or {:run, job}  # the new capability
\end{elixircode}
with one extra branch in \eelixir{loop} (to be added anywhere between lines \ref{bumpbranch} and \ref{total}):\footnote{The notation \eelixir{f.(e)} is Elixir's notation for applying a closure bound to a variable \eelixir{f}, the notation \eelixir{f(e)} being reserved for named functions, such as \eelixir{loop} on the same line.} 
\begin{elixircode}
      {:run, job}     -> job.(); loop(n)       !\phantomsection\label{job}!
\end{elixircode}

Note that the programmer writes a plain arrow \eelixir{(-> term())}, meaning \emph{any} zero-argument function (in Elixir, \eelixir{term()} is the top type). Nothing in
this type says what a job may or may not do to the mailbox of the process that runs
it. That information can be \emph{inferred}: from the branch \eelixir{job.()} in line~\ref{job} the
system learns that jobs execute inside the worker, so any \eelixir{self()}
occurring in a job denotes the worker, any \eelixir{receive} it performs reads
the worker's mailbox, and any self-send it performs lands there. How this is tracked (every arrow type silently carries mailbox information, inferred and never written by the programmer) is the subject of Section~\ref{sec:ann-inference}.

\paragraph{Two \texttt{self}'s, two processes.}
A client that has received \eelixir+{:ready, w}+ can now write:
\begin{elixircode}
send(w, {:get_total, self()})                   # self() = the client, bound here !\phantomsection\label{selfclient}!
send(w, {:run, fn -> send(self(), :bump) end})  # self() = the worker, bound there !\phantomsection\label{self3}!
\end{elixircode}

The two occurrences of \eelixir{self()}, one line apart, denote two different
processes. The first (line~\ref{selfclient}) is evaluated on the spot, by the
client: it is the client's pid, and the message typechecks only if the client
itself is ready to receive the \eelixir+{:total, integer()}+ reply, as its
interface (line~\ref{clientiface}) says it is. The second (line~\ref{self3}) sits
under \eelixir{fn}: it is evaluated only when the worker runs the job, so it
denotes the \emph{worker's} pid, and the self-send of \eelixir{:bump} is checked
against the \emph{worker's} interface (line~\ref{workerinterface}), in which it
duly appears. The function's typing owes nothing to the process in which it is defined.

Behind the check of the first occurrence lies interface subsumption: the
client's pid carries the client's full interface (line~\ref{clientiface}), which
is richer than \eelixir+{:total, integer()}+, yet it is accepted where a
\eelixir+pid({:total, integer()})+ is expected, because \eelixir{pid(t)} is
\emph{contravariant} in \eelixir{t}. Contravariance may surprise at first,
but it follows from what a pid grants. Since only the owner of a mailbox can
consume its messages (recall that \eelixir{receive} names no mailbox), holding a pid
confers a single capability: \emph{depositing} messages. Hence the more
messages a process accepts, the more contexts its pid can be used in: a pid
promising to accept every message of type \eelixir{t} is safely used
wherever only messages of some subtype of \eelixir{t} will be sent. This is
precisely the subtyping of output capabilities in the $\pi$-calculus with
i/o types of Pierce and Sangiorgi~\cite{PS96}, a pid being, in
$\pi$-calculus terms, a channel of which only the output capability is ever
handed out. Section~\ref{sec:types} derives, rather than
decrees, this contravariance from the set-theoretic interpretation of
interfaces.

The same subsumption was at work when the client connected: by sending
\eelixir|{:connect, self()}|, it passes its pid where the dispatcher's interface
(line~\ref{dispiface}) expects one that accepts only \eelixir{:ready} messages.
The client's pid thus has two incomparable supertypes, one per recipient, each recipient seeing only the
slice of the client's interface that it uses:
\eelixir+pid({:ready, pid(Server.worker)})+ for the dispatcher and
\eelixir+pid({:total, integer()})+ for the worker. The
worker's slice also types the reply: in the \eelixir|{:get_total, from}| branch
of \eelixir{loop} (line~\ref{total}), the worker's interface
(line~\ref{workerinterface}) gives \eelixir{from} the type
\eelixir+pid({:total, integer()})+, against which \eelixir|send(from, {:total, n})|
is checked. Request and reply are typed independently, in two processes, and agree
because both are checked against the same component of the worker's interface.

\paragraph{Two jobs that overstep, and one that does not.}
The job of line~\ref{self3} is accepted because the worker handles \eelixir{:bump}. A client shipping instead
\begin{elixircode}
send(w, {:run, fn -> send(self(), :tick) end})  # rejected         !\phantomsection\label{tick}!
\end{elixircode}
is stopped at this very \eelixir{send}: since jobs run inside the worker
\eelixir{w}, the \eelixir{self()} in this closure is the worker \eelixir{w},
its self-send would deposit \eelixir{:tick} in the worker's mailbox, and
\eelixir{:tick} is not among the messages the worker handles, which are specified in the type \eelixir{worker}  of
\eelixir{w} (line \ref{workerinterface}). 

A job can overstep on the read side as well: a client shipping
\begin{elixircode}
send(w, {:run, fn -> receive do x -> x + 1 end end})  # rejected         !\phantomsection\label{readjob}!
\end{elixircode}
is rejected too, this time because of what the closure \emph{reads}: its
\eelixir{receive} scans the mailbox of the worker that runs it, and the
worker's interface guarantees only that the messages found there are of type
\eelixir{worker}: the first message the job picks may well be
\eelixir{:bump}, and \eelixir{:bump + 1} does not typecheck. The first job
\emph{deposited} into the implicit mailbox a message its runner does not
handle; the second \emph{reads} from that mailbox under an assumption that
the runner's contract does not warrant.
Observe that all of this is
established without a single mailbox
annotation being written (\eelixir{job} is just \eelixir{(-> term())}) and
that the faulty code never travels: \textbf{the error is reported where the
closure is \emph{written}, in the client, not where it would have
\emph{misfired}, in the worker}. The stray self-send of line~\ref{reset} was at least the
worker's own code; these jobs would be injected by a third party, whose code the
worker can nevertheless police.\footnote{Interfaces do not police everything:
in the \texttt{\{:get\_total, from\}} branch of \texttt{loop}, it is \texttt{loop}'s
own \texttt{\color{anncol}\$}-prefixed annotation \texttt{integer() -> none()} in line~\ref{loopann} that guarantees that the
payload \texttt{n} of the reply is an integer.}

Let us finish with a third job, which both reads and writes the mailbox of its
runner and stays within the worker's contract:
\begin{elixircode}
send(w, {:run, fn -> receive do
                       {:add, 0} -> send(self(), :bump)  !\phantomsection\label{closurebump}!
                       {:add, n} -> send(self(), {:add, n+1}) !\phantomsection\label{closurework}!
                     end
               end})
\end{elixircode}
It turns a request to add 0 into a plain bump, and re-queues any other one with
its argument incremented. The job places two demands on whoever runs it, one per
capability. The \eelixir{n+1} of the second branch (line \ref{closurework}) \emph{reads}: it demands that
the runner's interface guarantee that the payload of an \eelixir{:add}
message is an integer. The first branch demands nothing on this side: its pattern
is made of literals alone, so it binds no variable, and the messages that reach
it are exactly those it matches, whatever the runner's interface may be. The two \eelixir|send(self(), ...)| \emph{write}: they
demand that the same interface accept both \eelixir{:bump} (line \ref{closurebump}) and
\eelixir|{:add, integer()}| (line \ref{closurework}). The \eelixir{worker} interface meets both
demands, so the job may be shipped to \eelixir{w}. These two demands, and nothing
else, are what the system silently records in the \emph{decorations} of the arrow type
it infers for the closure, that is, in the mailbox information that the arrow type carries. This is computed by the rely/guarantee discipline
presented in Section~\ref{sec:typing-rules}.

\paragraph{Legacy code: reconstructing the interface.}
The \kwinterface{} annotation is the recommended way to type a process,
but existing Elixir code carries none. For such code we \emph{reconstruct} the
interface of a spawned process from its body
(Section~\ref{sec:reconstruction}), from what its
\eelixir{receive}s do with the messages they select. The branch
\eelixir|loop(n + k)| (line~\ref{addbranch}) tells us that whatever arrives
under the tag \eelixir{:add} and is bound to \eelixir{k}, is added to \eelixir{n} and the result passed to \eelixir{loop}, which expects an integer; likewise, the branch
\eelixir|send(from, {:total, n})| (line~\ref{total}) tells us that the payload of
a \eelixir{:get_total} message, which is bound to \eelixir{from}, is the pid of a process that must accept \eelixir+{:total, integer()}+. Each branch
thus states a \emph{requirement} on the interface, and the interface is derived
as the most permissive one meeting them all: for the three branches of
\eelixir{loop} in lines~\ref{loopdef}--\ref{total}, exactly the \eelixir{worker}
of line~\ref{workerdef},
reconstructed with no annotation written. Being dictated by the code that reads
the mailbox, a reconstructed interface holds the code that writes to it to what
the body actually consumes, so the mistyped \eelixir|{:aadd, 12}| and the
ill-formed \eelixir|{:get_total, 42}| (lines~\ref{typo}--\ref{total42}) are rejected as
they would have been against a declaration. A typo in a receive pattern, on the
other hand, is not detected by the reconstruction, which simply derives an
interface that accepts the mistyped message: if the branch read
\eelixir|{:aadd, k}|, the reconstructed interface would contain
\eelixir+{:aadd, integer()}+ and not \eelixir+{:add, integer()}+, and the typo
would surface only through the rejection of the sends of \eelixir{:add}
messages to the worker.
This, like every interface in this paper, presupposes that all senders are
typed. When some are not, there are two fallbacks, both discussed in
Section~\ref{sec:elixir}. A process can be declared with interface
\eelixir{term()}, which accepts every message; in exchange, a branch may assume
about the message only what its own pattern guarantees, so the branch
\eelixir|{:add, k} -> loop(n + k)| is rejected unless a guard ensures that
\eelixir{k} is an integer. Or the process can be given a gradual interface,
which also accepts every message, and is what Elixir does today. The sharpest guarantees, and the
higher-order core of the system, rest on the single new annotation we require;
this is exactly what that annotation provides.

\behavior{
\paragraph{Behavioral types.} The system presented so far should cover the vast majority of the typing needs of Elixir programmers working with processes. Section~\ref{sec:behavioral-types} adds one last, optional layer, aimed at a much more specialized audience: the programmers who \emph{implement} concurrent protocols rather than just use them, and who are willing to trade heavier annotations for stronger guarantees.  

An interface declares which messages a mailbox may contain, but not how many, nor in which combinations: it is a set, not a protocol. A \emph{mailbox type} additionally prescribes multiplicities and alternatives. For instance, suppose a variant of our server whose dispatcher, when it is at capacity, turns a client away with a \eelixir{:busy} message instead of spawning a worker for it (its interface then declares the client's pid at \eelixir|pid({:ready, pid(worker)} or :busy)|). For a client of this server, which connects and, if it is served, asks for the total, one can declare at spawn time that its mailbox will conform to the mailbox type

\centerline{\eelixir|:busy + ({:ready, pid(worker)} . {:total, integer()})|}

\noindent
Where the interface promised only that \eelixir{:busy}, \eelixir{:ready} and \eelixir{:total} messages are acceptable, in any number and any combination, the mailbox type states that either the client is turned away, and its mailbox holds a single \eelixir{:busy}, or it is served, and then receives \emph{exactly one} \eelixir{:ready} and one \eelixir{:total}---no more, no less (`\eelixir{.}' joins contents, \eelixir{+} expresses alternatives, and \eelixir{*} ---see later--- unbounded repetition). The two alternatives are exclusive: no execution may deposit both a \eelixir{:busy} and a \eelixir{:ready} in that mailbox, and a client that still waited for a total after being turned away is rejected, since past the \eelixir{:busy} its mailbox type leaves nothing to read. Against such declarations the system checks a two-way \emph{balance}, made precise in Section~\ref{sec:behavioral-types}: whenever the programmer sends a message, somewhere there is a receive clause that can handle it and, vice versa, whenever the programmer writes a receive clause, somewhere there is a send of a message matching that clause. Although this falls short of deadlock freedom, it statically rules out entire classes of protocol errors: a process can no longer terminate leaving messages unread in its mailbox, a lock cannot be released twice---nor released by a process that never acquired it.
}

\section{Syntax}
\label{sec:syntax}
In order to define and study the type system for Elixir processes we first consider the following syntax, that we later extend to get closer to Elixir. 
\[
    \begin{array}{lrcl}
        \textbf{Variable} &
        x, y, z & \in & \VarSet
        \\
        \textbf{Process Id.} &
        \PID[a], \PID[b] & \in & \PIDSet
        \\
        \textbf{Atom} &
        \Atom & \in & \AtomSet
        \\
        \textbf{Constant} &
        \Const & ::= &
            n \mid \Atom
        \\
        \textbf{Type} &
        t, i, o & ::= &
            \Const \mid \TInt \mid \TAtom  \mid \TTuple \mid \TChan{t} \mid t \to_{o}^{i} t \mid \Tuple{\seqof{t}}
            \mid \Tany \mid t \lor t \mid \neg t
        \\    
        \textbf{Test Type} &
        \tau & ::= &
            \Const \mid \TInt \mid \TAtom \mid \TTuple \mid \TChan{\Tempty} \mid \TFun\mid \Tuple{\seqof{\tau}} \mid\Tany\mid             \tau \lor \tau \mid \neg \tau
        \\
        \textbf{Lambda annotations} &
        l & ::= &
            t \to^i_o t \mid l \land l
        \\
        \textbf{Expression} &
        e, f & ::= &
            \Const \mid \PID^t \mid x \mid \Self \mid \Fun^{l} x.~e \mid e~e \mid \Tuple{\seqof{e}} \mid
            \Case{e}{\seqof{\Branch{p}{e}}} \mid e+e
            \\
            & & \mid &
            \Send{e}{e} \mid \Recv{\seqof{\Branch{p}{e}}} \mid \Spawn{e}{t}
        \\
        \textbf{Pattern} & p & ::= & x \mid \tau \mid \Tuple{\seqof{p}} \mid p \land p
        \\
        \textbf{Value} &
        v, w & ::= & \Const \mid \PID^t \mid \Fun^l x.~e \mid \Tuple{\seqof{v}}
        \\
        \textbf{Eval. context} &
        \Eval & ::= &
            \Hole \mid \Eval~e \mid v~\Eval \mid \Tuple{\seqof{v},\Eval,\seqof{e}} \mid \Case{\Eval}{\seqof{\Branch{p}{e}}} 
            \mid\Eval+e \mid v+\Eval 
            \\
            & & \mid &
            \Send{\Eval}{e} \mid \Send{v}{\Eval}
        \\
        \textbf{Configuration} &
        P, Q & ::= & \Process\PID{t}{e}{\seqof{v}} \mid (P \parop Q)
    \end{array}
\] 

Types (ranged over by $t, s, u, i, o$) are the terms \emph{coinductively} generated by the
grammar above and which satisfy two additional constraints: (1)
\emph{regularity}: the term must have a finite number of distinct sub-terms; (2)
\emph{contractivity}: every infinite branch must contain an infinite number of
occurrences of type constructors (i.e., arrow, tuple, or pid types). All the other syntactic categories, test types
(ranged over by $\tau$) included, are generated \emph{inductively}: in particular test
types are finite terms, since Elixir guards are not allowed to perform recursive tests (in Elixir it is possible to test whether a value is a list, but not whether it is a list of, say, \eelixir|integers()|).

Types are essentially those of Core Elixir~\cite{CD26} with two notable exceptions: the addition of pid types $\TChan{t}$, and of \emph{decorations} $i$ and $o$ on arrow types. More precisely types may be singleton types $\Const$  (containing either an integer constant or an atom); the base types $\TInt$ (containing all integers),  $\TAtom$ (containing all atoms), $\TTuple$ (containing all tuples), or the top type $\Tany$ (containing all values); the constructors with type arguments are the tuple type $\Tuple{\seqof{t}}$, the pid type $\TChan{t}$ (containing all pids of processes accepting \emph{at least} the messages of type $t$), and the arrow type $s \to^i_o t$ of functions from $s$ to $t$ where $i$ and $o$ are input and output decorations: both are types that---as we explain in detail in Section~\ref{sec:types}---track the messages that the function \emph{assumes} it may consume from, and those it \emph{may produce} on, the local mailbox; the union $t \lor t$, and negation $\lnot t$ connectives complete the syntax. 

We use the abbreviations
$
  t_1 \land t_2 \eqdef \lnot (\lnot t_1 \lor \lnot t_2)
$, $
  t_1 \setminus t_2 \eqdef t_1 \land (\lnot t_2)
$, $
  \Tempty \eqdef \lnot \Tany$. We consider $\TBool$ as the union of the two atoms for true and false: $\TBool \eqdef \texttt{\color{teal}:true} \lor \texttt{\color{teal}:false}$.
We refer to $ \lor $, $ \lnot $, $ \land $, and $ \setminus $
as \emph{type connectives}, and to $\mkconstant{c}$, $ \TInt$, $\TAtom$, $\TTuple$, $\Tany$, $\Tuple\ldots$, $\TChan\ldots$, and $ \to $, that is, to everything but the connectives, as \emph{type constructors}. We call \emph{pid type} a type of the form $\TChan{t}$ and \emph{process type} any combination of pid types by union, intersection, and difference, just as an \emph{arrow type} is a single arrow and a \emph{function type} is a combination of arrow types. As customary, connectives have priority over constructors and negation has the highest priority---e.g., $\neg s \lor t\to^i_o u \land t'$ denotes  $((\neg s) \lor t)\to^i_o (u \land t')$. We use $\TFun$ to denote the top function type, that is $\Tempty\to^\Tempty_\Tany\Tany$, and use it in the definition of test types. Note that $\TFun$ is an \emph{atomic} symbol: although it denotes an arrow type, it is not itself a written arrow---a distinction that matters in Section~\ref{sec:ann-inference}, where the arrows written by the programmer carry no decorations and receive them from inference, while $\TFun$, being atomic, needs no decoration and denotes the same type throughout. 

Coinduction accounts for recursive types---e.g., to define the type of lists.
It is coupled with the contractivity condition, which excludes infinite terms
that have no meaningful interpretation as sets of values (e.g., the trees
satisfying $t = t \lor t$ or $t = \lnot t$), rules out infinite unions and
intersections, and provides an induction principle allowing the induction
hypothesis to be applied below type connectives (union and negation), though
not below non-atomic type constructors (tuples, pids, and arrows). The regularity condition is
necessary only to ensure that the subtyping relation is decidable~\cite{CD26}.

Expressions include variables, constants (integers $n$ and atoms $\Atom$), pids $\PID^{t}$ which are indexed by their interface type $t$, the self pid $\Self$, $\lambda$-abstractions that, as in the Elixir type system, are annotated with an intersection of arrow types, applications, tuples, case-expression (that can encode tuple projections). For the concurrent part of the language, there are expressions for message sending and receiving, and process spawning $\Spawn{e}{t}$, whose type argument $t$ is the interface of the spawned process. This type argument may of course contain decorated arrows, but in practice the programmer will only write interfaces without decorations, as in Section~\ref{sec:example}, and Section~\ref{sec:ann-inference} shows how the decorations can be inferred.
It is important to remark that pids are runtime-only values that cannot be constructed by the user, but are only generated when a process is spawned. 

Case-expressions match the value produced by the scrutinee against the patterns and execute the first matching branch. Receive-expressions select messages from the mailbox using the same patterns as case-expressions: the oldest message that matches some pattern is consumed and the first branch that matches it is executed; if no message matches, the process blocks until a new message arrives (this semantics is formalized in Section~\ref{sec:opsem}).

Patterns include capture variables (that match any value), the type test pattern $\tau$, which matches only the values that pass the test $\tau$ (Section~\ref{sec:opsem}), tuple patterns which match tuple values component-wise, and the intersection patterns which match values that match all the patterns in the intersection. 

Capture variables within a pattern are required to be pairwise distinct. This is necessary to ensure that the substitution generated by matching a value against a pattern is well-defined. Test types are essentially types in which arrows are replaced by $\TFun$, the top-function type, and pid types by $\TChan{\Tempty}$, the top process type (corresponding to Elixir's current \eelixir|pid()| type). The reason of their definition is that Elixir can observe whether a value is a pid or a function, but not which messages a pid accepts nor the domain and codomain of a function.  Intersection patterns partially simulate Elixir's patterns guards (e.g., \eelixir{x when is_integer(x)} is rendered as $x\land\TInt$); guards that are not type tests, such as the equality test for which the pin operator \eelixir{^x} of Elixir patterns is syntactic sugar, are not modeled (see Section~\ref{sec:related-elixir}). Disjunction patterns are not included since they can be expressed by multiple branches. 

Finally, expressions do not include recursion since it can be encoded by a fixpoint combinator, which is typable in our system~\cite[Section~5]{castagna2024inf}.

A \emph{mailbox} is a list of values written $\seqof{v}$. We write
$\EmptyMailbox$ for the empty mailbox. If $\seqof{v}$ and $\seqof{w}$ are
mailboxes, we write $\seqof{v}\seqof{w}$ for the concatenation of $\seqof{v}$
and $\seqof{w}$ (also written $\seqof{v}\cdot\seqof{w}$ when juxtaposition would
be ambiguous). A \emph{process} is a triple $\Process\PID{t}{e}{\seqof{v}}$ of a
pid $\PID^t$, an expression $e$, and a mailbox $\seqof{v}$. A
\emph{configuration} is a parallel composition of processes with pairwise
distinct pids.

Since run-time pattern matching tests values against (test) types, the reduction rules depend on subtyping: we thus define the subtyping relation first, followed by the operational semantics, and finally the typing relation.

\section{Types and Subtyping}
\label{sec:types}
Both the dynamic and static semantics of the calculus depend on the subtyping relation: the former to determine when a value matches a test type, the latter to determine when an expression can be used in a context expecting expressions of a different type. As for Elixir, we define the subtyping relation following the semantic subtyping approach \cite{FCB08}, which interprets types as  sets of elements of a model (intuitively representing the values of the type) and defines subtyping as set inclusion. Section~\ref{sec:channel-semantics} explains what the interpretation of the two new type constructors must represent and which ingredients this requires, Section~\ref{sec:arrow-semantics-pointed} gives the interpretation of the types of Section~\ref{sec:syntax}, by encoding them into the types of~\cite{FCB08}, and defines subtyping, and Section~\ref{sec:subtyping-laws} states the subtyping laws of the two new type constructors.

\subsection{Intuition}\label{sec:channel-semantics}
Most of the machinery to define the model and the interpretation of types is already present in \cite{FCB08}. The missing part is the interpretation of our two new type constructors, the pid type $\TChan{t}$ and the decorated arrow type $t_1 \to_o^i t_2$. In semantic subtyping there is no latitude in interpreting the type
\emph{connectives}: $\lor$, $\land$ and $\neg$ must denote the corresponding
set-theoretic operations. The modeling of the type \emph{constructors} is
instead constrained only by the semantics of the operations that introduce and
eliminate them, and by the trade-off between how faithfully a constructor is
modeled and how easily its containment can be decided. We thus let our intuition guide us in defining the interpretation of the two new type constructors.

\paragraph{Process types.} The interpretation of the pid type $\TChan{t}$ is straightforward since \citet{CDV07} already did it for us. A process of type $\TChan{t}$ is a buffer abstract data type exposing only a
\emph{write} operation and no \emph{read} one: a process may $\Send{\cdot}{\cdot}$
to a specific process, but it never receives \emph{from} a specific process: reception happens on
a process's own mailbox. Therefore, $\TChan{t}$ exactly corresponds to the type $\texttt{ch}^{-}(t)$ of write-only channels of \cite{CDV07}, whose interpretation is the set of all pids $\PID^s$ (``channels'' in \cite{CDV07}) such that $t\subt s$ (write-only channels are contravariant). \citet{CDV07} show that this corresponds to the set of all types that are supertypes of $t$,
\begin{equation}\label{eq:chan}
    \sem{\TChan{t}} = \{ \sem{s} \mid t\leq s \}
\end{equation}
They also show how to break the apparent circularity of the definition above. With this definition, the subtyping relation follows the contravariant pattern: if $\sem{s} \subseteq \sem{t}$, then $\sem{\TChan{t}} \subseteq \sem{\TChan{s}}$.

\paragraph{Arrow types.} The decorated arrow types are instead new, and their interpretation is subtler. The interpretation must account for the characteristics of the functions of our system. In general a function is a black box that, when executed in some environment, interacts with it by expecting an input and returning an output: from a typing viewpoint it says
``\emph{I expect an input of type $s$ and give you a result of type $t$}'', i.e.\
$s \to t$. In our calculus a function is a black box that has one further way to interact with the environment in which it runs: not only does it expect an input and produce an output, but it may also act on the mailbox of the process that runs it, by reading some messages from it and depositing new messages into it. Since a mailbox is never named by the code that uses it, the two operations must be \emph{provided} to the function by its runner. From a typing viewpoint the black box now says ``\emph{give me
an input of type $s$, a way to deposit messages of type $o$, and a way to read messages of type $i$; I return a result of type $t$.}'' So a function in $s\to^i_o t$ performs two transformations, one on values (the $s/t$ component), the other on the local mailbox (the $i/o$ component), the latter mediated by the two capabilities its runner hands it. It is then clear that the interpretation must behave contravariantly on $i$ and $s$ and covariantly on $o$ and $t$.
For instance, the type in \eqref{eq:rpc-type} for \eelixir{remote_procedure_call} given in the introduction will be decorated by the type in \eqref{eq:rpc-lower-bound} for its $o$ component and by the one in \eqref{eq:rpc-upper-bound} for its $i$ component. 

The subtle part is how these two transformations, $s/t$ and $i/o$, are related: are they independent or correlated? Within a \emph{single} arrow, as in \eelixir{remote_procedure_call}'s type, they are independent: as we will see, subtyping of decorated arrows factorizes componentwise, with no dependency between the value and the mailbox components (Proposition~\ref{prop:arrow-factor}). In systems without intersections this question is thus irrelevant, but in Elixir functions are typed by intersections of arrows to account for function overloading. 

As an example of overloading, consider the following definition of \eelixir{counter}, which when applied to an integer it builds a request by tagging its argument by \eelixir|:add|, and when applied to a reply \eelixir|{:total, n}| decodes its payload:
\begin{elixircode}
!\color{anncol}\$! (integer() -> {:add, integer()}) and ({:total, integer()} -> integer()) !\phantomsection\label{countertype}!
def counter(k) when is_integer(k), do: {:add, k}   # builds a request
def counter({:total, n}),          do: n           # decodes a reply
\end{elixircode}
\noindent
This overloaded behaviour is common in Elixir, and expressed by the intersection of arrows in line~\ref{countertype}. For these types, the question of correlation is essential and technically difficult. 

Now, let's modify the example so that the two clauses send the request and receive the reply, instead of building and decoding them. The result types do not change, because in Elixir the \eelixir{send} expression returns the message it sends.
\begin{elixircode}
!\color{anncol}\$! (integer() -> {:add, integer()}) and (:total -> integer()) !\phantomsection\label{intersection}!
def counter(k) when is_integer(k), do: send(self(), {:add, k})          # sends a request
def counter(:total),               do: receive do {:total, n} -> n end  # receives a reply
\end{elixircode}
\noindent
Each clause of the function now acts on the mailbox of the process that runs \eelixir{counter}, and does so in a different way: the first, by sending to \eelixir{self()}, bounds the interface of that process from below by \eelixir|{:add, integer()}|; the second, by its \eelixir{receive}, bounds it from above by requiring \eelixir{:total}-tagged messages to carry integers. Each bound must be recorded in the arrow of its clause: the type in line~\ref{intersection} can be decorated as follows:
\[
(\TInt \to^{\Tany}_{\Tuple{\Atom[{:}add],\TInt}} \Tuple{\Atom[{:}add],\TInt})
\;\land\;
(\Atom[{:}total] \to^{\neg\Tuple{\Atom[{:}total],\neg\TInt}}_{\Tempty} \TInt).
\]
(These decorations are actually the ones deduced by by the inference system of Section~\ref{sec:inference}).

Thus a worker of Section~\ref{sec:example}, whose interface contains \eelixir|{:add, integer()}|, can call \eelixir{counter(3)}, and a client, whose interface (line~\ref{clientiface}) contains \eelixir|{:total, integer()}| but not \eelixir|{:add, integer()}|, can call \eelixir{counter(:total)} but not \eelixir{counter(3)}. What the interpretation must record is thus the \emph{pairing} of the two transformations in each arrow: each arrow of an intersection must keep its own mailbox transformation, and it must not be derivable that the value transformation of one arrow can be recombined with the mailbox transformation of another. For instance, the arrow $\TInt \to^{\neg\Tuple{\Atom[{:}total],\neg\TInt}}_{\Tempty} \Tuple{\Atom[{:}add],\TInt}$, which pairs the value transformation of the first arrow above with the mailbox transformation of the second, would let the client call \eelixir{counter(3)} and deposit \eelixir|{:add, 3}| in its own mailbox, which its interface excludes. This is coherent with the typing rule for abstractions of Section~\ref{sec:typing}, which verifies each arrow of an intersection on the body of the function under that arrow's own mailbox decoration. We will say more about this in Section~\ref{sec:correlation}, where we show that a model validating the recombination (``mixing'') of decorations across the arrows of an intersection would break subject reduction.

\paragraph{The two capabilities.} We interpret our decorated arrows accordingly, as a
\emph{single} transformation: a function in this space receives its input
together with the two \emph{capabilities} on the mailbox of the process that
runs it:
\begin{itemize}
	\item the \emph{write} capability, of type $o \to ()$, which deposits into the mailbox: it identifies what the mailbox must be able to accept, and it is exactly the pid at which $\Self$ is exposed, since $o\to() $ is nothing but $\TChan{o}$ (a pid is its write operation, a reading due to~\citet{CDV06} and recalled below);
	\item the \emph{read} capability, of type $() \to i$, which extracts from the mailbox: it identifies what the mailbox may yield, that is, the type at which the $\mkkeyword{receive}$ instructions of the body are typed.
\end{itemize}
Here $()$ is the \emph{unit} type, the type of the empty tuple: a write
operation consumes a message and returns nothing, a read operation is invoked
on no argument and returns a message. Neither $o\to()$ nor $()\to i$ is a type of
the grammar of Section~\ref{sec:syntax}: they are notations for the two capabilities. This is the same reading of a channel as its write operation, due to \citet{CDV06} and recalled in Section~\ref{sec:arrow-semantics-pointed}, that underlies the pid types
above; only, this time, we keep both capabilities, and the two decorations of
an arrow are precisely the two types they carry.
The domain of the interpretation is
thus a \emph{triple} (a value, a write capability, and a read capability), while the codomain is a single value. This asymmetry calls for one adjustment: pointing the input.

\paragraph{Pointing the input.} Folding the input value naively into a product
$(\sem{s}\times\sem{o\to ()}\times\sem{() \to i})$ is degenerate: were $s$
empty, the whole domain would be empty and the arrow would be the top one
whatever $i$, $o$, and $t$ are, the contribution of the decorations and of
the result erased by a single empty factor. The repair is to \emph{point} the
input value component: we adjoin to it a distinguished element $\mho$ (i.e., we
make it a \emph{pointed set}), read ``no value'', and set
\begin{equation}\label{eq:arrow-encoding}
    \sem{s \to_o^i t} \;=\;
    \bigl((\sem{s}\uplus\{\mho\})\times\sem{o\to ()}\times\sem{()\to i}\bigr)
    \;\to\;
    \sem{t}.
\end{equation}
The other two domain components need no pointing: $\sem{o\to ()}$ is never
empty (it contains the empty capability, which accepts nothing), and neither is $\sem{()\to i}$, as it contains the empty capability, which yields nothing.
The domain is therefore never empty (no factor can collapse the others), and the degenerate clause of arrow
subtyping, an arrow with empty domain is above every arrow, never fires.
The point $\mho$ plays for the input value exactly the role that these two
empty capabilities play for the mailbox components.
The codomain, being a single value component, needs no such repair: nothing
there can erase anything. When $t$ is empty the codomain is empty too, but this
is harmless: the resulting arrow is simply the \emph{least} one over its domain, inhabited by the functions that return no result, and its place in
the subtyping order is fixed by its domain, so the decorations are preserved;
erroneous results are accounted for, as usual in semantic subtyping, by a distinguished
error element adjoined to the codomain (Section~\ref{sec:arrow-semantics-pointed}).
The decoration $o$ occurs \emph{once}, in the domain, in the argument position
of the write capability: since the domain is itself a contravariant
position, the two contravariances cancel and the interpretation is
\emph{covariant} in $o$, as expected. This single occurrence is also what ties
the write capability of each arrow of an intersection to that arrow's own
execution, exactly as the read capability ties to it the messages the body may
read.

\paragraph{What is forced.} Three features of this interpretation are in fact \emph{forced}: the pairing of
the value and of the two capabilities in a single input, the presence of the
write capability, and the reading of the read capability as an operation that
can be invoked repeatedly (Section~\ref{sec:arrow-semantics-pointed}); dropping
any of them makes the model validate a recombination law that breaks type
soundness, as we show in Section~\ref{sec:correlation}. The point $\mho$ is not
required by soundness, since it changes subtyping only for the arrows whose
argument type is empty, to which no value can be passed: it is required by the
non-degeneracy discussed above. We formalize all the intuitions above in the
next section, through an encoding into the types of Frisch, Castagna, and Benzaken~\cite{FCB08}.

\subsection{Interpretation of Types}\label{sec:arrow-semantics-pointed}\label{sec:fcb-encoding}

Following Frisch, Castagna, and Benzaken~\cite{FCB08}, one would now define a domain with one summand for each type
constructor and interpret the types of Section~\ref{sec:syntax} directly into it, as
Section~\ref{sec:channel-semantics} prescribes. This can be done
(Appendix~\ref{app:encoding-property}), but we give the interpretation \emph{through an
encoding} instead: we encode our types into the types of~\cite{FCB08}, and the
interpretation of a type is the interpretation of its encoding in the model
of Frisch, Castagna, and Benzaken~\cite{FCB08} (FCB). This has several advantages:
\begin{itemize}[noitemsep,topsep=2pt]
\item there is no new model to build: the domain, the interpretation of the type
connectives and of the type constructors, and the proof that they form a model are those
of~\cite{FCB08}; only the encoding is new, and it is a direct formalization of
Section~\ref{sec:channel-semantics};
\item subtyping %
is decided by the algorithm of~\cite{FCB08}, and
tallying\footnote{\emph{Tallying} is to subtyping what unification is to equality. It will be described in Section~\ref{sec:ann-inference}.}
	by the algorithm of~\citet{polyduce2}, both run on the encoded types, for
which efficient implementations already exist, notably in the Elixir compiler;
\item the encoding extends to type variables, by mapping each variable to itself, and it
commutes with type substitutions, so that the inference system of
Section~\ref{sec:ann-inference} reduces to tallying on encoded types, followed by a
readback of the solutions;
\item the ingredients identified in Section~\ref{sec:channel-semantics} become
constructions on types: each alternative interpretation rejected in
Section~\ref{sec:correlation} is an alternative encoding, and the unsound law it validates
is a subtyping fact of~\cite{FCB08};
\item the gradual typing of Elixir, discussed in Section~\ref{sec:elixir-gradual}, can be
added, at the level of the subtyping relations, by extending the encoding.
\end{itemize}

\paragraph{The FCB model of~\cite{FCB08}.} A set-theoretic interpretation maps each type to a
set of elements of a domain, so that the type connectives denote the corresponding
set-theoretic operations: union for $\lor$, complement for $\neg$, and the empty set for
$\Tempty$. The domain $\DomainFCB$ of~\cite{FCB08} contains constants, pairs, which
interpret products, and finite relations, which interpret functions. Intuitively, a
function type $t_1{\to}t_2$ should denote the set of the relations that map elements of
$\sem{t_1}_\textsf{FCB}$ into elements of $\sem{t_2}_\textsf{FCB}$, where
$\sem{\cdot}_\textsf{FCB}$ is the interpretation of~\cite{FCB08}. As explained
by~\citet[Section~4.2]{FCB08}, two adjustments are needed: outputs range over
$\DomainFCB\cup\Set{\Omega}$ for a fresh constant $\Omega$ marking the error (otherwise
$\Tany\to\Tany$ would denote the set of \emph{all} functions rather than the functions that
are defined for every value), and, since a domain cannot contain all the relations on itself for
cardinality reasons, relations are restricted to \emph{finite} approximations:
\[
\sem{t_1\to t_2}_\textsf{FCB} = \Pf(\oline{\sem{t_1}_\textsf{FCB}\times\oline{\sem{t_2}_\textsf{FCB}}}),
\]
where the over-line denotes set complement, the inner one taken within
$\DomainFCB\cup\Set{\Omega}$, so that a finite relation belongs to $\sem{t_1\to
t_2}_\textsf{FCB}$ when each of its pairs whose input is in $\sem{t_1}_\textsf{FCB}$ has
its output in $\sem{t_2}_\textsf{FCB}$, hence not $\Omega$. \citet{FCB08} prove that finite
approximations induce the same subtyping relation as arbitrary relations. We write
$\subt_\textsf{FCB}$ for the subtyping relation induced by this model, namely $t_1\subt_\textsf{FCB}t_2\iffdef\sem{t_1}_\textsf{FCB} \subseteq \sem{t_2}_\textsf{FCB}$, the inclusion of interpretations.

\paragraph{The encoding.} Let $\CPoint$, $\ell_{\Wch}$, $\ell_{\Wfun}$, and ${\Cnil}$ be pairwise distinct fresh constants, all distinct from the constants of Section~\ref{sec:syntax}, and let $\Wch$ and $\Wfun$  be unary constructors that tag their
argument so that pid types and decorated arrows remain mutually
disjoint. Each is encoded as a product with its constant tag:
\[
    \Wch(X) \;\triangleq\; (\ell_{\Wch},\, X),
    \qquad\qquad
    \Wfun(X) \;\triangleq\; (\ell_{\Wfun},\, X).
\]
The encoding of our types into the types of \cite{FCB08} is defined for the type constructors as:
\begin{align*}
\enc{t_1 \to_o^i t_2} & =
        \Wfun\Bigl(
            \bigl((\enc{t_1}\lor\CPoint)\times\enc{\TChan{o}}\times(\CPoint\to \enc{i})\bigr)
            \to
             \enc{t_2}
        \Bigr)\\
\enc{\TChan{t}} & = \Wch\bigl(\enc{t}\to\CPoint\bigr)\\
\enc{\Tuple{t_1,\dots,t_n}} & = \enc{t_1}\times\bigl(\enc{t_2}\times\cdots\times(\enc{t_n}\times{\Cnil})\cdots\bigr)
        \qquad\text{(in particular $\enc{\Tuple{}} = {\Cnil}$)}\\
\enc{\TTuple} & = T \quad\text{where $T$ is the (regular) type satisfying } T = {\Cnil} \lor (\Tany\times T)
\end{align*}
and homomorphically for the remaining types, that is, basic types, $\Tany$, and the type connectives. In the arrow clause, the ternary product abbreviates its right-nested binary form $A\times(B\times C)$ (no terminator is needed: the shape is the same, fixed one for every encoded arrow).

\begin{remark}[Well-foundedness]\label{rem:enc-corecursive}
Since types are coinductively defined, the equations above are a \emph{corecursive} definition of $\enc{\cdot}$. The clauses for the type constructors are guarded, since every occurrence of $\enc{\cdot}$ on their right-hand sides appears below at least one type constructor, hence productive and well-defined on infinite trees; the clauses for the type connectives are covered instead by the induction principle that contractivity provides. Moreover, the definition maps regular trees to regular trees and, since it inserts constructors and never erases them, it preserves contractivity. Therefore $\enc{t}$ is a type in the sense of~\cite{FCB08}.
\end{remark}

The encodings of arrows and pids are the formal counterpart of Section~\ref{sec:channel-semantics}. A pid is its write operation:
\citet{CDV06} show that, in the presence of only contravariant channels, a channel type
$\texttt{ch}^-(t)$ can be seen as the type of the functions from $t$ to the unit type
(that is, as write operations on the channel), and the encoding of $\TChan{t}$ is
accordingly the type of the functions from $\enc{t}$ to the fresh constant $\CPoint$, which
plays the role of the unit type and of the point $\mho$ of
Section~\ref{sec:channel-semantics}. An element of $\sem{\enc{\TChan{t}}}_\textsf{FCB}$ is thus a
pair $(\ell_{\Wch},R)$, where $R$, a \emph{pid approximation}, is a finite relation that
relates a message of type $t$, if at all, only to $\CPoint$ (the process accepts it), while
it may relate any other message to anything else, $\Omega$ included (the process may reject
it).
The write capability of an arrow is exactly this, $o\to()$ being $\TChan{o}$, so it needs
no new machinery. The read capability is the mirror image: it is encoded as a function type
whose domain is $\CPoint$, the unit argument on which the operation is invoked, and the
elements of $\sem{\CPoint\to\enc{i}}_\textsf{FCB}$ are the finite relations that relate
$\CPoint$ only to elements of $\sem{\enc{i}}_\textsf{FCB}$, the messages the capability may
yield (never to $\Omega$: a read capability that errs has no type). Note that the two
capabilities constrain \emph{opposite} components, the pid its input and the read
capability its output, which is precisely the contravariance of $o$ and the covariance of
$i$ within the arrow domain. The reading is that of a \emph{multi-shot} operation: the body
of a function may perform several $\mkkeyword{receive}$s, and successive invocations of the
same read capability may well return different messages, which is why a capability
\emph{relates} $\CPoint$ to several possible results, instead of recording just one.
Encoding it as a single (pointed) message, $\enc{i}\lor\CPoint$, would generate a
different, unsound subtyping relation (Section~\ref{sec:correlation}). The value component
of the domain of an arrow is the pointed type $\enc{t_1}\lor\CPoint$, so that no component
of the domain is ever empty: $\CPoint$, the empty relation tagged by $\ell_{\Wch}$, and the
empty relation inhabit them whatever the types are, and the degenerate clause of arrow
subtyping never fires. Tuples are encoded as the classic \emph{nil-terminated}
right-associative nestings of pairs. The terminator ${\Cnil}$ is not optional: a bare
right-associative encoding would map the pair $\Tuple{t_1,\Tuple{t_2,t_3}}$ and the triple
$\Tuple{t_1,t_2,t_3}$ to the same type, whereas the two types are disjoint. Notice that, contrary to $\CPoint$, the constant ${\Cnil}$ is denotable, being the interpretation of the empty tuple.
Finally, the
encodings of pid, arrow, and tuple types are pairwise disjoint even though all three are
encoded as products: the tags $\ell_{\Wch}$ and $\ell_{\Wfun}$ separate pids from arrows
(whose content is a function type, hence never empty), while the second component of a
tuple encoding is either ${\Cnil}$ or a product, never a function type. The disjointness of
pid, function, and tuple values on which pattern matching relies is therefore preserved by
the encoding.

\paragraph{Interpretation and subtyping.} We can now define the interpretation of our types
and the subtyping relation.
\begin{definition}[Interpretation and subtyping]\label{def:subtyping}
The \emph{interpretation} of a type $t$ is $\sem{t}\eqdef\sem{\enc{t}}_\textsf{FCB}$, and
the \emph{subtyping relation} $\leq$ is defined as
\[
  t_1 \leq t_2 \quad{\iffdef}\quad \sem{t_1} \subseteq \sem{t_2},
\]
that is, $\enc{t_1}\subt_\textsf{FCB}\enc{t_2}$.
\end{definition}
\noindent We write $s\simeq t$ when $s\leq t$ and $t\leq s$. Since the encoding commutes
with the type connectives, so does the interpretation: $\sem{t_1\lor
t_2}=\sem{t_1}\cup\sem{t_2}$, $\sem{\neg t}=\DomainFCB\setminus\sem{t}$, and
$\sem{\Tempty}=\varnothing$. For the two capabilities we write
$\sem{o\to()}\eqdef\sem{\enc{\TChan{o}}}_\textsf{FCB}$ and
$\sem{()\to i}\eqdef\sem{\CPoint\to\enc{i}}_\textsf{FCB}$, the sets of write and of read
capabilities, so that $\sem{t_1\to_o^i t_2}$ is the set of the pairs $(\ell_{\Wfun},R)$
with $R$ in
\[
\Pf\bigl(\oline{((\sem{t_1}\cup\Set{\CPoint})\times\sem{o\to()}\times\sem{()\to i})\times\oline{\sem{t_2}}}\bigr),
\]
which is equation~\eqref{eq:arrow-encoding} with finite approximations. Unlike the pointed
set of that equation, $\sem{t_1}\cup\Set{\CPoint}$ need not be a disjoint union:
$\DomainFCB$ contains elements that encode no value of our calculus, such as the constants
$\CPoint$, $\ell_{\Wch}$, $\ell_{\Wfun}$ and the untagged relations, which belong to the
interpretation of no type constructor other than $\Tany$, but to that of types built with
negation; for instance, $\CPoint\in\sem{\neg\TInt}$. These elements do not alter the
subtyping relation, as the direct interpretation shows.

\paragraph{The direct interpretation.} Appendix~\ref{app:encoding-property} follows~\cite{FCB08}
literally: it defines a domain $\Domain$ with one summand for each type constructor
(constants, tuples, and approximations of functions, pids, and read capabilities), and it
interprets the types of Section~\ref{sec:syntax} directly into it, as
equations~\eqref{eq:chan}, in its finite-approximation form, and~\eqref{eq:arrow-encoding}
prescribe. It shows that this \emph{direct interpretation}, written $\semD{\cdot}$, is a
model in the sense of~\citet[Definition~4.4]{FCB08}, and that it induces on our types the
same subtyping relation as Definition~\ref{def:subtyping}: $s\subt t$ if and only if
$\semD{s}\subseteq\semD{t}$ (Corollary~\ref{cor:same-subtyping}).

\subsection{Properties of Subtyping}\label{sec:subtyping-laws}
The subtyping relation satisfies the following properties, proved in Appendix~\ref{app:enc-laws}.

\paragraph{Properties of arrow types.}
\begin{restatable}[Arrow factorization]{proposition}{ArrowFactorProp}\label{prop:arrow-factor}
For all types $s,s',t,t',i,i',o,o'$:\quad
$s \to_o^i t \subt s' \to_{o'}^{i'} t'$ \ if and only if \
$s' \subt s$, $i' \subt i$, $t \subt t'$, and $o \subt o'$.
\end{restatable}
\noindent Single-arrow containment thus factorizes with no side condition, since
the domain of an arrow is never empty and the complement of its codomain always
contains $\Omega$. It is contravariant in the argument $s$ and in the input
decoration $i$, covariant in the result $t$ and in the output decoration $o$, with
no relation imposed between $i$ and $o$. The output decoration is covariant
through its single occurrence in the domain: there the two contravariances, of the
position and of the write capability, cancel out.
The interpretation also validates the following limit cases: $(a)$ $\Tany \to_{\Tempty}^{\Tany}\Tempty$ is the smallest arrow type:  its
    codomain $\sem{\Tempty}$ is empty, hence contained in every codomain,
    while its domain is the largest one: $\sem{\Tempty\to ()}$ is the largest
    write capability (it contains $(\ell_{\Wch},R)$ for every pid approximation $R$, since no message
    need be accepted) and $\sem{()\to\Tany}$ the largest read capability (every message it may
    yield is in $\Tany$); $(b)$
$\Tempty \to_{\Tany}^{\Tempty}\Tany$ is the largest arrow type, that is, $\TFun$: its domain is
    the smallest one (the value component is the single
    point $\CPoint$, the write capability $\sem{\Tany\to()}$ is contained in every
    write capability, and the read capability $\sem{()\to\Tempty}$, which yields no
    message, is contained in every read capability), and its
    codomain $\sem{\Tany}$ contains every output
    value, so every decorated arrow is below it; $(c)$ a function
$\TInt \to_{\Tempty}^{\Tany}\TBool$ that sends nothing to the process that runs it (output
decoration $\Tempty$) is below every $\TInt \to_{o}^{i}\TBool$ and is callable in
any environment (input decoration $\Tany$: it can read whatever the mailbox contains); $(d)$ $\TInt \to_{\Tany}^{\Tempty}\TBool$ is the largest
arrow mapping $\TInt$ into $\TBool$.

\paragraph{Properties of pid types.} For process types, the interpretation of $\TChan{t}$ by finite relations given above and the
one given by equation~\eqref{eq:chan} are different, but, as proved by Castagna, Dezani, and Varacca~\cite{CDV06}, they
induce the \emph{same} subtyping on process types, which is all that matters.
Both validate the following laws.
\begin{restatable}[Pid laws]{proposition}{PidLawsProp}\label{prop:pid-laws}
For all types $s,t,t_1,t_2$ and every finite family $(s_n)_{n\in N}$:
\begin{enumerate}[noitemsep,topsep=2pt]
    \item $\TChan{t}\subt\TChan{s}$ if and only if $s\subt t$;
    \item $\TChan{t_1}\land\TChan{t_2} \simeq \TChan{t_1\lor t_2}$ and
    $\TChan{t_1}\lor\TChan{t_2} \subt \TChan{t_1\land t_2}$, the latter strictly
    whenever $t_1$ and $t_2$ are incomparable;
    \item (union law) $\TChan{t} \subt \bigvee_{n\in N}\TChan{s_n}$ if and only if
    $s_n \subt t$ for some $n \in N$.
\end{enumerate}
\end{restatable}
\noindent The union law is the key case for the correspondence to extend to all
process types. With finite relations the ``only if'' direction is checked
directly: if $s_n \not\subt t$ for every $n$, pick
$d_n \in \sem{s_n}\setminus\sem{t}$; then the element
$(\ell_{\Wch},\Set{(d_n,\Omega) \mid n \in N})$ belongs to $\sem{\TChan{t}}$ but to no
$\sem{\TChan{s_n}}$.

\section{Dynamic Semantics}
\label{sec:opsem}

We define $\match{v}{p}$ as the substitution generated by matching a value $v$ against a pattern $p$\\[-2.5mm]
\begin{minipage}[t]{0.45\textwidth}
    \begin{align*}
    \match{v}{x} &= \subst{v}{x}  \\
    \match{\Tuple{v_1, \dots, v_n}}{\Tuple{p_1, \dots, p_n}} &= \textstyle\bigcup_{k=1}^n \match{v_k}{p_k} \\
    \match{v}{p_1 \land p_2} &= \match{v}{p_1} \cup \match{v}{p_2} 
\end{align*}
\end{minipage}\hfill
\begin{minipage}[t]{0.5\textwidth}
    \begin{align*}
    \match{v}{\tau} &= \emptyset & \text{if } v\in\tau  \\
    \match{v}{p} &= \textsf{fail} & \text{otherwise}
\end{align*}
\end{minipage}\\
where $\cup$ denotes the union of two substitutions with disjoint domains (condition satisfied since pattern must have distinct capture variables), and the convention that $\textsf{fail}$ is absorbing for $\cup$. When $\match{v}{p}\neq\textsf{fail}$ we write $e\subst{v}{p}$ for the result of applying this substitution to $e$: this extends to patterns the notation $e\subst{v}{x}$ for substituting a single variable, of which it is the particular case $p=x$. The relation $v\in \tau$ holds if and only if $\testof{v}\subt\tau$, where the \emph{test type} $\testof{v}$ of a value $v$---recording everything that a run-time type test can observe of $v$---is defined by induction on $v$:
\begin{align*}
    \testof{\Const} &= \Const &
    \testof{\PID^t} &= \TChan{\Tempty} &
    \testof{\Fun^{l} x.~e} &= \TFun &
    \testof{\Tuple{v_1,\dots,v_n}} &= \Tuple{\testof{v_1},\dots,\testof{v_n}}
\end{align*}
Membership is thus decided by a single subtyping check, with no recourse to
the type system (and can be efficiently computed by non-uniform automata \cite{frisch04}). Two aspects of the definition are deliberate. Constants are
tested \emph{exactly}---$\testof{\Const}$ is the singleton $\Const$, so the
literal patterns pervasive in Elixir (e.g., \eelixir{:bump}) match just the
intended value---whereas of pids and abstractions a test observes only
\emph{that} they are pids or functions, not their interfaces or their
domains.
We define a reduction semantics for expressions as the least relation $\red$
defined by the following rules:
\hfill\begin{align*}
    (\Fun^{l} x.~e)~v &\red e\subst{v}{x} \\
    n_1 + n_2 &\red n_3 & \text{if $n_3 = n_1 + n_2$} \\
    \hspace*{4cm}\Case{v}{\seqof{\Branch{p}{e}}} &\red e_j\subst{v}{p_j} & \text{if }\forall h<j.\; \match{v}{p_h}=\textsf{fail}\text{ and } \match{v}{p_j} \neq \textsf{fail} \\
    \Eval[e] &\red \Eval[e'] & \text{if $e \red e'$}
\end{align*}
As usual, in the reduction of $n_1 + n_2$ the symbol `$+$' is syntactic to the
left of $\red$ and stands for integer addition in the side condition, and the position in which values appear in the syntactic definition
of the evaluation context gives us the order in which they evaluate (for example, applications are evaluated from left to
right). A crucial point that may have passed unnoticed is that there is no evaluation context for $\Spawn{e}{t}$. The absence of $\Spawn{\Eval}{t}$  is exactly what makes the argument a thunk shipped to the child (and what makes $\Self$ inside it late-bound). In that we slightly depart from the semantics of Elixir, where spawn takes a zero-argument function whose body is spawn: that semantics is recovered by encoding \eelixir{spawn(fn -> e end)} as $\Spawn{((\Fun^{l}x.~e)())}{...}$, with $x$ fresh, $()$ the empty tuple, and $l$ a suitable annotation of domain $\Tuple{}$ (e.g., $\Tuple{}\to t$, where $t$ is the type of $e$).

We define the semantics of configurations using a combination of structural
congruence $\equiv$, which allows us to identify essentially equivalent terms,
and a reduction relation $\red$, which expresses computation. We use the same
symbol $\red$ already used for the reduction semantics of expressions since the
context always allows us to distinguish between the two.

Structural congruence is the smallest congruence relation satisfying the
following rules
\begin{mathpar}
    P \parop Q \equiv Q \parop P
    \qquad\qquad
    P \parop (Q \parop R) \equiv (P \parop Q) \parop R
\end{mathpar}
asserting that parallel composition is commutative and associative.

The reduction relation for configurations is the least relation $\red$ defined
by the following rules:
\[
    \begin{array}{@{}r@{~}c@{~}ll@{}}
        \Process{\PID}{t}{e}{\seqof{v}} & \red & \Process{\PID}{t}{e'}{\seqof{v}}
        & \text{if $e \red e'$}
        \\
        \Process\PID{t}{\Eval[\Spawn{e}{s}]}{\seqof{v}}
        & \red &
        \Process\PID{t}{\Eval[\PID[b]^s]}{\seqof{v}} \parop \Process{\PID[b]}{s}{e}{\EmptyMailbox} &
        \text{$\PID[b]$ fresh}
        \\
        \Process\PID{t}{\Eval[\Send{\PID[b]^s}{v'}]}{\seqof{v}} \parop \Process{\PID[b]}{s}{e_2}{\seqof{w}}
        & \red &
        \Process\PID{t}{\Eval[v']}{\seqof{v}} \parop \Process{\PID[b]}{s}{e_2}{\seqof{w}v'}
        \\
        \Process\PID{t}{\Eval[\Send{\PID^t}{v'}]}{\seqof{v}}
        & \red &
        \Process\PID{t}{\Eval[v']}{\seqof{v}v'}
        \\
        \Process\PID{t}{\Eval[\Self]}{\seqof{v}}
        & \red &
        \Process\PID{t}{\Eval[\PID^ {t}]}{\seqof{v}} \\
          \Process\PID{t}{\Eval[\Recv{\seqof{\Branch{p}{e}}}]}{\seqof{v}v'\seqof{w}}
        & \red &
        \Process\PID{t}{\Eval[e_j\subst{v'}{p_j}]}{\seqof{v}\seqof{w}} &\text{if }\left\{\small\begin{array}{l}
        \forall h.\;\forall v_k\in\seqof{v}.\;\match{v_k}{p_h}=\textsf{fail}\\
        \forall h<j.\;\match{v'}{p_h}=\textsf{fail}\\
        \match{v'}{p_j}\neq\textsf{fail}\end{array}\right.
      \\
        P \parop Q & \red & P' \parop Q & \text{if $P \red P'$}
        \\
        P & \red & P' & \text{if $P \equiv Q \red P'$}
    \end{array}
\]
A few rules deserve comments. Spawns create a pid that is fresh \emph{for the whole configuration}, not only for the component that reduces. The three side conditions of the receive rule implement the reduction strategy of the BEAM-based languages: the oldest message matching some pattern is removed from the mailbox, and the first pattern it matches selects the branch. Sends are asynchronous, and the message is appended to the mailbox of the recipient, except self-sends, which are delivered to the local mailbox. Finally, $\Self$ is reduced to the pid of the current process, which is late-bound since it is only known at runtime. We omit here the timeout clause $\mkkeyword{after}~n \to e$ that an Elixir $\mkkeyword{receive}$ may carry, because the system accounts for it as it stands: the clause consumes no message, so it constrains no interface; typing it only adds the type of $e$ to the union of the types of the branches; and it can only help liveness, since a receive with a timeout never blocks past it. This matters, since \eelixir{after 0} is the standard idiom for draining a mailbox.

\section{Static Semantics}
\label{sec:typing}
\subsection{Auxiliary definitions}
\begin{figure}[t]
\centering
\begin{minipage}[t]{.44\textwidth}
\centering
\begin{align*}
    \accept{x} &= \Tany \\
    \accept{\tau} &= \tau \\
    \accept{\Tuple{p_1, \dots, p_n}} &= \Tuple{\accept{p_1}, \dots, \accept{p_n}} \\
    \accept{p_1 \land p_2} &= \accept{p_1} \land \accept{p_2}
\end{align*}
\end{minipage}\hfill
\begin{minipage}[t]{.54\textwidth}
\centering
\begin{align*}
    \ptenv{t}{x} &= \{x : t\} \\
    \ptenv{t}{\tau} &= \emptyset \\
    \ptenv{t}{\Tuple{p_1, \dots, p_n}} &= \textstyle\bigcup_{k=1}^n \ptenv{\pi_k(t)}{p_k} \\
    \ptenv{t}{p_1 \land p_2} &= \ptenv{t}{p_1} \cup \ptenv{t}{p_2}
\end{align*}
\end{minipage}
\caption{The accepted type $\accept{p}$ of a pattern (left) and the pattern
environment $\ptenv{t}{p}$ (right).}\label{fig:pattern-aux}
\end{figure}
\noindent
We define the accepted type of a pattern $p$, denoted $\accept{p}$, as shown on
the left of Figure~\ref{fig:pattern-aux}, which is the type of all and only the values that match the pattern $p$.  The
environment $\ptenv{t}{p}$,  defined only for $t\leq\accept{p}$, extracts typing information from a pattern $p$
assuming the matched value has type $t$; it is defined on the right of the same
figure.
Here $\pi_k(t)$ is the standard projection type operator for tuples~\cite{FrischThesis}: every type smaller than the type $\Tuple{\Tany,\dots,\Tany}$ of the tuples of some fixed arity is equivalent to a finite union of non-empty tuple types of that arity, and $\pi_k$ returns the union of the $k$-th components of the tuples of that union.

\subsection{Typing rules}\label{sec:typing-rules}
The typing rules of Figure~\ref{fig:typing} derive a typing judgment of the form $\wte[i][o]{\Context}{e}{t}$ which carries two mailbox
decorations, and the way to read them is as a rely/guarantee pair: to type
$e$ we \emph{rely} on the assumption that the mailbox of the process
running $e$ contains messages of type $i$, and we \emph{guarantee}
that $e$ exposes its runner only at type $\TChan{o}$: every message sent
through the pid that $\Self$ returns in $e$, by $e$ itself or by any code that
obtains this pid from $e$, is of type $o$.
Together with the interface, three types thus describe how a piece of
code interacts with the mailbox of the process that runs it:
\begin{center}
\begin{tabular}{@{}llp{0.6\textwidth}@{}}
$s$ & \emph{interface} & attached to the \emph{process},
$\Process{\PID}{s}{e}{\seqof v}$: the type of the messages its mailbox may contain;\\[.5mm]
$i$ & \emph{input decoration} & attached to the \emph{code}: what it assumes
it may read---$\mkkeyword{receive}$ inspects the mailbox at type $i$;\\[.5mm]
$o$ & \emph{output decoration} & attached to the \emph{code}: what it may
itself produce---it publishes the identity of its runner as
$\Self : \TChan{o}$.\\
\end{tabular}
\end{center}
The decorations ``bracket'' the interface of any process that may safely execute
the code: reading requires $s \subt i$ (the code must be ready for whatever
the mailbox may contain) and exposing self requires $o \subt s$ (whatever is sent through $\Self$, the mailbox must accept),
whence $o \subt s \subt i$. This
relation underlies all the rules and statements that follow, so we name it.
\begin{definition}[Brackets and narrowing]\label{def:bracket}
A pair of decorations $[o,i]$ is called a \emph{bracket}. A bracket $[o',i']$
\emph{narrows} $[o,i]$ if $o \subt o'$ and $i' \subt i$. A bracket $[o,i]$ is \emph{coherent} if $o\subt i$.
\end{definition}
\noindent Thus a process of interface $s$ may safely execute code typed at
$[o,i]$ whenever $[s,s]$ narrows $[o,i]$ (which also implies that the bracket is coherent). The configuration rule
\rulename{t-thread} below types each process at the tightest bracket $[s,s]$,
and the mailbox-narrowing theorem (Theorem~\ref{thm:narrowing} in
Section~\ref{sec:properties}) states that
every judgment survives the narrowing of its bracket.
We do not impose coherence as a condition for typing, as we explain in Remark~\ref{rem:coherence}.

\subsubsection{Typing rules for expressions}\label{sec:typing-expr}
\begin{figure}[!ht]
\begin{mathpar}
    \inferrule[t-sub]{
        \wte{\Context}{e}{t}
    }{
        \wte{\Context}{e}{t'}
    }
    ~t \subt t'
    \and
    \inferrule[t-const]{~}{
        \wte{\Context}{\Const}{\Const}
    }
    \and
    \inferrule[t-pid]{~}{
        \wte{\Context}{\PID^t}{\TChan{t}}
    }
    \and
    \inferrule[t-self]{~}{
        \wte[i][o]{\Context}{\mkkeyword{self}}{\TChan{o}}
    }
    \and
    \inferrule[t-var]{~}{
        \wte{\Context, x : t}{x}{t}
    }
    \and
    \inferrule[t-abs]{
        \wte[i_j][o_j]{\Context, x : t_j}{e}{t'_j}
    }{
        \wte[i][o]{\Context}{\Fun^{\wedge t_j \to^{i_j}_{o_j} t'_j} x.~e}{\textstyle\bigwedge_j t_j \to_{o_j}^{i_j} t'_j}
    }
    \and
    \inferrule[t-app]{
        \wte{\Context}{e_1}{t \to_{o}^{i} t'}
        \\
        \wte{\Context}{e_2}{t}
    }{
        \wte{\Context}{e_1~e_2}{t'}
    }
    \and
    \inferrule[t-tuple]{
        \wte{\Context}{e_k}{t_k}
    }{
        \wte{\Context}{\Tuple{e_1, \dots, e_n}}{\Tuple{t_1, \dots, t_n}}
    }
    \and
    \inferrule[t-plus]{
        \wte{\Context}{e_1}{\TInt}
        \\
        \wte{\Context}{e_2}{\TInt}
    }{
        \wte{\Context}{e_1 + e_2}{\TInt}
    }
    \and
 \inferrule[t-case]{
        \wte{\Context}{e}{t}
        \and
        \wte{\Context, \ptenv{r_j}{p_j}}{e_j}{t'}
        \quad (j\in J)
    }{
        \wte{\Context}{\Case{e}{\seqof{\Branch{p}{e}}}}{t'}
    }
    {\;\footnotesize\begin{array}{@{}l@{}}
        t\leq \textstyle\bigvee_{j} \accept{p_j}\\
        r_j=(t\land\accept{p_j})\setminus\textstyle\bigvee_{h<j}\accept{p_h}\\
        J\supseteq\{\,j \mid r_j\not\simeq\Tempty\,\}
    \end{array}}
   \and
    \inferrule[t-spawn]{
        \wte[t][t]{\Context}{e}{t'}
    }{
        \wte{\Context}{\Spawn{e}{t}}{\TChan{t}}
    }
    \and
    \inferrule[t-send]{
        \wte[i][o]{\Context}{e}{\TChan{t}}
        \\
        \wte[i][o]{\Context}{e'}{t}
    }{
        \wte[i][o]{\Context}{\Send{e}{e'}}{t}
    }
    \and
    \inferrule[t-receive]{
        \wte{\Context, \ptenv{r_j}{p_j}}{e_j}{t}
        \quad (j\in J)
    }{
        \wte{\Context}{\Recv{\seqof{\Branch{p}{e}}}}{ t}
    }
    {\;\footnotesize\begin{array}{@{}l@{}}
        r_j=(i\land\accept{p_j})\setminus\textstyle\bigvee_{h<j}\accept{p_h}\\
        J\supseteq\{\,j \mid r_j\not\simeq\Tempty\,\}
    \end{array}}
\end{mathpar}
\caption{Declarative typing rules for expressions.}\label{fig:typing}
\end{figure}
In the rules of Figure~\ref{fig:typing}, a premise indexed by $j$ or $k$ stands for one premise for each value of the index.
Note that \rulename{t-sub} subsumes the type alone: one could let it narrow the
bracket as well, but such a rule is admissible
(Theorem~\ref{thm:narrowing}), and we omit it.
In both \rulename{t-case} and \rulename{t-receive} the $j$-th branch is typed
under the values that may actually reach it, of type $r_j$: those that match $p_j$ but none
of the earlier patterns, mirroring the first-match reduction of
Section~\ref{sec:opsem}. The side condition of \rulename{t-case} requires the
patterns to be exhaustive on the scrutinee (i.e., $t\leq \textstyle\bigvee_{j} \accept{p_j}$), so that matching cannot fail.
\rulename{t-receive} has no such condition: a message that matches no pattern
is not an error, it simply remains in the mailbox. These are the simple
versions of the rules, which do not perform occurrence typing (also called type narrowing, a notion
unrelated to the narrowing of brackets of Definition~\ref{def:bracket}); a naive
form of occurrence typing can be straightforwardly added by using the skeleton
technique of~\cite{CD26}, but it is not necessary for the purposes of this
paper, so we omit it. Likewise, the projections of Figure~\ref{fig:pattern-aux} lose
the correlation between the components of a tuple. It could be recovered by typing a
branch once per summand: given a finite union $r_j\simeq\bigvee_k r_{j,k}$, the premise
of branch $j$ is replaced by one premise for every $k$, with $r_{j,k}$ in place of
$r_j$. We expect this to be sound, but we do not adopt it
(it is an experimental option of our prototype).

The branches that the two rules type are those in $J$. The side condition on $J$ requires it to contain every branch that some value may reach, that is, every $j$ whose refined type $r_j$ is not empty, and lets it leave out the others: the premise typing a branch that no value reaches \emph{may} be omitted, and only such a premise. When the pattern binds some variable, the emptiness of $r_j$ amounts to its pattern environment assigning the empty type to some variable; the condition is stated on $r_j$ because of the patterns that bind no variable, such as literals. Omitting the premise is what makes functions with intersection types typable~\cite[Remark 1]{CD26}: since \rulename{t-abs} types the same code several times under different hypotheses, a branch that no value reaches under one arrow may be reached under another, and its body need not be typable under the hypotheses that make it unreachable. Such a branch is thus not necessarily dead code. It is dead code only when its refined type is empty in every instance of the rule typing its $\mkkeyword{case}$ or $\mkkeyword{receive}$: a compositional system cannot flag this, but a type checker can, by recording the branches whose refined type is not empty in at least one instance~\cite[Remark 1]{CD26}.
Finally, the remaining rules are mostly self-explanatory: \rulename{t-send} checks that the mailbox of the receiver is compatible with the message sent; \rulename{t-spawn} checks that the body of the spawned process is typed at the declared bracket; \rulename{t-self} types $\Self$ by the current output decoration; \rulename{t-pid} types a pid with the interface ascribed in it; the other rules are the standard rules of semantic subtyping enriched with the decorations (in particular notice that in \rulename{t-abs} lambdas are annotated and typed with intersection of arrows, each bringing its own decorations).

\subsubsection{Typing rules for configurations}\label{sec:typing-conf}

\begin{mathpar}
    \inferrule[t-thread]{
        \wte[s][s]{\emptyset}{e}{t}
        \\
        \wte[s][s]{\emptyset}{v_k}{s}
    }{
        \wtc{}{\Process{\PID}{s}{e}{\seqof{v}}}
    }
    \and
    \inferrule[t-par]{
        \wtc{P}
        \\
        \wtc{Q}
    }{
        \wtc{P \parop Q}
    }
\end{mathpar}

These rules work because pids are runtime-only values, generated by the $\mkkeyword{spawn}$ operator and therefore soundly annotated; the well-formedness invariant of Section~\ref{sec:properties} makes this global property precise. Were pids forgeable, a global environment $\Delta$ recording the interface of every pid in the system would be needed instead.

\subsubsection{The decorations at work}\label{sec:typing-decorations}
The two decorations play asymmetric roles, and the rules localize the
asymmetry. The input decoration is consulted by \rulename{t-receive} alone: it is
the type at which the mailbox is inspected, so it must over-approximate the
interface of every process that may run the code. The output decoration
is consulted by \rulename{t-self} alone: it is the type at which the code publishes
the identity of its runner, so it must under-approximate that interface.
In particular $o$ does \emph{not} bound the content of the mailbox: what a
send deposits is bounded, by \rulename{t-send}, by the interface of the pid
it targets, which \rulename{t-spawn}
fixes at the bracket at which it types the spawned body. The mailbox
invariant is therefore carried by the interface, and $o$ carries only the
weaker, local obligation of never handing out a capability larger than the
runner accepts---an obligation that binds not only the code itself, but
every process to which it forwards the $\Self$ it obtained.

For code running in \emph{another} process the bracket travels on arrow types:
the arrow $t_1 \to_o^i t_2$ records the bracket $[o,i]$
at which the \emph{body} of the abstraction is typed.
This bracket is determined in \rulename{t-abs} by the
annotation alone. In \rulename{t-abs} (as in every rule for values: \rulename{t-pid}, \rulename{t-const}, and, inductively, \rulename{t-tuple}) the bracket $[o,i]$ of the
conclusion is unconstrained, since creating a function is a pure act,
possible in any process; it is its \emph{application} that is constrained:
\rulename{t-app} requires (through \rulename{t-sub} and the variance of the
decorations) the ambient bracket to narrow that of the arrow, and any $\Self$
in the body then resolves to the pid of the process that \emph{runs} the application, typed $\TChan{o}$
for the output decoration $o$ of the ambient bracket of the application. This decoupling of creation site from execution
site is what makes closures shippable.

\begin{remark}[Coherence]\label{rem:coherence}
The rules do not
enforce coherence %
because every judgment that participates
in an execution is coherent by construction. Indeed, \rulename{t-thread} types
every process at $[s,s]$ and \rulename{t-spawn} types the spawned body at
$[t,t]$, both trivially coherent; and by \rulename{t-app} (through
\rulename{t-sub} and the variance of the decorations) an arrow
$t_1 \to_o^i t_2$ can be invoked in an ambient bracket $[o',i']$ only if
$[o',i']$ narrows $[o,i]$ (Definition~\ref{def:bracket}), so in a coherent
ambient ($o'\subt i'$) every applicable arrow satisfies
$o \subt o' \subt i' \subt i$, hence is itself coherent. An
abstraction annotated with an incoherent bracket is therefore not ill-typed: it
is merely \emph{inapplicable}---no coherent ambient can ever invoke it---so it
denotes a function that can be passed around but never called.\footnote{In the system of~\cite{FCB08} this is the case for any function typed by the top function type $\Tempty\to\Tany$. The same holds here for the top function type $\Tempty\to_{\Tany}^{\Tempty}\Tany$, which is also incoherent.}
\end{remark}

\subsection{Justifying the Model: Three Unsound Laws}
\label{sec:correlation}
With the typing rules now in hand, we can justify the three ingredients of
Section~\ref{sec:arrow-semantics-pointed}. Subtyping of decorated arrows is
delicate only on \emph{intersections}: an
intersection constrains one and the same function on the domain of each of its
members, and each such constraint bears on an input \emph{triple}, whose value,
write, and read components act together and cannot be recombined with those of
another branch. This is what the three ingredients of
Section~\ref{sec:arrow-semantics-pointed} secure: drop the correlation of the
components, the multi-shot reading, or the write capability, and the interpretation
validates one of the three recombination (``mixing'')
laws~\eqref{eq:mixing1}--\eqref{eq:mixing3} below, each of which
breaks subject reduction.

\paragraph{Correlation across components} Were the value and the mailbox
transformations independent, encoded say as a product of 3 separate arrows, then,
products distributing over intersections, one arrow's value transformation
could be paired with another's message transformation, validating
\begin{equation}\label{eq:mixing1}
(s_1 \to_{o_1}^{i_1} u_1) \;\land\; (s_2 \to_{o_2}^{i_2} u_2)
\;\subt\;
(s_1  \to_{o_2}^{i_2} u_1 ).
\end{equation}
The law is unsound. Take
$l_1 \eqdef (\TInt\to_{\Atom[{:}a]}^{\Atom[{:}a]}\TChan{\Atom[{:}a]}) \land
(\TBool\to_{\Atom[{:}b]}^{\Atom[{:}b]}\TChan{\Atom[{:}b]})$:
\rulename{t-abs} types $\Fun^{l_1}\x.\Self$, whose body, $\Self$ being
an expression of branch-indexed type $\TChan{o_j}$, types as
$\TChan{\Atom[{:}a]}$ in the first branch and $\TChan{\Atom[{:}b]}$ in the
second. Under~\eqref{eq:mixing1} the abstraction subsumes to
$\TInt\to_{\Atom[{:}b]}^{\Atom[{:}b]}\TChan{\Atom[{:}a]}$, so in a process of
interface $\Atom[{:}b]$ the expression
$\Send{(\Fun^{l_1}\x.\Self)\,42}{\Atom[{:}a]}$ is well-typed yet deposits an
$\Atom[{:}a]$-message in a mailbox read only at $\Atom[{:}b]$, breaking subject
reduction. In our interpretation the law fails: the triple pairing $42$ with a read
capability yielding $\Atom[{:}b]$-messages is in the domain of
$\TInt\to_{\Atom[{:}b]}^{\Atom[{:}b]}\TChan{\Atom[{:}a]}$ but in that of neither
member of $l_1$, the first expecting $\Atom[{:}a]$-messages and the second a
boolean, so neither member constrains the result on that triple: a function of
type $l_1$ may return there a value outside $\TChan{\Atom[{:}a]}$, and it is then
not in the interpretation of the right-hand side.

\paragraph{Why multi-shot read capabilities} Interpreting the read capability as
yielding a \emph{single} (pointed) message rather than a repeatedly invocable
operation (that is, encoding it as $\enc{i}\lor\CPoint$ instead of $\CPoint\to\enc{i}$) looks harmless, but it changes the subtyping relation: a single
message of type $i_1\lor i_2$ is an $i_1$-message or an $i_2$-message, so a
single-shot capability of type $i_1\lor i_2$ is one of type $i_1$ or one of
type $i_2$: $(()\to i_1)\lor(()\to i_2)\simeq()\to(i_1\lor i_2)$ in this
restricted reading. Since $i$ is the only component in which the two
conjuncts of the intersection differ, the simplified interpretation then
validates
\begin{equation}\label{eq:mixing2}
(s \to_{o}^{i_1} u)\;\land\;(s \to_{o}^{i_2} u)
\;\subt\;
s \to_{o}^{i_1\lor i_2} u.
\end{equation}
The law is unsound, because a mailbox can be read \emph{more than once}. Take
$e_r \eqdef \Recv{\Branch{y_1}{\Recv{\Branch{y_2}{\Tuple{y_1,y_2}}}}}$,
$u \eqdef \Tuple{\Atom[{:}a],\Atom[{:}a]}\lor\Tuple{\Atom[{:}b],\Atom[{:}b]}$,
and
$l_2 \eqdef (\Atom[{:}ok]\to_{\Tempty}^{\Atom[{:}a]}u)\land
(\Atom[{:}ok]\to_{\Tempty}^{\Atom[{:}b]}u)$: both \rulename{t-abs}
premises hold, since under input decoration $\Atom[{:}x]$ the two
\rulename{t-receive}s bind $y_1,y_2 : \Atom[{:}x]$ and the body types at
$\Tuple{\Atom[{:}x],\Atom[{:}x]} \subt u$. Under~\eqref{eq:mixing2} $l_2$ subsumes to
$\Atom[{:}ok]\to_{\Tempty}^{s}u$ with $s = \Atom[{:}a]\lor\Atom[{:}b]$,
which is invocable at the ambient bracket $[s,s]$, so the single-process
configuration
\[
\Process{\PID}{s}
{\Case{((\Fun^{l_2}\x.e_r)\;\Atom[{:}ok])}
  {\Branch{\Tuple{\Atom[{:}a],\Atom[{:}a]}}{\Atom[{:}ok]},\;
   \Branch{\Tuple{\Atom[{:}b],\Atom[{:}b]}}{\Atom[{:}ok]}}}
{\Atom[{:}a]\cdot\Atom[{:}b]}
\]
would be well-typed (the $\mkkeyword{case}$ is exhaustive on $u$). But the two
receives consume $\Atom[{:}a]$ and then $\Atom[{:}b]$, leaving a
$\mkkeyword{case}$ on the value $\Tuple{\Atom[{:}a],\Atom[{:}b]}$, which has
neither type $u$, so subject reduction fails, nor any of the two patterns, so
the $\mkkeyword{case}$ is stuck and progress fails too. In our interpretation the law
fails: a \emph{mixed} read capability, yielding one message in $i_1{\setminus}i_2$ and
one in $i_2{\setminus}i_1$, is in $\sem{()\to i_1\lor i_2}$ but in neither
$\sem{()\to i_1}$ nor $\sem{()\to i_2}$, so the intersection leaves its image
unconstrained (Lemma~\ref{lem:read-coverage} in Appendix~\ref{app:prop-arrow-inv}
generalizes the witness).

\paragraph{Why the write capability} Were $o$ absent from the domain and
recorded, if at all, only in the codomain, then arrows sharing argument type and
input decoration would have the same antecedent in the arrow clause, their
consequents would conjoin, and the interpretation would validate the collapse
\begin{equation}\label{eq:mixing3}
(s \to_{o_1}^{i} u_1)\;\land\;(s \to_{o_2}^{i} u_2)
\;\subt\;
s \to_{o_1\land o_2}^{i} (u_1 \land u_2),
\end{equation}
which recombines the output decorations and the results \emph{across} the
branches, a recombination of the same kind as~\eqref{eq:mixing1}. The law is unsound by the
very same $\Fun \x.\Self$ mechanism as above. Take
\[
l_3 \;\eqdef\;
(\Atom[{:}ok] \to_{\Atom[{:}a]\lor\Atom[{:}c]}^{\Atom[{:}a]\lor\Atom[{:}b]\lor\Atom[{:}c]}
   \TChan{\Atom[{:}a]\lor\Atom[{:}c]})
\;\land\;
(\Atom[{:}ok] \to_{\Atom[{:}b]\lor\Atom[{:}c]}^{\Atom[{:}a]\lor\Atom[{:}b]\lor\Atom[{:}c]}
   \TChan{\Atom[{:}b]\lor\Atom[{:}c]}):
\]
the abstraction $\Fun^{l_3}\x.\Self$ is well-typed by \rulename{t-abs}, both
premises being instances of \rulename{t-self}. Under~\eqref{eq:mixing3} and the
law $\TChan{t_1}\land\TChan{t_2} \simeq \TChan{t_1\lor t_2}$ we would get
$l_3 \simeq
\Atom[{:}ok] \to_{\Atom[{:}c]}^{\Atom[{:}a]\lor\Atom[{:}b]\lor\Atom[{:}c]}
\TChan{\Atom[{:}a]\lor\Atom[{:}b]\lor\Atom[{:}c]}$,
so the single-process configuration
\[
\Process{\PID}{\Atom[{:}b]\lor\Atom[{:}c]}
  {\Send{(\Fun^{l_3}\x.\Self)\;\Atom[{:}ok]}{\Atom[{:}a]}}
  {\EmptyMailbox}
\]
would be well-typed: the collapsed arrow is invocable at the ambient bracket
$[\Atom[{:}b]\lor\Atom[{:}c],\Atom[{:}b]\lor\Atom[{:}c]]$, its result attenuates by contravariance to
$\TChan{\Atom[{:}a]}$, and \rulename{t-send} concludes. But the application
reduces to $\Self$, hence to $\PID^{\Atom[{:}b]\lor\Atom[{:}c]}$, and the send
then deposits $\Atom[{:}a]$ in the process's own mailbox, yielding
$\Process{\PID}{\Atom[{:}b]\lor\Atom[{:}c]}{\Atom[{:}a]}{\Atom[{:}a]}$, which
\rulename{t-thread} cannot type: the promise
$\TChan{\Atom[{:}a]\lor\Atom[{:}b]\lor\Atom[{:}c]}$ was about a pid whose
true interface is $\Atom[{:}b]\lor\Atom[{:}c]$. In our interpretation the collapse fails:
the domains of the two branches of $l_3$ \emph{differ} in their write component,
since by the union law of Proposition~\ref{prop:pid-laws}, for
incomparable $o_1,o_2$ some element of
$\sem{\TChan{o_1\land o_2}}$ lies in neither $\sem{\TChan{o_j}}$ (here
$(\ell_{\Wch},\Set{(\Atom[{:}a],\Omega),(\Atom[{:}b],\Omega)})$, whose pid approximation rejects both an
$\Atom[{:}a]$ and a $\Atom[{:}b]$), so the corresponding images stay
unconstrained.
Lemma~\ref{lem:arrow-inv} in Appendix~\ref{app:prop-arrow-inv} turns this per-arrow
correlation into the decomposition of arrow subtyping that the $\beta$ case of
subject reduction consumes.

\section{Properties}
\label{sec:properties}

Soundness of the type system of Section~\ref{sec:typing} is, as customary,
the combination of two theorems on configurations: \emph{subject reduction}
(typing is preserved by reduction) and \emph{progress} (a well-typed
configuration is never stuck: the only processes that cannot advance are
those that have returned a value and those waiting for a message that has
not arrived yet). Each theorem comes in two versions: one for expressions, and one for configurations, the latter being the one that matters for the whole system.

Subject reduction, however, is stated for an \emph{enriched} typing
judgment, written $\wtee{\Context}{e}{t}$ for expressions and $\wtce{P}$
for configurations: the system of Section~\ref{sec:typing} extended with
two additional typing rules for \emph{values}---the negated-arrow rule of
\cite{FCB08} and its analogue for pids---defined in
Appendix~\ref{app:prop-enriched} (Definition~\ref{def:enriched}). The
enriched system is a proof device: programs are typed by the system of
Section~\ref{sec:typing}, and neither the inference of
Section~\ref{sec:ann-inference} nor the reconstruction of
Section~\ref{sec:reconstruction} ever manipulates $\Vdash$. Every
$\vdash$ derivation is in particular a $\Vdash$ derivation, and the two
judgments derive exactly the same types for every value that contains
neither abstractions nor pids (Lemma~\ref{lem:vee-principal}); the
discussion in the appendix explains why the detour through $\Vdash$ is not an
option but a necessity. In short: preservation holds for $\Vdash$,
soundness is delivered for $\vdash$.

A first result concerns brackets: every judgment survives the narrowing of its
bracket (Definition~\ref{def:bracket}), so that code demanding little and
promising much can run in any process whose interface fits. It is also a lemma
of subject reduction, since the body of an applied abstraction, typed at the
bracket of its arrow, runs after $\beta$-reduction at the ambient one.

\begin{restatable}[Mailbox narrowing]{theorem}{NarrowingThm}\label{thm:narrowing}
If $\wte[i][o]{\Context}{e}{t}$ and $[o',i']$ narrows $[o,i]$, then
$\wte[i'][o']{\Context}{e}{t}$.
\end{restatable}

\begin{restatable}[Subject reduction, expressions]{theorem}{SRExprThm}\label{thm:sr-expr}
If $\wtee[i][o]{\Context}{e}{t}$ and $e \red e'$, then
$\wtee[i][o]{\Context}{e'}{t}$.
\end{restatable}

\begin{restatable}[Subject reduction]{theorem}{SRThm}\label{thm:sr}
If $\wtce{P}$ and $P \red P'$, then $\wtce{P'}$. Moreover, if
$\Process{\PID}{s}{e}{\seqof{v}}$ is a process of $P$ with
$\wtee[s][s]{\emptyset}{e}{t}$, then $P'$ contains a process
$\Process{\PID}{s}{e'}{\seqof{w}}$ with $\wtee[s][s]{\emptyset}{e'}{t}$.
\end{restatable}

\noindent Neither statement of subject reduction requires coherence of the bracket (cf.\
Remark~\ref{rem:coherence}).

\begin{restatable}[Progress, expressions]{theorem}{ProgressExprThm}\label{thm:progress-expr}
If\ \  $\wtee[i][o]{\emptyset}{e}{t}$, then one of the following holds:
\begin{enumerate}[noitemsep,topsep=2pt]
    \item $e$ is a value;
    \item $e \red e'$ for some $e'$;
    \item $e = \Eval[f]$ where $f$ is one of $\Self$,\; $\Spawn{e'}{s}$,\;
    $\Send{\PID^s}{v}$,\; or $\Recv{\seqof{\Branch{p}{e}}}$.
\end{enumerate}
\end{restatable}

\noindent The expressions in \emph{(3)} are the communication redexes: they reduce
only at the level of configurations.

Progress for configurations relies on a
well-formedness invariant---every pid value names an existing process and
carries its interface---formally defined as follows and preserved by
reduction.%

\begin{restatable}[Well-formed configuration]{definition}{WFConfigDef}\label{def:wf-config}
A configuration $P$ is \emph{well-formed} if its processes have pairwise
distinct pids and, for every pid value $\PID^s$ occurring in $P$ (in an
expression, in a mailbox, or as the identity of a process), $P$ contains a
process $\Process{\PID}{s}{e}{\seqof{v}}$.
\end{restatable}
\begin{restatable}[Progress]{theorem}{ProgressThm}\label{thm:progress}
Let $P$ be well-formed with $\wtce{P}$, and let $\mathcal{P}=\Process{\PID}{t}{e}{\seqof{v}}$ be a
process of $P$. Then one of the following holds:
\begin{enumerate}[itemsep=0pt,topsep=2pt]
    \item $e$ is a value (the process $\mathcal{P}$ has terminated);
    \item $e=\Eval[\Recv{\seqof{\Branch{p}{e'}}}]$ and $\match{w}{p_j} = \textsf{fail}$
    for every message $w$ of $\seqof{v}$ and every pattern $p_j$ (the process $\mathcal{P}$ waits
    for a message that has not arrived);
    \item $P \red P'$ for some $P'$ by applying to $\mathcal{P}$ one of the rules of
    Section~\ref{sec:opsem}. 
\end{enumerate}
\end{restatable}

\noindent A receive whose mailbox holds no matching message is the one
legitimately irreducible redex: ruling out that a process waits forever
(deadlock, orphan messages) is beyond this set-of-messages layer%
\behavior{{} and is the
object of the behavioral mailbox types of
Section~\ref{sec:behavioral-types}, which rule out some of these situations}.

Subject reduction and progress combine into the final soundness
statement, whose shape deserves emphasis: its hypothesis is $\wtc{P}$, in
the negation-free, programmer-visible system of Section~\ref{sec:typing};
the enriched system appears only in the invariant maintained along the
execution, which Theorem~\ref{thm:progress} turns into the absence of
run-time errors.

\begin{restatable}[Soundness]{corollary}{SoundnessCor}\label{cor:soundness}
Let $P$ be a well-formed configuration with $\wtc{P}$, and let $Q$ be
reachable from $P$, that is, $P \red^* Q$. Then:
\begin{enumerate}[noitemsep,topsep=2pt]
    \item $Q$ is well-formed and $\wtce{Q}$; in particular, for every process
    $\Process{\PID}{s}{e}{\seqof{v}}$ of $Q$ and every message $w$ of $\seqof{v}$,
    $\wtee{}{w}{s}$, and every pid value $\PID[b]^{s'}$ occurring in $Q$ names a
    process of $Q$ of interface $s'$;
    \item every process of $Q$ has terminated on a value, waits for a message that
    has not arrived, or can take a step (Theorem~\ref{thm:progress});
    \item if $\Process{\PID}{s}{e}{\seqof{v}}$ is a process of $P$ with
    $\wte[s][s]{\emptyset}{e}{t}$, then $Q$ contains a process
    $\Process{\PID}{s}{e'}{\seqof{w}}$ with $\wtee[s][s]{\emptyset}{e'}{t}$; in
    particular, if $e'$ is a value $v$, then $\wtee{}{v}{t}$, and $\wte{}{v}{t}$ when
    $v$ contains neither abstractions nor pids.
\end{enumerate}
\end{restatable}

For programs, this is the familiar statement that well-typed programs do
not go wrong. Pids are runtime-only values (Section~\ref{sec:syntax}), so every source
program is pid-free, and for a program $e$ with $\wte[t][t]{\emptyset}{e}{s}$ the
initial configuration $\Process{\PID}{t}{e}{\EmptyMailbox}$ is well-typed (by
\rulename{t-thread}) and well-formed (the only pid occurring in it is $\PID$ itself).
Every configuration reachable from it is thus well-formed and well-typed, none of its
processes is stuck, and if the main process terminates, it does so on a value of type
$s$.

Item 1 is the invariant carried along the execution: enriched
typing, the only typing preserved by reduction, plus well-formedness, so that
no pid ever dangles, a send always finds its target, and no mailbox holds a message
that the interface of its process rejects. The last guarantee comes from typing, not
from a failure of the semantics, since a delivery is not checked at run time. Item 2
is the absence of run-time errors, process by process: no application applies a
non-function, no $\mkkeyword{case}$ fails to match, no send targets a value that is
not a pid, whatever the other processes do. 
Item 3 is type
preservation for the result of a process, and it shows that the detour through the
enriched system leaves no trace on the results that contain neither abstractions nor
pids.

The proofs of the five theorems and of the corollary, together
with the auxiliary results they rely on (minimal types, canonical forms,
matching, substitution, the inversion lemmas, intersection introduction, and the
enriched system itself) are in Appendix~\ref{app:properties}.

\section{Closing the Gap with Elixir: Annotations and Inference}
\label{sec:ann-inference}
The calculus defined in~\Cref{sec:syntax} is convenient for the study of the
meta-theory, but it still needs a crucial modification to serve as reference for
Elixir: arrow types carry the mailbox decorations $i$ and $o$, and we cannot
reasonably expect a programmer, even more a programmer of a dynamic language, to
systematically write them---the whole point of our design is that decorations
remain invisible to the programmer. In this section we address the problem of
decoration inference (Section~\ref{sec:inference}), we argue that inferred
decorations are as good as those that a the programmer could write
(Section~\ref{sec:requirement-inference}) and, in fact, are the best ones in
most cases. We therefore separate decorated \emph{types} from undecorated
\emph{annotations}: annotations (written in \textcolor{anncol}{\sf bordeaux})
are types written by the programmer in $\lambda$-abstractions and in
$\mkkeyword{spawn}$, the only two places where the syntax mentions types. Only
for code whose body is not available for inference, such as a precompiled
foreign module, will we require decorations to be written
(Section~\ref{sec:elixir-decorations}).

\subsection{Annotations}
\label{sec:annotations}
We add a syntactic category of \emph{annotations} $\ta,\sa$, that is,
types with \emph{undecorated arrows}:
\[
    \begin{array}{lrcl}
        \textbf{Annotation} &
        \ta, \sa & ::= &
            \ann{\Const} \mid \ann{\TInt} \mid \ann{\TAtom} \mid \ann{\TTuple} \mid \ann{\TChan{\ta}} \mid \ann{\TFun} \mid \ann{\ta \to \ta} \mid \ann{\Tuple{\seqof{\ta}}}  \mid \ann{\Tany}  \mid
            \ann{\ta \lor \ta} \mid \ann{\neg \ta}
    \end{array}
\]
coinductively defined and satisfying the well-formedness conditions of regularity and contractivity. Note that test types are a subset of annotations: their only function and pid components are the atomic constructors $\TFun$ and $\TChan{\Tempty}$, so no test type mentions a decoration.
Test types are thus types and annotations at once, and every occurrence of a test type in the
rules---in the accepted types $\accept{p}$ as well as in the constraints
generated in Section~\ref{sec:inference}---is well formed as it stands.\lucam{Perché si parla di well formedness qui? Le annotazioni sono induttive o coinduttive?}\beppem{Coinduttive (sono i tipi senza decorazioni). Aggiunta una frase per chiarire. }

That the annotation $\TFun$, an alias for the top function type
$\Tempty\to^{\Tempty}_{\Tany}\Tany$ of Section~\ref{sec:syntax}, leaves no decoration to supply is reasonable: a value known only to be \emph{some} function
may be passed around but never applied, since no coherent ambient\lucam{Definito?}\beppem{coherent, sì: definizione 6.1 remark 6.3 ... ambient è sinonimo di bracket (per la funzione)} narrows the
bracket $[\Tany,\Tempty]$ of $\TFun$.
Similarly, a pid of type $\TChan{\Tempty}$ cannot be used in any meaningful way.

Inference is performed on a calculus whose syntax is the one of Section~\ref{sec:syntax}, except for the two
annotated constructs which now carry annotations instead of types:
\[
    \begin{array}{lrcl}
        \textbf{Expression} &
        e, f & ::= &
            \cdots \mid \Fun^{\ann{\wedge \overline{\ta \to \ta}}} x.~e \mid \Spawn{e}{\ta}
    \end{array}
\]

Note that abstractions now carry an \emph{intersection} of arrow annotations
$\ann{\wedge\,\overline{\ta\to\ta}}$, instead of an intersection of arrow types. To relate annotations and types we use two functions:
\begin{itemize}
    \item $\erase t$ erases all the decorations in a type $t$, yielding an
    annotation;
    \item $\decorate{\ta}$ supplies the decorations missing from $\ta$,
    yielding a type.
\end{itemize}

Decorations are supplied through type variables: from now on types may contain
type variables, ranged over by $\alpha,\beta,\gamma,\delta$; a type without type variables is \emph{ground}. A type substitution $\sigma$ maps type variables to types,
$t\sigma$ denoting its application to $t$. The interface $t$ that a pid $\PID^{t}$ carries
is always ground: type variables serve inference only, and pids appear only at run time.
\begin{definition}[Decorating annotations]\label{def:decorate}
Fix a family $(\alpha_{\sa},\beta_{\sa})_{\sa}$ of pairwise distinct type
variables indexed by the arrow annotations $\sa$. The function $\decorate{\cdot}$
is defined on annotations \emph{coinductively}, by the equation
\[
    \decorate{\ann{\sa_1 \to \sa_2}} \;=\;
        \decorate{\sa_1}\to_{\beta_{\ann{\sa_1\to\sa_2}}}^{\alpha_{\ann{\sa_1\to\sa_2}}}\decorate{\sa_2}
\]
and homomorphically for the remaining types, that is, basic types, $\Tany$,
$\TFun$, pids, tuples and the type connectives. A type $t$ is a \emph{decoration} of
$\ta$ if $t=\decorate{\ta}\sigma$ for some type substitution $\sigma$.
\end{definition}
\noindent
The variables are indexed by the arrow annotation, instead of being chosen afresh
at each position of the tree where that annotation occurs, so that
$\decorate{\ta}$ is a regular tree, hence a type, even when $\ta$ is recursive.
It is the \emph{program} that chooses the families: each occurrence in it of one
of the two annotated constructs, that is, each $\lambda$-abstraction and each
$\mkkeyword{spawn}$, has a family of its own, disjoint from all the others. Two
functions that happen to carry the same annotation (e.g., integer sum and difference) are thus
decorated independently of each other. The decorations of $\ta$ are therefore the types
that erase to $\ta$, save that equal arrow annotations---e.g., all the unfoldings
of a recursive one---must have the same bracket
(see Lemma~\ref{lem:decorate-wd} and Remark~\ref{rem:decorate}).

The rules for typing the new expressions are the same as the declarative rules of Section~\ref{sec:typing}, except for the two constructs that
changed. They both now ``guess'' the decorations that their annotation leaves open,
that is, they take any decoration of it:
\[
    \inferrule[t-abs]{
        \wte[i_j][o_j]{\Context, x : s_j}{e}{t_j}
    }{
        \wte[i][o]{\Context}{\Fun^{\ann{\wedge \sa_j \to \ta_j}} x.~e}{\textstyle\bigwedge s_j \to_{o_j}^{i_j} t_j}
    }{\footnotesize\;\begin{array}{l}
        s_j \text{ decoration of } \sa_j\\[-.1mm]
        t_j \text{ decoration of } \ta_j\end{array}}
    \qquad\qquad
    \inferrule[t-spawn]{
        \wte[t][t]{\Context}{e}{t'}
    }{
        \wte{\Context}{\Spawn{e}{\ta}}{\TChan{t}}
    }{\footnotesize\;t \text{ decoration of } \ta}
\]

Like the families, the decorations are chosen once per occurrence
in the program of an annotated construct: a single type substitution decorates all the
annotations of an occurrence, the brackets of its arrows included, and every premise
that types this occurrence (for instance, once for each arrow of an enclosing
abstraction) uses the same decorations for it.

The operational semantics discards annotations and decorations
alike (Section~\ref{sec:typerase}): as before, annotations play no role at runtime.

\subsection{Type inference}
\label{sec:inference}

Simplifying the syntax is only the first step to reduce the gap with Elixir. The next step is to show how to decide the typing relation described by the previous rules, by defining an algorithmic deduction system. The system deduces judgments of the form $\tinf{\Context}{e}{t}[C]$, which reads
``\emph{under $\Context$, the expression $e$ has type $t$ provided the constraint set
$C$ is satisfiable}''. Intuitively $\Context$, $i$, $o$, and $e$ are the inputs of the algorithm, while the type $t$ and the constraints $C$ are the result. 

A constraint set $C$ is a finite set of \emph{constraints} $s\issubt t$; a type
substitution $\sigma$ \emph{satisfies} $C$ if $s\sigma\subt t\sigma$ holds for
every $s\issubt t \in C$, and $C$ is \emph{satisfiable} if some $\sigma$
satisfies $C$. 
A constraint set $C$ \emph{entails} a constraint $c$, written
$C\models c$, when every substitution that satisfies $C$ satisfies $c$. The type
variables that occur in constraints are the \emph{inference variables}: they are
introduced by the rules and never by the programmer, who writes only annotations,
and they can occur in the types that inference returns.

Constraints are solved once, at the end of the inference process, by the tallying
algorithm of Castagna et al.~\cite{polyduce2}, which applies as it stands, since our types are encoded into those
of~\cite{polyduce2} (Lemma~\ref{lem:tally-readback}). The tallying algorithm returns a finite set of type
substitutions, written $\textsf{Solutions}(C)$, which is \emph{sound}, that is, each
of its elements satisfies $C$, and \emph{complete}, that is, every substitution that
satisfies $C$ is, on the variables of $C$, an instance of one of its elements.

In order to give the rules (Section~\ref{sec:inference-rules}), we must
adapt to inference the pattern environments through which $\mkkeyword{case}$
and $\mkkeyword{receive}$ bind capture variables.

\subsubsection{Constraint-generating pattern environments}
\label{sec:cg-penv}

The declarative pattern environment $\ptenv{t}{p}$ of Figure~\ref{fig:pattern-aux} uses the projection operators
$\pi_k(t)$: for a tuple pattern it requires $t$ to denote a set of tuples (i.e.,
to be a subtype of $\TTuple$); this can always be expressed as a finite union of
tuple constructors and the projection returns the union of the projection of
each type. These projections are not computable when $t$ contains type
variables, which is the common case during inference. We therefore replace
$\ptenv{t}{p}$ by a relation $\ptrel{t}{p}{\Gamma}{C}$ that produces a set of
bindings $\Gamma$
together with a set of constraints $C$, both possibly mentioning fresh
variables, and that expresses each projection by a fresh variable deferred to
tallying:
\begin{mathpar}
    \inferrule[pe-var]{~}{
        \ptrel{t}{x}{\{x : t\}}{\varnothing}
    }
    \and
    \inferrule[pe-test]{~}{
        \ptrel{t}{\tau}{\varnothing}{\{t \issubt \tau\}}
    }
    \and
    \inferrule[pe-and]{
        \ptrel{t}{p_1}{\Gamma_1}{C_1}
        \\
        \ptrel{t}{p_2}{\Gamma_2}{C_2}
    }{
        \ptrel{t}{p_1 \land p_2}{\Gamma_1 \cup \Gamma_2}{C_1 \cup C_2}
    }
    \\
    \hspace*{6mm}\inferrule[pe-tuple]{
        \ptrel{\gamma_k}{p_k}{\Gamma_k}{C_k}
        \qquad (1 \le k \le n)
    }{
        \ptrel{t}{\Tuple{p_1, \dots, p_n}}{\textstyle\bigcup_{k} \Gamma_k}{\{\, t \issubt \Tuple{\gamma_1, \dots, \gamma_n} \,\} \cup \textstyle\bigcup_{k} C_k}
    }{\;\gamma_1, \dots, \gamma_n \text{ fresh}}
\end{mathpar}
In \rulename{pe-tuple} the constraint $t \issubt \Tuple{\gamma_1,\dots,\gamma_n}$
forces $t$ to be product-shaped and makes each $\gamma_k$ an upper bound of the
$k$-th projection: every solution $\sigma$ satisfies
$\pi_k(t\sigma) \subt \gamma_k\sigma$, and the least solution gives
$\gamma_k\sigma = \pi_k(t\sigma)$. Binding a variable to such a $\gamma_k$ instead
of to $\pi_k(t)$ is sound, because the matched sub-value has type
$\pi_k(t) \subt \gamma_k$; and complete, because the least solution recovers the
declarative environment exactly.

\subsubsection{Inference rules}
\label{sec:inference-rules}

The algorithmic rules are named after their declarative counterparts, with the
prefix \textsc{i-} in place of \textsc{t-} and are given in Figure~\ref{fig:inference}. 
\begin{figure}[t]
\begin{mathpar}
    \inferrule[i-const]{~}{
        \tinf{\Context}{\Const}{\Const}
    }
    \and
    \inferrule[i-pid]{~}{
        \tinf{\Context}{\PID^t}{\TChan{t}}
    }
    \and
    \inferrule[i-self]{~}{
        \tinf[i][o]{\Context}{\mkkeyword{self}}{\TChan{o}}
    }
    \and
    \inferrule[i-var]{~}{
        \tinf{\Context}{x}{\Context(x)}
    }
\end{mathpar}
\begin{mathpar}
    \inferrule[i-abs]{
        \tinf[\alpha_j][\beta_j]{\Context, x : s_j}{e}{t'_j}[C_j]
    }{
        \tinf[i][o]{\Context}{\Fun^{\ann{\wedge\sa_j \to \ta_j}} x.~e}{\textstyle\bigwedge_j (s_j \to_{\beta_j}^{\alpha_j} t_j)}[\bigcup_j(C_j\cup\{t'_j\issubt t_j\})]
    }{\footnotesize\begin{array}{l}
        s_j=\decorate{\sa_j} \\
        t_j=\decorate{\ta_j}\\
        \alpha_j,\beta_j\text{ fresh}
    \end{array}}
    \and
    \inferrule[i-app]{
        \tinf{\Context}{e_1}{t_1}[C_1]
        \\
        \tinf{\Context}{e_2}{t_2}[C_2]
    }{
        \tinf{\Context}{e_1~e_2}{\gamma}[C_1\cup C_2\cup\{t_1\issubt t_2 \to_{o}^{i}\gamma\}]
    }{\gamma\text{ fresh}}
\end{mathpar}
\begin{mathpar}
    \inferrule[i-tuple]{
        \tinf{\Context}{e_j}{t_j}[C_j]
    }{
        \tinf{\Context}{\Tuple{e_1, \dots, e_n}}{\Tuple{t_1, \dots, t_n}}[\textstyle\bigcup_{j=1}^n C_j]
    }
    \and
    \inferrule[i-plus]{
        \tinf{\Context}{e_1}{t_1}[C_1]
        \\
        \tinf{\Context}{e_2}{t_2}[C_2]
    }{
        \tinf{\Context}{e_1 + e_2}{\TInt}[C_1\cup C_2\cup\{t_1\issubt\TInt,\; t_2\issubt\TInt\}]
    }
    \\
    \inferrule[i-spawn]{
        \tinf[\alpha][\beta]{\Context}{e}{t'}[C]
    }{
        \tinf{\Context}{\Spawn{e}{\ta}}{\TChan{t}}[C\cup\{ \beta\issubt t,\; t\issubt \alpha\}]
    }{\footnotesize\begin{array}{l}
        t=\decorate{\ta}\\
        \alpha,\beta\text{ fresh}
    \end{array}}
    \and
    \inferrule[i-send]{
        \tinf[i][o]{\Context}{e}{t}[C]
        \\
        \tinf[i][o]{\Context}{e'}{t'}[C']
    }{
        \tinf[i][o]{\Context}{\Send{e}{e'}}{t'}[C\cup C'\cup\{t\issubt\TChan{t'}\}]
    }
\end{mathpar}
\begin{mathpar}
    \inferrule[i-case]{
        \tinf{\Context}{e}{t}[C]
        \and
        \ptrel{r_j}{p_j}{\Gamma_j}{D_j}
        \and
        \tinf{\Context, \Gamma_j}{e_j}{t_j}[C_j]
        \quad (j\in J)
    }{
        \tinf{\Context}{\Case{e}{\seqof{\Branch{p}{e}}}}{\textstyle\bigvee_{j\in J} t_j}[C\cup\{t \issubt \textstyle\bigvee_{j}\accept{p_j}\}\cup \textstyle\bigcup_{j\in J} (C_j \cup D_j)]
    }{\;\footnotesize\begin{array}{@{}l@{}}
        r_j=(t\land\accept{p_j})\setminus\textstyle\bigvee_{h<j}\accept{p_h}\\
        J=\{\,j \mid r_j\not\simeq\Tempty\,\}
    \end{array}}
\and%
    \inferrule[i-receive]{
    \ptrel{r_j}{p_j}{\Gamma_j}{D_j}
    \\
    \tinf{\Context, \Gamma_j}{e_j}{t_j}[C_j]
    \quad (j\in J)
}{
    \tinf{\Context}{\Recv{\seqof{\Branch{p}{e}}}}{\textstyle\bigvee_{j\in J} t_j}
        [\textstyle\bigcup_{j\in J} (C_j \cup D_j)]
}{\;\footnotesize\begin{array}{@{}l@{}}
        r_j=(i\land\accept{p_j})\setminus\textstyle\bigvee_{h<j}\accept{p_h}\\
        J=\{\,j \mid r_j\not\simeq\Tempty\,\}
    \end{array}}
\end{mathpar}
\caption{Algorithmic typing rules.}\label{fig:inference}
\end{figure}
\noindent
A few rules deserve a comment.

\rulename{i-pid} is there only so that the algorithmic system covers every expression
that the declarative one types: inference is run on programs, which contain no pid,
since pids appear only at run time. The rule reads the interface that the pid carries,
which, as stated in Section~\ref{sec:annotations}, is ground.%

\rulename{i-abs} types the body once for each annotation arrow $\ann{\sa_j\to\ta_j}$ of the
declared intersection, at fresh mailbox decorations $\alpha_j,\beta_j$; the
result is the intersection of the corresponding arrow types, and the only
constraints emitted are the expected $t'_j\issubt t_j$. As in the declarative rules,
the variables that decorate an occurrence, that is, the $\alpha_j,\beta_j$ of
\rulename{i-abs}, the $\alpha,\beta$ of \rulename{i-spawn} and the families of
$\decorate\cdot$, are minted once per occurrence
of the construct in the program and shared by all the deductions that type it, such
as the premises of an enclosing \rulename{i-abs}; the other fresh variables, those of
\rulename{i-app} and \rulename{pe-tuple}, are minted anew at each
application of their rule. No
constraint $\beta_j\issubt\alpha_j$ is emitted: coherence is not required here
(\Cref{rem:coherence}), and an abstraction
whose inferred decorations turn out incoherent is simply one that no coherent
ambient can ever invoke.

\rulename{i-app} does not decompose $t_1$ syntactically. It posits a fresh result
variable $\gamma$ and requires $t_1$ to be below the arrow $t_2\to_{o}^{i}\gamma$,
where $t_2$ is the type of the argument and $[o,i]$ is the bracket of the judgment, as
in the declarative \rulename{t-app}. A function created at $[\beta,\alpha]$ may thus
be called at $[o,i]$ when $[o,i]$ narrows $[\beta,\alpha]$: by
Proposition~\ref{prop:arrow-factor}, $s\to_{\beta}^{\alpha}t\subt
t_2\to_{o}^{i}\gamma$ holds if and only if $t_2\subt s$, $t\subt\gamma$,
$\beta\subt o$, and $i\subt\alpha$, and tallying derives these four constraints when
it decomposes the constraint of the rule. The constraint is well defined whatever the
shape of $t_1$ (a variable, a union or an intersection of arrows or their
negations).

\rulename{i-plus} proceeds like \rulename{i-app}:\lucam{Quale stile? perché è in italico?}\beppem{Lo "stile" era quello di \rulename{i-app}: non ispezionare il tipo degli operandi, ma emettere un vincolo di sottotipo.} it does not inspect the types of
its operands, it emits a constraint on each, and it is the template for
every primitive that constrains its operands (comparisons, arithmetic, etc.): it
types the operands at whatever types inference gives them and bounds each from
above, $\{t_1\issubt\TInt,\;t_2\issubt\TInt\}$. An operand whose type is still a
\emph{variable}, such as a value just extracted from the mailbox or the result of
an application, thus records its demand as a constraint, which is what requirement
inference (\S\ref{sec:requirement-inference}) collects.

\rulename{i-spawn} types the body at fresh decorations $[\beta,\alpha]$ and
brackets the declared interface between them, $\beta\issubt t\issubt\alpha$; the
declarative typing at $[t,t]$ is then recovered by mailbox narrowing. Note that
$t=\decorate{\ta}$ carries the variables that $\decorate\cdot$ minted for the
arrows written under the interface, so an annotation fixes an interface only
relative to the derivation that solves them. This is immaterial as long as the
whole program is checked at once, which is the setting of this section, and it is
what a client checked against the annotation alone would lack;
Section~\ref{sec:elixir-decorations} takes up the point. Note also that the rule
uses $t$ only through the two constraints it emits: the rule, and the results of
Section~\ref{sec:sound-complete}, apply as they stand to a $\Spawn{e}{t}$ that
carries a ground type $t$ instead of an annotation $\ta$, as those of
Section~\ref{sec:reconstruction} do.\footnote{\label{fn:spawn-variants}Formally, [\textsc{i-spawn}] comes in two variants: for $\Spawn{e}{\ta}$ the side condition is $t=\decorate{\ta}$; for $\Spawn{e}{t}$, which carries a type $t$, as in Section~\ref{sec:syntax}, there is none. The two rules can be easily mixed to handle partially decorated annotations.}\beppem{DO NOT REMOVE: this footnote is important for Section~\ref{sec:reconstruction} and \ref{sec:elixir}.}

\rulename{i-case} and \rulename{i-receive} obtain their branch environments
through $\rightsquigarrow$, over the same per-branch refinement as their
declarative counterparts: in \rulename{i-case} the scrutinee type, and in
\rulename{i-receive} the type $i\land\accept{p_j}$ of the messages that the
mailbox (bounded by $i$) may contain, are both stripped of the previously
matched patterns, mirroring the first-match reduction of
Section~\ref{sec:opsem}. Both rules type exactly the branches whose refined type
$r_j$ is not empty whatever the inference variables stand for, which is what
$r_j\not\simeq\Tempty$ means for a type with variables: where the declarative rules
only require $J$ to contain these branches, the algorithmic ones take $J$ to be
exactly this set, so that the deduction is determined by the types.

The exhaustiveness constraint $t\issubt\bigvee_j\accept{p_j}$ of \rulename{i-case}
is the side condition of \rulename{t-case}, which guarantees that matching cannot
fail; the sets $D_j$ do not entail it, since each constrains only the values that
match $p_j$. No such constraint accompanies \rulename{i-receive}: a message that matches no pattern is not
an error, it simply remains in the mailbox. But \rulename{i-receive} hides a
subtler point, which Section~\ref{sec:requirement-inference} develops.

\subsubsection{Soundness, Completeness, and Decidability}
\label{sec:sound-complete}

The algorithmic inference system of Figure~\ref{fig:inference} is sound and complete with respect to the declarative
system of Section~\ref{sec:annotations}, and it terminates. Soundness states that
every solution of the generated constraints yields a declarative derivation.
Completeness states that every declarative derivation is captured by some
solution, up to instantiating the fresh decorations at which inference works and
up to subsumption on the result type.

Completeness needs one assumption. The declarative system may skip the typing of the
$\mkkeyword{case}$ and $\mkkeyword{receive}$ branches that no value can reach, which is necessary to type functions with
intersection types (see the discussion in Section~\ref{sec:typing-expr}). Inference, which solves its variables only
at the end, skips a branch only when its refined type is empty whatever these
variables stand for, so completeness assumes that the declarative derivation skips
no other branch. We call such derivations \emph{compatible with inference}
(see Definition~\ref{def:compatible} for the formal definition). In particular, a compatible derivation must type
the dead code that inference does not recognize as such, which a compositional
system cannot flag.

\begin{restatable}[Soundness of inference]{theorem}{InfSoundThm}\label{thm:inf-sound}
If $\tinf{\Context}{e}{t}[C]$ and $\sigma$ satisfies $C$, then
$\wte[i\sigma][o\sigma]{\Context\sigma}{e}{t\sigma}$, by a derivation that skips
only the branches that the deduction skips.
\end{restatable}

\noindent In particular the conclusion holds for every
$\sigma\in\textsf{Solutions}(C)$, since $\textsf{Solutions}(C)$ is sound. When $i$
and $o$ are fresh variables and $\Context$ contains no inference variable, the
derivation is compatible with inference.

\begin{restatable}[Completeness of inference]{theorem}{InfCompleteThm}\label{thm:inf-complete}
Let $\wte[i][o]{\Context}{e}{t}$ be derivable by a derivation \emph{compatible with
inference}. Then $\tinf[\alpha][\beta]{\Context}{e}{t'}[C]$, with $\alpha,\beta$
fresh, and there exist $\sigma\in\textsf{Solutions}(C)$ and a substitution $\rho$
such that $i\subt\rho\sigma(\alpha)$, $\rho\sigma(\beta)\subt o$, and
$t'\rho\sigma \subt t$.
\end{restatable}

\noindent The instantiation $\rho$ is what completeness of tallying provides.
Modulo $\rho$, the inferred decorations bracket the declarative ones; the
declarative judgment is then recovered by mailbox narrowing, since $[o,i]$
narrows $[\rho\sigma(\beta),\rho\sigma(\alpha)]$
(Definition~\ref{def:bracket}).

Both theorems are proved in Appendices~\ref{app:elab-soundness} and~\ref{app:elab-completeness} on the
\emph{elaboration} of the deduction, that is, the term of the calculus of
Section~\ref{sec:syntax} obtained from $e$ by decorating its annotations
with the variables that the deduction introduces.
Finally, the rules of Figure~\ref{fig:inference} define 
an algorithm, whose constraints tallying solves.

\begin{restatable}[Termination and decidability]{proposition}{InferenceTerminatesProp}\label{prop:inference-terminates}
Let $\Context$, $i$, $o$ and $e$ be given such that $\tinf{\Context}{e}{t}[C]$ is
derivable. Then
\begin{enumerate}[noitemsep,topsep=2pt]
    \item there is exactly one such derivation, up to the renaming of the
    variables it mints; the derivation is finite, and $C$ is a finite set of
    constraints between regular types;
    \item tallying applies to $C$, and $\textsf{Solutions}(C)$ is the finite,
    sound, and complete set of regular solutions that it returns;
    \item let $\Context$ contain no inference variable, and $i$ and $o$ be fresh
    variables. Then $\textsf{Solutions}(C)\neq\varnothing$ if and only if
    $\wte[i'][o']{\Context}{e}{t'}$ is derivable, for some $i',o',t'$, by a
    derivation compatible with inference; and, for all ground types $i'$ and $o'$,
    $\textsf{Solutions}(C\cup\{i'\issubt i,\;o\issubt o'\})\neq\varnothing$ if and only if
    $\wte[i'][o']{\Context}{e}{t'}$ is derivable, for some $t'$, by a
    derivation compatible with inference.
\end{enumerate}
\noindent Whether an expression is typable under an environment with no inference
variable by a derivation compatible with inference, at some bracket or at a given
ground bracket, is decidable.
\end{restatable}

\subsection{Requirement inference, or why annotations carry no decorations}
\label{sec:requirement-inference}
\lucam{C'è qualcosa che non mi torna in questa sezione. Si parte dicendo che \textbf{in generale} specificare le decorazioni non si può. Va bene. Poi però si fa un esempio (discusso per due pagine) per il quale la decorazione migliore di tutte esiste. Semmai l'esempio andrebbe presentato come caso in cui \textbf{si potrebbe} usare un'unica decorazione (per altro con benefici). Ma allora servirebbe anche un altro esempio che illustri il caso in cui l'unica decorazione non c'è, mentre ora è solo detto a parole (pagina successiva).}
Specifying in a program the decorations in the annotation of a function definition would be, in general (but not always: cf.\ \S\ref{sec:elixir-decorations}), a bad
idea, since inference does this job at least as well as the programmer could. A function that reads the mailbox has no interface of its own: its
decorations record what the process \emph{running} it must accept, and, being
named, the function may be run by several processes whose interfaces differ,
directly or shipped to a $\mkkeyword{spawn}$. What its $\mkkeyword{receive}$s
expect places a constraint on those runners without singling one out. Inference
accepts exactly the programs that some decoration of their annotations makes well typed by a derivation compatible with
inference (Theorems~\ref{thm:inf-sound} and~\ref{thm:inf-complete}), and in most
cases it deduces the
\emph{best} decoration of a function, the one that accepts every runner that some
decoration accepts. A programmer, instead, would most likely write the interface of
the runner at hand, and lose the (possibly unforeseen) others. Consider a private
function spawned under a declared interface:
\begin{elixircode}
!\kwinterface! {:half, integer()}
spawn(fn -> halve() end)

!\color{anncol}\$! (-> integer())
defp halve() do
  receive do {:half, x} -> div(x, 2) end
end
\end{elixircode}
\noindent
\eelixir|halve| is well typed, since it runs at the declared
interface \eelixir|{:half, integer()}|. But its name may be reused by
another $\mkkeyword{spawn}$ at a wider
interface---say \eelixir|{:half, integer()} or {:triple, integer()}|---which is an
equally legal runner. A programmer who had to decorate \eelixir|halve| would most
likely write the interface at hand, \eelixir|{:half, integer()}|, which rejects the
second runner. A decoration that accepts both exists, and we do not want to
re-typecheck the body of \eelixir|halve| at every $\mkkeyword{spawn}$ that uses it
in order to find it.
\beppem{Risposta alla nota di Luca a inizio sezione. Giusto: per \texttt{halve} la decorazione migliore {$[o_0,i_0]$} esiste. Il testo diceva il contrario di quel che intendevo (l'inferenza fa almeno quanto il programmatore, e di solito trova la decorazione migliore). Riscritto: \texttt{halve} è il caso in cui la decorazione migliore esiste, \texttt{run\_job} (fine sezione) quello in cui non esiste.}

The inference system of Section~\ref{sec:inference-rules} finds it without doing
so, because its \rulename{i-receive} keeps the ambient interface \emph{symbolic}.
Type the body of \eelixir|halve| at a variable interface $\alpha$: the
single branch is matched against the scrutinee
$\alpha\land\Tuple{\Atom[{:}half],\Tany}$ and, since this type is not a
syntactic tuple, \rulename{pe-tuple} defers its two projections to fresh
variables. The branch generates
\[
D \;=\; \{\,\alpha\land\Tuple{\Atom[{:}half],\Tany}\issubt\Tuple{\gamma_1,\gamma_2},\;\gamma_1\issubt\Atom[{:}half]\,\}
\qquad
C \;=\; \{\,\gamma_2\issubt\TInt\,\},
\]
the second constraint of $D$ coming from the test pattern \eelixir{:half}
(\rulename{pe-test}), and the constraint of $C$ from the use of \eelixir{x} in
\eelixir|div(x, 2)|, as \rulename{i-plus} does for \eelixir|x + 2|.
Nothing here commits $\alpha$ to a value. Eliminating the deferred variables
($\gamma_1$ and $\gamma_2$ can be taken equal to their upper bounds
$\Atom[{:}half]$ and $\TInt$) leaves a single \emph{constraint} on the
interface,
\begin{equation}\label{eq:requirement}
\alpha\land\Tuple{\Atom[{:}half],\Tany}\;\issubt\;\Tuple{\Atom[{:}half],\TInt},
\end{equation}
which is the \emph{requirement} of \eelixir|halve|: the payload of a
$\Atom[{:}half]$ message (i.e., the second component of all pairs in $\alpha$ whose first component is $\Atom[{:}half]$) must be an integer, and the function demands nothing
else of whoever runs it.

\begin{definition}[Requirement]\label{def:requirement}
Let $\tinf[\alpha][\beta]{\Context}{e}{t}[C]$ with $\alpha,\beta$ fresh. The
\emph{requirement} of $e$ is the set of the constraints of $C$ in which either $\alpha$ or
$\beta$ occurs, and the brackets it \emph{admits} are the ground brackets
$[\beta\sigma,\alpha\sigma]$ for $\sigma$ satisfying $C$. The requirement is
\emph{uniform} if the brackets it admits are exactly the ground narrowings
of a single ground bracket $[o_0,i_0]$, its \emph{best
decoration}; it is \emph{non-uniform} otherwise.
\end{definition}

\noindent The requirement of $e$ is \emph{uniform} when whether a process running $e$
may accept a given message depends only on that message, and not on the other messages
the process accepts. For \eelixir|halve|, a process may accept \eelixir|{:half, v}| if
and only if $v$ is an integer, whatever its other messages are. Likewise, the messages
that $e$ sends to the process running it, which that process must accept, are the same
whichever process runs $e$.

The requirement is \emph{non-uniform} otherwise, for instance when $e$ receives a
function and runs it. The function then runs in the process that received it, so that
process may accept a message carrying the function only if the function can run at its
interface, hence depending on the other messages it accepts. The function
\eelixir|run_job|, at the end of this section, is a detailed example showing this phenomenon. But a non-uniform requirement may also happen without
closures:
\begin{elixircode}
receive do {:whoami, from} -> send(from, self()) end  !\phantomsection\label{whoami}!
\end{elixircode}
\noindent Here a process may accept a message \eelixir|{:whoami, from}| only if
\eelixir{from} can receive the pid of that process, whose type depends on the messages
the process accepts.

When the requirement of $e$ is uniform, the processes
that can run $e$ are exactly those whose interface $s$ satisfies
$o_0\subt s\subt i_0$ as for Definition~\ref{def:requirement} (see Proposition~\ref{prop:requirement} below). So, among the interfaces of
the processes that can run $e$, $i_0$ is the \emph{greatest} (provided $o_0\subt i_0$, without which no process can run $e$). The requirement~\eqref{eq:requirement} of \eelixir|halve| is
equivalent to
$\alpha\subt\Tuple{\Atom[{:}half],\TInt}\lor\neg\Tuple{\Atom[{:}half],\Tany}$; so
$i_0$ is this type (which contains every message but the $\Atom[{:}half]$ pairs whose
payload is not an integer), and $o_0=\Tempty$. The pair $[o_0,i_0]$ is the
best decoration of \eelixir|halve| in a precise sense: arrows are contravariant in the input
decoration and covariant in the output one,
so $\Tuple{}\to_{\Tempty}^{i_0}\TInt$ is a subtype of the arrow that any decoration
admitted by the requirement gives to \eelixir|halve|, and, written in its annotation,
it would accept every runner that some decoration accepts. Inference deduces it,
and the programmer need not even know it. A non-uniform requirement has no best
decoration: no bracket it admits has all the others as narrowings.
Remark~\ref{rem:uniform} in Appendix~\ref{app:elab-uniform} proves this, shows how $i_0$
and $o_0$ are computed from $C$, and shows that a greatest interface can exist even when
the requirement is non-uniform, as for the \eelixir{:whoami} receive of line~\ref{whoami}, whose greatest interface is the unique solution of
$x=\neg\Tuple{\Atom[{:}whoami],\Tany}\lor\Tuple{\Atom[{:}whoami],\TChan{\TChan x}}$.

A requirement is thus not a new form of type and calls for no addition
to the system: the judgment reads it as it stands, ``$e$ has type $t$
\emph{provided $C$ is satisfiable}'', inference variables being free to occur in
$t$. Here \eelixir|halve| has the ordinary inferred type
$\Tuple{}\to_{\beta}^{\alpha}\TInt$, and~\eqref{eq:requirement} says which
$\alpha$ make that typing valid. A single derivation therefore speaks about every
admissible runner at once, which is exactly what the soundness of the inference
system (Theorem~\ref{thm:inf-sound}) states: every solution
$\sigma$ of $C$ yields a declarative typing of the body at the
decorations that $\sigma$ assigns. The requirement is carried by the judgment,
not by the type.

\paragraph{One check, at the join of the runners.}
The requirement is computed once, from the body, and this loses nothing with
respect to the per-use re-typechecking we ruled out. All the uses of a function
share its decorations: if \eelixir|halve| is run by processes of interfaces
$s_1,\dots,s_n$, the bracket that \rulename{i-spawn} emits for each declared
$s_k$, together with the constraint of the call (\rulename{i-app}), entails
$\beta\issubt s_k$ and $s_k\issubt\alpha$ on the same $\alpha$ and $\beta$, so a
single solution must satisfy
$\beta\issubt\bigwedge_k s_k$ and $\bigvee_k s_k\issubt\alpha$: the requirement is
checked once, at the \emph{join} of the runners' interfaces. For a uniform
requirement, this single check accepts exactly the runners that separate checks
would accept: all the runners face the same bounds $o_0$ and $i_0$, and a join is
below $i_0$, and a meet above $o_0$, exactly when all interfaces are.

\begin{restatable}[Uniform requirements]{proposition}{RequirementProp}\label{prop:requirement}
Let $\tinf[\alpha][\beta]{\Context}{e}{t}[C]$ with $\alpha,\beta$ fresh, let the
requirement of $e$ be uniform, with $[o_0,i_0]$ as in
Definition~\ref{def:requirement}, and let $s_1,\dots,s_n$ ($n\geq1$) be ground
types. Some $\sigma$ satisfying $C$ has $\beta\sigma\subt\bigwedge_k s_k$ and
$\bigvee_k s_k\subt\alpha\sigma$ if and only if $o_0\subt s_k\subt i_0$ for every
$k$.
\end{restatable}

\noindent The left-hand side of the equivalence is the single check that the shared decorations
impose; the right-hand side is a conjunction of checks, one per runner. Sharing a function with a uniform requirement among several processes thus
costs nothing, as long as the variables of $C$ other than $\alpha$ and $\beta$ occur
nowhere else in the program. This excludes parameters of arrow type, whose decorations
are shared with the callers. For \eelixir|halve|, the wider interface
\eelixir+{:half, integer()} or {:triple, integer()}+ is below $i_0$ (recall that, for \eelixir|halve|, $i_0 =\Tuple{\Atom[{:}half],\TInt}\lor\neg\Tuple{\Atom[{:}half],\Tany}$). A runner whose payload for
\eelixir{:half} is not contained in \eelixir{integer()} is not below $i_0$ and is
rejected, on its payload rather than on its tag, whatever the other runners of \eelixir|halve| are.  Thus, \eelixir+{:half, term()}+ is rejected, and so is
\eelixir+{:half, float()}+ (rightly so since \eelixir|div/2|, the integer division,\footnote{In Elixir,
  $f/n$ denotes the function $f$ of arity $n$, and $M.f/n$ the function $f/n$ of
  module $M$.} is defined on integers, neither on \eelixir{term()} nor on floats). A decoration written by hand in
the type of \eelixir|halve|, such as \eelixir+{:half, integer()}+, would replace this
join by a single fixed interface, and the second, wider $\mkkeyword{spawn}$ would be
rejected though nothing in the body objects to it. 
{Keeping the interface
symbolic is what lets the annotated system find by itself the decoration that types
\emph{modularly} a function shared by processes with different interfaces.}

\paragraph{When no decoration is best.}
A best decoration need not exist. Consider a function that runs the jobs it
receives:
\begin{elixircode}
!\color{anncol}\$! (-> term())
defp run_job() do
  receive do {:run, job} -> job.() end
end
\end{elixircode}
\noindent and two \emph{separate} programs, each made of this definition and of
one of the two fragments below. The first spawns a worker and sends it a
job that reads an \eelixir{:add} message and halves its payload:
\begin{elixircode}
!\kwinterface! {:run, (-> term())} or {:add, integer()}
w = spawn(fn -> run_job() end)
send(w, {:run, fn -> receive do {:add, n} -> div(n, 2) end end})
\end{elixircode}
\noindent The second program spawns a process that accepts every \eelixir{:add}
payload, and sends it nothing:
\begin{elixircode}
!\kwinterface! {:run, (-> term())} or {:add, term()}
spawn(fn -> run_job() end)
\end{elixircode}
\noindent The application \eelixir|job.()| demands of \eelixir|job| an arrow that
can run at the decorations $[\beta,\alpha]$ of \eelixir|run_job|, and eliminating
the other variables as for \eelixir|halve| leaves the requirement
\[
\alpha\land\Tuple{\Atom[{:}run],\Tany}\;\issubt\;\Tuple{\Atom[{:}run],\;\Tuple{}\to_{\beta}^{\alpha}\gamma},
\]
where $\gamma$ is the type of the result of \eelixir|job.()|. Here $\alpha$ occurs
on both sides: the jobs that a process accepts must be able to run in it, so whether a
process may accept \eelixir|{:run, job}| depends on the other messages it accepts, and
the requirement is non-uniform. In both
programs, the bracket of the $\mkkeyword{spawn}$ bounds $\alpha$ from below by the
interface of the runner. In the first program, the job also bounds it from above.
Let $[\delta,\delta']$ be the decorations of the arrow in the \eelixir{:run} slot of
the worker's interface. Since that interface is
below $\alpha$, the requirement puts this arrow below
$\Tuple{}\to_{\beta}^{\alpha}\gamma$, whence $\alpha\issubt\delta'$, arrows being
contravariant in the input decoration. The
job sent to \eelixir|w| must have this arrow type, so its input decoration
contains $\delta'$, and the job reads the payloads of \eelixir{:add} as integers.
The two programs thus impose, respectively,
\[
\alpha\land\Tuple{\Atom[{:}add],\Tany}\;\issubt\;\Tuple{\Atom[{:}add],\TInt}
\qquad\text{and}\qquad
\Tuple{\Atom[{:}add],\Tany}\;\issubt\;\alpha .
\]
Each program alone is well typed, with \eelixir|run_job| decorated at
$[\Tempty,u_1]$ in the first and at $[\Tempty,u_2]$ in the second, where
\[
u_1=\Tuple{\Atom[{:}add],\TInt}\lor\Tuple{\Atom[{:}run],\Tuple{}\to_{\Tempty}^{u_1}\Tany}
\qquad
u_2=\Tuple{\Atom[{:}add],\Tany}\lor\Tuple{\Atom[{:}run],\Tuple{}\to_{\Tempty}^{u_2}\Tany},
\]
the arrow in the \eelixir{:run} slot of the interface being decorated in the same
way, so that the interface is $u_1$ in the first program and $u_2$ in the second.
The two constraints, however, are incompatible, which has two consequences. First,
no decoration written in the annotation of \eelixir|run_job| serves both programs.
Which one is right depends on the rest of the program, not on the body of
\eelixir|run_job|, and inference chooses it together with that program
(Section~\ref{sec:elixir-decorations} considers what happens when the rest of the
program is not at hand). Second, the
program made of the definition of \eelixir|run_job| and of \emph{both} fragments is
rejected: the two $\mkkeyword{spawn}$s share the decorations of
\eelixir|run_job|, so $\alpha$ is subject to both constraints. The conflict is
between the job sent to \eelixir|w| and the second runner, not between the two
runners: without the $\mkkeyword{send}$, both run \eelixir|run_job| at
$[\Tempty,u_2]$. The requirement being non-uniform, the message carrying that job may
be accepted by the first runner and not by the second. That program is
nevertheless safe: the worker receives only integer \eelixir{:add} payloads, and
nothing is ever sent to the second process, whose pid is discarded. The rejection is
the price of decorations shared by all the runners of a function: every runner of
\eelixir|run_job| accepts only the jobs that can run at the decorations of
\eelixir|run_job|, and these bracket the interfaces of all its runners. Defining
\eelixir|run_job| twice, once for each runner, makes the program well
typed.
So does declaring \eelixir|run_job| with the duplicated annotation
\eelixir|(-> term()) and (-> term())|: \rulename{i-abs} gives each arrow of the
intersection fresh decorations of its own, the call in each runner selects one arrow,
and the two arrows are decorated at $[\Tempty,u_1]$ and $[\Tempty,u_2]$ respectively.
\footnote{
Inference could build this intersection itself by \emph{expansion}, as in
intersection type systems, that is, by duplicating the typing of the body with renamed
decoration variables. Neither our system nor our prototype does: unrestricted expansion
is what makes type inference undecidable for intersection types, and here it would at
least degrade the performance of inference only to serve this corner case. We leave its
study for future work.
}

\subsection{Type erasure}
\label{sec:typerase}

It still remains to prove that the
annotations serve static typing alone and have no effect at run time, so that a
typed program runs on the stock BEAM. Let $\typeerase{e}$ denote the term
obtained by erasing every type from $e$, replacing $\Fun^{l}x.~e$ by
$\Fun x.~\typeerase{e}$, $\Spawn{e}{\ta}$ by $\mkkeyword{spawn}(\typeerase{e})$
and $\PID^{t}$ by $\PID$, and let $\typeerase{\cdot}$ act on a configuration by
erasing each process, $\Process\PID{t}{e}{\seqof v}$ becoming
$(\PID,\typeerase e,\typeerase{\seqof v})$. Erased terms reduce by the rules of
Section~\ref{sec:opsem} read on erased syntax, with one adjustment: the two send
rules locate the recipient by its \emph{name}, matching $\Send{\PID[b]}{v'}$
against the process of pid $\PID[b]$, there being no longer an annotation to
match. We write $\red$ for this relation too.

\begin{restatable}[Type erasure]{theorem}{TypeEraseThm}\label{thm:typerase}
Let $P$ be a well-formed configuration (Definition~\ref{def:wf-config}). Then
\begin{enumerate}[noitemsep,topsep=2pt]
    \item if $P \red P'$, then $\typeerase{P} \red \typeerase{P'}$;
    \item if $\typeerase{P} \red Q$, then $P \red P'$ for some $P'$ with
    $\typeerase{P'} = Q$.
\end{enumerate}
\end{restatable}

\noindent So $\typeerase\cdot$ is a strong bisimulation, and a well-typed program
and its erasure are indistinguishable at run time. The proof
(Appendix~\ref{app:elab-erasure}) is a case analysis on the reduction rules: only
the send rules consult an annotation, and well-formedness makes the annotation of
the target pid coincide with the interface of the recipient. Composed with inference and
with the soundness results of Section~\ref{sec:properties}, this gives what the
programmer needs.

\begin{restatable}[Safety of inferred programs]{corollary}{InferredSafetyCor}\label{cor:inferred-safety}
Let $e$ be a program, that is, a closed pid-free expression, with
$\tinf[\alpha][\beta]{\emptyset}{e}{t}[C]$ for fresh $\alpha,\beta$; let $s$ be a
ground type, $\sigma$ a type substitution that satisfies
$C\cup\{\beta\issubt s,\;s\issubt\alpha\}$, and $\rho$ a type substitution that maps
every type variable to a ground type. Then every
configuration $Q$ reachable from $(\PID,\typeerase{e},\EmptyMailbox)$ satisfies:
\begin{enumerate}[noitemsep,topsep=2pt]
    \item every process of $Q$ has terminated on a value, waits for a message that
    matches none of its patterns, or can take a step;
    \item if the process of pid $\PID$ has terminated on a value $v$ that contains
    neither abstractions nor pids, then $\wte{}{v}{t\sigma\rho}$.
\end{enumerate}
\end{restatable}

\noindent The ground type $s$ is the interface of the main process, which the
program declares by a top-level annotation \kwselfinterface~$s$. A program accepted
by inference at some interface thus runs on the stock BEAM without run-time type
errors, and if its main process terminates on data, the
data have the inferred type.

\section{Reconstruction of the Interface}
\label{sec:reconstruction}
The system of Section~\ref{sec:ann-inference} is our type system for Elixir, and we
take it as the default. It still departs from current Elixir on one point,
though: the programmer must annotate every $\mkkeyword{spawn}$ with the interface
of the process it creates. For legacy code that carries no such annotation we
show how to \emph{reconstruct} an interface from the code of the spawned
process.\footnote{The current release of Elixir does not support type annotations
for functions either---they are planned for a future release~\cite{Elixir-1.20}---and the system copes with their absence by resorting to gradual typing.
We ignore this aspect in this work.}

The reconstruction algorithm has four steps: generation, derivation, solving, and checking.
\begin{enumerate}[noitemsep,topsep=2pt]
    \item \emph{Generation} (Section~\ref{sec:recon-bracket}). We generate the constraints of the whole program, with the
    inference rules of Section~\ref{sec:inference-rules} and one new rule for the
    unannotated $\mkkeyword{spawn}$s. This rule leaves the interface of the spawned process
    open. Nothing is solved at this stage.
    \item \emph{Derivation} (Section~\ref{sec:recon-derivation}). From these constraints we derive, for each unannotated
    $\mkkeyword{spawn}$, an upper bound on its interface, from what the code of the process does
    with the messages \emph{it reads}: as anticipated in Section~\ref{sec:example}, each branch of
    a $\mkkeyword{receive}$ states a requirement on the interface. Nothing is solved at this stage
    either: the bound may still mention type variables.
    \item \emph{Solving} (Section~\ref{sec:recon-solving}). We solve each bound: the interface is the ``most permissive'' type that
    satisfies the requirements of the code the process runs. When no such type can be singled
    out, this step fails and the interface must be declared.
    \item \emph{Checking} (Section~\ref{sec:recon-check}). We replace the new rule by the rule \rulename{i-spawn} of
    Section~\ref{sec:inference-rules} at the interface so obtained, and we check the whole
    program against this interface, exactly as against a declared one.
\end{enumerate}
Solving and checking are the only steps in which the reconstruction may fail. The derivation and solving steps are what recover payloads,
and with them the received capabilities on which the higher-order core of the
system rests; the checking step is what makes a reconstructed interface a contract like
a declared one. ``Most permissive'' must be taken with care: when the requirement of the code
that the process runs is uniform (Definition~\ref{def:requirement}), among the types that satisfy
it and accept only messages that the code can consume there is a \emph{greatest} one, but when it
is non-uniform none need exist. The solving step yields a type \emph{maximal} among those that
respect the bound it solves (Section~\ref{sec:recon-guarantees}), and the greatest of them when
this bound does not mention the interface, as for the worker of Section~\ref{sec:example}.

The approach has thus an inherent dissymmetry: the $\mkkeyword{receive}$s of a process determine
its interface, and the rest is just checked against it. The new rule of the generation step makes
this possible, by separating the code that reads a mailbox, which bounds the interface from above,
from the code that writes to it, which bounds it from below: the derivation and solving steps use
only the upper bounds, and the checking step verifies the lower ones.
Sections~\ref{sec:recon-guarantees} and~\ref{sec:recon-limits} state what the reconstruction
guarantees and what it misses, respectively. As we discuss in Section~\ref{sec:elixir-tallying} the overhead of reconstruction  is negligible with respect to the cost of performing the inference discussed in the previous section.

\subsection{Generation: the split bracket}\label{sec:recon-bracket}

Reconstruction adds a new form of expression to the syntax of Section~\ref{sec:annotations},
the \emph{unannotated spawn} $\mkkeyword{spawn}(e)$, which creates a process of body $e$ without
stating its interface; a program may contain both annotated and unannotated spawns. We must
therefore extend to this new expression both the declarative system of
Section~\ref{sec:annotations} (\textsc{t-}prefixed rules) and the inference system of
Section~\ref{sec:inference} (\textsc{i-}prefixed rules). Each gains one rule, distinguished
by the suffix \textsc{-alt}, in which the annotation gives way, respectively, to a type to be
guessed and to a fresh variable; all the other rules are unchanged:
\begin{mathpar}
    \inferrule[t-spawn-alt]{
        \wte[t][t]{\Context}{e}{t'}
    }{
        \wte{\Context}{\mkkeyword{spawn}(e)}{\TChan{t}}
    }{\;\text{for some type } t}
    \and
    \inferrule[i-spawn-alt]{
        \tinf[\alpha][\beta]{\Context}{e}{t'}[C]
    }{
        \tinf{\Context}{\mkkeyword{spawn}(e)}{\TChan{\gamma}}[C]
    }{\;\alpha,\beta,\gamma\text{ fresh}}
\end{mathpar}
The declarative rule \rulename{t-spawn-alt} is the \emph{specification} of the
reconstruction: a valid interface for the spawned process is any type $t$ at which $e$ types at the
coherent bracket $[t,t]$, as \rulename{t-spawn} requires of a declared one, and
against whose pid type the rest of the program types. The derivation and solving steps compute a
candidate for $t$, and the checking step verifies it.

The algorithmic rule \rulename{i-spawn-alt} is the rule that the generation step adds to
those of Section~\ref{sec:inference-rules}. It differs from \rulename{i-spawn} in two
respects: a fresh variable $\gamma$, the \emph{pid variable} of the $\mkkeyword{spawn}$, replaces
the annotation $t$ in the pid type of the $\mkkeyword{spawn}$, and the constraints $\beta\issubt t\issubt\alpha$, which tie the
bracket of $e$ to the annotation, are \emph{not} emitted. Like the variables that
decorate a construct (Section~\ref{sec:inference-rules}), $\alpha$, $\beta$ and
$\gamma$ are minted once per occurrence of the $\mkkeyword{spawn}$. The two ends of the
bracket thus stay \emph{apart} until the interface $t$ of this $\mkkeyword{spawn}$ is found;
then $\beta\issubt t\issubt\alpha$ is added and $\gamma$ is replaced by $t$.
They must stay apart until then: had \rulename{i-spawn-alt} emitted
$\beta\issubt\gamma\issubt\alpha$ at once, every lower bound on $\beta$ or $\gamma$
would also be a lower bound on $\alpha$, and the code of the clients would constrain
the interface.

Kept apart, the two ends are bounded from opposite sides, as the division of labor between the
two decorations prescribes (Section~\ref{sec:typing-decorations}).

\paragraph{Readers bound $\alpha$ from above.}
Only two rules use the input decoration of their judgment, so $\alpha$ occurs in the constraint set $C$ of
\rulename{i-spawn-alt} in two ways only. \rulename{i-receive} types the messages of a
branch at a type below $\alpha$, and the use that the code makes of them bounds this
type from above. \rulename{i-app} places $\alpha$ below the input decoration of the
function called. This is again an upper bound, and since the body of the function is
typed at that decoration, its $\mkkeyword{receive}$s and calls bound the decoration,
hence $\alpha$, in the same two ways. For instance, spawn the worker of
Section~\ref{sec:example} without annotation (line~\ref{spawnclosure}): its body calls
\eelixir{serve}, which calls \eelixir{loop}, so the $\alpha$ minted for this spawn is below the
input decoration of \eelixir{loop}, whose branch \eelixir|{:add, k} -> loop(n + k)|
(line~\ref{addbranch}) requires an integer \eelixir{k}, whence
$\alpha\land$\eelixir|{:add, term()}|$\;\issubt\;$\eelixir|{:add, integer()}|.

\paragraph{Writers bound $\beta$ and $\gamma$ from below.}
Likewise, only two rules use the output decoration, so $\beta$ occurs in $C$ in two ways
only. \rulename{i-self} gives $\Self$ the type $\TChan{\beta}$, and since a pid type is
contravariant, what the code sends through this pid ---or lets another process send by
forwarding the pid--- is a lower bound on $\beta$. \rulename{i-app} places the output
decoration of the function called below $\beta$. This is again a lower bound, and the
occurrences of $\Self$ and the calls in the body of the function bound that decoration,
hence $\beta$, in the same two ways. For instance, consider, as before, the spawn of a worker without specifying the interface (line~\ref{spawnclosure}) and the $\beta$ minted for it: the self-send in the body of \eelixir{serve}
(line~\ref{self2}) gives $\Atom[{:}bump]\issubt\beta$. The variable $\gamma$ does not
occur in $C$, but in the constraints of the code that uses the pid returned by the
$\mkkeyword{spawn}$, where it is bounded as $\beta$ is: with \eelixir{w} this pid,
\eelixir|send(w, :bump)| gives $\Atom[{:}bump]\issubt\gamma$.

\medskip

\noindent What a \emph{client sends} bounds $\beta$ or $\gamma$, depending on how it obtained the
pid of the process, and never $\alpha$, so the derivation and solving steps depend on the code
that the process runs, not on its clients. A client cannot narrow the interface of a process: it can only conform
to the interface or not, and a client that does not is rejected at the checking step, at
its own $\mkkeyword{send}$. Such a client is not detected earlier: what it sends does not bound
$\alpha$, so the constraints that the generation step produces may have a solution even when no
type is a valid interface for the process.

Finally, one process is created by no $\mkkeyword{spawn}$: a program $e$ is itself the body of
a process, the main process, and is typed at a bracket of its own, the \emph{root
bracket}, a pair of fresh variables $[\beta_0,\alpha_0]$
(Corollary~\ref{cor:inferred-safety}). If the interface of the main process is declared by
a top-level annotation \kwselfinterface~$t_0$, the constraints $\beta_0\issubt t_0$ and
$t_0\issubt\alpha_0$ are added, as \rulename{i-spawn} adds them for an annotated
$\mkkeyword{spawn}$; otherwise
$e$ is treated as the body of an unannotated $\mkkeyword{spawn}$ without a
$\gamma$, since there is no $\mkkeyword{spawn}$ to return its pid: what its clients send bounds
$\beta_0$ from below.

\subsection{Derivation: two upper bounds to reconstruct the interface}\label{sec:recon-derivation}

Fix an unannotated $\mkkeyword{spawn}(e')$ of a program $e$, and call 
\emph{the spawned process}  the process that it creates (or the \emph{spawned processes}, when the $\mkkeyword{spawn}$ is evaluated several
times, as the one of the dispatcher, which creates a worker per client: they all get the same
interface). The derivation step bounds from above the interface
$t$ to be found for it. Let
$[\beta,\alpha]$ be the bracket at which \rulename{i-spawn-alt} types $e'$, and $\gamma$ the pid
variable of the $\mkkeyword{spawn}$. The type variable $\alpha$ is not the interface, but the checking step places the
interface $t$ below it, $\beta\issubt t\issubt\alpha$, so every upper bound on $\alpha$
is an upper bound on $t$. The derivation step therefore
\begin{enumerate}[noitemsep,topsep=2pt]
    \item finds the $\mkkeyword{receive}$s that the spawned process runs (Section~\ref{sec:recon-runs});
    \item extracts from them an upper bound on $\alpha$ (Section~\ref{sec:recon-demands}).
\end{enumerate}
The solving step then solves this bound, which yields the interface $t$ (Section~\ref{sec:recon-solving}).

\subsubsection{The receives that the spawned process runs}\label{sec:recon-runs}

The upper bounds on $\alpha$ come from the $\mkkeyword{receive}$s that the spawned process runs: those
syntactically in $e'$ and those of the functions that $e'$ calls, directly or through other calls. Let the
fixed $\mkkeyword{spawn}$ be that of the worker (line~\ref{spawnclosure}) without its
annotation. The worker runs the $\mkkeyword{receive}$ of \eelixir{loop}, since its body calls
\eelixir{serve}, which calls (only) \eelixir{loop} (line~\ref{loopzero}) without performing any receive. The worker does not run the
$\mkkeyword{receive}$ of \eelixir{dispatcher}, which the server runs, although its
$\mkkeyword{spawn}$ is in a branch of that $\mkkeyword{receive}$.

Which function a call reaches is not always visible in the syntax: it may be a parameter, or a
recursive function, encoded with a fixpoint combinator. However, it \emph{is} visible in the \emph{constraint
set} of $e$, the set $C$ that the generation step computes for the whole program $e$ from its
\emph{deduction}, that is, its unique derivation by the rules of Figure~\ref{fig:inference}. A $\mkkeyword{receive}$ is
typed by \rulename{i-receive} at the input decoration of its judgment: in $e'$ this is $\alpha$,
except in the body of a function or of another $\mkkeyword{spawn}$, which are typed at
decorations of their own. A call places the input decoration of the caller below that of the
function called (\rulename{i-app}). So along a chain of calls, $C$ places $\alpha$ below the
input decoration $\delta$ of each $\mkkeyword{receive}$ reached, and entails
$\alpha\issubt\delta$ even when the syntax does not show the function called.
\begin{definition}[Receives run by a process]\label{def:recon-runs}
Let $C$ be the constraint set of a program $e$ and $\alpha$ a type variable. The set
$\mathsf{rcv}_e(\alpha)$ of the \emph{receives run at} $\alpha$ is the set of the uses of
\rulename{i-receive} in the deduction of $e$ whose input decoration $\delta$ satisfies
$C\models\alpha\issubt\delta$. We write $\rcv\alpha$ when $e$ is clear.
\end{definition}

\noindent When $\alpha$ is the input decoration
of the fixed $\mkkeyword{spawn}$, which is minted for this $\mkkeyword{spawn}$ alone, it identifies
the spawned process, and $\rcv\alpha$ contains the $\mkkeyword{receive}$s of $e'$ and of the
functions that $e'$ calls. The set $\rcv\alpha$ is finite, since the
deduction is. For the worker example, the calls give
$\alpha\issubt\delta_{\mathtt{serve}}\issubt\delta_{\mathtt{loop}}$ (where $\delta_f$ denotes the
input decoration of function $f$), and $\rcv\alpha$ contains
the $\mkkeyword{receive}$ of \eelixir{loop} only.

The definition misses, however, the $\mkkeyword{receive}$s of a closure that the spawned process
receives in a message and then calls, such as a job sent to the extended \eelixir{loop} of
Section~\ref{sec:example} (line~\ref{job}): the process knows such a closure only through its
interface, which is being computed. We return to this limit in Section~\ref{sec:recon-limits}.

\subsubsection{What the spawned process reads: demands and coverage}\label{sec:recon-demands}

Each branch $j$ of a $\mkkeyword{receive}$ in $\rcv\alpha$, of pattern $p_j$, selects the
messages $q_j=\accept{p_j}\setminus\bigvee_{h<j}\accept{p_h}$, and contributes to two upper
bounds on $\alpha$, the \emph{demands} and the \emph{coverage}. Each excludes one way in which a
message can fail to be served. To illustrate them, let the fixed $\mkkeyword{spawn}$ be again
that of the worker in line~\ref{spawnclosure}. The ill-typed sends of lines~\ref{typo}--\ref{total42}, which
Elixir accepts today, send the worker two messages that $\alpha$ must \emph{not} contain:
\begin{enumerate}[noitemsep,topsep=2pt]
    \item the demands: $\alpha$ may contain only messages that the branch selecting
    them can use. This excludes \eelixir|{:get_total, 42}|, which is selected by a branch
    whose body cannot use this payload (since it expects a pid), and crashes the worker.
    This bound is forced by the typing: the constraints of the program entail it;
    \item the coverage: $\alpha$ may contain only messages that some branch selects.
    This excludes \eelixir|{:aadd, 12}|, which is selected by no branch, and stays in the
    mailbox forever. This bound is a choice: no rule emits it, and a declared interface may
    ignore it, but a reconstructed interface must not contain such a message.
\end{enumerate}
In other terms, given an unannotated $\mkkeyword{spawn}$ and the receives it runs, the demands give the bound coming from the code that uses the messages, and the coverage gives the bound coming from the receive patterns.
The demands of all the branches are therefore intersected, and their patterns joined.
\paragraph{First upper bound: what the code demands of the messages.}
Let $j$ be a branch of a $\mkkeyword{receive}$ in $\rcv\alpha$, and $\delta$ the input decoration
at which this $\mkkeyword{receive}$ is typed. Let $m_j$, the type of the messages that the branch
supports, be the pattern $p_j$ read as a type, with each capture variable replaced by the
meet of the upper bounds that the program places on its type: in the body of $j$ and, after the
$\mkkeyword{receive}$, on the value that $j$ returns. The constraint set $C$ entails
$\delta\land q_j\issubt m_j$ and, since $C\models\alpha\issubt\delta$, it entails also the \emph{demand}
$\alpha\land q_j\issubt m_j$. This is the
requirement inference of Section~\ref{sec:requirement-inference}, solved for an
interface instead of checked against an annotation. The demand is equivalent to
$\alpha\issubt m_j\lor\neg q_j$, an upper bound on $\alpha$.

Let us show it on an example: take the third branch (i.e., $j=3$) of \eelixir{loop} (line~\ref{total}), reproduced here\lucam{Commentato strani comandi perché danno errori e \LaTeX\ fallisce}\beppem{argh, no, era per avere lo stesso numero di riga della versione originale: salvo il counter, metto la linea giusta, poi riporto il counter a posto: protetto con un ifminted}
\ifminted
\edef\savedline{\the\value{FancyVerbLine}}%
\setcounter{FancyVerbLine}{\numexpr\getrefnumber{total}-1\relax}%
\fi\begin{elixircode}
   {:get_total, from} -> send(from, {:total, n}); loop(n)   # return the counter's value
\end{elixircode}
\ifminted\setcounter{FancyVerbLine}{\savedline}\fi%
This $\mkkeyword{receive}$ is in $\rcv\alpha$, with $\alpha$ identifying the spawn of the worker. The messages that the pattern $p_3=$\eelixir|{:get_total, from}| selects are $q_3$ =
\eelixir|{:get_total, term()}|. Its body sends \eelixir|{:total, n}| to \eelixir{from}.
Since \eelixir|n|  is the argument of \eelixir{loop}, then it must be an integer, and so this branch bounds the type of \eelixir{from} from above
by \eelixir|pid({:total, integer()})|.
Hence $m_3$ is
\eelixir|{:get_total, pid({:total, integer()})}| and the demand %
$\alpha\land q_3\issubt m_3$ becomes
    $\alpha\land$\eelixir|{:get_total,!\,!term()}|$\;\issubt\;$\eelixir|{:get_total,!\,!pid({:total,!\,!integer()})}|.
This demand excludes the \eelixir|{:get_total, 42}| of line~\ref{total42},
which is in $q_3$ but not in $m_3$.

The upper bounds that define $m_j$ have one exception. Consider a spawned process running an $e'$ that contains a receive that re-queues every
message it reads 
\begin{elixircode}
receive do other -> send(self(), other) end !\phantomsection\label{requeue}!
\end{elixircode}
The single branch ($j=1$) of this receive selects every message
($q_1=\Tany$), and since $\Self$ has type $\TChan\beta$ in $e'$ (\rulename{i-self}), the
$\mkkeyword{send}$ bounds the type of the capture variable
\eelixir{other} from above by $\beta$. So $m_1$ would be $\beta$, and the demand
$\alpha\land q_1\issubt m_1$ would be $\alpha\issubt\beta$. This bound is not needed for the
branch to use the message: it states that the spawned process accepts what it sends to itself, a
condition on what the spawned process writes, which the checking step verifies as it verifies the
sends of the clients (Section~\ref{sec:recon-check}). We therefore leave out of
$m_j$ the upper bounds placed by every \emph{self-send}, a $\mkkeyword{send}$ whose target may
be a pid that the spawned process obtained from $\Self$ while running $e'$ (see
Definitions~\ref{def:recon-selfsend} and~\ref{def:recon-demand} for the formal definitions of
\emph{self-send}, of \emph{demand} and, in particular, of $m_j$). For the re-queue, $m_1$ is
then $\Tany$ and the demand is trivial.

The first upper bound is thus $\alpha\issubt\dem\alpha$, where
\begin{equation}\label{eq:demands}
    \mathsf{dem}_e(\alpha)\;=\;\bigwedge\{\,m_j\lor\neg q_j\;\mid\;j\text{ a branch of a }
        \mkkeyword{receive}\text{ in }\mathsf{rcv}_e(\alpha)\,\},
\end{equation}
written $\dem\alpha$ when $e$ is clear.

\paragraph{Second upper bound: what the receives of the spawned process can consume.}
Demands alone do not suffice, since each demand constrains only the messages that its own branch
selects: $\alpha\issubt m_j\lor\neg q_j$ leaves $\alpha$ free outside $q_j$, so a message that no
branch selects, such as the mistyped \eelixir|{:aadd, 12}| of line~\ref{typo}, satisfies all the
demands. It is, however, excluded by the \emph{coverage} bound $\alpha\issubt\cov\alpha$, where
\begin{equation}\label{eq:coverage}
    \mathsf{cov}_e(\alpha)\;=\;\bigvee\{\,\accept{p_j}\;\mid\;j\text{ a branch of a }
        \mkkeyword{receive}\text{ in }\mathsf{rcv}_e(\alpha)\,\},
\end{equation}
written $\cov\alpha$ when $e$ is clear.
No rule emits the coverage bound, because it concerns \emph{all} the $\mkkeyword{receive}$s that
the spawned process runs, at once: emitted by each $\mkkeyword{receive}$, it would bound $\alpha$ by
the intersection of their patterns instead of their union. For the worker, $\cov\alpha$ is the
union of the accepted types of the patterns of \eelixir{loop}, that is, the type
\eelixir|:bump or {:add, term()} or {:get_total, term()}|, which contains no \eelixir{:connect}
message, since the worker does not run the $\mkkeyword{receive}$ of \eelixir{dispatcher}.

To sum up, $\alpha$ has two complementary upper bounds, $\dem\alpha$ and $\cov\alpha$, each
excluding a message that the other admits, and the derivation step yields the single bound
\begin{equation}\label{eq:recon-bound}
    \alpha\;\issubt\;\cov\alpha\;\land\;\dem\alpha.
\end{equation}

\subsection{Solving: from the bound(s) to the interface}\label{sec:recon-solving}

We still consider the unannotated $\mkkeyword{spawn}(e')$ of the program $e$ fixed at the
beginning of Section~\ref{sec:recon-derivation}, with its type variables $\alpha$, $\beta$ and
$\gamma$, and the constraint set $C$ of $e$. The solving step solves the bound
$\alpha\issubt\cov\alpha\land\dem\alpha$ of~\eqref{eq:recon-bound} into a ground type $t$, the
\emph{reconstructed interface} of this $\mkkeyword{spawn}$.

We want $t$ as large as possible, since, by contravariance of $\TChan\cdot$, the larger the
interface, the more senders its pid type admits. The bound~\eqref{eq:recon-bound} does not
give $t$ directly, because the types $m_j$ in $\dem\alpha$ may mention, besides $\alpha$ and
$\beta$, other type variables (from the types of capture variables, the results of calls, the
decorations of functions and closures, other unannotated $\mkkeyword{spawn}$s).
We first eliminate the type variables other than $\alpha$ and $\beta$,
which turns the right-hand side of~\eqref{eq:recon-bound} into a type $F(\alpha,\beta)$ in
which $\alpha$ and $\beta$ are the only type variables that occur. Then, since \rulename{t-spawn-alt} types $e'$ at the
coherent bracket $[t,t]$, we make $\alpha$ and $\beta$ coincide: $t$ solves the
equation $x=F(x,x)$.

To weaken the bound as much as possible, each type variable $\zeta$ other than $\alpha$ and
$\beta$ that occurs in $\dem\alpha$ takes its \emph{extremum} $\zeta^\ast$: the meet of its upper
bounds in $C$, the largest type that they allow, if $\zeta$ occurs covariantly; the join of its
lower bounds, the smallest, if it occurs contravariantly. These bounds, read after decomposing
the constraints,\footnote{Tallying rewrites constraints into equivalent ones with intersections of
pid (or arrow) types on the left and unions on the right, which decompose into constraints on their
components. For example, $\TChan{t_1}\land\TChan{t_2}\issubt\TChan{\xi}$ yields the upper bound
$\xi\issubt t_1\lor t_2$. Decomposition is formally defined in Definition~\ref{def:recon-dec}.} may mention other type variables, which are treated in the
same way: the extrema are the solution of a system of equations, which may be recursive. Computed
from bounds only, the extrema---like the types $m_j$---miss what a constraint says of several types
at once; the bound may then
admit messages that the process cannot handle, and the checking step rejects such an interface
(see the third category of limits in Section~\ref{sec:recon-limits}).

A type variable that occurs in $\dem\alpha$ with both polarities has no extremum, since the value
that weakens one occurrence strengthens the other. This is the case of the \emph{reply server}, a
process with the branch \eelixir|{:run, from, job} -> send(from, job.())|. The type $u$ of the result
of the application \eelixir|job.()| occurs covariantly as the codomain of the arrow demanded of \eelixir{job},
but contravariantly inside the pid type demanded of \eelixir{from}, which must accept that result.
Indeed no most permissive interface exists: for every
type $u$ the process can serve \eelixir|{:run, pid(!$u$!), (-> !$u$!)}| but, by the rules of
Figure~\ref{fig:typing}, not the union of two such types. The solving step therefore checks the
\emph{polarity hypothesis}; when it does not hold, reconstruction fails and the interface must be declared:
\begin{enumerate}[label=(PH),ref=(PH),noitemsep,topsep=2pt,leftmargin=5em,labelindent=2em,labelsep=0.8em,rightmargin=3em]
    \item\label{h:polarity} every type variable other than $\alpha$ and $\beta$ that occurs in
    $\dem\alpha$, or in the bounds from which the extrema are computed, has a single polarity.
\end{enumerate}

One bound is missing only because it is not known yet. Writers bound the pid variable of another
unannotated $\mkkeyword{spawn}$ only from below (Section~\ref{sec:recon-bracket}); what bounds it
from above is the interface of that $\mkkeyword{spawn}$. Its extremum would be $\Tany$: with
\eelixir{w} the pid of such a $\mkkeyword{spawn}$, a process with the branch
\eelixir|{:fwd, x} -> send(w, {:val, x})| would accept any payload. The solving step therefore
computes the interface of that $\mkkeyword{spawn}$ first and treats it as declared at this type:
its pid variable is replaced by this type and the constraints of \rulename{i-spawn} are added. In
general, a $\mkkeyword{spawn}$ \emph{depends} on another unannotated $\mkkeyword{spawn}$ when a type
variable minted for the latter occurs in the constraints from which the demands of the process
created by the former are computed. The solving step takes the $\mkkeyword{spawn}$s in dependency
order and fails on a cycle, as between two processes that each forward to the other: one of the
two interfaces must then be declared.

Like $\mathsf{rcv}$, $\mathsf{dem}$ and $\mathsf{cov}$, the resulting type $F$ is indexed by the
program $e$ and by $\alpha$, since the extrema depend on the rest of the program: on the other
runners of the functions that the process runs, and on the interfaces of the $\mkkeyword{spawn}$s
on which it depends. Definitions~\ref{def:recon-cstar} to~\ref{def:recon-elim} give the constraints, the polarities and the
extrema formally.
\begin{definition}[The type $F$]\label{def:recon-bound}
Assume \ref{h:polarity}, let $\alpha$ and $\beta$ be the input and output
decorations of an unannotated $\mkkeyword{spawn}$ of a program $e$, the $\mkkeyword{spawn}$s on
which it depends being treated as declared at their interfaces, and let
$\mathsf{dem}^\ast_e(\alpha)$ be $\mathsf{dem}_e(\alpha)$ in which every type variable $\zeta$ other
than $\alpha$ and $\beta$ is replaced by its extremum $\zeta^\ast$. The type $F_{e,\alpha}(a,b)$ is
\begin{equation}\label{eq:recon-req}
    F_{e,\alpha}(a,b)\;=\;\bigl(\mathsf{cov}_e(\alpha)\;\land\;
        \mathsf{dem}^\ast_e(\alpha)\bigr)[\alpha:=a,\,\beta:=b].
\end{equation}
We write $F(a,b)$ when $e$ and $\alpha$ are clear.
\end{definition}

The equation $x=F(x,x)$ has a unique solution, because $\alpha$ and
$\beta$ occur in $F(\alpha,\beta)$ only below a type constructor. This is where the exclusion
of self-sends matters: the re-queue of Section~\ref{sec:recon-demands} (line~\ref{requeue}), would contribute the
demand $\alpha\issubt\beta$, an unguarded $\beta$, and the equation would in general
have infinitely many solutions: all types, for a process whose only receive is the re-queue in line~\ref{requeue}.
The interface must satisfy $t\subt F(t,t)$, and no type that satisfies it is strictly above this
solution (Proposition~\ref{prop:recon-canonical}).
\begin{definition}[Reconstructed interface]\label{def:recon-interface}
Under \ref{h:polarity}, the interface reconstructed for the unannotated
$\mkkeyword{spawn}$ of $e$ of input decoration $\alpha$ is the unique solution of the equation
$x=F_{e,\alpha}(x,x)$.
\end{definition}

\noindent
When $F(\alpha,\beta)$ depends neither on $\alpha$ nor on $\beta$, which is the case when the
process runs no function received in a message and sends to other processes neither its own
pid nor any part of a message it receives, the requirement is, self-sends aside, uniform, and $t$ is $F(\alpha,\beta)$ itself. For the
worker of Section~\ref{sec:example} it is
    \eelixir|:bump or {:add, integer()} or {:get_total, pid({:total, integer()})}|,
\noindent precisely the \eelixir{worker} interface a programmer would have declared.

A higher-order branch makes the requirement non-uniform and the equation recursive. Let the
worker run the extended \eelixir{loop} of Section~\ref{sec:example}, whose branch
\eelixir|{:run, job}| (line~\ref{job}) applies the function it receives, at the bracket
$[\beta_\ell,\alpha_\ell]$ of the body of \eelixir{loop}. Then
$m_{\mathtt{run}}=\Tuple{\Atom[{:}run],\,\Tuple{}\to_{\beta_\ell}^{\alpha_\ell}\gamma'}$, with
$\gamma'$ the result of the call. The type variable $\gamma'$ is covariant and has no upper bound,
so its extremum is $\Tany$. The calls leading to \eelixir{loop} bound $\alpha_\ell$, which is
contravariant, from below by $\alpha$, and $\beta_\ell$, which is covariant, from above by
$\beta$, so their extrema are $\alpha$ and $\beta$, since no other process runs
\eelixir{loop}. The interface is therefore the solution of
\begin{equation}\label{eq:recon-run}
    t\;=\;\Atom[{:}bump]\;\lor\;\Tuple{\Atom[{:}add],\TInt}
               \;\lor\;\Tuple{\Atom[{:}get\_total],\TChan{\Tuple{\Atom[{:}total],\TInt}}}
               \;\lor\;\Tuple{\Atom[{:}run],\;\Tuple{}\to_{t}^{t}\Tany},
\end{equation}
an arrow decorated at the interface's own bracket, since a job is run by the very process whose
interface this is.

\subsection{Checking: replacing the new rule}\label{sec:recon-check}

The solving step gives every unannotated $\mkkeyword{spawn}$ of the program $e$ its
reconstructed interface. We number these $\mkkeyword{spawn}$s: the $k$-th ($k\geq1$) is
$\mkkeyword{spawn}(e_k)$, $\alpha_k,\beta_k,\gamma_k$ are its variables and $t_k$ is its
reconstructed interface. When the interface of the main process is not declared, the index $0$
stands for this process: $[\beta_0,\alpha_0]$ is its root bracket, $t_0$ is its reconstructed
interface, and there is no $\gamma_0$.

The checking step replaces \rulename{i-spawn-alt} by \rulename{i-spawn} at these interfaces, and
checks the resulting program as if they had been declared. We write $e[\vec{t}\,]$ for this
program: $e$ in which each $\mkkeyword{spawn}(e_k)$ is replaced by $\Spawn{e_k}{t_k}$, typed
by \rulename{i-spawn} with $t=t_k$ in place of $t=\decorate{\ta}$, which the rule allows
(see Footnote~\ref{fn:spawn-variants} in Section~\ref{sec:inference-rules}).

No new deduction is needed: the two rules have the same premise, so the deduction of
$e[\vec{t}\,]$ is that of $e$ with each use of \rulename{i-spawn-alt} replaced by a use of
\rulename{i-spawn}, and its constraint set is the one displayed in item~4 of
Definition~\ref{def:reconstruction}. This is
where what the derivation and solving steps left aside is verified: the sends aimed at each process, the closures
shipped to it and its own self-sends must deposit in its mailbox only messages of type $t_k$,
exactly as \rulename{t-thread} requires. A message that does not conform to $t_k$ makes this
step fail at the $\mkkeyword{send}$ of that message, which is what the compiler reports.
\begin{definition}[Reconstruction]\label{def:reconstruction}
Let $e$ be a program. The \emph{reconstruction} of $e$ consists of four steps:
\begin{enumerate}[noitemsep,topsep=2pt]
    \item \emph{(generation)} the deduction of $e$ yields its constraint set $C$;
    \item \emph{(derivation)} for each $k\geq0$, the bound~\eqref{eq:recon-bound} at $\alpha_k$ is
    derived from $C$;
    \item \emph{(solving)} for each $k\geq0$, in dependency order, the interface $t_k$ is
    computed as Definition~\ref{def:recon-interface} prescribes;
    \item \emph{(checking)} tallying is run on the constraint set
    \[
        C[\gamma_k:=t_k]_{k\geq1}\;\cup\;\{\,\beta_k\issubt t_k,\ t_k\issubt\alpha_k\;\mid\;k\geq0\,\}.
    \]
\end{enumerate}
If the interface of $e$ is declared, use $k\geq1$ instead of $k\geq0$: the constraints on $t_0$ are
already in $C$. The program $e$ is \emph{accepted by the reconstruction} when the solving step
succeeds for every $k$ (i.e, \ref{h:polarity} holds and the dependencies have no
cycle), and the checking step returns a solution.
\end{definition}

\begin{restatable}[Computability of the reconstruction]{proposition}{ReconComputableProp}\label{prop:recon-computable}
For every program $e$, whether $e$ is accepted by the reconstruction is decidable, and its
interfaces $t_k$ are computed when it is.
\end{restatable}

\subsection{Guarantees}\label{sec:recon-guarantees}

The reconstruction gives each unannotated $\mkkeyword{spawn}$ of an accepted program an interface
$t_k$. Two questions remain: whether the program is safe, which the checking step straightforwardly settles, and
how permissive these interfaces are, which the solving step settles.

\begin{restatable}[Soundness of the reconstruction]{proposition}{ReconSoundnessProp}\label{prop:recon-sound}
Let $e$ be accepted by the reconstruction, with the interface $t_0$ for $e$ itself
and $t_k$ for its $k$-th unannotated $\mkkeyword{spawn}$. Then inference accepts
$e[\vec{t}\,]$ at the interface
$t_0$, so Corollary~\ref{cor:inferred-safety} applies to $\typeerase{e}$, which is
$\typeerase{e[\vec{t}\,]}$. Moreover, in every configuration reachable from the
initial one, a process created by the $k$-th unannotated $\mkkeyword{spawn}$ has
interface $t_k$, and every message in its mailbox has type $t_k$.
\end{restatable}

\noindent
Acceptance yields a typing derivation in which \rulename{t-spawn-alt} is instantiated at the
reconstructed types. Reading an unannotated $\mkkeyword{spawn}$ at run time as
$\Spawn{e_k}{t_k}$ is legitimate by Theorem~\ref{thm:typerase}, and how $t_k$ was computed
decides how permissive the interface is, not whether it is sound.

To measure how permissive it is, we compare the interface with the types that respect the
bound solved by the solving step, however these types were obtained.
\begin{definition}[Servable type]\label{def:recon-servable}
Assume \ref{h:polarity} and let $\alpha$ be the input decoration of an unannotated
$\mkkeyword{spawn}$ of a program $e$. A ground type $s$ is \emph{servable} when
$s\subt F_{e,\alpha}(s,s)$.
\end{definition}

\noindent
That is, a type $s$ is servable if it respects the bound that the solving step yields: this over-approximates the requirements of the code run by the process that the $\mkkeyword{spawn}$ with input decoration $\alpha$ creates, ignoring the messages sent to this process and accepting only messages that this code can consume.

\begin{restatable}[Canonical interface]{proposition}{ReconCanonicalProp}\label{prop:recon-canonical}
Assume \ref{h:polarity} and let $t$ be the interface reconstructed for an unannotated
$\mkkeyword{spawn}$ of input decoration $\alpha$ of a program $e$.
\begin{enumerate}[noitemsep,topsep=2pt]
    \item $t$ is servable, and no servable type is strictly above it: it is maximal.
    \item Let $s\subt\cov\alpha$. If inference accepts $e[\vec{t}\,]$ with $s$ as the interface of
    this $\mkkeyword{spawn}$, then $s$ is servable.
    \item If $F(\alpha,\beta)$ does not depend on $\alpha$, then $t$ is also the greatest servable
    type; if it does not depend on $\beta$ either, then $t$ is $F(\alpha,\beta)$ itself.
\end{enumerate}
\end{restatable}

\noindent
Items 1 and 2 together say that no annotation below the coverage at which the program stays
typable is strictly above $t$: a declaration below the coverage cannot enlarge the reconstructed
interface. Without the hypothesis of item 3, nothing stronger than maximality holds: the
extended \eelixir{loop} of equation~\eqref{eq:recon-run} has no greatest interface. Writing $u_0$
for the union of the first three summands of that equation, every type
$s_u=u\lor\Tuple{\Atom[{:}run],\Tuple{}\to_{s_u}^{s_u}\Tany}$ with $u\subt u_0$ is
servable and maximal, and the reconstructed interface is $s_{u_0}$, the canonical choice among
these pairwise incomparable types.

\subsection{Limits}\label{sec:recon-limits}

Proposition~\ref{prop:recon-sound} has no converse: a program that is typable with
\rulename{t-spawn-alt} may be rejected by the reconstruction. The main cases for such a program can be classified in four categories.
In the first three the checking step rejects the program. Since the $\mkkeyword{receive}$s of a
process determine its interface and the sends to it are only checked, the checking step rejects a
sender in the first two categories, where the program needs messages that the reconstructed
interface does not contain, and a receiver in the third, where the interface computed from the
$\mkkeyword{receive}$s is too large. In the fourth the solving step fails. Let us examine in detail the four cases in which a program typable with
\rulename{t-spawn-alt} may be rejected by the reconstruction.

First, the program is typable only with an interface that is not below the coverage, whereas
the reconstructed interface always is. The interface that such a program requires contains messages that
no $\mkkeyword{receive}$ handles, or that are handled only by a $\mkkeyword{receive}$ that
coverage does not reach. A body that first sends
itself a message with \eelixir|send(self(), :later)| and then only runs
\eelixir|receive do :go -> :ok end| is well typed under the annotation
\eelixir{:go or :later} (even if the \eelixir{:later} message remains in the mailbox),
whereas the reconstruction computes $\Atom[{:}go]$ and the checking step rejects the self-send.
Likewise, coverage does not reach a $\mkkeyword{receive}$ performed by a closure that arrives in
a message (which could thus consume the previous \eelixir{:later} message left in the mailbox), and the checking step rejects the send of a message that only this closure handles.

Second, the program is typable only with an interface that is below the coverage but
incomparable with the reconstructed one: the reconstruction computes a single interface among
the maximal ones of Proposition~\ref{prop:recon-canonical}, whereas a programmer may declare
another one. A program that sends the extended \eelixir{loop} a job that can run at some $s_u$ as defined right after Proposition~\ref{prop:recon-canonical}, but not at
the reconstructed interface is rejected, whereas it may be typable when the
$\mkkeyword{spawn}$ is annotated with $s_u$.

Third, the program is typable but the reconstructed interface is too large: it contains
messages that the process cannot handle, so the checking step rejects the code of the process,
not a send to it. The reason is that $F$ retains only part of the constraints: the demands of the
$\mkkeyword{receive}$s in $\rcv\alpha$, in which every other type variable is taken at its
extremum, computed from bounds only (Section~\ref{sec:recon-solving}). The checking step uses
every constraint of $C$, and a constraint that $F$ does not retain may exclude messages that $F$
admits. This happens in three ways.
\begin{itemize}[noitemsep,topsep=2pt]
    \item \emph{A constraint says something of several types at once}, which the extrema miss.
    The $\mkkeyword{receive}$s of a closure that the process sends constrain the input decoration
    of the closure together with what they read, but do not bound it, so this decoration is taken
    at its extremum. For instance, a process with the branch\par
    \centerline{\eelixir|{:job, p} -> send(p, fn -> receive do {:int, n} -> n + 1 end end)|}\par
    \noindent gets an interface in which \eelixir{p} is the pid of a process that accepts closures
    able to run in any process, whereas this closure can run only in a process whose
    \eelixir{:int} messages carry integers. Likewise, the branch \eelixir|{:to, p, x} -> send(p, x)|
    relates the types of two capture variables: the demand cannot express that \eelixir{p} accepts
    \eelixir{x}, and the interface accepts any pid as \eelixir{p}.
\item \emph{A decoration in the demand has no bound.} A type variable that no bound constrains is taken at the extremum that weakens the demand, $\Tany$ or $\Tempty$ according to its polarity. In one case these are the decorations of an arrow of a declared interface: the annotation does not write them and nothing bounds them in the constraints from which the extrema are computed. For instance, if \eelixir{runner} is declared at \eelixir|{:run, (-> term())}| and the spawned process has the branch\par
\centerline{\eelixir|{:fwd, job} -> send(runner, {:run, job})|}\par
\noindent the demand on \eelixir{job} is that arrow, whose decorations take the extrema that make it the largest arrow: the interface accepts any closure as \eelixir{job}, whereas \eelixir{runner} accepts only the closures that can run in \eelixir{runner}. In the other case a bound is lost: the sends to its runner of a closure that the process both sends and runs are self-sends, left out of the demands, so the output decoration of the closure is taken at $\Tempty$.
    \item \emph{The interface has a bound that is not a demand.} When the process calls a function
    whose written input decoration (Section~\ref{sec:elixir-decorations}) is smaller than what its
    branches demand, this decoration bounds the interface, but $F$ ignores it
    (Remark~\ref{rem:recon-written}).
\end{itemize}

Fourth, the solving step fails: the polarity hypothesis \ref{h:polarity} does not hold, as for
the reply server, or the dependencies form a cycle (Section~\ref{sec:recon-solving}).

In all these cases, the interface has to be declared.

\behavior{%
\input{behavioral-types}
}

\section{Integration into Elixir}
\label{sec:elixir}
Sections~\ref{sec:typing} to~\ref{sec:reconstruction} type a closed program,
written in a core calculus, whose every process is typed. An Elixir program is
type-checked in units, and it runs beside code that our system never sees:
libraries, other languages of the BEAM, the virtual machine. This section discusses
what an interface annotation means outside its type-checking unit, and where decorations must
then be written (Section~\ref{sec:elixir-decorations}); untyped senders and
Elixir's gradual type \eelixir{dynamic()} (Section~\ref{sec:elixir-dynamic}); and the
library loops and virtual-machine messages that own or fill a mailbox
(Section~\ref{sec:elixir-ecosystem}). This is a design discussion: nothing in it is proved.
Section~\ref{sec:elixir-tallying} ends the section with how our prototype modifies the tallying algorithm implemented in Elixir, to efficiently solve the inference
constraints, and with what inference and reconstruction cost.

\subsection{Decorations at the unit boundary}
\label{sec:elixir-decorations}

\paragraph{The type-checking unit.}
Inference (Section~\ref{sec:ann-inference}) collects the constraints of a whole
program in a single set $C$, solved once, at the end, by $\textsf{Solutions}(C)$.
An Elixir compiler applies inference to one part of the program at a time, which we
call a \emph{type-checking unit}: the code whose constraints are collected in the
same set and solved by the same call of $\textsf{Solutions}(\cdot)$. What the unit
is (a module, a file, a set of modules that depend on each other) is for the
compiler to decide; we only assume that it is \emph{bounded}, that is, not the
whole program. Inference leaves free the decorations of the arrows written in an
interface annotation: it supplies them as type
variables, which only the code of the unit where the annotation occurs
constrains. Within a unit, free decorations are right, since inference chooses them
at least as well as the programmer could.
Across units they are not.

\paragraph{Why a crossing name must be closed.}
Suppose a client in another unit holds a pid \eelixir{w} of type
\eelixir{pid(Server.worker)} and ships the job of line~\ref{tick},

\centerline{\eelixir|send(w, {:run, fn -> send(self(), :tick) end})|}

\noindent which must be rejected, the worker not handling \eelixir{:tick}. The
client's own occurrence of \eelixir{worker} decorates the arrow type of \eelixir{:run} payload
$\Tuple{}\to_{\beta}^{\alpha}\Tany$ with a fresh pair, and the client constrains
$\beta$ only by $\Atom[{:}tick]\subt\beta$, which $\beta=\Atom[{:}tick]$
satisfies; symmetrically, the job

\centerline{\eelixir|send(w, {:run, fn -> receive do x -> x + 1 end end})|}

\noindent only demands $\alpha\subt\TInt$, which $\alpha=\Tempty$ satisfies.
Both jobs are caught when client, dispatcher and worker are solved together,
since $\alpha$ and $\beta$ are then tied to the worker's interface through the
\eelixir|{:ready, w}| of the dispatcher and the \eelixir{job.()} of
\eelixir{loop}; a client in another unit has no such chain. What an annotation
leaves open is exactly the mailbox contract of the closures that travel.

\paragraph{Default closing.}
The compiler knows where its units end, so it can close such names itself, by
assuming that a function that travels in a message is run by the process that
receives it. Write $t_N$ for the interface annotation named $N$ with every arrow
decorated at $[t,t]$, where $t$ is the interface of that process: $t=t_N$, so that
only a process of interface $t_N$ can run the arrow, except for an arrow under a pid
type of $N$, for which $t$ is the type in the innermost such pid, closed in the same
way. For the worker,
$t_{\mathtt{worker}}=\cdots\lor\Tuple{\Atom[{:}run],\,\Tuple{}\to_{t_{\mathtt{worker}}}^{t_{\mathtt{worker}}}\Tany}$,
a regular type, contractive since its recursive occurrences sit in the
decorations of an arrow. The \emph{default closing} is the rule: a name resolved
outside the unit that defines it denotes $t_N$; inside, it keeps the free pair of
Definition~\ref{def:decorate}; and the $\mkkeyword{spawn}$ annotated with an exported
$N$ is checked against $t_N$, as Section~\ref{sec:reconstruction} checks a program
against a reconstructed interface. A client in another unit then rejects the two
jobs above and accepts that of line~\ref{self3}, as when the whole program is
checked. Exporting instead the decorations that inference finds in the unit would
make the interface change with every edit of the unit, whereas an interface
annotation is a contract.

\paragraph{When the default is wrong.}
A function run by the process that receives it is a common case, but not the only one. Let \eelixir{runner} name the interface
\eelixir|{:run, (-> term())}|, and let a \eelixir{proxy} accept \eelixir{:stop}
and \emph{forward} the jobs it receives to a runner. Under the default closing
these jobs are decorated at the proxy's bracket, and the forwarding
$\mkkeyword{send}$, which needs them at the runner's, requires
$t_{\mathtt{proxy}}=t_{\mathtt{runner}}$ (the input decoration is
contravariant, the output one covariant): impossible, since only the proxy
accepts \eelixir{:stop}. So the programmer must provide the decorations of arrow of the job. For that we add to Elixir types the syntax to add decorations to an arrow type. We write \eelixir{(!$s$! -> !$t$!)!@![!$o$!,!$i$!]} where $o$ and $i$ are the output and input decorations, respectively, and use \eelixir{(!$s$! -> !$t$!)!@$u$!} as a shorthand for \eelixir{(!$s$! -> !$t$!)!@![!$u$!,!$u$!]}. In our example, the proxy declares
\eelixir|:stop or {:run, (-> term())!@!runner}|, that is,
$t_{\mathtt{proxy}}=\Atom[{:}stop]\lor\Tuple{\Atom[{:}run],\,\Tuple{}\to_{t_{\mathtt{runner}}}^{t_{\mathtt{runner}}}\Tany}$,
and the default closing of the worker is the declaration
\begin{elixircode}
!\color{anncol}\$type! worker = :bump or {:add, integer()} or {:get_total, pid({:total, integer()})}
                or {:run, (-> term())!@!worker}
\end{elixircode}
so the suffix writes what the compiler does on its own. Qualifiers also tell apart
two equal arrow annotations, which Definition~\ref{def:decorate} decorates at the
same bracket. The inference of Section~\ref{sec:ann-inference} must be extended to deal with annotations that can be partially decorated: this is straightforward (cf.\ Footnote~\ref{fn:spawn-variants}) and our prototype already implements this.

\paragraph{Foreign modules.}
Both closings need an annotation or a body to derive from. A foreign module
(a precompiled \eelixir{.beam} file, an Erlang module, untyped code) supplies
neither, and leaving its decorations unwritten is worse than a rejection.
\eelixir{Task.async/1} runs the closure it is given in a process whose interface
the library fixes, but the arrow of the closure reaches the importing unit with
free decorations: the call typechecks at decorations chosen to suit the caller,
against code never checked at them. Such types must carry their decorations,
and with no body to infer them from, writing them loses nothing
(Section~\ref{sec:requirement-inference}). There are three forms, in decreasing
order of what is known: \eelixir{!@!}$\,N$, when the runner's interface is
named; \eelixir{(s -> t) !@! [o, i]}, when the bracket is known, in particular the
least arrows \eelixir{!@! [none(), term()]} for a first-order function that
touches no mailbox (\eelixir{Kernel.+/2}, \eelixir{String.length/1}); and
\eelixir{dynamic()} (Section~\ref{sec:elixir-dynamic}) in the decorations, when
nothing is known: a decoration variable would be solved, that is, some value picked
for it, whereas \eelixir{dynamic()} states that what the function reads from and
writes to its runner's mailbox is unknown. Section~\ref{sec:typing} already types
decorated arrows. \eelixir{Enum.map/2} does not get the least arrows, since it
runs its closure in the caller's process, and qualifying the arrow of the closure,
although sound, rejects the callbacks that pass their own pid on for a reply, as
does \eelixir|fn w -> send(w, {:job, self()}) end|.

\paragraph{Polymorphic decorations.}
Decorations are monomorphic: inference mints them once per program occurrence
and solves them once, so all the callers of a function share its interval
$[o,i]$ of admissible runners, and a library function such as
\eelixir{Enum.map/2}, which runs its argument in the caller's process, has no
decoration valid for every caller. When parametric
polymorphism is added to Elixir's type system, decorations should be quantified
together with the other type variables and instantiated at each call, as in
effect systems~\cite{TJ92}: \eelixir{Enum.map/2} would get
$\forall\gamma,\delta,\alpha,\beta.\,\Tuple{\mathtt{list}(\gamma),\,\gamma\to^{\alpha}_{\beta}\delta}\to^{\alpha}_{\beta}\mathtt{list}(\delta)$,
whose instances follow the bracket of each caller. Decorations being types, they
would be generalized and instantiated by polymorphic inference for set-theoretic
types~\cite{castagna2024inf}, and still never written. The soundness argument of
Section~\ref{sec:properties} would have to be revisited, since that inference
instantiates a function with a set of substitutions. This should be harmless for
\eelixir{Enum.map/2}, whose instances all run in the caller's process, but not for
a variable in the interface of a spawned process, since the process has one mailbox
for all the instances: a function that spawns its argument, instantiated at the two
input decorations of~\eqref{eq:mixing2} and applied to the body of the process that
refutes this law, would create that very process, breaking subject reduction without
using the law. The point is that process add effects to the pure functional core of Elixir. It is the same problem as with the type of \texttt{ref\,[]} in ML, where all the
instances share one cell, but the value restriction does not solve it: a function
that spawns is a value, like \texttt{fun () -> ref\,[]}, and generalizing it is sound
in ML only because each call chooses one instance, whereas a set of substitutions
gives the call the intersection of all of them. Such a variable must therefore be
instantiated once for each spawn.

\subsection{Interoperating with untyped code}
\label{sec:elixir-dynamic}

Every interface, declared or reconstructed, bounds a mailbox only because every
$\mkkeyword{send}$ of the program is typed against the pid type of its
recipient: the mailbox premise of \rulename{t-thread}, on which
Corollary~\ref{cor:soundness} rests. A single untyped sender defeats it. There
are two answers: a defensive one, available as the system stands, and a gradual
one, which is what Elixir does today.

\paragraph{Defensive typing.}
\label{sec:elixir-defensive}
If the messages in a mailbox cannot be trusted, give the process the
annotation \kwinterface~\eelixir|term()|. At $i=\Tany$, \rulename{t-receive} types branch $j$
at $\accept{p_j}\setminus\bigvee_{h<j}\accept{p_h}$: a branch assumes of its
message only what its pattern tested. The exported $\TChan\Tany$ is the
\emph{smallest} pid type, so every $\mkkeyword{send}$ to the process is
admitted, and the mailbox premise of \rulename{t-thread},
$\wte[\Tany][\Tany]{\emptyset}{v_k}{\Tany}$, holds for every value whose
abstractions are well typed: by Corollary~\ref{cor:soundness}, the process never
gets stuck and never uses a value at a type it does not have, whoever filled its
mailbox with such values. The check becomes a discipline of
defensive programming: every use a branch makes of its message must be justified
by its pattern, as in the \eelixir{loop} of Section~\ref{sec:example}, whose
\eelixir{k} is now bound at $\Tany$:

\begin{center}
    \eelixir|{:add, k} when is_integer(k) -> loop(n + k)|
\end{center}

\noindent A body is defensive exactly when its requirement
(Section~\ref{sec:requirement-inference}) is satisfied by $\Tany$, that is, when
it demands nothing of the messages it reads.

The limit of defensive typing is that a test recovers data, never capabilities: patterns establish only the
\emph{test types} of Section~\ref{sec:syntax}, where every arrow is $\TFun$ and
every pid type $\TChan\Tempty$, so a received pid sharpens at best to a pid to
which nothing can be sent, and a received closure to a function that can be
applied to nothing. The reply branch
\eelixir|{:get_total, from} -> send(from, {:total, n})| (line~\ref{total}) and
\eelixir|{:run, job} -> job.()| (line~\ref{job}) cannot be typed at
\eelixir{term()}, however rewritten. Yet these are the shapes of every request carrying
its reply address and of every closure shipped for execution, that is, of
\eelixir{GenServer.call/2} and \eelixir{Task.async/1}. Typing them means refining
a binding by its use, as requirement inference does, which is sound only
because senders are held to the interface. Untyped senders are held to nothing,
hence the second answer.

\paragraph{Gradual typing.}
\label{sec:elixir-gradual}
The gradual answer adds Elixir's \eelixir{dynamic()}, written $\TDyn$, to the
types, and extends the encoding $\enc\cdot$ of Section~\ref{sec:fcb-encoding}
homomorphically, so that its target becomes the gradual types of
Elixir~\cite{CD26,castagna2019gradual,lanvin2021semantic}. Since $\enc\cdot$
preserves polarities, it commutes with the minimal and maximal materializations
that characterize gradual subtyping, precision, and consistent
subtyping~\cite{lanvin2021semantic}. Thus the three relations on our types are, like
subtyping, those of Elixir on the encoded types. No result of
Sections~\ref{sec:syntax}--\ref{sec:reconstruction} is claimed for this
unformalized extension, and our statements about run time rest on the safe
erasure of Elixir~\cite{CD26}, not re-proved with mailboxes. It gives two
things.

First, \kwinterface~\eelixir|dynamic()| accepts every receive that
\eelixir{term()} refuses. Since $\TDyn\land\accept{p_j}$ materializes to any
subtype of $\accept{p_j}$, the \eelixir{k} of \eelixir|{:add, k}| may be
taken at \eelixir{integer()} and \eelixir|loop(n + k)| typechecks as written, as
Elixir does today. A received pid arrives at $\TDyn$ as well, and materializing
it to $\TChan{t}$ lets the reply branch typecheck; we return below to the
soundness of both steps. The two declarations agree on defensive
branches, and elsewhere \eelixir{term()} rejects where \eelixir{dynamic()}
defers. The price to pay is that the types deduced by gradual typing may be less precise (e.g. \eelixir{dynamic()} instead of \eelixir{integer()} for \eelixir|k|) than with defensive programming and, thus, they capture fewer errors.

Second, \eelixir{pid(dynamic())} types a pid whose mailbox is unknown, making
the system incrementally deployable: such a pid is otherwise \eelixir{pid()},
that is, $\TChan\Tempty$, to which every send is rejected, and
Section~\ref{sec:reconstruction}, which needs the spawned body, offers no middle
ground. A send to it is admitted, with the checks, if any, that the gradual
discipline of~\cite{CD26} decides.

No amount of \eelixir{dynamic()} checks the interface itself. Gradual typing is
sound on the BEAM~\cite{CD26} because every materialization is backed by a check
that the virtual machine performs anyway, and because a function applied outside
its domain either fails at such a check or returns a result typed $\TDyn$.
Processes break both conditions. A $\mkkeyword{send}$ is not checked: if the
\eelixir|{:add, k}| branch reported \eelixir{k} to a process of interface
\eelixir|{:sq, integer()}| before adding it, a message
\eelixir|{:add, "x"}| from an untyped sender would place
\eelixir|{:sq, "x"}| in the mailbox of that process, whose code and interface do
not mention $\TDyn$, before any check on \eelixir{"x"}. And capabilities travel
inside payloads: materializing a received pid, or sending to a pid of type
\eelixir{pid(dynamic())}, chooses the interface of another process, which nothing
at run time records. Which materializations remain sound (presumably those of
data, whose uses the virtual machine checks, and not those of capabilities), and
what this does to the mailbox invariant and to requirement inference once the
input decoration of a spawn may be instantiated by a gradual type, is left for
future work.

\subsection{Library loops and the virtual machine}
\label{sec:elixir-ecosystem}

\paragraph{Library loops.}
A process is usually written as callbacks for a library loop (an OTP
\emph{behaviour} such as \eelixir{GenServer}) that owns the
$\mkkeyword{receive}$, dispatches each message to a callback and sends the
reply. The interface is the union of the messages the callbacks handle (plus a few the
loop handles), and a request carries its reply address, the higher-order pattern
of this paper and the one that defensive typing cannot type. The interface of such a loop cannot be
reconstructed: its single branch sends to the reply address the result of an
overloaded callback, a constraint between the types of two capture variables that
no demand expresses (Section~\ref{sec:recon-limits}). It
must therefore be declared, but it need not be written by hand: \citet{Cas26}
computes it from the intersection type of
the callback and types the loop at it for each module that adopts the behaviour, by
the typing per summand of Section~\ref{sec:typing-rules}, which we do not adopt.
Typing the loop once and for all remains to be done.

\paragraph{Messages from the virtual machine.}
A process that \emph{monitors} or is \emph{linked} to another learns of its
death by a message, \eelixir|{:DOWN, ...}| or \eelixir|{:EXIT, ...}|, that the
virtual machine deposits and no $\mkkeyword{send}$ produces. Such messages must
enter the interface at which receives are typed (declared, added by the compiler
at each monitor or link, or reconstructed from the handling clause), but must not be
demanded of senders. So the type system cannot check that a declared interface
lists them: if it omits \eelixir|{:DOWN, ...}|, the handling clause binds its
variables at the empty type and is silently skipped
(Section~\ref{sec:typing-rules}). The interface thus plays two roles that would
be better separated, what receives rely on and what senders are held to; we
leave this open. 
\behavior{%
    The behavioral layer of Section~\ref{sec:behavioral-types}
must also exempt these clauses from the balance of sends and receives.}

\subsection{Solving the constraints: performance of inference and reconstruction}
\label{sec:elixir-tallying}

The prototype mentioned in the introduction relies on the API of
the Elixir compiler for the type representation and the subtyping checks, and
on the algorithm in the development version of Elixir for tallying. The latter
follows the three phases of \citet{polyduce2}, normalize, saturate,
and solve, implemented as detailed by \citet{LN25}. With tallying unmodified, inference did not finish within two minutes on the program of
Section~\ref{sec:example} without any decoration. We thus made three changes to tallying, none of which changes the set of accepted
programs, and the prototype now infers every decoration of that program in less than a
second. We measured the impact of each change by removing it from the final implementation.
\begin{enumerate}[label=\textbf{\arabic*.},leftmargin=*,itemsep=2pt,topsep=2pt]
\item \textbf{One system of equations.} The solving phase turns the saturated set into
one equation per variable. The standard tallying implementation solved them one after the other,
substituting each recursive solution into the remaining ones, and the subtyping checks
that follow blew up on the resulting nest of recursive types. Since normalization
ensures that every cycle of the equations goes through a type constructor, we solve
them at once, as one system of mutually recursive types. This change is decisive:
without it, inference lags, with or without the other two optimizations.

\item \textbf{Elimination of bound variables.} Most variables that the rules
introduce, such as those that \rulename{pe-tuple} gives to the components of
tuple patterns, are constrained in one of the two directions only by bare bounds
$\gamma\issubt u$ or $l\issubt\gamma$. Before tallying, such a variable is
replaced by the intersection of its upper bounds (the union of its lower bounds)
when the other constraints in which it occurs can only become easier to satisfy
as it grows (as it shrinks), which preserves satisfiability
(Proposition~\ref{prop:elimination}). This drastically reduces the time (in our
experiments, up to a factor of more than a hundred), and makes it
stable: without it, the time for the same program varies with the internal names
of the type variables.

\item \textbf{One solution.} Tallying computed every solution, whereas typability only
needs $C$ to be satisfiable (Proposition~\ref{prop:inference-terminates}). The choices
of the saturation phase are now explored depth first, with backtracking, and the search
stops at the first saturated set, so it fails only if $C$ is unsatisfiable; the
solution need not be principal, which Corollary~\ref{cor:inferred-safety} does not
require. This saves a factor of about three, but only when $C$ has several principal
solutions.
\end{enumerate}
Rejecting a program costs more, since locating its errors solves many prefixes of $C$:
with the two jobs of Section~\ref{sec:example} that overstep and two ill-typed messages,
it takes about ten seconds when the decorations of the jobs are inferred. Every other
variant of our examples, obtained by removing any subset of their decorations, is
checked in less than half a second.

The prototype also implements the reconstruction of Section~\ref{sec:reconstruction}, at
almost no additional cost. Its generation step is the constraint generation of inference, with
\rulename{i-spawn-alt} for the unannotated $\mkkeyword{spawn}$s. Its derivation and solving
steps never call tallying: they only extract bounds from the constraints and compute with them.
On all our examples these two steps take between 3 and 25
milliseconds. The checking step is the tallying of inference, run on the constraints already
generated, in which the reconstructed interfaces replace the pid variables. The time of a check
is thus the time of tallying, which depends on the decorations left to infer, not on the
interfaces left to reconstruct: the program of Section~\ref{sec:example} is checked in about a
tenth of a second when its decorations are written and in about half a second when they are not,
whether the interface of its worker is declared or reconstructed
(see Appendix~\ref{app:measurements} for measurements).

\section{Related Work}
\label{sec:related}

The positioning in one sentence: several systems type actor mailboxes, and some
type them much more finely than we do, but none of those we discuss below types a \emph{higher-order
function whose late-bound \(\Self\) is resolved, and checked, at the process that runs it} without fixing in advance the exact interface of that process, and none obtains its guarantees as we do, by reusing an
existing set-theoretic subtyping and tallying machinery unchanged.

\paragraph{Late-bound \(\Self\) and the actor model.}
We borrowed the term ``late binding'' from object-oriented dispatch, although
the classical actor literature never uses it. The phenomenon it names, however, is
the model's own, not a BEAM idiosyncrasy. From its inception the actor model
separates an actor's \emph{identity} (its mail address) from its reusable
\emph{behavior} (its script): in Hewitt's original
formulation~\cite{Hewitt1973,Hewitt1977} as in Agha's~\cite{Agha1986}, one
behavior may be instantiated into many actors and re-adopted across turns
(Agha's \textsf{become}), so the self-reference available inside a behavior (the classical \textsf{SELF} primitive) cannot be resolved when the
behavior is defined: it denotes whichever actor is currently executing the
script. Implementations differ in \emph{when} they bind it. Erlang and
Elixir bind latest: \textsf{self()} is re-evaluated at each occurrence (the operational semantics of concurrent Core Erlang gives it its own reduction rule~\cite{Bereczky2024}), so a closure can capture a still-unresolved self
and ship it. Akka binds once per actor instance, at creation
(\textsf{ctx.self})~\cite{AkkaTyped}, and the C++ Actor Framework passes self
to the function defining an actor as an explicit first
argument~\cite{CAF16}, so in those languages a shipped closure carries an
identity that is already bound, and already typed. Our
decorated arrows thus type the general situation that the model itself
mandates: they are meaningful for actor languages whose behaviors are
first-class, reusable values, and they matter most where binding happens last,
as on the BEAM.

\paragraph{Set-theoretic types for Elixir.}
\label{sec:related-elixir}
This work is an extension of the type system that Elixir is currently
adopting~\cite{CDV24,CD26,duboc-phd}, which types the sequential part of the
language with semantic subtyping: types denote sets of values, and union,
intersection, and negation types denote the corresponding set-theoretic
operations~\cite{FCB08,castagna2022programming}. That system deliberately
leaves process creation and message passing untyped; this paper adds that
layer. The addition is conservative by design: interfaces $\TChan{t}$ and
decorated arrows $t\to_o^i t'$ are \emph{encoded} into the constructors of
\cite{FCB08} (Section~\ref{sec:fcb-encoding}), so the subtyping and tallying
algorithms of \cite{polyduce1,polyduce2}, already implemented for Elixir, apply verbatim, and the inference of Section~\ref{sec:ann-inference}
resolves to a single tallying call. Nothing in the sequential system needs to
be re-proved or re-implemented. This is also why our calculus omits the features
that this system already handles: multiarity functions and guards that test
arities and equality~\cite{CD26,duboc-phd}. Equality includes the pin
\eelixir{^x}, which matches exactly the current value of \eelixir{x}: a pinned
pid is compared by identity, and the guard establishes at most the type of
\eelixir{x} intersected with a test type, so it recovers no capability
(Section~\ref{sec:elixir-defensive}).

\paragraph{Semantic subtyping for the \(\pi\)-calculus.}
The interpretation of our channel types descends directly from Castagna, De
Nicola, and Varacca~\cite{CDV07}, who interpret channel types set-theoretically
and, in particular, show how to make sense of the apparently circular
``sets of types'' interpretation that we reuse in
Section~\ref{sec:channel-semantics}. The analogy stops at the communication
model: their calculus is synchronous and channel-based, with read and write
capabilities as separate constructors in the tradition of the i/o types of
Pierce and Sangiorgi~\cite{PS96}, whereas here reception happens only on a
process's own mailbox, which is why our $\TChan{t}$ keeps only the (write)
capability, contravariant in $t$, and the read side migrates to the turnstile
and arrow decorations. Their calculus has no functions, so the question of
late-bound $\Self$, our central concern, does not arise.

\paragraph{Mailbox types.}
De'Liguoro and Padovani~\cite{DP18} introduce mailbox types, which describe the
\emph{contents} of a mailbox as a commutative regular expression, tracking
multiplicities (a message that must arrive exactly once) and combinations of
messages; Fowler et al.~\cite{FAD23} bring this discipline to a programming
language, Pat, with an algorithmic, co-contextual type system tamed by
quasi-linear typing. On the mailbox itself these systems are strictly finer
than ours: they detect orphan messages and certain deadlocks, which a
set-of-messages interface cannot express. In exchange they give up what we
focus on: there is no subtyping of the set-theoretic kind, no negation or
intersection of interfaces, no interface reconstruction for unannotated code,
and closures are not shipped and re-typed at their runner. The two
disciplines are complementary, 
\behavior{
    and Section~\ref{sec:behavioral-types} layers
exactly this kind of behavioral information over the present set-of-messages
foundation.}
\ifbehavior\else
    and we are currently working on a system that combines the two, with a behavioral layer overlaid on the present set-of-messages foundation.
\fi

\paragraph{Session types for actors.}
Mostrous and Vasconcelos~\cite{MV11} type a featherweight Erlang with binary
sessions, correlating requests and replies through references; Neykova and
Yoshida~\cite{NY17} implement multiparty session types over actors, assigning
roles to processes and checking protocol conformance dynamically. Closest in
target language, ElixirST~\cite{ElixirST} brings this discipline to Elixir
itself: a public function of a module is annotated with the endpoint protocol it
must observe, its sends and receives are checked against it, and session
fidelity is proved against an LTS semantics for the supported fragment. It is,
as far as we know, the only other static type system for Elixir's concurrency,
and it is complementary to ours almost exactly: it disciplines the \emph{order}
of a two-party conversation, which our interfaces leave free, where we type the
\emph{mailbox} of a process open to arbitrarily many senders. Session
disciplines are \emph{order-sensitive}: they guarantee that interactions follow
a protocol, which is beyond our interfaces. But they type conversations, not
mailboxes: the mailbox of a process participating in several sessions has no
type of its own, self-sends are not an object of study, and the systems are
first-order with respect to process identities: nothing corresponds to typing
a function against the interface of whatever process will run it.

\paragraph{Typed actors in Scala.}
The academic precursor is He, Wadler, and Trinder's TAkka~\cite{TAkka}, which
retrofits Akka with type-parameterized references \textsf{ActorRef[M]}, contravariant in the accepted messages exactly as our
$\TChan{t}$, and typed supervision; the industrial Akka Typed~\cite{AkkaTyped} later made the same
discipline standard, typing the actor's behavior as \textsf{Behavior[T]}. In
both, the typing lives inside Scala's nominal variance: heterogeneous message
shapes are encoded, in the documented idiom, as sealed class hierarchies fixed
in advance, there are no negations, and the interface is always declared, never reconstructed from the receives. Moreover
\textsf{ctx.self} is an eagerly evaluated \emph{value}: a closure that
captures it ships the reference, correctly typed, of its \emph{creator}, so
a late-bound self is not left untyped there, it is inexpressible. Elixir's
\textsf{self()} is instead a dynamic call, so any system for existing Elixir
code must type it at its runner: that is what our decorated arrows do.

\paragraph{Comparing actors and channels.}
Fowler, Lindley, and Wadler~\cite{FLW17} define $\lambda_{\mathsf{act}}$, a
concurrent $\lambda$-calculus with actors, to compare the actor and channel
metaphors through translations in both directions: it is a vehicle for that
comparison, not a type system for a programming language. Its arrows are
annotated with a mailbox type $C$, $A\to^{C}B$, at which the $\mathsf{self}$
and $\mathsf{receive}$ of the function body are typed. But the calculus has
neither subsumption nor polymorphism, and a function applies only in an actor
whose mailbox type is \emph{exactly} $C$: rather than typing the function at
its runner, $\lambda_{\mathsf{act}}$ fixes the type of the runner when it
types the function. A function run by several actors can then be typed only by
giving all of them one mailbox type, which is what the authors themselves call
the type pollution problem, and not even then if
two of them read the same tag at different payload types. Adding subsumption
would not suffice: the same $C$ is read by $\mathsf{receive}$, which needs it
contravariant, and by $\mathsf{self}$, which needs it covariant, so any sound
subtyping would leave it invariant. This is why our arrows carry two decorations
of opposite variances (Section~\ref{sec:typing-decorations}), which delimit the
interval of processes that can run the function. Nor do sums replace union
types, even for tagged messages, which are the
idiom of Erlang and Elixir. Without
subtyping, a message must have exactly the mailbox type of its recipient:
forwarding a received message to a process that accepts more requires taking the message apart and injecting its
payload into the sum of the recipient; and a client that sends its own address
for the reply sends a reference with its \emph{whole} mailbox type, where our contravariant \eelixir{pid(t)} needs only the reply component of the mailbox type.

\paragraph{Actors typed by their message set.}
Closest to us are the systems that type a process directly by the \emph{set} of
messages it accepts. Colaço, Pantel, and Sallé~\cite{CPS97} give a
set-constraint analysis (in the style of Aiken and Wimmers) for a primitive
actor calculus whose actors are typed by the messages they can handle (a set that may even change dynamically), and use it to flag \emph{orphan} messages
that no actor will ever consume, which is the complement of what our
reconstruction checks: it holds the senders to the interface the receives
justify, and an orphan message is one that no interface so derived contains
(Section~\ref{sec:reconstruction}). On the BEAM itself
the idea is embodied by Gleam~\cite{Gleam}, a statically typed language
compiling to Erlang whose actors communicate through a typed handle
$\mathsf{Subject}(\mathit{msg})$, checked by Hindley--Milner
inference. The parallel is strong (an actor is typed by what it accepts, and every send is checked against it), but Gleam is
instructive for what it must give up. Its interfaces are \emph{nominal} sums
fixed per actor, whereas ours are set-theoretic unions with genuine subtyping,
intersection, and negation, decided by tallying and reusing Elixir's existing
machinery rather than a fresh Hindley--Milner core; it does not reconstruct an
interface from the receives of unannotated code; and it cannot type a $\Self$
shipped inside a closure, because in Gleam the type of a message comes from an
explicit handle, not from the process: $\mathsf{process.self()}$ returns an
untyped $\mathsf{Pid}$, and a typed receive names a $\mathsf{Subject}$ (or a
$\mathsf{Selector}$ of subjects) that the process holds.
Attaching types to explicit handles rather than to the ambient operations is indeed \emph{a} way to make actors typable,
and it is the one Gleam, a new language, has taken; it is not available to us, whose object is existing Elixir,
where those operations are the ones programmers actually write.

\paragraph{Type-and-effect systems.}
The decorations are recognizably a type-and-effect discipline in the sense of
Lucassen and Gifford~\cite{LG88}: the turnstile carries an effect, arrows
carry a latent effect, and application discharges the latter against the
former. The closest precedent is Nielson and Nielson's type system for
Concurrent ML~\cite{NN94}, whose function types carry latent
\emph{behaviors} describing the communications the body may perform; and it
is Talpin and Jouvelot~\cite{TJ92} who established that such annotations can
be \emph{inferred} rather than declared, which is what Section~\ref{sec:ann-inference}
does for mailboxes. The
differences are the point of this paper: our effects are ordinary set-theoretic types, decided by the same subtyping and inferred by the same tallying, with no separate effect algebra, and they are anchored to the
\emph{identity} of the runner: no CML behavior can say that
$\Self$ denotes whichever process runs the closure, which is precisely what
the decorations transport. Whether that holds beyond mailboxes (for exceptions, for the process dictionary, for the tables a process reads and writes) is the general study that remains.

\paragraph{Success typings.}
Dialyzer's success typings~\cite{LS06} are the established static analysis for
Erlang: types are inferred so as to never reject a program that can succeed,
warnings are emitted only for code that certainly fails, and message passing is
essentially transparent to the analysis. Christakis and Sagonas~\cite{CS11}
carry that line into concurrency, adding to Dialyzer's infrastructure an
analysis that flags messages no receive can ever match and receive clauses no send can ever satisfy, \behavior{%
  the same two-way mismatch that the behavioral layer of
   Section~\ref{sec:behavioral-types} rules out by typing, and 
}
   the closest existing
tool to what interface reconstruction detects. Being a bug-finder rather than a
type system, it reports only the mismatches it can prove, where we reject every
send outside the interface. Our reconstruction
(Section~\ref{sec:reconstruction}) is a sound counterpart for the same
no-annotation setting, and it differs from that tradition by where it puts the
blame. A bug-finder derives from the receives the \emph{least} interface they
justify and reports only what provably fails, so a mistyped send is left alone;
we derive the \emph{greatest} interface the receive bodies can serve and hold the
senders to it. The least interface is in any case not worth deriving: a process
that is to accept whatever is sent to it is simply declared at \eelixir{term()},
which asks nothing of its senders and, as Section~\ref{sec:elixir-dynamic}
shows, requires no machinery of its own. Where a success typing stays silent and
we do not is on the use a branch makes of its message: that use becomes a demand
on the interface (Section~\ref{sec:elixir-defensive}), so a sender that
violates it is rejected instead of merely suspected. The same no-annotation setting is targeted by the industrial Erlang
checkers descended from this tradition, such as eqWAlizer~\cite{eqWAlizer} and
Gradualizer~\cite{Gradualizer}, with its Elixir front end Gradient,
which type the sequential language and leave pids generic, exactly the gap
this paper fills. Remarkably, eqWAlizer uses the theory of gradual typing for semantic
subtyping of \cite{castagna2019gradual,lanvin2021semantic}, the very theory
Elixir has adopted. Extending our system in that direction is the next step for
this work.

\paragraph{Early Erlang typing.}
Marlow and Wadler~\cite{MW97} gave the first subtyping-based type system for
Erlang, with inference based on constraint solving over a lattice of types
including atoms and unions of tagged tuples, recognizably the same data
vocabulary as our interfaces. Their system covered the sequential language;
processes and message passing were left to future work, and the line of work
was not pursued in practice. It is pleasant to note that the constraint-based
architecture they advocated is, thirty years later, the shape of our
Section~\ref{sec:ann-inference}.

\section{Conclusion}
\label{sec:conclusion}
We have shown how to type the actor model in the presence of first-class
functions. A process is typed by \texttt{pid($t$)}, where its interface $t$ is
the type of the messages it accepts, and an arrow is decorated by a bracket, the
interval of the interfaces of the processes that can run the function. This is
what makes it possible to type code that travels: a closure that evaluates
\eelixir{self()} or performs a \eelixir{receive} is typed once, where it is
written, and can then be run by every process whose interface lies in the closure's
bracket. The system is sound: every message sent
belongs to the interface of its recipient, and every \eelixir{receive} can rely
on the interface of the process that runs it. The errors it catches are, for
the most part, ordinary ones: a misspelled tag, a payload of the wrong type, a
job that deposits in the mailbox of its runner a message that the runner does
not handle. Elixir accepts all of them today, and at run time they leave a
message in a mailbox forever, or crash a process far from where the error is. Although not fancy, these errors are the ones that Elixir programmers encounter most often, and they are the ones that our system prevents.

Throughout this work we stayed as close as possible to Elixir as it is today:
to its types, its compiler, its virtual machine, and the way its programmers
write processes. Interfaces and
decorations are encoded into the set-theoretic types that Elixir already uses,
so that the subtyping and constraint resolution of Elixir apply to them as they
are. The only new annotation is the interface of a spawned process: decorations
are inferred and (mostly) never written, and an
interface that is not declared is reconstructed, when possible, from the code of
the spawned process, which makes the system
applicable to existing code. Annotations are erased at compilation, and typed
programs run on the BEAM as it is. The system can thus be integrated into Elixir
as an extension of its current type system, and we analyzed what this requires
beyond our core calculus: type-checking units, untyped code, explicit decorations, library loops, and
the messages that the virtual machine deposits.

We tested the type system, the inference of decorations and the reconstruction of
interfaces on a prototype that type-checks a subset of Elixir. It runs
on the current release of Elixir and builds on the type representation,
subtyping and tallying of the development version of the Elixir compiler; with
three optimizations of tallying, which do not change the set of accepted
programs, it infers every decoration of the running example of
Section~\ref{sec:example} in a reasonable amount of time, and reconstructing its
interfaces adds almost nothing to it.

Future work includes three extensions. First, behavioral types: an interface
states which messages a process accepts, but neither how many nor whether they
are ever handled. Mailbox types in the style of~\cite{DP18,FAD23} would
check that sends and receives balance; the question is how much of this can be
obtained without asking programmers to write more than an interface. Second, parametric polymorphism: once it
is added to Elixir's type system, decorations should be generalized and
instantiated together with the other type variables. But processes add effects
to the pure functional core that the types of Elixir have covered so far, and,
as with references in ML, effects restrict generalization: a call that spawns a
process creates one mailbox, whatever the number of instances at which the call
is typed, so a type variable in the interface of a spawned process must be
instantiated once for each spawn. Third, gradual typing: Elixir's
\eelixir{dynamic()} can already be given to interfaces and pids, but which of
its materializations remain sound in the presence of processes is still to be
determined.

\bibliographystyle{ACM-Reference-Format}
\bibliography{main}

\clearpage

\appendix
\section{Types and Subtyping: Direct Interpretation and Proofs}
\label{app:encoding-property}
This appendix gives the direct interpretation of the types of Section~\ref{sec:syntax}
announced in Section~\ref{sec:arrow-semantics-pointed}: a domain with one summand per type
constructor (Appendix~\ref{app:enc-domain}), in which types are interpreted as
Section~\ref{sec:channel-semantics} describes (Appendix~\ref{app:enc-interpretation}). It shows
that this interpretation is a model (Appendix~\ref{app:enc-model}), that it induces the same
subtyping relation as the encoding (Theorem~\ref{thm:encoding} and
Corollary~\ref{cor:same-subtyping}, Appendix~\ref{app:enc-subtyping}), and it proves, in this
model, the properties of arrow and pid types stated in Section~\ref{sec:subtyping-laws}
(Appendix~\ref{app:enc-laws}). It ends with two extensions needed from
Section~\ref{sec:ann-inference} on: type variables (Appendix~\ref{app:enc-variables}), and the
readback of types of~\cite{FCB08} into ours (Appendix~\ref{app:enc-readback}).

\danielem{I really don't understand why all this. We build a new model and then we prove that we can encode using FCB. This is very technical, and not very interesting. We just \emph{define} the interpretation of the decorated arrow types to be the encoding, that's it.}\beppem{Done: Section~\ref{sec:arrow-semantics-pointed} now defines the interpretation through the encoding. The direct interpretation stays here because it shows that the encoding induces the intended relation (it is the interpretation that Section~\ref{sec:channel-semantics} describes), and because the proofs of Appendices~\ref{app:properties} and~\ref{app:reconstruction}, and the behavioral extension, reason on its elements (the latter needs it for its basic types).}

\subsection{Domain}\label{app:enc-domain}
The direct interpretation follows Section~\ref{sec:channel-semantics} literally. As
in~\cite{FCB08}, functions are interpreted by finite relations whose outputs range over
$\Domain_\Omega\eqdef\Domain\cup\Set{\Omega}$, for a constant $\Omega\notin\Domain$ that
marks the error; but an input is now a \emph{triple} of a (pointed) value, a write
capability, and a read capability. So functions are interpreted in
$\Pf((\Domain_\mho{\times}\Domain{\times}\Domain)\times\Domain_\Omega)$, where
$\Domain_\mho\eqdef\Domain\cup\Set{\mho}$ is the pointed domain, for a constant
$\mho\notin\Domain$ distinct from $\Omega$: only the inputs change shape with respect
to~\cite{FCB08}, $\Domain$ becoming $\Domain_\mho{\times}\Domain{\times}\Domain$, while the
outputs remain $\Domain_\Omega$. Following~\cite{CDV06}, a pid is its write operation, a
function from messages to the unit type, here rendered by $\mho$: a \emph{pid
approximation} is a finite set of pairs $(d,\omega)$, where $d$ is a message sent to the
process and $\omega$ says whether the process accepts it ($\mho$) or not ($\Omega$); the
write capability of an arrow is a pid approximation. Symmetrically, a \emph{read
approximation} is a finite set of pairs $(\mho,\domega)$, which relate the unit argument
to an outcome $\domega$, either a message yielded or the error $\Omega$; it may relate the
argument to several results, the read capability being multi-shot.

\begin{definition}[Interpretation domain]\label{def:interpretation}
 The \emph{interpretation domain} $ \Domain $ is the set of finite terms $ d $
produced inductively by the following grammar
\begin{align*}
  d & ::= c \:\mid\: \Tuple{d,\dots,d} \:\mid\: \Set{((\dagemo,d,d) , \domega), \dots, ((\dagemo,d,d) , \domega)} \:\mid\: \Set{(d,\omega), \dots, (d,\omega)} \:\mid\: \Set{(\mho,\domega), \dots, (\mho,\domega)}
    \\
    \domega & ::= d \mid \Omega \qquad\qquad \dagemo ::= d \mid \mho \qquad\qquad \omega ::= \mho \mid \Omega
\end{align*}
where $ c $ ranges over the set $ \Constants $ of constants,
and $ \Omega $ and $\mho$ are constants such that $ \Omega \notin \Domain $, $ \mho \notin \Domain $, and $ \Omega \neq \mho $. The three set alternatives of the grammar are considered \emph{disjoint}: every finite set carries the alternative that produced it.
\end{definition}
 In other terms, the domain is the solution of
$\Domain = \Constants + \sum_{n\geq 0}\Domain^{n} + \Pf((\Domain_\mho{\times}\Domain{\times}\Domain)\times\Domain_\Omega)+\Pf(\Domain\times\Set{\mho,\Omega})+\Pf(\Set{\mho}\times\Domain_\Omega)$, where $+$ denotes disjoint union; the elements of the second summand represent tuples, those in the third summand represent (approximations of) functions (the components of an input triple being the argument, the write capability, and the read capability, and the output being the result), those in the fourth summand represent (approximations of) pids, that is, of write capabilities, and those in the fifth summand represent (approximations of) read capabilities. The last two summands are the two halves of the same picture: a pid relates the messages it is sent to their acceptance, a read capability relates the unit argument to the messages it yields. The disjointness requirement in the definition is not redundant: without it the empty set would inhabit every arrow type, every pid type, and every read capability, making for instance $\TChan{\Tempty}\land\TFun$ nonempty and thus breaking the disjointness of pid and function types on which pattern matching relies. With it, every finite set belongs to exactly one alternative, and each clause for sets in the membership predicate that defines the interpretation (below) applies only to the elements of the corresponding alternative.

\subsection{Definition of the interpretation}\label{app:enc-interpretation}
The direct interpretation $\semD{.}:\Types\to\Pd(\Domain)$ must satisfy the following
equalities:
\begin{gather*}
\begin{array}{r@{~=~}l@{\qquad}r@{~=~}l@{\qquad}r@{~=~}l}
\semD{\Tempty} & \varnothing &
\semD{\neg t} & \Domain\setminus\semD{t} &
\semD{\Tuple{t_1,\dots,t_n}} & \semD{t_1}\times\dots\times \semD{t_n}\\
\semD{b} & \ConstantsInBasicType({b}) &
\semD{t_1\lor t_2} & \semD{t_1}\cup \semD{t_2} &
\semD{\TTuple} & \textstyle\bigcup_{n\geq 0}\Domain^{n}\\
\semD{\Tany} & \Domain &
\semD{\TChan{t}} & \Pf(\oline{\semD{t}\times\Set{\Omega}}) &
\semD{()\to i} & \Pf(\oline{\Set{\mho}\times\oline{\semD{i}}})
\end{array}
\\[1ex]
\semD{t_1\to^i_o t_2} = \Pf(\oline{((\semD{t_1}\cup\Set{\mho}){\times}\semD{o\to ()}{\times}\semD{()\to i}){\times}\oline{\semD{t_2}}})
\end{gather*}
where $\semD{o\to()}$ is $\semD{\TChan{o}}$, since the write capability of an arrow is a
pid approximation; $b$ ranges over the \emph{basic types}, that is, the singleton types $c$, $\TInt$, and $\TAtom$, and $\ConstantsInBasicType(b)$ denotes the set of constants of the basic type $b$: $\ConstantsInBasicType(c)=\Set{c}$, $\ConstantsInBasicType(\TInt)$ is the set of all integers, and $\ConstantsInBasicType(\TAtom)=\AtomSet$;
the complements are taken with respect to the corresponding interpretation spaces: for the arrow, the inner complement is taken within $\Domain_\Omega$ (so it always contains $\Omega$ and is never empty) and the outer one within $(\Domain_\mho{\times}\Domain{\times}\Domain)\times\Domain_\Omega$; for pid types, the complement is taken within $\Domain\times\Set{\mho,\Omega}$; for the read capability, the inner complement is taken within $\Domain_\Omega$ (so it contains $\Omega$: a read capability that errs has no type) and the outer one within $\Set{\mho}\times\Domain_\Omega$. Notice that no factor of the \emph{domain} of an arrow is ever empty, since $\mho$ and the two empty approximations $\varnothing$, of the write and of the read capability, inhabit them whatever the types are (Section~\ref{sec:channel-semantics}).

Since types are coinductive, these equalities cannot be taken as an inductive
definition of $\semD{\cdot}$. Giving a well-founded definition that validates them is a
standard technique of semantic subtyping: we define $\semD{t} = \{ d\in\Domain\mid
(d:t)\}$, where $(d:t)$ is the following binary predicate, defined by induction on the
pair $(d,t)$ ordered lexicographically, using the induction principle on types recalled
in Section~\ref{sec:syntax} and structural induction on $\Domain$:\vspace{-.9mm}
\bgroup\setlength{\jot}{0ex}
\begin{align*}
  (d : \Tany) & = \mathsf{true} \\
  (c : b) & = c \in \ConstantsInBasicType(b) \\
  (\Tuple{d_1,\dots, d_n} : \TTuple) & = \mathsf{true} \\
  (\Tuple{d_1,\dots, d_n} : \Tuple{t_1,\dots, t_n} ) & =
    (d_1 : t_1) \mathrel{\mathsf{and}} \dots \mathrel{\mathsf{and}} (d_n : t_n) \\
  (\Set{((\dagemo_1, d_1, \mu_1),\domega_1), \dots, ((\dagemo_n, d_n, \mu_n),\domega_n)} : t_1 \to^i_o t_2) & =
    \forall j \in \Set{1, \dots, n} . \\
    &\hspace*{-5mm}\mathsf{if} \: (\dagemo_j :_\mho t_1) \mathrel{\mathsf{and}} (d_j : o\to()) \mathrel{\mathsf{and}} (\mu_j: ()\to i) \\
    &\hspace*{-5mm}{\mathsf{then}}\ \domega_j = d'_j \mathrel{\mathsf{and}} (d'_j : t_2) \\
  (\Set{(d_1,\omega_1),\dots,(d_n,\omega_n)} : o \to ()) & = \forall j \in \Set{1, \dots, n} . \: \mathsf{if}\ (d_j:o)\ \mathsf{then}\ \omega_j = \mho\\
  (\Set{(\mho,\domega_1),\dots,(\mho,\domega_n)} : () \to i) & = \forall j \in \Set{1, \dots, n} . \: \domega_j = d'_j \mathrel{\mathsf{and}} (d'_j : i)\\
  (\Set{(d_1,\omega_1),\dots,(d_n,\omega_n)} : \TChan{t}) & = \forall j \in \Set{1, \dots, n} . \: \mathsf{if}\ (d_j:t)\ \mathsf{then}\ \omega_j = \mho\\
(d : t_1 \lor t_2) & = (d : t_1) \mathrel{\mathsf{or}} (d : t_2) \\
  (d : \lnot t) & = \mathsf{not} \: (d : t) \\
  (d : t) & = \mathsf{false}  \hspace*{42mm}\text{otherwise}\\[-5mm]
\end{align*}\egroup
where, in the clause for arrows, $d_j$ and $\mu_j$ are the write and the read capability of the $j$-th input triple, and $(\dagemo :_\mho t)$ is short for $\dagemo = \mho \mathrel{\mathsf{or}} (\dagemo : t)$, that is, membership in the pointed set $\semD{t}\cup\Set{\mho}$. The clause for $o\to()$ is by definition the one for $\TChan{o}$ (the two denote the same set), and the clause for $()\to i$ is the only new one: $o\to()$ and $()\to i$ being notations rather than types, the predicate is extended to them by these two dedicated clauses. All the checks of the arrow clause concern components of the function element at hand, which are structurally smaller elements, so the lexicographic induction on $(d,t)$ goes through even though these two are not subterms of the arrow type. The error $\Omega$ satisfies no consequent, so an element that maps some input of the (pointed) domain of an arrow type to $\Omega$ belongs to no such arrow type: this is what prevents $\Tany\to^{\Tany}_{\Tany}\Tany$ from denoting the set of all function approximations.

\subsection{The interpretation is a model}\label{app:enc-model}
The pair $\Domain$ and $\semD{.}$ is a \emph{model} of types in the sense of
\citet[Definition 4.4]{FCB08}: it induces the same subtyping relation as its
\emph{extensional interpretation} $\extint{\cdot}$~\cite[Definition 4.3]{FCB08},
that is, $\semD{t}=\varnothing$ if and only if $\extint{t}=\varnothing$, for every
type $t$. The extensional interpretation is the set-theoretic interpretation that
coincides with $\semD{.}$ on all type constructors but arrows and pids, which it
interprets by arbitrary, rather than finite, relations:
\begin{gather*}
\extint{t_1\to^i_o t_2} = \Pd(\oline{((\semD{t_1}\cup\Set{\mho}){\times}\semD{o\to ()}{\times}\semD{()\to i}){\times}\oline{\semD{t_2}}})
\\
\extint{\TChan{t}} = \Pd(\oline{\semD{t}\times\Set{\Omega}})
\end{gather*}
with the complements taken as above.
Following \citet[Definition 6.28]{FCB08}, an interpretation is \emph{finitely
extensional} if its domain is the restriction to finite relations of the domain of
its extensional interpretation, and if $\semD{a}=\extint{a}\cap\Domain$ for every
type constructor $a$. The interpretation $\semD{.}$ is finitely extensional by construction:
$\extint{\cdot}$ ranges over the domain obtained from the equation defining
$\Domain$ by replacing the finite powersets of the function and pid summands with
full powersets, it coincides with $\semD{.}$ on the other type constructors, and on
arrows and pids $\semD{.}$ keeps exactly the finite relations of $\extint{\cdot}$.
\citet[Lemma 6.32]{FCB08} prove that every finitely extensional interpretation is a
model. Their proof splits a Boolean combination of type constructors according to
the summand of the domain on which it lies. The statement is trivial for the
summands that $\Domain$ shares with the extensional domain (here constants, tuples,
and read capabilities), and for the function summand it rests on three
lemmas~\cite[Lemmas 6.6, 6.7, and 6.30]{FCB08} that are properties of the relation
spaces $X\to Y=\Pd(\oline{X\times\oline{Y}})$ over arbitrary sets of inputs and of
outputs. They therefore apply unchanged to our function summand, whose inputs range
over $\Domain_\mho{\times}\Domain{\times}\Domain$, and to our pid summand, whose
outputs range over $\Set{\mho,\Omega}$ with $Y=\Set{\mho}$. Hence
$\semD{t}=\varnothing$ if and only if $\extint{t}=\varnothing$, for every type $t$.

\subsection{Same subtyping}\label{app:enc-subtyping}
The direct interpretation and the encoding induce the same subtyping relation on the
types of Section~\ref{sec:syntax}. Since both interpretations commute with the type
connectives, and a subtyping statement is in both cases the emptiness of a difference,
it suffices to show that a type is empty in the direct interpretation if and only if its
encoding is empty in the model of~\cite{FCB08}.

\begin{theorem}[Soundness and completeness of the encoding]\label{thm:encoding}
For every type $t$, $\semD{t} = \varnothing$ if and only if
$\sem{\enc{t}}_\textsf{FCB} = \varnothing$, that is, $\sem{t}=\varnothing$.
\end{theorem}
\begin{proof}
Let $\Domain$ be the domain of Definition~\ref{def:interpretation} and $\DomainFCB$ the
domain of \cite{FCB08} built over the set of constants extended with the fresh
constants $\CPoint$, $\ell_{\Wch}$, $\ell_{\Wfun}$, ${\Cnil}$: its elements are
constants, pairs $(x,y)$, and finite relations
$\Set{(x_1,y_1),\dots,(x_n,y_n)}$ with each $y_j$ an element or $\Omega$. We write
$(x:T)$ also for membership in the interpretation of \cite{FCB08}. Both directions
relate memberships in the two domains, by induction on pairs (element, type) ordered
lexicographically, the principle used to define $(d:t)$: an inductive call may be made
on a strictly smaller element at any type, or on the same element at a type below a
connective. Since the encoding commutes with the connectives and with $\Tany$, these
cases are immediate in both directions and we omit them.

\emph{Embedding.} Define $\varphi:\Domain\to\DomainFCB$ by structural induction, for
constants, tuples, pid approximations, read capabilities, and function approximations:
\[
\begin{array}{r@{\;}c@{\;}l}
\varphi(c) &=& c\\
\varphi(\Tuple{d_1,\dots,d_n}) &=& (\varphi(d_1),(\dots,(\varphi(d_n),{\Cnil})))
\qquad\text{(so that $\varphi(\Tuple{})={\Cnil}$)}\\
\varphi(\Set{(d_j,\omega_j)}_j) &=& (\ell_{\Wch},\Set{(\varphi(d_j),\hat\omega_j)}_j)
\qquad\text{with $\hat\mho=\CPoint$ and $\hat\Omega=\Omega$}\\
\varphi(\Set{(\mho,\domega_j)}_j) &=& \Set{(\CPoint,\hat\varphi(\domega_j))}_j\\
\varphi(\Set{((\dagemo_j,d_j,\mu_j),\domega_j)}_j) &=&
(\ell_{\Wfun},\Set{((\hat\varphi(\dagemo_j),(\varphi(d_j),\varphi(\mu_j))),\hat\varphi(\domega_j))}_j)
\end{array}
\]
where $\hat\varphi$ extends $\varphi$ with $\hat\varphi(\mho)=\CPoint$ and
$\hat\varphi(\Omega)=\Omega$. By construction $\varphi(d)\neq\CPoint$ for every
$d\in\Domain$, $\varphi(d)$ is an untagged relation if and only if $d$ is a read
capability, and $\varphi(d)$ is a pair with first component $\ell_{\Wch}$ (resp.\
$\ell_{\Wfun}$) and a relation as second component if and only if $d$ is a pid (resp.\
function) approximation.

\emph{Claim 1: $(d:t)$ if and only if $(\varphi(d):\enc{t})$; and $(d:()\to i)$ if and
only if $(\varphi(d):\CPoint\to\enc{i})$.} By cases on the type.
\begin{itemize}[leftmargin=*,itemsep=1pt,topsep=2pt]
\item Basic type $b$: $\varphi(d)$ is a constant only if $d$ is a constant or the empty
tuple, and ${\Cnil}$, being fresh, belongs to no basic type.
\item $\TTuple$: $\enc{\TTuple}=T$ with $T={\Cnil}\lor(\Tany\times T)$, whose elements are
the ${\Cnil}$-terminated nestings of pairs; $\varphi(d)$ is one if and only if $d$ is a
tuple, since a tagged pair has a relation, neither a pair nor ${\Cnil}$, as second
component.
\item $\Tuple{t_1,\dots,t_n}$: $\varphi(d)$ belongs to
$\enc{t_1}\times(\dots\times(\enc{t_n}\times{\Cnil}))$ if and only if it is a
${\Cnil}$-terminated nesting of exactly $n$ pairs whose components belong to the
$\enc{t_k}$, that is, if and only if $d=\Tuple{d_1,\dots,d_n}$ with
$(\varphi(d_k):\enc{t_k})$ for every $k$, hence, by induction on the smaller $d_k$, if
and only if $(d:\Tuple{t_1,\dots,t_n})$.
\item $\TChan{t}$, encoded as $\ell_{\Wch}\times(\enc{t}\to\CPoint)$: if $d$ is not a pid
approximation, both memberships fail. Otherwise $(\varphi(d):\enc{\TChan t})$ if and only
if, for every $j$, $(\varphi(d_j):\enc{t})$ implies $\hat\omega_j=\CPoint$, that is, by
induction on the smaller $d_j$, if and only if $(d_j:t)$ implies $\omega_j=\mho$ for every
$j$, which is $(d:\TChan{t})$.
\item $()\to i$: if $d$ is not a read capability, $(d:()\to i)$ is false and $\varphi(d)$ is
not an untagged relation. Otherwise, since $\CPoint$ inhabits $\sem{\CPoint}_\textsf{FCB}$,
$(\varphi(d):\CPoint\to\enc{i})$ if and only if, for every $j$,
$\hat\varphi(\domega_j)\neq\Omega$ and $(\hat\varphi(\domega_j):\enc{i})$, that is, by
induction, if and only if $\domega_j=d'_j$ with $(d'_j:i)$ for every $j$.
\item $s\to_o^i u$, encoded as $\ell_{\Wfun}\times(D\to\enc{u})$ with
$D=(\enc{s}\lor\CPoint)\times(\enc{\TChan{o}}\times(\CPoint\to\enc{i}))$: if $d$ is not a
function approximation, both memberships fail. Otherwise
$(\varphi(d):\enc{s\to_o^i u})$ if and only if, for every $j$, whenever
$(\hat\varphi(\dagemo_j):\enc{s}\lor\CPoint)$, $(\varphi(d_j):\enc{\TChan o})$, and
$(\varphi(\mu_j):\CPoint\to\enc{i})$, then $\hat\varphi(\domega_j)\neq\Omega$ and
$(\hat\varphi(\domega_j):\enc{u})$. By induction on the smaller elements $d_j$, $\mu_j$,
and $\domega_j$, the last three conditions are $(d_j:o\to())$, $(\mu_j:()\to i)$, and
$\domega_j=d'$ with $(d':u)$; the first is $(\dagemo_j:_\mho s)$, because
$\hat\varphi(\mho)=\CPoint$ and $\varphi(\dagemo)\neq\CPoint$ for $\dagemo\in\Domain$.
This is the clause of $(d:s\to_o^i u)$.
\end{itemize}
Consequently $\semD{t}\neq\varnothing$ implies $\sem{\enc{t}}_\textsf{FCB}\neq\varnothing$.

\emph{Junk.} Call an element of $\DomainFCB$ \emph{regular} if it is a constant of the
source language, a ${\Cnil}$-terminated nesting of pairs, or a pair $(\ell_{\Wch},R)$ or
$(\ell_{\Wfun},R)$ with $R$ a finite relation, and \emph{junk} otherwise (the fresh
constants other than ${\Cnil}$, untagged relations, and the other pairs). A junk element
inhabits the encoding of no type \emph{constructor} other than $\Tany$: not that of a basic type, since
the fresh constants belong to none; not that of a tuple type, which is contained in
$\sem{T}_\textsf{FCB}$; not that of a pid or arrow type, which contains only tagged relations. Yet
junk inhabits the encodings of types built with negation or $\Tany$, for instance
$\sem{\enc{\neg\TInt}}_\textsf{FCB}=\DomainFCB\setminus\sem{\TInt}_\textsf{FCB}$. The domain
$\Domain$ has an element in the same situation,
$\jzero=\Set{((\mho,\varnothing,\varnothing),\Omega)}$, the function approximation that
maps to the error the input made of no value, the empty write approximation (in every
$\semD{o\to()}$), and the empty read approximation (in every $\semD{()\to i}$); the two
empty approximations are distinct elements, since the set alternatives of the domain
are disjoint. Indeed $\jzero$ is neither a constant, nor a tuple, nor a pid
approximation, and it belongs to no arrow type, not even to $\TFun$, since its input
belongs to the domain of every arrow type. \emph{Claim 2: for every junk element $e$
and every type $t$, $(e:\enc{t})$ if and only if $(\jzero:t)$}, by induction on $t$
(contractivity): both memberships are false at the type constructors, true at $\Tany$,
and commute with the connectives.

\emph{Readback.} Define $\psi:\DomainFCB\to\Domain$ by structural induction. On regular
elements: $\psi(c)=c$ for a constant of the source language; $\psi({\Cnil})=\Tuple{}$ and
$\psi((x,y))=\Tuple{\psi(x),d_2,\dots,d_n}$ when $(x,y)$ is a ${\Cnil}$-terminated nesting
and $\psi(y)=\Tuple{d_2,\dots,d_n}$;
$\psi((\ell_{\Wch},R))=\Set{(\psi(x),\omega(y))\mid (x,y)\in R}$, with
$\omega(\CPoint)=\mho$ and $\omega(y)=\Omega$ for every other $y$, $\Omega$ included; and
\[
\psi((\ell_{\Wfun},R))=\Set{((\hat\psi(x_1),\psi(x_2),\rho(x_3)),\psi_\Omega(y))\mid
((x_1,(x_2,x_3)),y)\in R,\ x_3\text{ a finite relation}},
\]
where
$\hat\psi(\CPoint)=\mho$ and $\hat\psi(x)=\psi(x)$ otherwise,
$\psi_\Omega(\Omega)=\Omega$ and $\psi_\Omega(y)=\psi(y)$ otherwise, and
$\rho(x_3)=\Set{(\mho,\psi_\Omega(y'))\mid (\CPoint,y')\in x_3}$. On junk,
$\psi(e)=\jzero$. The readback thus drops the pairs of a tagged relation whose input does
not have the shape $(x_1,(x_2,x_3))$ with $x_3$ a relation, and the pairs of $x_3$ whose
first component is not $\CPoint$: such inputs belong to the domain of no encoded arrow,
respectively to $\sem{\CPoint}_\textsf{FCB}$, so they constrain no membership.
\emph{Claim 3: $(e:\enc{t})$ if and only if $(\psi(e):t)$; and, for every finite relation
$x$, $(x:\CPoint\to\enc{i})$ if and only if $(\rho(x):()\to i)$.} The junk case is
Claim 2. For regular elements the cases mirror those of Claim 1, since $\psi$ maps
constants, nestings, and $\ell_{\Wch}$- and $\ell_{\Wfun}$-tagged relations to constants,
tuples, pid approximations, and function approximations, respectively; when the kind of
$e$ does not match the constructor, both memberships fail. For $\TChan{t}$ and
$e=(\ell_{\Wch},R)$: $(e:\enc{\TChan t})$ if and only if, for every $(x,y)\in R$,
$(x:\enc{t})$ implies $y=\CPoint$, that is, by induction on the smaller $x$, if and only
if $(\psi(x):t)$ implies $\omega(y)=\mho$, which is $(\psi(e):\TChan t)$; distinct pairs
with the same image are harmless, since the clause quantifies universally. For a
relation $x$: $(x:\CPoint\to\enc{i})$ if and only if every $(\CPoint,y')\in x$ has
$y'\neq\Omega$ and $(y':\enc{i})$, that is, by induction, $\psi_\Omega(y')=d'$ with
$(d':i)$, which is $(\rho(x):()\to i)$. For $s\to_o^i u$ and $e=(\ell_{\Wfun},R)$: the
pairs dropped by $\psi$ constrain nothing; for every other pair
$((x_1,(x_2,x_3)),y)$, $(x_1:\enc{s}\lor\CPoint)$ if and only if
$(\hat\psi(x_1):_\mho s)$ (both hold if $x_1=\CPoint$; otherwise $\psi(x_1)\neq\mho$ and
the induction hypothesis applies), $(x_2:\enc{\TChan o})$ if and only if
$(\psi(x_2):o\to())$, $(x_3:\CPoint\to\enc{i})$ if and only if $(\rho(x_3):()\to i)$, and
$y\neq\Omega$ with $(y:\enc{u})$ if and only if $\psi_\Omega(y)=d'$ with $(d':u)$. So
$\psi(e)$ satisfies the clause of $s\to_o^i u$ if and only if $e$ satisfies that of its
encoding. Consequently $\sem{\enc{t}}_\textsf{FCB}\neq\varnothing$ implies
$\semD{t}\neq\varnothing$.
\end{proof}

\begin{corollary}[Same subtyping]\label{cor:same-subtyping}
For all types $s$ and $t$: \ $s \subt t$ if and only if $\semD{s}\subseteq\semD{t}$.
\end{corollary}
\begin{proof}
Both $\semD{\cdot}$ and $\enc{\cdot}$ commute with the type connectives, so
$\semD{s\land\neg t}=\semD{s}\setminus\semD{t}$ and
$\enc{s\land\neg t}=\enc{s}\land\neg\enc{t}$. By Definition~\ref{def:subtyping}, $s\subt t$
if and only if $\sem{\enc{s\land\neg t}}_\textsf{FCB}=\varnothing$, and
Theorem~\ref{thm:encoding} applied to $s\land\neg t$ concludes.
\end{proof}

\subsection{Properties of arrow and pid types}\label{app:enc-laws}
By Corollary~\ref{cor:same-subtyping}, it suffices to prove the two propositions of
Section~\ref{sec:subtyping-laws} for the direct interpretation.

\ArrowFactorProp*

\danielem{Why this name?}
\begin{proof}
Write $X_{s,o,i}$ for $(\semD{s}\cup\Set{\mho})\times\semD{o\to()}\times\semD{()\to i}$, so that
$\semD{s\to^i_o t}=\Pf(\oline{X_{s,o,i}\times\oline{\semD{t}}})$, and
$\varnothing_{\mathsf w}$ and $\varnothing_{\mathsf r}$ for the empty approximations
of a write and of a read capability, which belong to every $\semD{o\to()}$ and to
every $\semD{()\to i}$, respectively.

For the ``if'' direction: $s'\leq s$ gives $(\semD{s'}\cup\Set{\mho})\subseteq(\semD{s}\cup\Set{\mho})$;
$o\leq o'$ gives $\semD{\TChan{o'}}\subseteq\semD{\TChan{o}}$, since an approximation
that maps no element of $\semD{o'}$ to $\Omega$ maps no element of
$\semD{o}\subseteq\semD{o'}$ to $\Omega$; and $i'\leq i$ gives
$\semD{()\to i'}\subseteq\semD{()\to i}$, since a read approximation that yields only
elements of $\semD{i'}$ yields only elements of $\semD{i}$. Hence
$X_{s',o',i'}\subseteq X_{s,o,i}$ and, with $t\leq t'$,
$X_{s',o',i'}\times\oline{\semD{t'}}\subseteq X_{s,o,i}\times\oline{\semD{t}}$, so that
$\oline{X_{s,o,i}\times\oline{\semD{t}}}\subseteq\oline{X_{s',o',i'}\times\oline{\semD{t'}}}$,
an inclusion that carries over to finite subsets.

For the ``only if'' direction, whenever one of the four inequalities fails we exhibit
a singleton relation in $\semD{s\to^i_o t}\setminus\semD{s'\to^{i'}_{o'}t'}$, that is,
a pair whose input is not in $X_{s,o,i}$ or whose output is in $\semD{t}$, and whose
input is in $X_{s',o',i'}$ while its output is not in $\semD{t'}$:
\begin{itemize}
\item if $s'\not\leq s$, take $d\in\semD{s'}\setminus\semD{s}$ and the pair
$((d,\varnothing_{\mathsf w},\varnothing_{\mathsf r}),\Omega)$, whose input is not in
$X_{s,o,i}$ since $d\neq\mho$;
\item if $i'\not\leq i$, take $d\in\semD{i'}\setminus\semD{i}$ and the pair
$((\mho,\varnothing_{\mathsf w},\Set{(\mho,d)}),\Omega)$, the read approximation
$\Set{(\mho,d)}$ being in $\semD{()\to i'}$ but not in $\semD{()\to i}$;
\item if $o\not\leq o'$, take $d\in\semD{o}\setminus\semD{o'}$ and the pair
$((\mho,\Set{(d,\Omega)},\varnothing_{\mathsf r}),\Omega)$, the write approximation
$\Set{(d,\Omega)}$ being in $\semD{\TChan{o'}}$ but not in $\semD{\TChan{o}}$;
\item if $t\not\leq t'$, take $d\in\semD{t}\setminus\semD{t'}$ and the pair
$((\mho,\varnothing_{\mathsf w},\varnothing_{\mathsf r}),d)$, whose input is in both
domains.\qedhere
\end{itemize}
\end{proof}

\PidLawsProp*
\begin{proof}
We give the proof for the direct interpretation. For the set-builder form of
equation~\eqref{eq:chan}, which interprets $\TChan t$ as the set of the interpretations of
the supertypes of $t$, each step is immediate, the interpretation of $t$ itself replacing
the witnesses below.

\emph{Item 1.} If $s\leq t$, an approximation that maps no element of $\semD{t}$ to
$\Omega$ maps no element of $\semD{s}\subseteq\semD{t}$ to $\Omega$, so
$\semD{\TChan t}\subseteq\semD{\TChan s}$. If $s\not\leq t$, take
$d\in\semD{s}\setminus\semD{t}$: the approximation $\Set{(d,\Omega)}$ is in
$\semD{\TChan t}$ but not in $\semD{\TChan s}$.

\emph{Item 3.} The ``if'' direction follows from item 1. For the ``only if''
direction, if $s_n\not\leq t$ for every $n\in N$, pick $d_n\in\semD{s_n}\setminus\semD{t}$
for each $n$: the approximation $\Set{(d_n,\Omega)\mid n\in N}$ is in
$\semD{\TChan t}$, since no $d_n$ is in $\semD{t}$, and in no $\semD{\TChan{s_n}}$, since
it maps $d_n\in\semD{s_n}$ to $\Omega$.

\emph{Item 2.} A finite set of pairs belongs to both $\semD{\TChan{t_1}}$ and
$\semD{\TChan{t_2}}$ if and only if it maps to $\Omega$ no element of $\semD{t_1}$ and
no element of $\semD{t_2}$, that is, no element of $\semD{t_1\lor t_2}$: this is the
first law. For the second, $t_1\land t_2\leq t_k$ gives
$\TChan{t_k}\leq\TChan{t_1\land t_2}$ by item 1, for $k=1,2$. Conversely,
$\TChan{t_1\land t_2}\leq\TChan{t_1}\lor\TChan{t_2}$ would give, by item 3,
$t_1\leq t_1\land t_2$ or $t_2\leq t_1\land t_2$, that is, $t_1\leq t_2$ or
$t_2\leq t_1$: the inclusion is strict whenever $t_1$ and $t_2$ are incomparable.
\end{proof}

\danielem{What is the point of the next two paragraphs?}
\subsection{Type variables}\label{app:enc-variables}
From Section~\ref{sec:ann-inference} on, types may contain type variables
$\alpha,\beta,\gamma,\delta$, which are leaves of the type terms and do not count as type
constructors for contractivity. Subtyping extends to them in the standard
way~\cite[Section~3.3]{castagna2022programming}. Every element $d$ of $\DomainFCB$ and of
$\Domain$ carries a finite set of type variables, its \emph{tags} $\mathsf{tags}(d)$ (the
symbols $\mho$ and $\Omega$, which are not elements of $\Domain$, carry none); in both
domains a type variable $\alpha$ is interpreted as the set of the elements $d$ with
$\alpha\in\mathsf{tags}(d)$, so that the membership predicate of the direct interpretation
gains the clause $(d:\alpha)=\alpha\in\mathsf{tags}(d)$; all the other clauses ignore tags,
which are therefore irrelevant on ground types. The encoding is extended with
$\enc{\alpha}=\alpha$, and Definition~\ref{def:subtyping} is read with these tagged
interpretations. The proof of Theorem~\ref{thm:encoding} carries
over to the tagged domains: $\varphi$ gives $\varphi(d)$ the tags of $d$ (and empty tags
to the auxiliary pairs and constants it introduces), $\psi$ gives $\psi(e)$ the tags of
$e$, a junk element tagged $L$ being read back as $\jzero$ tagged $L$, so that the clause
for type variables is preserved and reflected. Hence Theorem~\ref{thm:encoding} and
Corollary~\ref{cor:same-subtyping} hold for types with type variables. Moreover
$\enc{\cdot}$ commutes with type substitutions, $\enc{t\sigma}=\enc{t}\enc{\sigma}$ where
$\enc{\sigma}$ maps each $\alpha$ to $\enc{\sigma(\alpha)}$. Subtyping is therefore
preserved by type substitutions: if $s\subt t$ then $\enc{s}\subt\enc{t}$, hence
$\enc{s}\enc{\sigma}\subt\enc{t}\enc{\sigma}$ by the same property for the types
of~\cite{FCB08}~\cite[Section~3.3]{castagna2022programming}, that is, $s\sigma\subt t\sigma$.

\subsection{Reading back types of~\texorpdfstring{\cite{FCB08}}{FCB08}}\label{app:enc-readback}
The inference of Section~\ref{sec:ann-inference} solves its constraints on their
encodings, and the substitutions it obtains may have images that are not encodings, such
as $\Tany\setminus\CPoint$ (Appendix~\ref{app:elab-tallying}). These images must be read back
as types of Section~\ref{sec:syntax}, and the readback cannot be defined on all types
of~\cite{FCB08}. Let $T$ be the type $\enc{\TTuple}$ of Section~\ref{sec:fcb-encoding}. A
readback that commutes with the connectives, is monotone, and maps $\enc{t}$ to $t$ would
send $\TInt\times T$, which contains $\enc{\Tuple{\TInt,\Tany,\dots,\Tany}}$ at every arity
and is disjoint from $\enc{\Tuple{}}=\Cnil$ and from
$\enc{\Tuple{\neg\TInt,\Tany,\dots,\Tany}}$ at every arity, to a type
whose tuples are exactly those of positive arity with an integer first component. No type
of ours denotes that set: a regular type mentions finitely many arities, and above them it
contains either all tuples or none. We therefore define the readback on a class of types
that contains the encodings.

Write $\mathsf{In}(X_1,X_2,X_3)$ for
$(X_1\lor\CPoint)\times\bigl(\Wch(X_2\to\CPoint)\times(\CPoint\to X_3)\bigr)$, the shape of
the domain of an encoded arrow. Let $\EncTypes$ be the greatest set of contractive regular
types of~\cite{FCB08} in which every element has one of the following forms, each $X$,
$X_k$ on the right being again in $\EncTypes$ and $n\geq1$:
\[
\begin{array}{r@{\;\;}c@{\;\;}l}
X &::=& \alpha \mid b \mid \Cnil \mid T \mid \Tempty \mid X\lor X \mid \neg X
  \mid X_1\times(X_2\times\cdots\times(X_n\times\Cnil)) \mid \Tany\times T\\
  &\mid& \Wch(X\to\CPoint) \mid \Wfun\bigl(\mathsf{In}(X_1,X_2,X_3)\to X_4\bigr)\\
  &\mid& \CPoint \mid \ell_{\Wch} \mid \ell_{\Wfun} \mid \mathsf{In}(X_1,X_2,X_3)
\end{array}
\]
where $b$ ranges over the basic types. The forms of the third line are the \emph{junk}
forms. The variables of a type in $\EncTypes$ occur only in the positions of an $X$, which
we call \emph{$X$-positions}.

\begin{definition}[Readback]\label{def:readback}
The \emph{readback} $\dec{\cdot}:\EncTypes\to\Types$ is defined by the clauses
\[
\begin{array}{r@{\;=\;}l@{\qquad}r@{\;=\;}l@{\qquad}r@{\;=\;}l}
\dec{\alpha} & \alpha & \dec{b} & b & \dec{\Cnil} & \Tuple{}\\
\dec{T} & \TTuple & \dec{X_1\lor X_2} & \dec{X_1}\lor\dec{X_2} & \dec{\neg X} & \neg\dec{X}\\
\multicolumn{6}{c}{\dec{X_1\times(\cdots\times(X_n\times\Cnil))} = \Tuple{\dec{X_1},\dots,\dec{X_n}}
\qquad\qquad \dec{\Tany\times T} = \TTuple\setminus\Tuple{}}\\
\multicolumn{6}{c}{\dec{\Wch(X\to\CPoint)} = \TChan{\dec{X}}}\\
\multicolumn{6}{c}{\dec{\Wfun(\mathsf{In}(X_1,X_2,X_3)\to X_4)} = \dec{X_1}\to^{\dec{X_3}}_{\dec{X_2}}\dec{X_4}}
\end{array}
\]
and $\dec{X}=\Tempty$ when $X$ is $\Tempty$ or has a junk form. A type substitution
$\hat\theta$ whose images are in $\EncTypes$ is read back as the substitution
$\dec{\hat\theta}$ that maps each variable $\alpha$ to $\dec{\hat\theta(\alpha)}$.
\end{definition}

\noindent The forms are pairwise distinct as trees, except that $T$ is also the union $\Cnil\lor(\Tany\times T)$; the clause
for $T$ takes precedence, and the two readings give equivalent results anyway. The clauses
for the constructors are guarded, the others are covered by contractivity, and a regular tree
is mapped to a regular tree, so $\dec{X}$ is a type. The set $\EncTypes$ contains $\enc{t}$
for every type $t$, since every clause of $\enc{\cdot}$ produces one of these forms. It is
closed under the substitutions $\hat\theta$ whose images are in $\EncTypes$, since
substituting at an $X$-position does not change the form of any node. We call such
$\hat\theta$ \emph{$\EncTypes$-substitutions}. Two syntactic facts follow by coinduction:
\begin{equation}\label{eq:dec-syntactic}
\dec{\enc{t}}=t
\qquad\qquad
\dec{X\hat\theta}=\dec{X}\,\dec{\hat\theta}
\quad\text{for every $\EncTypes$-substitution $\hat\theta$.}
\end{equation}

\begin{lemma}[Readback]\label{lem:readback}
For every $X\in\EncTypes$ and every element $d$ of the tagged domain, $(d:\dec{X})$ if and
only if $(\varphi(d):X)$. Consequently, for all $X,Y\in\EncTypes$, $X\subt_\textsf{FCB}Y$
implies $\dec{X}\subt\dec{Y}$.
\end{lemma}

\begin{proof}[Proof sketch]
The first statement is proved as Claim~1 of the proof of Theorem~\ref{thm:encoding}, by the
same induction, with arbitrary components in $\EncTypes$ in place of the $\enc{t_k}$. Only
four cases are new.
\begin{itemize}[leftmargin=*,itemsep=1pt,topsep=2pt]
\item Junk constants: $\varphi(d)$ is never $\CPoint$, $\ell_{\Wch}$, or $\ell_{\Wfun}$.
\item $\mathsf{In}(X_1,X_2,X_3)$: no $\varphi(d)$ belongs to it. That would require a pair
whose second component is a pair whose second component is a relation. The second component
of $\varphi(d)$ is a nesting or $\Cnil$ when $d$ is a tuple, whose own second component is
again a nesting or $\Cnil$, and a relation when $d$ is a pid or function approximation.
\item $\Tany\times T$: $\varphi(d)$ belongs to it if and only if $d$ is a tuple of positive
arity.
\item Variables: at an $X$-position the element met is $\varphi(d')$, which carries the tags
of $d'$, or the untagged $\CPoint$ that stands for $\mho$ in $X_1\lor\CPoint$, which belongs to
that union whatever $X_1$ is.
\end{itemize}
The second statement follows, since both subtyping relations are inclusions in the tagged
domains and $\varphi^{-1}$ is monotone.
\end{proof}

\section{Soundness of the Type System}
\label{app:properties}

This appendix proves the properties stated in Section~\ref{sec:properties},
restated with their original numbers, together with the auxiliary results the
main text omits. It is organized by dependence in five parts. The first (Appendix~\ref{app:prop-values}) proves
the properties of pattern environments, values and matching that culminate in the
substitution lemma. The second (Appendix~\ref{app:prop-enriched}) introduces the enriched system $\Vdash$ announced in
Section~\ref{sec:properties} and the results that make its negative information
usable (generators, dichotomy, $\lor$-selection). The third (Appendix~\ref{app:prop-arrow-inv}) proves abstraction
inversion, the non-standard ingredient of subject reduction, from the
decomposition of arrow subtyping in the model of \cite{FCB08}. The fourth (Appendix~\ref{app:prop-contexts}) collects
context replacement, mailbox narrowing and intersection introduction. The fifth (Appendix~\ref{app:prop-sr})
proves the theorems of Section~\ref{sec:properties}: subject reduction, progress
and soundness. Throughout, the subtyping facts we use (emptiness, projections,
arrow decomposition) are facts about the \emph{encoded} types, since subtyping is
defined on them (Definition~\ref{def:subtyping}); where an argument is simpler on the
elements of the direct interpretation of Appendix~\ref{app:encoding-property}, we use
that interpretation instead, which induces the same subtyping relation
(Corollary~\ref{cor:same-subtyping}).

\subsection{Values, patterns and substitution}\label{app:prop-values}
\danielem{this sentence is mysterious}
The chain is: projections and pattern environments are monotone
(Lemma~\ref{lem:penv-mono}), judgments survive a smaller environment
(Lemma~\ref{lem:env-sub}), every value has a minimal type that decides its
run-time tests (Lemma~\ref{lem:min-types2}), whence canonical forms and the
correspondence between matching and typing (Corollary~\ref{cor:canonical},
Lemma~\ref{lem:matching}), which the substitution lemma
(Lemma~\ref{lem:substitution}) consumes.

\paragraph{Conventions} Three conventions are used below.
\begin{itemize}
	\item \emph{Inversion.}
Every expression form other than a value has exactly one typing rule besides
\rulename{t-sub}, so a derivation for it ends with that rule followed by instances of
\rulename{t-sub}; we invert derivations in this way without further comment (for values
in the enriched system the inversion is Lemma~\ref{lem:vee-principal}). 
\item \emph{Omitted
branches.} A derivation may omit the premise of a branch of \rulename{t-case} or
\rulename{t-receive} only when the refined type of the branch is empty
(Section~\ref{sec:typing-rules}); when a proof builds a derivation, the branches it omits
are always such branches, and it says which. 
\item \emph{Type variables.} From
Section~\ref{sec:ann-inference} on, types may contain type variables, and
Appendix~\ref{app:elaboration} applies the results up to Theorem~\ref{thm:narrowing} to
such types. They hold there by the same proofs, which use only subtyping facts that
Appendix~\ref{app:enc-variables} extends to types with variables, the projections of
Lemma~\ref{lem:penv-proj} included.
\end{itemize}

\begin{lemma}[Projections]\label{lem:penv-proj}
For every type $t\subt\Tuple{\Tany,\dots,\Tany}$ ($n$ components) and every
$k\le n$:\; $\semD{\pi_k(t)} = \Set{d_k \mid \Tuple{d_1,\dots,d_n}\in\semD{t}}$.
Consequently $\pi_k$ is monotone and invariant under $\simeq$,
$t \subt \Tuple{\pi_1(t),\dots,\pi_n(t)}$, and $\pi_k(t\sigma)\subt\pi_k(t)\sigma$
for every type substitution $\sigma$.
\end{lemma}
\begin{proof}
	 Write $t$ as a finite union of non-empty tuple types
	$\Tuple{t^r_1,\dots,t^r_n}$, $r\in R$ (Section~\ref{sec:typing}). Then
	$\semD{t}=\bigcup_{r\in R}\semD{t^r_1}\times\dots\times\semD{t^r_n}$ and, every
	$\semD{t^r_c}$ being non-empty, the $k$-th projection of $\semD{t}$ is
	$\bigcup_{r\in R}\semD{t^r_k}=\semD{\pi_k(t)}$. The projection of a set is monotone and
	depends only on the set, and $\semD{t}$ is contained in the product of its
	projections. For types with type variables (Appendix~\ref{app:enc-variables}),
	$\pi_k(t)$ is computed in the same way on the disjunctive normal form of $t$,
	discarding the literals on type variables and the empty tuple types, and the
	characterization still holds, since a literal on a type variable constrains only
	the tags of a tuple, not its components. Then $t\subt\Tuple{\pi_1(t),\dots,\pi_n(t)}$
	and the stability of subtyping under type substitutions give
	$t\sigma\subt\Tuple{\pi_1(t)\sigma,\dots,\pi_n(t)\sigma}$, whence
	$\pi_k(t\sigma)\subt\pi_k(t)\sigma$ by projecting (if some $\pi_c(t)\sigma$ is empty,
	so is $t\sigma$, and $\pi_k(t\sigma)\simeq\Tempty$).

\end{proof}

\begin{lemma}[Pattern environments]\label{lem:penv-mono}
If $t \subt t' \subt \accept{p}$, then $(\ptenv{t}{p})(x) \subt
    (\ptenv{t'}{p})(x)$ for every capture variable $x$ of $p$; in particular,
    $\ptenv{t}{p}$ depends on $t$ only up to $\simeq$.
\end{lemma}

\begin{proof}
By induction on $p$. For $p=x$ the two environments bind $x$ to $t$
and to $t'$; for $p=\tau$ they are empty; for $p=p_1\land p_2$ the induction
hypothesis applies to $p_1$ and to $p_2$, since $t\subt t'\subt\accept{p_i}$. For
$p=\Tuple{p_1,\dots,p_n}$, a capture variable $x$ of $p$ occurs in exactly one
$p_k$, and $(\ptenv{t}{p})(x)=(\ptenv{\pi_k(t)}{p_k})(x)$; by Lemma~\ref{lem:penv-proj}
$\pi_k(t)\subt\pi_k(t')$, and $\pi_k(t')\subt\accept{p_k}$ since $t'$ is below
$\Tuple{\accept{p_1},\dots,\accept{p_n}}$; the induction hypothesis applies. The
last claim is the first one applied in both directions.
\end{proof}

\noindent Lemma~\ref{lem:penv-proj}  is what the proofs below use whenever a branch is re-typed at a
type smaller than the original one.

\begin{definition}[Environment order]
We define $\Context' \subt \Context$ when
\begin{enumerate}
	\item $\dom\Context \subseteq \dom{\Context'}$
	\item $\Context'(x) \subt \Context(x)$ for every $x \in \dom\Context$.	
	
\end{enumerate} 
\end{definition}

And environment refines another if it defines more variables, and for the common variables it assigns smaller types.

\begin{lemma}[Environment subsumption]\label{lem:env-sub}
 If
$\wte[i][o]{\Context}{e}{t}$ and $\Context' \subt \Context$, then
$\wte[i][o]{\Context'}{e}{t}$.
\end{lemma}

\begin{proof}[Proof sketch]
By induction on the derivation, simultaneously for all $\Context'\subt\Context$. An
instance of \rulename{t-var} is followed by one of \rulename{t-sub}, since
$\Context'(x)\subt\Context(x)$; the rules that extend the environment
(\rulename{t-abs}, \rulename{t-case}, \rulename{t-receive}) do so with bindings that
override those of $\Context$ and of $\Context'$ alike, so the induction hypothesis
applies to their premises; no side condition mentions the environment, so every rule
can be reapplied.
\end{proof}

\noindent This lemma subsumes weakening. It is
consumed by reduction and narrowing, which re-type branch bodies under environments
\emph{smaller} than those of the original derivation: narrowing the input decoration
from $i$ to $i' \subt i$ turns the environment
$\ptenv{((i\land\accept{p_j})\setminus\bigvee_{h<j}\accept{p_h})}{p_j}$ of a $\Recv{}$
branch into
$\ptenv{((i'\land\accept{p_j})\setminus\bigvee_{h<j}\accept{p_h})}{p_j}$, which is
pointwise smaller by Lemma~\ref{lem:penv-mono}, and the lemma lets the original
branch derivations be reused.

The next lemma ties the run-time matching of Section~\ref{sec:opsem} to
derivability. It attaches to every value $v$ its \emph{minimal type}
$\mathsf{min}(v)$, defined by induction on $v$:
\[
\mathsf{min}(\Const)=\Const \quad
\mathsf{min}(\PID^t)=\TChan t \quad
\mathsf{min}(\Fun^{l}x.e)=l \quad
\mathsf{min}(\Tuple{v_1,\dots,v_n})=\Tuple{\mathsf{min}(v_1),\dots,\mathsf{min}(v_n)}
\]
(where the annotation $l$ is read as the intersection type it spells).

\begin{restatable}[Minimal types for values]{lemma}{MinTypesLemma1}\label{lem:min-types1}
For every value $v$, 
\begin{enumerate}[noitemsep,topsep=2pt]
    \item $\mathsf{min}(v) \not\simeq \Tempty$
    \item if $v$ is well-typed under $\Context$ (i.e., $\wte{\Context}{v}{s}$ for some
    $s$), then for every type $t$:\; $\wte{\Context}{v}{t}$ if and only if
    $\mathsf{min}(v) \subt t$;
    \item for every $t_1,t_2$, if $\wte{\Context}{v}{t_1}$ and $\wte{\Context}{v}{t_2}$, then $\mathsf{min}(v) \subt t_1\wedge t_2$ and $\wte{\Context}{v}{t_1\wedge t_2}$;
\end{enumerate}
\end{restatable}

Note that the statement cannot be made simpler: the minimal type is defined also for functions that are ill-typed.

\begin{proof}[Proof sketch]
	\emph{Item 1.} By induction on $v$: constant and pid types are nonempty
	by definition of the interpretation, an intersection of decorated arrow types
	always contains the empty approximation (i.e., the function diverging on all inputs), and a tuple of nonempty types is
	nonempty. This is what makes the dichotomy of Lemma~\ref{lem:min-types2} \emph{exclusive}. 
	
	\emph{Item 2.} A derivation for a value that does not end with
	\rulename{t-sub} assigns it exactly the shape of $\mathsf{min}(v)$ (by
	inspection of the value rules \rulename{t-const}, \rulename{t-pid},
	\rulename{t-abs}, \rulename{t-tuple}), so every derived type is above
	$\mathsf{min}(v)$; conversely, the same rules derive
	$\wte{\Context}{v}{\mathsf{min}(v)}$, the premises of \rulename{t-abs} being
	available since $v$ is well-typed under $\Context$, and \rulename{t-sub} closes upward.
	
	\emph{Item 3.} By item 2, $\mathsf{min}(v) \subt t_1$ and $\mathsf{min}(v) \subt t_2$, thus $\mathsf{min}(v) \subt t_1\wedge t_2$. And, always by item 2,  $\wte{\Context}{v}{t_1\wedge t_2}$.
\end{proof}

\begin{restatable}[Minimal type dychotomy]{lemma}{MinTypesLemma2}\label{lem:min-types2}
	For every value $v$, for every test type $\tau$:
	\begin{enumerate}[noitemsep,topsep=2pt]
		\item $v\in\tau$ if and only if
		$\mathsf{min}(v)\subt\tau$;
		\item either $\mathsf{min}(v)\subt\tau$
		or $\mathsf{min}(v)\subt\neg\tau$ 
	\end{enumerate}
\end{restatable}

In the second item, the disjunction is exclusive since $\mathsf{min}(v)$ is never
empty.

\begin{proof}[Proof sketch]
Membership is defined in Section~\ref{sec:opsem} as
$v\in\tau \iff \testof{v}\subt\tau$, and the item rests on the following claim, which
says that \emph{test types cannot split values}:
\[
\text{for every value $v$ and every test type $\tau$,}\qquad
\testof{v}\subt\tau \quad\text{or}\quad \testof{v}\land\tau\simeq\Tempty.
\]
The claim is proved by induction on the (finite) test type $\tau$, for all values at
once. A test type built with a constructor either contains $\testof{v}$ or is disjoint
from it: a singleton or basic type contains the constant $\testof{c}=c$ or is disjoint
from it, and is disjoint from every other shape; in the direct interpretation
(Corollary~\ref{cor:same-subtyping}), $\TChan\Tempty$ is the whole pid
summand of the domain, so it contains $\testof{\PID^t}=\TChan\Tempty$ and is disjoint
from every other shape; $\TFun$ is the only test constructor that meets the function
summand, so a test type contains all of $\semD{\TFun}$ or none of it (that $\semD{\TFun}$
is a proper subset of the summand is immaterial); $\TTuple$ contains every tuple; and
for a tuple test $\Tuple{\tau_1,\dots,\tau_n}$, distinct arities are disjoint, while at
equal arity the induction hypothesis at the components $v_k$ gives either
$\testof{v_k}\subt\tau_k$ for every $k$, whence $\testof{v}\subt\tau$, or
$\testof{v_k}\land\tau_k\simeq\Tempty$ for some $k$, whence
$\testof{v}\land\tau\simeq\Tempty$. Union and negation preserve the two alternatives.
Next, $\mathsf{min}(v)\subt\testof{v}$: by equality for constants, by
Proposition~\ref{prop:pid-laws}(1) for pids, by Proposition~\ref{prop:arrow-factor} for
abstractions (every decorated arrow is below $\TFun=\Tempty\to^{\Tempty}_{\Tany}\Tany$),
and componentwise for tuples. Hence if $\mathsf{min}(v)\subt\tau$ then
$\testof{v}\subt\tau$, since otherwise $\mathsf{min}(v)\subt\testof{v}\land\tau\simeq\Tempty$
contradicts nonemptiness; the converse is transitivity. This is the first half. For the
dichotomy, if $\mathsf{min}(v)\not\subt\tau$ then $\testof{v}\not\subt\tau$, so the claim
gives $\testof{v}\land\tau\simeq\Tempty$, that is, $\mathsf{min}(v)\subt\testof{v}\subt\neg\tau$.
\end{proof}

\noindent Subject reduction needs both lemmas. By dychotomy, the outcome of
every run-time match is derivable, positively \emph{and} negatively: indeed a failed
match yields $\wte{}{v}{\neg\accept{p}}$. %
By Lemma~\ref{lem:min-types1} the types of a value are closed under
intersection, which no typing rule provides, and which lets the type of a
scrutinee be combined with the outcomes of the matches into the refined type
$(s\land\accept{p_j})\setminus\bigvee_{h<j}\accept{p_h}$ of \rulename{t-case} and
\rulename{t-receive} (in the enriched system, through
Lemma~\ref{lem:vee-principal}(3)). For abstractions and pids Lemma~\ref{lem:min-types1}  speaks of
\emph{derivability}, not of the model, a gap that is harmless for
$\mkkeyword{case}$ and $\Recv{}$ but not for applications
(Definition~\ref{def:enriched} below).

An immediate consequence of Lemma~\ref{lem:min-types1}  is that the top-level shape of a type
determines the top-level shape of its values, the standard ingredient of progress.

\begin{corollary}[Canonical forms]\label{cor:canonical}
Let $\wte{\Context}{v}{t}$.
\begin{enumerate}[noitemsep,topsep=2pt]
    \item If $t \subt \TInt$, then $v$ is an integer constant $n$; if
    $t \subt \TAtom$, then $v$ is an atom $\Atom$.
    \item If $t \subt \TChan{\Tempty}$, then $v = \PID^s$ for some $s$ with
    $\TChan{s} \subt t$.
    \item If $t \subt \TFun$, then $v = \Fun^{l}x.e$ for some $l \subt t$.
    \item If $t \subt \Tuple{\Tany,\dots,\Tany}$ ($n$ components), then
    $v = \Tuple{v_1,\dots,v_n}$ with $\wte{\Context}{v_k}{\pi_k(t)}$ for every $k$.
\end{enumerate}
\end{corollary}

\begin{proof}
By Lemma~\ref{lem:min-types1}, $\mathsf{min}(v)\subt t$, and $\mathsf{min}(v)$ is
non-empty and lies, by its definition, in the summand of the domain determined by
the shape of $v$ (constants, pids, functions, tuples of the arity of $v$); the
summands being pairwise disjoint, $v$ has the shape of $t$'s summand, which gives
the four shape claims, with $\TChan{s}=\mathsf{min}(v)\subt t$ in item 2 and
$l=\mathsf{min}(v)\subt t$ in item 3. For item 4,
$\mathsf{min}(v_k)=\pi_k(\mathsf{min}(v))\subt\pi_k(t)$ by
Lemma~\ref{lem:penv-proj}, the components being non-empty, and
Lemma~\ref{lem:min-types1} types $v_k$ at it, $v_k$ being well-typed under
$\Context$ as a subterm of $v$.
\end{proof}

\noindent Progress uses it in every elimination position (the function of an
application is an abstraction, the operands of $+$ are integers, the target
of a send is a pid), and subject reduction for sends uses item 2: the
annotation on the pid value bounds, contravariantly, the type of the message
in flight.

\begin{restatable}[Matching]{lemma}{MatchingLemma}\label{lem:matching}
Let $v$ be a value well-typed under $\Context$. Then:
\begin{enumerate}[noitemsep,topsep=2pt]
	\item $\match{v}{p} \neq \textsf{fail}$ if and only if $\wte{\Context}{v}{\accept{p}}$.
	\item If $\wte{\Context}{v}{s}$ with $s \subt \accept{p}$, then
$\wte{\Context}{(\match{v}{p})(x)}{(\ptenv{s}{p})(x)}$ for every capture variable
$x$ of $p$.
\end{enumerate}
\end{restatable}

\begin{proof}[Proof sketch]
Both items are inductions on $p$. The first item reduces, at the leaves, to
Lemma~\ref{lem:min-types1}: the reduction of Section~\ref{sec:opsem} tests
$v\in\tau$ as $\testof{v}\subt\tau$, and $\accept{p}$ is a test type for
every $p$. For a tuple pattern $p=\Tuple{p_1,\dots,p_n}$, both sides fail unless $v$ is
a tuple $\Tuple{v_1,\dots,v_n}$ of the same arity, since $\mathsf{min}(v)$ is non-empty
and lies in the tuple summand of the arity of $v$; and then
$\mathsf{min}(v)\subt\Tuple{\accept{p_1},\dots,\accept{p_n}}$ if and only if
$\mathsf{min}(v_k)\subt\accept{p_k}$ for every $k$, the $\mathsf{min}(v_k)$ being
non-empty. The second item rests, in the tuple case, on the semantic
projections: $\wte{\Context}{v}{s}$ with $s\subt\Tuple{\Tany,\dots,\Tany}$ implies
$\wte{\Context}{v_k}{\pi_k(s)}$ (Corollary~\ref{cor:canonical}(4)), and
$\pi_k(s)\subt\accept{p_k}$ (Lemma~\ref{lem:penv-proj}).
\end{proof}

As customary, the proofs of subject reduction, for expressions but also for configurations, rely on a substitution lemma, which in our case is
stated in terms of pattern matching rather than single-variable
substitution.

\begin{lemma}[Substitution]\label{lem:substitution}
If $\wte[i][o]{\Context, \ptenv{s}{p}}{e}{t}$, \;$\wte[i'][o']{\Context}{v}{s}$
with $s \subt \accept{p}$, and $\match{v}{p} \neq \textsf{fail}$, then
$\wte[i][o]{\Context}{e\subst{v}{p}}{t}$.
\end{lemma}

\noindent The proof uses two auxiliary facts. First, the typing of a \emph{value}
does not depend on the bracket: if $\wte[i][o]{\Context}{v}{t}$, then
$\wte[i'][o']{\Context}{v}{t}$ for every bracket $[o',i']$, by an immediate
induction on the derivation, since the rules that can conclude a judgment for a
value (\rulename{t-const}, \rulename{t-pid}, \rulename{t-abs}, \rulename{t-tuple},
and \rulename{t-sub}) read no bracket; in particular, $\Self$ is not a value. Second,
the following simultaneous-substitution property, where $\sigma$ ranges over
substitutions of values for variables, and bound and capture variables are chosen
distinct from the variables of $\dom\sigma$ and from the free variables of its
images: \emph{if $\wte[i][o]{\Context, \ContextD}{e}{t}$ and
$\wte{\Context}{\sigma(x)}{\ContextD(x)}$ for every
$x \in \dom\ContextD = \dom\sigma$, then $\wte[i][o]{\Context}{e\sigma}{t}$.}

\begin{proof}
We prove the simultaneous-substitution property by induction on the derivation of
$\wte[i][o]{\Context, \ContextD}{e}{t}$, for all brackets, environments, and
substitutions at once, by cases on the last rule.
\begin{itemize}[leftmargin=*,itemsep=1pt,topsep=2pt]
\item \rulename{t-sub}: the induction hypothesis, then \rulename{t-sub}.
\item \rulename{t-const}, \rulename{t-pid}, \rulename{t-self}: $e\sigma=e$, and the
rule, which reads no environment, concludes under $\Context$.
\item \rulename{t-var}: $e=y$. If $y\in\dom\ContextD$, then $e\sigma=\sigma(y)$ and
$t=\ContextD(y)$, and the hypothesis on $\sigma$, moved to the bracket $[o,i]$ by the
first fact, concludes. Otherwise $e\sigma=y$ and $t=\Context(y)$, and
\rulename{t-var} concludes.
\item \rulename{t-abs}: $e=\Fun^{l}x.e'$ with
$l=\bigwedge_j t_j\to_{o_j}^{i_j}t'_j$ and premises
$\wte[i_j][o_j]{\Context,\ContextD,x:t_j}{e'}{t'_j}$. Since $x\notin\dom\ContextD$,
the environment $\Context,\ContextD,x:t_j$ is $(\Context,x:t_j),\ContextD$, and
$e\sigma=\Fun^{l}x.(e'\sigma)$. The hypothesis on $\sigma$ holds under
$\Context,x:t_j$ by Lemma~\ref{lem:env-sub}, so the induction hypothesis, at the
bracket $[o_j,i_j]$, gives $\wte[i_j][o_j]{\Context,x:t_j}{e'\sigma}{t'_j}$ for every
$j$, and \rulename{t-abs} concludes.
\item \rulename{t-app}, \rulename{t-tuple}, \rulename{t-plus}, \rulename{t-send}: the
premises are typed under $\Context,\ContextD$ at the ambient bracket; the induction
hypothesis applies to each of them, and the same rule concludes.
\item \rulename{t-spawn}: $e=\Spawn{e'}{s'}$, with premise
$\wte[s'][s']{\Context,\ContextD}{e'}{t''}$; the induction hypothesis applies at the
bracket $[s',s']$.
\item \rulename{t-case}: $e=\Case{e_0}{\seqof{\Branch{p}{e}}}$, with premises
$\wte[i][o]{\Context,\ContextD}{e_0}{u}$ and, for every branch $j$ that the derivation
does not omit, $\wte[i][o]{\Context,\ContextD,E_j}{e_j}{t}$ with $E_j=\ptenv{r_j}{p_j}$
and $r_j=(u\land\accept{p_j})\setminus\bigvee_{h<j}\accept{p_h}$, and the side
condition $u\subt\bigvee_j\accept{p_j}$. The capture variables of $p_j$ are not in
$\dom\ContextD$, so $e\sigma=\Case{e_0\sigma}{\seqof{\Branch{p}{e\sigma}}}$ and
$\Context,\ContextD,E_j=(\Context,E_j),\ContextD$. The induction hypothesis gives
$\wte[i][o]{\Context}{e_0\sigma}{u}$ and, after weakening the hypothesis on $\sigma$
to $\Context,E_j$ (Lemma~\ref{lem:env-sub}), $\wte[i][o]{\Context,E_j}{e_j\sigma}{t}$.
The type $u$, the refined types, and the side condition depend on types only and are
unchanged, so \rulename{t-case} concludes, omitting the same branches.
\item \rulename{t-receive}: as for \rulename{t-case}, with
$r_j=(i\land\accept{p_j})\setminus\bigvee_{h<j}\accept{p_h}$ and no scrutinee.
\end{itemize}
For the lemma, let $\ContextD\eqdef\ptenv{s}{p}$ and $\sigma\eqdef\match{v}{p}$, so
that $e\subst{v}{p}=e\sigma$. The capture variables of $p$ being pairwise distinct,
$\dom\sigma=\dom\ContextD$. By the first fact $\wte{\Context}{v}{s}$, so $v$ is
well-typed under $\Context$ and the second half of Lemma~\ref{lem:matching} gives
$\wte{\Context}{\sigma(x)}{\ContextD(x)}$ for every $x\in\dom\sigma$. The
simultaneous-substitution property concludes.
\end{proof}

\noindent Subject reduction uses the lemma at $p = x$ for $\beta$-reduction (where
$\accept{x} = \Tany$ imposes nothing) and, for $\mkkeyword{case}$ and $\Recv{}$, at
the refined branch type $s = (t\land\accept{p_j})\setminus\bigvee_{h<j}\accept{p_h}$
at which the branch environment is computed ($t$ the scrutinee type, respectively
the mailbox bound $i$).

\begin{remark}[Test types and negative information]\label{rem:test-negative}
Restricting patterns to test types is what makes the Substitution lemma
usable at the refined branch types. Reduction is ordered: the value selected by a $\mkkeyword{case}$ or
a $\Recv{}$ has \emph{failed} all earlier patterns, and the environment of
the selected branch accordingly subtracts their accepted types (rules
\rulename{t-case} and \rulename{t-receive}). Re-typing the branch after
reduction thus requires the \emph{negative} judgments
$\wte{}{v}{\neg\accept{p_h}}$ for all the failed $p_h$. Were arbitrary types
allowed in patterns, these judgments could be underivable:
$\Fun^{\TInt\to\TInt}\y.\y$ fails the test $\TInt\to\TBool$, and it does
belong to $\neg(\TInt\to\TBool)$ in the model, but no typing rule derives it: this is the classical negative-arrow gap of semantic subtyping, which
\cite{FCB08} closes with an ad hoc typing rule for abstractions. Test types
close it for free: since $\accept{p}$ is a test type for every pattern $p$,
the dichotomy of Lemma~\ref{lem:min-types2} makes every failed match
derivably negative, and the offending test above, which would split the
functions strictly below $\TFun$, is simply not expressible. Applications,
where the types confronted with the value are arbitrary, are another
matter: see Definition~\ref{def:enriched}.
\end{remark}

\subsection{The enriched system}\label{app:prop-enriched}
The results so far concern the judgment of Section~\ref{sec:typing}. For
subject reduction that system lacks one ingredient: no rule of
Section~\ref{sec:typing} derives \emph{negative} information for a value (which arrow types an abstraction does \emph{not} inhabit, which pid types a pid does not), and this information is needed when an application
confronts a value with a \emph{union} of domains: it is the gap of
Remark~\ref{rem:test-negative} resurfacing in \rulename{t-app}, where test
types cannot close it. The gap defeats subject reduction for $\vdash$
outright. Consider
$v \eqdef \Fun^{\TInt\to_{\Tempty}^{\Tany}(\Atom[{:}t]\lor\Atom[{:}f])}\y.\Atom[{:}t]$,
whose minimal type $m$ overlaps $\TInt\to_{\Tempty}^{\Tany}\Atom[{:}t]$
without being contained in it. Let
$t_1 \eqdef \TInt\to_{\Tempty}^{\Tany}\Atom[{:}t]$ and
$t_2 \eqdef \TFun\land\neg t_1$, so that $v$ has type
$\TFun = t_1\lor t_2$, but neither $t_1$ (the codomains clash) nor,
derivably, $t_2$; and let
$l \eqdef (t_1\to_{\Tempty}^{\Tany}\Tuple{t_1,t_1})
\land(t_2\to_{\Tempty}^{\Tany}\Tuple{t_2,t_2})$. Then
$\wte[\Tany][\Tempty]{}{(\Fun^{l}x.\Tuple{x,x})\,v}
{\,\Tuple{t_1,t_1}\lor\Tuple{t_2,t_2}}$, since
$l \subt \TFun\to_{\Tempty}^{\Tany}(\Tuple{t_1,t_1}\lor\Tuple{t_2,t_2})$;
but the reduct $\Tuple{v,v}$ has \emph{no} type below
$\Tuple{t_1,t_1}\lor\Tuple{t_2,t_2}$ in the system of
Section~\ref{sec:typing}: its minimal type is $\Tuple{m,m}$, which
contains the mixed pairs, with one component in $m{\setminus}t_1$ and the other in $m{\setminus}t_2$, so subject reduction \emph{fails} for $\vdash$.
Nothing wrong can happen at run time: every element of $\sem{m}$ lies in
$t_1$ or in $t_2$; what fails is that $\vdash$, which knows $v$ only
through $m$, cannot say on which side $v$ falls, nor can any pattern
observe the split, since test types see of $v$ only $\TFun$. (The pid
analogue is $t_1 = \TChan{\Atom[{:}u]}$,
$t_2 = \TChan{\Tempty}\land\neg t_1$, and $v = \PID^{\Atom[{:}w]}$ with
$\Atom[{:}u]$ and $\Atom[{:}w]$ incomparable.) \cite{FCB08} recover such
negative judgments by extending their
abstraction rule with ad hoc negated arrows. We do not add such rules to
the system of Section~\ref{sec:typing}: we confine them, together with a
pid analogue, to an enriched system used only in the statements and proofs
of soundness.

\begin{definition}[Enriched typing]\label{def:enriched}
The judgments $\wtee[i][o]{\Context}{e}{t}$ and $\wtce{P}$ are derived by
the rules of Section~\ref{sec:typing}, read with $\Vdash$ in place of
$\vdash$ throughout (including \rulename{t-sub}, \rulename{t-thread}, and
\rulename{t-par}), plus the two rules
\begin{mathpar}
    \inferrule[t-abs-neg]{
        \wtee[i_j][o_j]{\Context, x : t_j}{e}{t'_j}
    }{
        \wtee[i'][o']{\Context}{\Fun^{\wedge t_j \to^{i_j}_{o_j} t'_j} x.~e}{\textstyle\bigwedge_j (t_j \to_{o_j}^{i_j} t'_j) \land \bigwedge_{h\in H}\neg a_h}
    }
    \and
    \inferrule[t-pid-neg]{~}{
        \wtee{\Context}{\PID^t}{\textstyle\TChan{t} \land \bigwedge_{h\in H}\neg\TChan{s_h}}
    }
\end{mathpar}
where $H$ is finite (possibly empty), the $a_h$ are decorated arrow types,
and both rules carry the side condition that the type in their conclusion
is not empty.
\end{definition}

\noindent At $H = \varnothing$ the new rules are \rulename{t-abs} and
\rulename{t-pid}, so every $\vdash$ derivation is a $\Vdash$ derivation.
Like \rulename{t-abs} and \rulename{t-pid}, the new rules ignore the
bracket and (for pids) the environment, and the premises of
\rulename{t-abs-neg} are exactly those of \rulename{t-abs}: consequently
Lemmas~\ref{lem:env-sub}, \ref{lem:matching}, \ref{lem:substitution},
\ref{lem:context} and Theorem~\ref{thm:narrowing}, as well as the
bracket-independence of the typing of values stated before
Lemma~\ref{lem:substitution}, hold verbatim with
$\Vdash$ in place of $\vdash$, by the same proofs (every induction over derivations acquires two base cases that are settled exactly like \rulename{t-abs} and \rulename{t-pid}; for Lemma~\ref{lem:matching}, Lemma~\ref{lem:vee-principal}(4) below replaces Lemma~\ref{lem:min-types1}), and it is these $\Vdash$
instances that the proofs below consume silently.

The enriched system settles the counterexample that motivated it:
$\wtee{}{v}{t_2}$ is derived by negating $t_1$ in rule
\rulename{t-abs-neg}, whence $\wtee{}{\Tuple{v,v}}{\Tuple{t_2,t_2}}$ and
preservation resumes. The design seems to us an improvement over
\cite{FCB08}, where the negated-arrow rule sits in the programmer-facing
system even though their inference algorithm never uses it: here even the
declarative system stays negation-free, and the rule is relegated only to the
proofs. What is given up is only the preservation of $\vdash$-typability
along reduction, a property no execution observes, while every guarantee
with observable content survives: reachable configurations are
$\Vdash$-typed and never stuck (Corollary~\ref{cor:soundness}), and values
that are pure data (abstraction- and pid-free) have exactly the same types
in the two systems.

The next lemma is the basic tool of the enriched system: the two summands
touched by the new rules are \emph{prime} with respect to unions of their
atoms, which is what makes the negative information of
Definition~\ref{def:enriched} consistent and, ultimately, what turns the
disjunctions that defeat $\vdash$ into derivable selections.

\begin{lemma}[Primality]\label{lem:atomic-prime}
Let $l$ be a finite non-empty intersection of decorated arrow types, $N$ a
finite (possibly empty) set of decorated arrow types, and $(s_h)_{h\in H}$
a finite family of types. Then:
\begin{enumerate}[noitemsep,topsep=2pt]
    \item $l \subt \bigvee_{a\in N} a$\; if and only if\; $l \subt a$ for
    some $a \in N$;
    \item $\TChan{t} \subt \bigvee_{h\in H}\TChan{s_h}$\; if and only if\;
    $s_h \subt t$ for some $h \in H$.
\end{enumerate}
\end{lemma}

\begin{proof}
Item 2 is the union law of Proposition~\ref{prop:pid-laws}. For
item 1 the ``if'' direction is trivial; for ``only if'', unfold
Definition~\ref{def:subtyping}: the encoding of a decorated arrow is
$\ell_{\Wfun}\times A$ for an arrow $A$ of \cite{FCB08}
(Section~\ref{sec:fcb-encoding}), and since
$\sem{\ell\times B}_\textsf{FCB}=\Set{\ell}\times\sem{B}_\textsf{FCB}$, products with the same singleton
first component satisfy $\bigwedge_j(\ell\times B_j)\simeq\ell\times\bigwedge_j B_j$,
$\bigvee_j(\ell\times B_j)\simeq\ell\times\bigvee_j B_j$ and
$\ell\times B\subt\ell\times A$ iff $B\subt A$. The statement thus becomes a
containment of an intersection of arrows of \cite{FCB08} in a finite union
of such arrows, and by \cite[Lemma~6.8]{FCB08}, whose characterization is an existential over the arrows of the union, such a containment holds if
and only if it holds for a single member, i.e.\ (transporting back), iff
$l \subt a$ for some $a \in N$.
\end{proof}

We will use the following easy consequence

\begin{corollary}\label{lem:atomic-prime-cor}
	Let $l$ be a finite non-empty intersection of decorated arrow types, $N$ a
	finite (possibly empty) set of decorated arrow types, and $(s_h)_{h\in H}$
	a finite family of types. Then:
	\begin{enumerate}[noitemsep,topsep=2pt]
		\item $l\land\bigwedge_{a\in N}\neg a$ is empty if and only if $l \subt a$ for some $a\in N$.
		\item $\TChan{t}\land\bigwedge_{h\in H}\neg\TChan{s_h}$ is empty if and only if
		$s_h \subt t$ for some $h\in H$
	\end{enumerate}
\end{corollary}
This follows from the fact that  $u\land\bigwedge_h\neg u_h \simeq \Tempty$ iff
$u \subt \bigvee_h u_h$.

The derivable $\Vdash$-types of a value are then completely described by a
set of \emph{generators} playing, jointly, the role that the single
minimal type plays for $\vdash$: there is no least element (the generators of a pid or an abstraction are incomparable), which is why
Lemma~\ref{lem:min-types1} has no verbatim $\Vdash$ analogue and why the
enriched system contains no intersection-introduction rule for values
(item 3 below shows the rule is nevertheless admissible, thanks to
primality).
\begin{definition}[Generators]
For a value $v$ define the set $\gen(v)$ of its \emph{generators} by
\[
\begin{array}{c}
	\gen(\Const)=\Set{\Const}
	\qquad
	\gen(\PID^t)=\Set{\TChan t\land\textstyle\bigwedge_{s\in N}\neg\TChan s
		\;\bigm|\; N \text{ finite}}
	\\[1mm]
	\gen(\Fun^{l}x.e)=\Set{l\land\textstyle\bigwedge_{a\in N}\neg a
		\;\bigm|\; N \text{ finite set of decorated arrows}}\\
	\gen(\Tuple{v_1,\dots,v_n})=
	\Set{\Tuple{u_1,\dots,u_n} \bigm| u_k\in\gen(v_k)}
\end{array}
\]
with the empty types discarded from each set.	
\end{definition}

\begin{lemma}[Properties of generators]\label{lem:vee-principal}
 Let $v$ be a value
$\Vdash$-well-typed under $\Context$ (i.e., $\wtee{\Context}{v}{s}$ for some $s$). Then:
\begin{enumerate}[noitemsep,topsep=2pt]
    \item every $u \in \gen(v)$ is non-empty and satisfies
    $u \subt \mathsf{min}(v)$; moreover $\mathsf{min}(v) \in \gen(v)$;
    \item $\wtee{\Context}{v}{t}$ if and only if $u \subt t$ for some
    $u \in \gen(v)$;
    \item for all $u_1, u_2 \in \gen(v)$ there is $u \in \gen(v)$ with
    $u \subt u_1\land u_2$; consequently $\wtee{\Context}{v}{t_1}$ and
    $\wtee{\Context}{v}{t_2}$ imply $\wtee{\Context}{v}{t_1\land t_2}$;
    \item for every test type $\tau$:\; $\wtee{\Context}{v}{\tau}$ iff
    $\mathsf{min}(v)\subt\tau$, and, when $v$ is also well-typed under $\Context$ in
    the system of Section~\ref{sec:typing}, iff $\wte{\Context}{v}{\tau}$; if $v$
    contains neither abstractions nor pids, then $\gen(v) = \Set{\mathsf{min}(v)}$,
    so $\Vdash$ and $\vdash$ derive exactly the same types for $v$.
\end{enumerate}
\end{lemma}

\begin{proof}
\emph{Item 1.} Non-emptiness is by construction; $u \subt \mathsf{min}(v)$
because negated atoms only shrink the positive part, componentwise for
tuples; $N = \varnothing$ (in every position) yields $\mathsf{min}(v)$,
which is non-empty by (the proof of) Lemma~\ref{lem:min-types1}.

\emph{Item 2.} ($\Leftarrow$) Each generator is derivable: by
\rulename{t-pid-neg}, respectively \rulename{t-abs-neg} (whose premises
are available since $v$ is $\Vdash$-well-typed under $\Context$), respectively, for tuples, by induction and \rulename{t-tuple}; one \rulename{t-sub} closes
up to $t$. ($\Rightarrow$) By induction on the derivation: if the last
rule is \rulename{t-sub}, the induction hypothesis and transitivity; if it
is a value rule, its conclusion is itself a generator (non-empty by the
side conditions), except for \rulename{t-tuple}, where the induction
hypothesis provides componentwise generators $u_k \subt t_k$ and
$\Tuple{u_1,\dots,u_n} \in \gen(v)$ is below $\Tuple{t_1,\dots,t_n}$.

\emph{Item 3.} For a pid, take the union $N_1\cup N_2$ of the two negation
sets: it is admissible by Lemma~\ref{lem:atomic-prime}(2), since a member
$s$ with $s \subt t$ would already make its own generator empty; likewise
for an abstraction by Lemma~\ref{lem:atomic-prime}(1); for tuples,
componentwise; constants are trivial. The consequence follows by item 2.

\emph{Item 4.} If $\mathsf{min}(v)\subt\tau$, then $\wtee{\Context}{v}{\tau}$ by
items 1 and 2, since $\mathsf{min}(v)\in\gen(v)$. Conversely let $u \subt \tau$ with
$u\in\gen(v)$: by the dichotomy of Lemma~\ref{lem:min-types2}, which does not
depend on typing, either $\mathsf{min}(v)\subt\tau$, as desired, or
$\mathsf{min}(v)\subt\neg\tau$, and then
$u \subt \mathsf{min}(v)\land\tau \subt \neg\tau\land\tau$ contradicts the
non-emptiness of $u$. When $v$ is well-typed under $\Context$ in the system of
Section~\ref{sec:typing}, $\mathsf{min}(v)\subt\tau$ iff $\wte{\Context}{v}{\tau}$ by
Lemma~\ref{lem:min-types1}. This hypothesis cannot be dropped: an abstraction
whose body is typable only in $\Vdash$, such as
$\Fun^{\Atom[{:}ok]\to_{\Tempty}^{\Tany}\Tuple{t_2,t_2}}z.\Tuple{v,v}$ with $v$ and
$t_2$ as in the counterexample above, is $\Vdash$-well-typed but not
$\vdash$-well-typed. For the second half, if $v$ contains neither abstractions nor
pids, no clause of $\gen$ introduces negations, so $\gen(v)=\Set{\mathsf{min}(v)}$;
such a $v$ is well-typed in both systems (by \rulename{t-const} and
\rulename{t-tuple}), and item 2 collapses to Lemma~\ref{lem:min-types1}.
\end{proof}

\noindent Item 4 is what lets all the $\mkkeyword{case}$ and $\Recv{}$
bookkeeping of subject reduction and progress (accepted types of patterns are test types) transfer to $\Vdash$ unchanged. Canonical forms survive
in the following form.

\begin{corollary}[Canonical forms, enriched]\label{cor:canonical-vee}
Let $\wtee{\Context}{v}{t}$. Then the shape claims of
Corollary~\ref{cor:canonical} hold unchanged. Moreover, if $t$ is a pid
type $\TChan{t'}$ then $v = \PID^s$ with $t' \subt s$, and if $t$ is a
decorated arrow type then $v = \Fun^{l}x.e$ with $l \subt t$.
\end{corollary}

\begin{proof}
For the shape claims, a generator is non-empty and its interpretation lies
within a single summand of the model, which determines the top-level shape of
$v$; for item 4 of Corollary~\ref{cor:canonical}, a generator
$\Tuple{u_1,\dots,u_n}\subt t$ has non-empty components, so
$u_k\subt\pi_k(t)$ by Lemma~\ref{lem:penv-proj} and
Lemma~\ref{lem:vee-principal}(2) types $v_k$ at $u_k$. For the positive bounds, Lemma~\ref{lem:vee-principal}(2) gives a
non-empty generator below $t$, and Lemma~\ref{lem:atomic-prime} lets its
negated atoms be discarded: from
$u = \TChan{s}\land\bigwedge_h\neg\TChan{s_h} \subt \TChan{t'}$, i.e.\
$\TChan{s} \subt \TChan{t'}\lor\bigvee_h\TChan{s_h}$, primality yields
$t' \subt s$ or some $s_h \subt s$, and the latter contradicts
$u \not\simeq\Tempty$; identically for arrows.
\end{proof}

\noindent This localized positive bound is exactly what the send case of
subject reduction and the transport step of the abstraction-inversion
proof below need.

\begin{lemma}[Value dichotomy]\label{lem:vee-dichotomy}
Let $v$ be a value $\Vdash$-well-typed under $\Context$. For every type $t$, either
$\wtee{\Context}{v}{t}$ or $\wtee{\Context}{v}{\neg t}$.
\end{lemma}

\begin{proof}
By induction on $v$; throughout, Lemma~\ref{lem:vee-principal}(2) reduces
the claim to exhibiting a generator below $t$ or below $\neg t$.

\emph{Constants.} $\sem{\mathsf{min}(v)}$ is the singleton $\Set{v}$, so
according to whether its element belongs to $\sem{t}$ or not,
$\mathsf{min}(v) \subt t$ or $\mathsf{min}(v) \subt \neg t$.

\emph{Pids, $v = \PID^s$.} Every generator is below $\TChan\Tempty$, so
only $t\land\TChan\Tempty$ matters. Put it in disjunctive normal form over
the atoms of the pid summand and collapse the positive literals by the
law $\TChan{t_1}\land\TChan{t_2} \simeq \TChan{t_1\lor t_2}$ of
Proposition~\ref{prop:pid-laws}:
\[
t\land\TChan\Tempty \;\simeq\;
\bigvee_{k\in K}\Bigl(\TChan{c_k}\land\bigwedge_{h\in H_k}\neg\TChan{d_{k,h}}\Bigr).
\]
Suppose first that some $k$ has $c_k \subt s$ and $d_{k,h} \not\subt s$
for every $h \in H_k$. Then
$u \eqdef \TChan{s}\land\bigwedge_{h}\neg\TChan{d_{k,h}}$ is a non-empty
generator (Lemma~\ref{lem:atomic-prime}(2)) with
$u \subt \TChan{c_k}\land\bigwedge_h\neg\TChan{d_{k,h}} \subt t$: so
$\wtee{\Context}{v}{t}$. Otherwise every $k$ has $c_k \not\subt s$ or some
$d_{k,h} \subt s$; let $N$ collect one $c_k$ of the former kind for each
such $k$. The generator
$u \eqdef \TChan{s}\land\bigwedge_{c\in N}\neg\TChan{c}$ is non-empty
(again primality: every $c \in N$ has $c \not\subt s$) and disjoint from
every disjunct: if the disjunct has a chosen $d_{k,h} \subt s$ then
$\TChan{s} \subt \TChan{d_{k,h}}$ clashes with the literal
$\neg\TChan{d_{k,h}}$; otherwise $c_k \in N$ and
$\TChan{s}\land\TChan{c_k} \subt \TChan{c_k}$ clashes with $\neg\TChan{c_k}$.
Hence $u \subt \neg t$ and $\wtee{\Context}{v}{\neg t}$.

\emph{Abstractions, $v = \Fun^{l}x.e$.} The same argument one summand
over, with $\TFun$ for $\TChan\Tempty$ and decorated arrows for the atoms
(no collapse of the positive literals is available, nor needed), the
non-emptiness checks now resting on Lemma~\ref{lem:atomic-prime}(1):
writing $t\land\TFun \simeq \bigvee_k(\bigwedge_m a_{k,m}\land
\bigwedge_h\neg b_{k,h})$, either some disjunct has $l \subt a_{k,m}$ for
every $m$ and $l \not\subt b_{k,h}$ for every $h$, and then $l\land\bigwedge_h\neg b_{k,h}$ is a generator below $t$, or from every
disjunct we may pick a failing positive atom (collected in $N$) or a
negated atom above $l$; then $l\land\bigwedge_{a\in N}\neg a$ is a
generator disjoint from every disjunct, hence below $\neg t$.

\emph{Tuples, $v = \Tuple{v_1,\dots,v_n}$.} Only the $n$-tuple summand of
$t$ matters, and it is a finite union of rectangles,
$t\land\Tuple{\Tany,\dots,\Tany} \simeq
\bigvee_{k\in K}\Tuple{t_{k,1},\dots,t_{k,n}}$: intersections of $n$-ary
products are products of intersections, and the difference of two products
is the union of the $n$ products obtained by complementing one coordinate.
By the induction hypothesis, for every $k$ and coordinate $c$ either
$\wtee{\Context}{v_c}{t_{k,c}}$ or $\wtee{\Context}{v_c}{\neg t_{k,c}}$. If some $k$ is
positive at every coordinate, Lemma~\ref{lem:vee-principal}(2) provides
generators $u_c \subt t_{k,c}$, and $\Tuple{u_1,\dots,u_n} \subt t$: so
$\wtee{\Context}{v}{t}$. Otherwise pick for each $k$ a negative coordinate
$c(k)$; by Lemma~\ref{lem:vee-principal}(3) each $v_c$ has a generator
$u_c$ below $\bigwedge\Set{\neg t_{k,c} \mid c(k) = c}$ (below $\Tany$ if
there is no such $k$), and $\Tuple{u_1,\dots,u_n}$ is disjoint from every rectangle, at coordinate $c(k)$, hence below $\neg t$.
\end{proof}

\begin{corollary}[Selection]\label{cor:vee-selection}
If $\wtee{\Context}{v}{\bigvee_{j\in K}s_j}$ with $K$ finite and non-empty, then
$\wtee{\Context}{v}{s_j}$ for some $j \in K$.
\end{corollary}

\begin{proof}
Induction on $|K|$; for a singleton there is nothing to prove. Otherwise
pick $j_0 \in K$: either $\wtee{\Context}{v}{s_{j_0}}$, and we are done, or
Lemma~\ref{lem:vee-dichotomy} gives $\wtee{\Context}{v}{\neg s_{j_0}}$, whence by
Lemma~\ref{lem:vee-principal}(3)
$\wtee{\Context}{v}{(\bigvee_{j\in K}s_j)\land\neg s_{j_0}}$, a type below
$\bigvee_{j\in K\setminus j_0}s_j$: apply the induction hypothesis.
\end{proof}

\noindent This is the union law \cite[Lemma~6.23]{FCB08} transported to
$\Vdash$: it supplies the selection step of abstraction inversion, to which we now
turn. No rule of Section~\ref{sec:typing} could pick the served arrow for the value
$v$ of the counterexample that motivated Definition~\ref{def:enriched}.

\subsection{Abstraction inversion}\label{app:prop-arrow-inv}
The $\beta$ case of subject reduction types $(\Fun^{l}x.e)\,v$ through
\rulename{t-app} at an arrow $t\to_o^i t'$ obtained from the annotation $l$ by
subsumption, and must recover, from $l\subt t\to_o^i t'$, which arrows of $l$
served the argument $v$, so as to reuse the corresponding premises of
\rulename{t-abs} on the body. Lemma~\ref{lem:arrow-inv} below does this. Its proof
runs on the encoded types, on which subtyping is defined (Definition~\ref{def:subtyping}), in
four steps: strip the $\Wfun$ tags, decompose
the containment between arrows (Lemma~\ref{lem:model-decomp}), decompose the
coverage of the ternary domain rectangles coordinatewise
(Lemma~\ref{lem:rect-coverage}) and read each coordinate back as a subtyping between
source types (Lemmas~\ref{lem:noval-purge} and~\ref{lem:read-coverage}, and the union
law of pid types), which gives Lemma~\ref{lem:folded-decomp}; then, in the proof of
Lemma~\ref{lem:arrow-inv} itself, turn the typing hypotheses into subtyping ones
(Corollary~\ref{cor:canonical-vee}, Lemma~\ref{lem:vee-principal}) and select the
served arrows by the $\lor$-selection of Corollary~\ref{cor:vee-selection}: this
is where the enriched system is consumed.
\daniele{if we follow my idea, this should become simpler, as we use the encoding as the definition}\beppem{Done: subtyping is now defined on the encodings (Definition~\ref{def:subtyping}), so the transport step is gone.}

\paragraph{Notation}
Throughout, $J$ is finite and non-empty,
$l = \bigwedge_{j\in J} t_j \to_{o_j}^{i_j} t'_j$, and we abbreviate the
encoded domains and codomains as
\[
D_j \;\eqdef\; (\enc{t_j}\lor\CPoint)\times\enc{\TChan{o_j}}\times(\CPoint\to\enc{i_j}),
\qquad
C_j \;\eqdef\; \enc{t'_j},
\]
and likewise $D$, $C$ for a target arrow $t \to_o^i t'$, so that
$\enc{s_1\to_o^i s_2} = \Wfun(D\to C)$ with
$\Wfun(X) = \ell_{\Wfun}\times X$ (Section~\ref{sec:fcb-encoding}; ternary
products are right-nested). Inside the lemmas below we write $\sem{\cdot}$
for the interpretation in the \emph{universal model} of
\cite[\S\S6.8--6.9]{FCB08} (over the constants extended as in
Section~\ref{sec:fcb-encoding}), in which subtyping coincides with
containment of interpretations, and $\subt$ between encoded types for the
subtyping it induces; elsewhere $\subt$ is the relation of
Definition~\ref{def:subtyping} on source types, so that $u\subt u'$ if and only if
$\enc{u}\subt\enc{u'}$. Recall from \cite[\S2.7]{FCB08}
that the empty relation belongs to the interpretation of every arrow type,
and note that no coordinate of the \emph{domain} rectangles above is ever
empty: $\CPoint$ inhabits the value coordinate, the (tagged) empty relation
the write coordinate, and the empty relation the read coordinate; the codomain
$C_j = \enc{t'_j}$ is a single value component, possibly empty.

\begin{lemma}[Purging the point]\label{lem:noval-purge}
For every source type $u$:\; $\enc{u} \subt \CPoint$ if and only if
$u \simeq \Tempty$.
\end{lemma}

\begin{proof}
If $u \simeq \Tempty$ then $\sem{\enc{u}} = \varnothing$ by
Definition~\ref{def:subtyping}, and the containment is trivial. Conversely, if
$u \not\simeq \Tempty$, pick $d \in \semD{u}$, which is nonempty by
Theorem~\ref{thm:encoding}, and let
$\varphi : \Domain \to \DomainFCB$ be the embedding of
Appendix~\ref{app:enc-subtyping}: membership is preserved,
$\varphi(d) \in \sem{\enc{u}}$, and $\varphi(d) \neq \CPoint$ ($\varphi$
maps every element to a constant other than $\CPoint$, namely itself or ${\Cnil}$
for the empty tuple, to a pair, or to a finite relation). Hence $\sem{\enc{u}} \not\subseteq \sem{\CPoint} = \Set{\CPoint}$.
\end{proof}

\begin{lemma}[Coverage of read capabilities]\label{lem:read-coverage}
For source types $w$ and $(i_j)_{j\in K}$, $K$ finite (possibly empty):
\[
\sem{\CPoint\to\enc{w}} \subseteq \bigcup_{j\in K}\sem{\CPoint\to\enc{i_j}}
\quad\iff\quad
\exists j\in K.\; w \subt i_j.
\]
\end{lemma}

\begin{proof}
The interpretation of $\CPoint\to X$ is the set of the finite relations whose
pairs with first component $\CPoint$ have their second component in
$\sem{X}$ (Section~\ref{sec:fcb-encoding}).
($\Leftarrow$) If $w \subt i_j$ then $\enc{w} \subt \enc{i_j}$
(Definition~\ref{def:subtyping}) and the containment is elementwise.
($\Rightarrow$) Contrapositive: assume $w \not\subt i_j$ for every
$j \in K$ and pick, by Definition~\ref{def:subtyping}, a witness
$x_j \in \sem{\enc{w}}\setminus\sem{\enc{i_j}}$ for each $j$. The capability
$\Set{(\CPoint,x_j) \mid j \in K}$, which may yield any of the $x_j$ (for
$K = \varnothing$, the empty relation), belongs to $\sem{\CPoint\to\enc{w}}$
but, being able to yield $x_j$, to no $\sem{\CPoint\to\enc{i_j}}$; so the
union does not cover.
\end{proof}

\noindent It is here that read capabilities must be \emph{multi-shot}: the
witness is a capability that may yield a different message at each
invocation, and it is what makes the read coordinate \emph{disjunctive}
with no auxiliary hypothesis, exactly what the two-receive
counterexample of Section~\ref{sec:correlation} demands. The write
coordinate is disjunctive for a different reason: by
Definition~\ref{def:subtyping} and the union law of pid types
(Proposition~\ref{prop:pid-laws}),
\begin{equation}\label{eq:pid-coverage}
\sem{\enc{\TChan{o'}}} \subseteq \bigcup_{j\in K}\sem{\enc{\TChan{o_j}}}
\quad\iff\quad
\TChan{o'} \subt \textstyle\bigvee_{j\in K}\TChan{o_j}
\quad\iff\quad
\exists j\in K.\; o_j \subt o'.
\end{equation}
Only the value coordinate, whose types are arbitrary, resists: its coverage
is genuinely a union, whence the $\lor$-selection step of the proof (and
the enriched system behind it).

\begin{lemma}[Coverage of rectangles]\label{lem:rect-coverage}
Let $A = A^1\times\dots\times A^m$ and, for $j$ in a finite set $J'$,
$A_j = A^1_j\times\dots\times A^m_j$ be $m$-ary products of types of
\cite{FCB08} with $\sem{A^c} \neq \varnothing$ for every coordinate $c$.
Then $\sem{A} \subseteq \bigcup_{j\in J'}\sem{A_j}$ if and only if for
every map $\kappa : J' \to \Set{1,\dots,m}$ there is a coordinate $c$ with
$\sem{A^c} \subseteq \bigcup_{j\in\kappa^{-1}(c)}\sem{A^c_j}$.
\end{lemma}

\begin{proof}
This is \cite[Lemma~6.4]{FCB08} for $m$-ary products, and its proof applies verbatim: a
point of $\sem{A}$ avoiding every $\sem{A_j}$ selects for each $j$ a failing coordinate
$\kappa(j)$, and conversely a $\kappa$ for which no coordinate is covered yields such a
point coordinatewise, the coordinates being non-empty.
\end{proof}

\begin{lemma}[Arrow decomposition, single target]\label{lem:model-decomp}
Let $(D_j)_{j\in J}$, $(C_j)_{j\in J}$, $D$, $C$ be types of \cite{FCB08}.
Then $\bigwedge_{j\in J}(D_j\to C_j) \subt D \to C$ if and only if for
every $J' \subseteq J$ at least one of the following holds:
\begin{enumerate}[noitemsep,topsep=2pt]
    \item[(a)] $\sem{D} \subseteq \bigcup_{j\in J'}\sem{D_j}$;
    \item[(b)] $J' \neq J$ and
    $\bigcap_{j\in J\setminus J'}\sem{C_j} \subseteq \sem{C}$.
\end{enumerate}
\end{lemma}

\begin{proof}
This is \cite[Lemma~6.8]{FCB08} instantiated with
$P = \Set{D_j\to C_j \mid j\in J}$ and the singleton $N = \Set{D\to C}$. When the map
$j \mapsto D_j\to C_j$ is not injective, a subset of $P$ corresponds to several $J'$
with the same union in (a) and the same intersection in (b), so the statement over
indices is equivalent to the statement over arrows.
\end{proof}

\begin{lemma}[Decomposition of the folded arrow]\label{lem:folded-decomp}
$l \subt t\to_o^i t'$ if and only if for every $J' \subseteq J$ at least
one of the following holds:
\begin{enumerate}[noitemsep,topsep=2pt]
    \item[(a)] for every $\kappa : J' \to \Set{1,2,3}$:\quad
    $\bigl(\kappa^{-1}(1) \neq \varnothing$ and
    $t \subt \bigvee_{j\in\kappa^{-1}(1)} t_j\bigr)$\; or\;
    $\bigl(\exists j\in\kappa^{-1}(2).\; o_j \subt o\bigr)$\; or\;
    $\bigl(\exists j\in\kappa^{-1}(3).\; i \subt i_j\bigr)$;
    \item[(b)] $J' \neq J$ and
    $\bigwedge_{j\in J\setminus J'} t'_j \subt t'$.
\end{enumerate}
\end{lemma}

\begin{proof}
By Definition~\ref{def:subtyping}, $l \subt t\to_o^i t'$ iff
$\enc{l} \subt \enc{t\to_o^i t'}$. Since the encoding commutes with the
connectives, $\enc{l} = \bigwedge_{j\in J}(\ell_{\Wfun}\times(D_j\to C_j))$
and $\enc{t\to_o^i t'} = \ell_{\Wfun}\times(D\to C)$. Products with the
same singleton first component satisfy
$\sem{\ell\times B} = \Set{\ell}\times\sem{B}$, so
$\bigwedge_j(\ell\times B_j) \simeq \ell\times\bigwedge_j B_j$ and
$\ell\times B \subt \ell\times A \iff B \subt A$ (both directions
pointwise). Hence the statement reduces to
$\bigwedge_{j\in J}(D_j\to C_j) \subt D\to C$, which
Lemma~\ref{lem:model-decomp} decomposes, for every $J'\subseteq J$, into
the coverage of $\sem{D}$ by the $\sem{D_j}$, $j\in J'$, or the codomain
clause.

\emph{Coverage is condition (a).} By Lemma~\ref{lem:rect-coverage} (the
coordinates of the rectangles are never empty), coverage holds iff for
every $\kappa: J'\to\Set{1,2,3}$ some coordinate is covered by its
$\kappa$-branches. For the value coordinate, since $\CPoint$ belongs to
every $\sem{\enc{t_j}\lor\CPoint}$ and to $\sem{\enc{t}\lor\CPoint}$,
coverage by $K \eqdef \kappa^{-1}(1)$ requires $K \neq \varnothing$ and
then, as the encoding commutes with the connectives,
\[
\sem{\enc{t}\lor\CPoint}\setminus
\bigcup_{j\in K}\sem{\enc{t_j}\lor\CPoint}
= \sem{\enc{t\setminus\textstyle\bigvee_{j\in K}t_j}}\setminus\Set{\CPoint},
\]
which is empty iff $t \subt \bigvee_{j\in K}t_j$ by
Lemma~\ref{lem:noval-purge}. The write coordinate is
equation~\eqref{eq:pid-coverage}, and the read coordinate is
Lemma~\ref{lem:read-coverage}.

\emph{The codomain clause is condition (b).} The codomain is a single value
component, and the encoding commutes with intersection:
\[
\bigcap_{j\in J\setminus J'}\sem{C_j}
= \sem{\textstyle\bigwedge_{j\in J\setminus J'}\enc{t'_j}}
= \sem{\enc{\textstyle\bigwedge_{j\in J\setminus J'} t'_j}}.
\]
Containment in $\sem{C} = \sem{\enc{t'}}$ is therefore
$\bigwedge_{j\in J\setminus J'} t'_j \subt t'$, by
Definition~\ref{def:subtyping}.
\end{proof}

\noindent Two families of instances are worth singling out. At
$J' = \varnothing$, condition (a) is vacuously false (the empty $\kappa$
has $\kappa^{-1}(c) = \varnothing$ for every $c$), so (b) yields the
\emph{unconditional} consequence
$\bigwedge_{j\in J} t'_j \subt t'$,
with no emptiness side condition: the classical escape clause for empty
domains never fires, because the point $(\CPoint,\varnothing,\varnothing)$
inhabits the domain of every encoded arrow. At $J' = J$, condition (b) is
unavailable, so (a) holds; its constant instances $\kappa \equiv 1,2,3$
give $t \subt \bigvee_{j\in J}t_j$, $o_j \subt o$ for some $j$, and
$i \subt i_j$ for some $j$. In particular, for a singleton $J$ the
lemma yields exactly $t \subt t_1$, $i \subt i_1$, $o_1 \subt o$, and $t'_1 \subt t'$, the componentwise factorization of
Proposition~\ref{prop:arrow-factor}.

\begin{lemma}[Abstraction inversion]\label{lem:arrow-inv}
Let $\wtee{\Context}{\Fun^{l}x.e}{t \to_o^i t'}$ with
$l = \bigwedge_{j\in J} t_j \to_{o_j}^{i_j} t'_j$, and let $v$ be a value
with $\wtee{\Context}{v}{t}$. Then there exists a non-empty $P \subseteq J$ such
that:
\begin{enumerate}[itemsep=1pt,topsep=2pt]
    \item $\wtee{\Context}{v}{t_j}$ for every $j \in P$;
    \item $\bigwedge_{j\in P} t'_j \subt t'$;
    \item $[o,i]$ narrows $[o_j,i_j]$ for every $j \in P$
    (Definition~\ref{def:bracket}).
\end{enumerate}
\end{lemma}

\begin{proof}
By Corollary~\ref{cor:canonical-vee} applied to the abstraction,
$l \subt t\to_o^i t'$ (the negated arrows of its generator are discarded
by primality); by Lemma~\ref{lem:vee-principal}(2) applied to $v$, there
is a non-empty generator $u \in \gen(v)$ with $u \subt t$. From
$u \subt t$ and the componentwise factorization of single arrows,
$t\to_o^i t' \subt u\to_o^i t'$, hence $l \subt u\to_o^i t'$.

Let $P \eqdef \Set{j\in J \mid \wtee{\Context}{v}{t_j}$, $o_j \subt o$, and
$i \subt i_j}$, that is, the branches whose domain the value derivably inhabits and whose bracket is narrowed by $[o,i]$ (Definition~\ref{def:bracket}), and apply
Lemma~\ref{lem:folded-decomp} to $l \subt u\to_o^i t'$ at
$J' \eqdef J\setminus P$. We claim condition (a) fails at $J'$. If
$J' = \varnothing$ this is vacuous (see above); otherwise define
$\kappa^* : J' \to \Set{1,2,3}$ by picking, for each $j \in J'$, one of
its failing conjuncts: $\kappa^*(j) \eqdef 1$ if $\wtee{\Context}{v}{t_j}$ is not
derivable, else
$2$ if $o_j \not\subt o$, else $3$ (and then $i \not\subt i_j$, since
$j \notin P$). No coordinate can satisfy (a) for $\kappa^*$: for the value
coordinate, the clause fails outright if $\kappa^{*-1}(1)=\varnothing$, and
otherwise $u \subt \bigvee_{j\in\kappa^{*-1}(1)} t_j$ would give
$\wtee{\Context}{v}{\bigvee_{j\in\kappa^{*-1}(1)} t_j}$ by
Lemma~\ref{lem:vee-principal}(2), hence, by
Corollary~\ref{cor:vee-selection}, $\wtee{\Context}{v}{t_j}$ for some $j$ with
$\kappa^*(j) = 1$,
contradicting the choice of $\kappa^*$; for the write coordinate,
$o_j \subt o$ for some $j$ with $\kappa^*(j) = 2$ contradicts it directly;
likewise for the read coordinate. So condition (b) holds at $J'$:
$J' \neq J$, i.e.\ $P \neq \varnothing$, and
$\bigwedge_{j\in P} t'_j \subt t'$, which is item 2. Items 1 and 3 hold by
the construction of $P$.
\end{proof}

\noindent In the $\beta$ case of subject reduction (Theorem~\ref{thm:sr-expr}), item 1
feeds the substitution lemma on the premises
$\wtee[i_j][o_j]{\Context, x:t_j}{e}{t'_j}$ of \rulename{t-abs}, item 3
(with the narrowing theorem) brings the results back to the ambient
bracket $[o,i]$, and item 2 delivers the expected type; when $P$ is not a
singleton, deriving the intersection $\bigwedge_{j\in P} t'_j$ for $e\subst{v}{x}$, which is not a value, requires the admissibility of
intersection introduction (Lemma~\ref{lem:inter-intro}). Note how the model pays for each item: without
the write component the per-branch $o_j \subt o$ of item 3 is false
(the codomain collapse of Section~\ref{sec:correlation}), and without the
multi-shot read capabilities so is the per-branch $i \subt i_j$ (the two-receive counterexample
there); in the unrepaired model both halves of item 3, and subject
reduction with them, fail outright. Note also that only the value coordinate consumes
the enriched system, the other two being disjunctive outright, and only for abstraction
and pid arguments: for a value $v$ built from constants and tuples,
$\sem{\mathsf{min}(v)}$ is a singleton, contained in a finite union only if contained in
one of its members, so the selection step is free and the lemma, with
$\mathsf{min}(v)\subt t_j$ in item 1, holds already for $\vdash$. It is precisely the
values whose run-time tests are capped, at $\TFun$ and $\TChan\Tempty$, that require the
negative information of Definition~\ref{def:enriched}.

\subsection{Contexts, narrowing and intersections}\label{app:prop-contexts}
Three more results that subject reduction consumes: a redex may be replaced in its
evaluation context by any expression of the same type (Lemma~\ref{lem:context}),
a judgment survives the narrowing of its bracket (Theorem~\ref{thm:narrowing},
the result of Section~\ref{sec:properties} that brings the body of an applied
abstraction back to the ambient bracket), and intersection introduction is
admissible for arbitrary expressions (Lemma~\ref{lem:inter-intro}, needed when
abstraction inversion selects several arrows).

\begin{lemma}[Context replacement]\label{lem:context}
If $\wte[i][o]{\Context}{\Eval[e]}{t}$, then there exists a type $s$ such
that $\wte[i][o]{\Context}{e}{s}$, and $\wte[i][o]{\Context}{\Eval[e']}{t}$
for every $e'$ such that $\wte[i][o]{\Context}{e'}{s}$.
\end{lemma}

\begin{proof}
By induction on $\Eval$. For $\Eval=\Hole$ take $s\eqdef t$. Otherwise $\Eval$ is built
from a smaller context $\Eval_1$ by one of the productions of Section~\ref{sec:syntax}
(an application, a tuple, a $\mkkeyword{case}$, an addition, or a send). After removing
the trailing instances of \rulename{t-sub}, the derivation of
$\wte[i][o]{\Context}{\Eval[e]}{t}$ ends with the rule of that construct, one of whose
premises types $\Eval_1[e]$ at some type $s_1$; the other premises and the side
conditions do not mention $\Eval_1[e]$, and for $\Case{\Eval_1}{\cdots}$ the branch
environments and the omitted branches depend on $\Eval_1[e]$ only through $s_1$. The
induction hypothesis on $\Eval_1$ yields $s$, and types $\Eval_1[e']$ at $s_1$ for every
$e'$ of type $s$; reapplying the rule, with the same premises otherwise and the same
omitted branches, and then the removed instances of \rulename{t-sub}, concludes.
\end{proof}

\noindent Note that the bracket does not change between the context and the
hole: evaluation contexts never enter the body of an abstraction (the
reduction strategy is weak), so the redex is typed at the ambient bracket.
The lemma is used by subject reduction both for the context rule of
expression reduction and for the configuration rules, all of which fire
inside an evaluation context.

\NarrowingThm*

\begin{proof}[Proof sketch]
Induction on the derivation, patching the rules that read the bracket:
\rulename{t-self} yields $\TChan{o'} \subt \TChan{o}$ by contravariance
($o \subt o'$), so one subsumption restores the original type; in
\rulename{t-app} the arrow type of the function is subsumed from
$t\to_o^i t'$ to $t\to_{o'}^{i'} t'$ (contravariant in $i$, covariant in
$o$); the branch environments of \rulename{t-receive}, computed at the refined types
$(i'\land\accept{p_j})\setminus\bigvee_{h<j}\accept{p_h}$ instead of
$(i\land\accept{p_j})\setminus\bigvee_{h<j}\accept{p_h}$, are pointwise smaller
(Lemma~\ref{lem:penv-mono}); the new derivation omits exactly the branches the given
one omits, which is allowed since their refined types have shrunk and were already empty,
and Lemma~\ref{lem:env-sub} re-types the others. The brackets occurring in
the premises of \rulename{t-abs} belong to the arrows of the annotation, and the premise
of \rulename{t-spawn} is typed at the bracket fixed by its annotation: both are
untouched.
\end{proof}

\noindent The statement does not require coherence of either bracket (cf.\
Remark~\ref{rem:coherence}).

\begin{lemma}[Intersection introduction]\label{lem:inter-intro}
If $\wtee[i][o]{\Context}{e}{t_1}$ and $\wtee[i][o]{\Context}{e}{t_2}$, then
$\wtee[i][o]{\Context}{e}{t_1\land t_2}$.
\end{lemma}

\begin{proof}[Proof sketch]
By induction on $e$ and, for a given $e$, on the number of instances of
\rulename{t-sub} that end the two derivations, simultaneously for all environments and
types. If a derivation ends with \rulename{t-sub}, say from $s_1\subt t_1$, the
induction hypothesis gives the type $s_1\land t_2$, below $t_1\land t_2$, as
in~\cite[Lemma~6.15]{FCB08}. Otherwise both derivations end with the rule
determined by the shape of $e$.
\begin{itemize}[leftmargin=*,itemsep=1pt,topsep=2pt]
\item \rulename{t-var}, \rulename{t-const}, \rulename{t-self}, \rulename{t-spawn},
\rulename{t-plus}: the two conclusions assign the same type.
\item \rulename{t-tuple}: the induction hypothesis componentwise, since
$\Tuple{t_1,\dots,t_n}\land\Tuple{t'_1,\dots,t'_n}\simeq
\Tuple{t_1\land t'_1,\dots,t_n\land t'_n}$.
\item \rulename{t-abs}, \rulename{t-abs-neg}: the conclusions are
$l\land\bigwedge_{a\in N_1}\neg a$ and $l\land\bigwedge_{a\in N_2}\neg a$, with the same
premises; \rulename{t-abs-neg} with $N_1\cup N_2$ concludes, its side condition holding
by Lemma~\ref{lem:atomic-prime}(1), since no arrow of $N_1\cup N_2$ is above $l$.
Likewise \rulename{t-pid}, \rulename{t-pid-neg}, with
Lemma~\ref{lem:atomic-prime}(2).
\item \rulename{t-app}: $e=e_1\,e_2$ with $e_1$ of type $s_k\to_o^i t_k$ and $e_2$ of
type $s_k$ ($k=1,2$), the two derivations sharing the ambient bracket. By the
induction hypothesis $e_2$ has type $s_1\land s_2$ and $e_1$ has type
$(s_1\to_o^i t_1)\land(s_2\to_o^i t_2)$, which is below
$(s_1\land s_2)\to_o^i(t_1\land t_2)$: every input in the domain of the latter lies in
the domains of both conjuncts, so a relation in both conjuncts maps it into
$\sem{t_1}\cap\sem{t_2}$. \rulename{t-app} concludes.
\item \rulename{t-send}: $e=\Send{e'}{e''}$ with $e'$ of type $\TChan{t_k}$ and $e''$
of type $t_k$ ($k=1,2$). By the induction hypothesis and
Proposition~\ref{prop:pid-laws}, $e'$ has type
$\TChan{t_1}\land\TChan{t_2}\simeq\TChan{t_1\lor t_2}\subt\TChan{t_1\land t_2}$, and
$e''$ has type $t_1\land t_2$; \rulename{t-send} concludes.
\item \rulename{t-receive}: the branch environments depend only on $i$ and on the
patterns, so they coincide in the two derivations. Let the conclusion omit the branches
that either derivation omits, whose refined type is empty; every other branch is typed
by both derivations, and the induction hypothesis applies to it.
\item \rulename{t-case}: $e=\Case{e_0}{\seqof{\Branch{p}{e}}}$ with scrutinee types
$u_1$ and $u_2$. By the induction hypothesis $e_0$ has type $u_1\land u_2$, which
satisfies the exhaustiveness condition. Its refined types
$r_j=(u_1\land u_2\land\accept{p_j})\setminus\bigvee_{h<j}\accept{p_h}$ are below those
of both derivations, so a branch omitted by either derivation has $r_j\simeq\Tempty$ and
may be omitted by the conclusion; every other branch is typed by both derivations, and
both of its derivations can be re-typed under $\Context,\ptenv{r_j}{p_j}$
(Lemmas~\ref{lem:penv-mono}(2) and~\ref{lem:env-sub}); the induction hypothesis applies
to it.\qedhere
\end{itemize}
\end{proof}

\subsection{Subject reduction and progress}\label{app:prop-sr}
Subject reduction for expressions (Theorem~\ref{thm:sr-expr}) is proved first; its
$\beta$ case is where abstraction inversion, substitution, narrowing and
intersection introduction combine. Subject reduction for configurations
(Theorem~\ref{thm:sr}) adds the communication rules, and needs the
well-formedness invariant (Definition~\ref{def:wf-config},
Lemma~\ref{lem:config-inv}), which progress (Theorems~\ref{thm:progress-expr}
and~\ref{thm:progress}) needs as well; soundness (Corollary~\ref{cor:soundness})
iterates both along a reduction sequence. Every lemma is used in its $\Vdash$
form.

\SRExprThm*

\begin{proof}
By induction on the derivation of $e\red e'$ by the rules of
Section~\ref{sec:opsem}, in the enriched system throughout. By the Conventions, the derivation of
$\wtee[i][o]{\Context}{e}{t}$ ends with the rule of the redex followed by instances
of \rulename{t-sub}, so it suffices to type $e'$ at the type $t_0\subt t$ in the
conclusion of that rule.

\emph{Addition.} \rulename{t-plus} gives $t_0=\TInt$, and the reduct is an integer
constant.

\emph{$\beta$-reduction, $(\Fun^{l}x.e_1)\,v\red e_1\subst{v}{x}$}, with
$l=\bigwedge_{j\in J}t_j\to_{o_j}^{i_j}t'_j$. The rule is \rulename{t-app}, with
premises $\wtee[i][o]{\Context}{\Fun^{l}x.e_1}{s\to_o^i t_0}$ and
$\wtee[i][o]{\Context}{v}{s}$ for some $s$. The Abstraction Inversion Lemma~\ref{lem:arrow-inv} applied to
these two judgments yields a non-empty $P\subseteq J$ such that
$\wtee{\Context}{v}{t_j}$ and $[o,i]$ narrows $[o_j,i_j]$ for every $j\in P$, and
$\bigwedge_{j\in P}t'_j\subt t_0$. Independently, the derivation of the first
premise ends, below its instances of \rulename{t-sub}, with \rulename{t-abs} or
\rulename{t-abs-neg} for the annotation $l$, whose premises are in both cases
$\wtee[i_j][o_j]{\Context,x:t_j}{e_1}{t'_j}$ for every $j\in J$. Fix $j\in P$. Since
the typing of a value does not depend on the bracket (the first fact before
Lemma~\ref{lem:substitution}), $\wtee[i_j][o_j]{\Context}{v}{t_j}$; the
substitution lemma at $p=x$ (Lemma~\ref{lem:substitution}, with
$\ptenv{t_j}{x}=\{x:t_j\}$, $\accept{x}=\Tany$, and $\match{v}{x}$ never failing) gives
$\wtee[i_j][o_j]{\Context}{e_1\subst{v}{x}}{t'_j}$; and Theorem~\ref{thm:narrowing},
since $[o,i]$ narrows $[o_j,i_j]$, gives
$\wtee[i][o]{\Context}{e_1\subst{v}{x}}{t'_j}$. Lemma~\ref{lem:inter-intro} applied
to these judgments for all $j\in P$ gives
$\wtee[i][o]{\Context}{e_1\subst{v}{x}}{\bigwedge_{j\in P}t'_j}$, and
\rulename{t-sub} with $\bigwedge_{j\in P}t'_j\subt t_0$ concludes.

\emph{Branch selection, $\Case{v}{\seqof{\Branch{p}{e}}}\red e_j\subst{v}{p_j}$.}
The rule is \rulename{t-case}, with the scrutinee typed at some $u$. The side
condition of the reduction rule gives $\match{v}{p_j}\neq\textsf{fail}$ and
$\match{v}{p_h}=\textsf{fail}$ for every $h<j$; since accepted types are test types,
Lemmas~\ref{lem:min-types2} and~\ref{lem:matching} apply to the $\Vdash$-typed
$v$ through Lemma~\ref{lem:vee-principal}(4) and give
$\wtee{\Context}{v}{\accept{p_j}}$ and $\wtee{\Context}{v}{\neg\accept{p_h}}$ for
$h<j$, hence, by Lemma~\ref{lem:vee-principal}(3),
$\wtee{\Context}{v}{r_j}$ with $r_j=(u\land\accept{p_j})\setminus\bigvee_{h<j}\accept{p_h}$,
the type at which \rulename{t-case} computed the environment of branch $j$. The
type $r_j$ is not empty, being above a non-empty generator of $v$
(Lemma~\ref{lem:vee-principal}(1--2)), so the rule did not omit the premise
$\wtee[i][o]{\Context,\ptenv{r_j}{p_j}}{e_j}{t_0}$, and Lemma~\ref{lem:substitution}
at $p=p_j$, $s=r_j$ concludes.

\emph{Context, $\Eval[e]\red\Eval[e']$ with $e\red e'$.} Lemma~\ref{lem:context}
gives $\wtee[i][o]{\Context}{e}{s}$ for some $s$; by the induction hypothesis on
the derivation of $e\red e'$, $\wtee[i][o]{\Context}{e'}{s}$, and the lemma
re-types $\Eval[e']$ at $t$.
\end{proof}

\paragraph{Soundness for configurations}
The typing rules \rulename{t-thread} and \rulename{t-par} check each process
in isolation; they cannot (and do not) check that the pid values scattered
in expressions and mailboxes actually name processes of the configuration.
This global invariant is instead maintained by reduction itself, because
pids are runtime-only: every pid is created by a $\mkkeyword{spawn}$, which
annotates it with the interface of the process it creates, and no construct
ever rewrites the annotation. We package the
invariant as a well-formedness condition: it is exactly what would
otherwise be recorded by a global pid environment $\Delta$, as discussed in
Section~\ref{sec:typing-conf}.

\WFConfigDef*

\begin{lemma}[Configuration invariants]\label{lem:config-inv}
Let $P \equiv Q$. Then $\wtce{P}$ if and only if $\wtce{Q}$ (and likewise
for $\wtc{\cdot}$), and $P$ is
well-formed if and only if $Q$ is. Moreover, if $P$ is well-formed and
$P \red P'$, then $P'$ is well-formed.
\end{lemma}

\begin{proof}[Proof sketch]
Structural congruence only permutes and reassociates the leaves of a derivation made of
\rulename{t-par} instances, and well-formedness is a property of the set of processes of
the configuration, which $\equiv$ does not change. Preservation of well-formedness holds
because processes are never removed (a terminated process stays in the configuration as
$\Process{\PID}{s}{v}{\seqof{v}}$) and reduction only moves pid values
around (sends copy them into mailboxes, receives and cases substitute them
into branches, $\Self$ reduces to the pid of the executing process, which is
in $P$); new pids appear only at $\mkkeyword{spawn}$, where the freshly
minted $\PID[b]^s$ carries by construction the interface of the freshly
created process.
\end{proof}

\SRThm*

\begin{proof}[Proof sketch]
The expression case is Theorem~\ref{thm:sr-expr}, and the closure under
structural congruence is Lemma~\ref{lem:config-inv}; the remaining cases are
the communication rules of Section~\ref{sec:opsem}. For $\Self$, the redex is
typed $\TChan{s}$ by \rulename{t-self} at the top-level bracket $[s,s]$ of
\rulename{t-thread}, and the reduct $\PID^s$ receives the same type by
\rulename{t-pid}; Lemma~\ref{lem:context} concludes. For $\mkkeyword{spawn}$,
inverting \rulename{t-spawn} yields $\wte[s][s]{\emptyset}{e}{t'}$ for the
shipped body, which is precisely the premise \rulename{t-thread} needs for
the new process (with empty mailbox), while the parent resumes with
$\PID[b]^s$ at type $\TChan{s}$ (Lemma~\ref{lem:context}). For a send, the
sender's derivation types the target at some $\TChan{t}$ and the message at
$t$; by canonical forms (Corollary~\ref{cor:canonical-vee}) the target value
is $\PID[b]^s$ with $t \subt s$,
so the delivered message has the type $s$ at which
\rulename{t-thread} checks the recipient's mailbox; self-sends are the
instance where recipient and sender coincide and the bracket is $[s,s]$. For
a receive, the mailbox invariant of \rulename{t-thread} types the selected
message $v'$ at the interface $s$ (which is the input decoration at top
level), the side conditions of the message-major reduction rule ($v'$ fails all patterns preceding $p_j$, older messages fail every pattern), together
with Lemmas \ref{lem:min-types1} and \ref{lem:matching} refine it to
$(s\land\accept{p_j})\setminus\bigvee_{h<j}\accept{p_h}$, and the
substitution lemma types the selected branch; the messages skipped or left
in place keep their types, so the residual mailbox is still typed at $s$.
In every case the expression that the step rewrites keeps the type it had in
\rulename{t-thread}, by Theorem~\ref{thm:sr-expr} or Lemma~\ref{lem:context}, and the
expressions of the other processes are unchanged (the recipient of a send only
receives a message), which is the second half of the statement.
\end{proof}

\ProgressExprThm*

\begin{proof}[Proof sketch]
The expressions of item 3 are the communication redexes: they reduce only at
the level of configurations, where the executing process (for $\Self$ and
$\mkkeyword{spawn}$), the recipient (for a send), or the mailbox (for a
receive) become available. The proof is by induction on the typing
derivation, with canonical forms (Corollary~\ref{cor:canonical-vee})
providing the required shapes in every elimination position. The $\mkkeyword{case}$
construct is where the exhaustiveness side condition of \rulename{t-case}
pays: from $\wtee{}{v}{t}$ and $t \subt \bigvee_j \accept{p_j}$, whence $\mathsf{min}(v) \subt \bigvee_j \accept{p_j}$ by
Lemma~\ref{lem:vee-principal}(4), unions of test types being test types, the dichotomy of Lemma~\ref{lem:min-types2}, with $\mathsf{min}(v)$ non-empty, forces
$\mathsf{min}(v) \subt \accept{p_j}$ for some $j$, and the least such $j$ is,
by the Matching lemma, a branch that fires: a well-typed $\mkkeyword{case}$
never fails to match.
\end{proof}

\ProgressThm*

\begin{proof}[Proof sketch]
Let $\Process{\PID}{t}{e}{\seqof{v}}$ be a process of $P$. Inverting \rulename{t-par}
and \rulename{t-thread} gives $\wtee[t][t]{\emptyset}{e}{t''}$ for some $t''$, and
Theorem~\ref{thm:progress-expr} applies to $e$. If $e$ is a value, this is case 1. If
$e\red e'$, the first rule for configurations gives a step of this process. Otherwise
$e=\Eval[f]$ with $f$ a communication redex, and well-formedness is what makes it fire.
The redexes $\Self$ and $\Spawn{e'}{s}$ reduce unconditionally. The target
$\PID[b]^s$ of a send names, by well-formedness, a process of $P$ whose interface is
$s$: if $\PID[b]=\PID$ it is the process itself, $s=t$, and the rule for self-sends
applies; otherwise the delivery rule applies, after rearranging $P$ by structural
congruence. Finally, a receive reduces if some message of $\seqof{v}$ matches some of
its patterns, the oldest such message selecting the branch; otherwise it is case 2 of
the statement.
\end{proof}

\SoundnessCor*

\begin{proof}[Proof sketch]
By induction on the reduction sequence: $P$ satisfies $\wtc{P}$, hence
$\wtce{P}$, and is well-formed by hypothesis, and both properties are
preserved by every reduction step
(Theorem~\ref{thm:sr} for typing, Lemma~\ref{lem:config-inv} for
well-formedness), which is item 1; its second part is the mailbox premise of
\rulename{t-thread}, reached by inverting \rulename{t-par}, together with
Definition~\ref{def:wf-config}. Item 2 is Theorem~\ref{thm:progress} applied to $Q$.
For item 3, the process of pid $\PID$ is never removed, and iterating the second half
of Theorem~\ref{thm:sr} along the reduction sequence gives
$\wtee[s][s]{\emptyset}{e'}{t}$ for its expression $e'$ in $Q$; if $e'$ is a value
$v$, then $\wtee{}{v}{t}$, and $\wte{}{v}{t}$ when $v$ contains neither abstractions
nor pids, by Lemma~\ref{lem:vee-principal}(4). For programs, the initial configuration
$\Process{\PID}{t}{e}{\EmptyMailbox}$ is well-typed by \rulename{t-thread}
(its mailbox is empty and $\wte[t][t]{\emptyset}{e}{s}$) and well-formed
because, $e$ being pid-free, the only pid occurring in it is $\PID$ itself.
\end{proof}

\section{Supplement to \Cref{sec:ann-inference}}
\label{app:elaboration}
This appendix proves the results of Section~\ref{sec:ann-inference}, restated with
their original numbers, in the order of that section: decorations
(Appendix~\ref{app:elab-decorating}); the elaboration of a deduction and the derivations
compatible with it (Appendix~\ref{app:elab-elaboration}); tallying on the encoded constraints
(Appendix~\ref{app:elab-tallying}, Lemma~\ref{lem:tally-readback}); soundness and completeness
of inference, both proved on elaborations (Appendices~\ref{app:elab-soundness}
and~\ref{app:elab-completeness}, Lemmas~\ref{lem:elab-sound} and~\ref{lem:elab-complete});
termination (Appendix~\ref{app:elab-termination}); uniform requirements
(Appendix~\ref{app:elab-uniform}); type erasure and safety (Appendix~\ref{app:elab-erasure}). It
ends with the elimination of variables that Section~\ref{sec:elixir-tallying} describes
(Appendix~\ref{app:elab-elimination}).

\subsection{Decorating annotations}\label{app:elab-decorating}

\begin{lemma}[Decorations are types]\label{lem:decorate-wd}
For every annotation $\ta$ and every family $(\alpha_{\sa},\beta_{\sa})_{\sa}$, the equations
of Definition~\ref{def:decorate} have exactly one solution $\decorate{\ta}$, and it is a
type, that is, a regular and contractive tree.
\end{lemma}

\begin{proof}
\emph{Existence and uniqueness.} As for $\enc\cdot$ (Remark~\ref{rem:enc-corecursive}). The
clauses for the type constructors are guarded, each recursive call occurring below the
constructor that its clause produces. The clauses for the type connectives are covered by
induction on the connectives above a constructor, which is well founded since annotations
are contractive (Section~\ref{sec:syntax}).

\emph{Regularity.} By the equations, a subtree of $\decorate{\ta}$ is either a variable of
the family or $\decorate{\ta'}$ for a subtree $\ta'$ of $\ta$. Since the family is indexed by
the arrow annotations, $\decorate{\ta'}$ depends only on $\ta'$; and $\ta$ is regular, so it
has finitely many subtrees.

\emph{Contractivity.} The equations erase no constructor, so every infinite branch of
$\decorate{\ta}$ crosses infinitely many constructors, as the corresponding branch of $\ta$
does.
\end{proof}

\begin{remark}[What decorations identify]\label{rem:decorate}
The decorations of $\ta$ are not all the types that erase to $\ta$: equal arrow annotations
receive the same pair of variables, hence the same bracket in every decoration.
\begin{itemize}[leftmargin=*,itemsep=1pt,topsep=2pt]
\item For the unfoldings of a recursive annotation, this is forced by regularity. On
$\sa=\ann{\TInt\to\sa}$, a fresh pair at each position of the tree would introduce infinitely
many variables and pairwise distinct subtrees, hence no type; indexed by the annotation, the
decoration is the regular type $t=\TInt\to_\beta^\alpha t$.
\item For equal arrow annotations at different positions it is a loss: no derivation gives
different brackets to the two components of $\ann{(\TInt\to\TInt,\,\TInt\to\TInt)}$,
although a type that does so erases to this annotation. Completeness of
inference (Theorem~\ref{thm:inf-complete}) is to be read up to this identification, which
the declarative system of Section~\ref{sec:annotations} makes as well. The programmer, who
writes no decoration, does not see it. No definition whose results are types avoids it
altogether: a regular decoration of $\ann{\TInt\to\sa}$ has, from some depth on, the same pair
of variables at depths $k$ and $k+P$ for some $P\geq1$, so none of its instances has
brackets that cycle through $P+1$ distinct types.
\end{itemize}
Two things are not identified. The occurrences of one annotation at two places of the
program draw their variables from different families (Section~\ref{sec:annotations}), so
the inner arrows of two abstractions annotated $\ann{(\TInt\to\TInt)\to\TInt}$ receive
unrelated brackets.
And an arrow written under a $\TChan\cdot$ receives a pair unrelated to the interface:
tying the pair to the interface would make the interface of a process that \emph{forwards}
the functions it receives to another process the decoration of no annotation
(Section~\ref{sec:elixir-decorations} returns to this point at a module boundary).
\end{remark}

\subsection{Elaboration and compatible derivations}\label{app:elab-elaboration}

\begin{definition}[Decoration, elaboration]\label{def:elaboration}
A \emph{decoration} of an expression $e$ is a term $f$ of the calculus of
Section~\ref{sec:syntax}, over types that may contain type variables, obtained from $e$ by
replacing each abstraction $\Fun^{\ann{\wedge_j\sa_j\to\ta_j}}x.\,e'$ by
$\Fun^{\wedge_j s_j\to_{o_j}^{i_j}t_j}x.\,f'$ and each $\Spawn{e'}{\ta}$ by $\Spawn{f'}{t}$,
where $f'$ is a decoration of $e'$ and, for a type substitution $\sigma$ and types $i_j,o_j$
chosen for the occurrence, $s_j=\decorate{\sa_j}\sigma$, $t_j=\decorate{\ta_j}\sigma$ and
$t=\decorate{\ta}\sigma$, the family of $\decorate\cdot$ being that of the occurrence.

The \emph{elaboration} of a deduction $\tinf{\Context}{e}{t}[C]$ is the decoration $f$ of $e$
in which, for each occurrence, $\sigma$ is the identity and $i_j,o_j$ are the variables
$\alpha_j,\beta_j$ minted for it (Section~\ref{sec:inference-rules}). We write
$\telab{\Context}{e}{t}[C]{f}$.
\end{definition}

\noindent The deduction of an expression is unique up to the renaming of the variables it
mints (Proposition~\ref{prop:inference-terminates}(1)), and so is its elaboration. Every
subterm of $e$ is elaborated, including the branches whose premise the deduction omits.
Patterns are unchanged, test types being types and annotations at once
(Section~\ref{sec:annotations}); so, extending $\erase\cdot$ homomorphically from types to
terms, $\erase f=e$ for every decoration $f$ of $e$. The types that $e$
itself carries, in its pids (Section~\ref{sec:annotations}) and in a $\mkkeyword{spawn}$
that carries a type (Section~\ref{sec:inference-rules}), are ground. So, the families of distinct occurrences
being disjoint, the decorations of $e$ are the terms $f\theta$, for $f$ the elaboration and
$\theta$ a type substitution. Since the variables that decorate an occurrence are minted
once for that occurrence (Section~\ref{sec:inference-rules}), the premises of
\rulename{i-abs} all type the same elaborated body; in the proofs below, each rule of
Figure~\ref{fig:inference} is read together with the elaborated subterms of its premises.

\begin{lemma}[Decorations]\label{lem:decoration}
A derivation of $\wte[i][o]{\Context}{e}{t}$ in the system of
Section~\ref{sec:annotations} is a derivation of $\wte[i][o]{\Context}{f}{t}$ in the
system of Section~\ref{sec:typing} for a decoration $f$ of $e$, and conversely.
\end{lemma}

\begin{proof}
The two systems differ only in \rulename{t-abs} and \rulename{t-spawn}, which in
Section~\ref{sec:annotations} take a decoration of the annotation where those of
Section~\ref{sec:typing} read the type that the term carries. The decorations of an
occurrence being the same in all the premises that type it
(Section~\ref{sec:annotations}), they determine a single decoration $f$ of $e$, and each
instance of a rule of one system is the instance of the same rule of the other.
\end{proof}

\begin{definition}[Compatible derivation]\label{def:compatible}
Let $\telab{\Context}{e}{t'}[C]{f}$ be a deduction and $\theta$ a type substitution. A
derivation of $\wte[i\theta][o\theta]{\Context\theta}{f\theta}{t}$ in the system of
Section~\ref{sec:typing} is \emph{compatible with the deduction} if every branch of a
$\mkkeyword{case}$ or of a $\mkkeyword{receive}$ that it omits, at a point that the
deduction reaches, has at the corresponding point of the deduction a refined type $r$ with
$r\simeq\Tempty$. The points of the derivation and of the deduction correspond once the
instances of \rulename{t-sub} are removed from the derivation, the $j$-th premise of
\rulename{t-abs} corresponding to the $j$-th premise of \rulename{i-abs}.

A derivation of $\wte[i][o]{\Context}{e}{t}$ in the system of
Section~\ref{sec:annotations} is \emph{compatible with inference} if, as a derivation for
the decoration $f\theta$ of $e$ (Lemma~\ref{lem:decoration}), it is compatible with the
deduction $\telab[\alpha][\beta]{\Context}{e}{t'}[C]{f}$ of $e$ at fresh $\alpha,\beta$,
where $\theta(\alpha)=i$ and $\theta(\beta)=o$.
\end{definition}

\begin{remark}[Omitted branches]\label{rem:compatible}
\rulename{i-case} and \rulename{i-receive} omit a branch exactly when its refined type $r$
has $r\simeq\Tempty$ (Appendix~\ref{app:encoding-property}), that is, when it is empty
whatever the variables stand for. Since $r\simeq\Tempty$ implies $r\sigma\simeq\Tempty$, a
derivation may omit every branch that the deduction omits.
\begin{enumerate}[leftmargin=*,itemsep=1pt,topsep=2pt]
\item It may omit more, and then inference is not complete. The abstraction
\[
\Fun^{(\TInt\to\TInt)\land(\TAtom\to\TAtom)}x.\Recv{\Branch{\TInt}{1},\;\Branch{\TAtom}{\Atom[{:}a]}}
\]
is typable at $(\TInt\to_{\Tempty}^{\TInt}\TInt)\land(\TAtom\to_{\Tempty}^{\TAtom}\TAtom)$,
each branch being omitted under one of the two arrows (this type states that a process can apply this function to an integer only if it accepts only integers, and the same for atoms: inference cannot discover this).
 In the deduction, the refined types
$\alpha_1\land\TInt$ and $(\alpha_1\land\TAtom)\setminus\TInt$ of the first arrow are not
$\simeq\Tempty$, so both branches are typed and \rulename{i-abs} emits
$1\lor\Atom[{:}a]\issubt\TInt$, which no substitution satisfies.
\item A derivation need not omit a branch whose refined type is empty
(Section~\ref{sec:typing-rules}). Were the omission mandatory, the abstraction
$\Fun^{\TInt\to\TInt}x.\Recv{\Branch{y\land\TInt}{y},\;\Branch{z\land\TAtom}{z}}$ would have
no compatible derivation: under an input decoration $i$, typing the second branch requires
$(i\land\TAtom)\setminus\TInt\subt\TInt$, hence this type to be empty and the branch to be
omitted, whereas its refined type $(\alpha_1\land\TAtom)\setminus\TInt$ in the deduction is
not $\simeq\Tempty$. Yet inference accepts the abstraction: its constraint
$(\alpha_1\land\TInt)\lor((\alpha_1\land\TAtom)\setminus\TInt)\issubt\TInt$ is satisfied by
$\alpha_1\eqdef\neg\TAtom$. The omission being optional, the derivation that types the second
branch at $z:\Tempty$ is compatible.
\end{enumerate}
\end{remark}

\subsection{Tallying through the encoding}\label{app:elab-tallying}
Tallying is run on the encoded constraints
$\enc{C}=\Set{\enc{s}\issubt\enc{t}\mid (s\issubt t)\in C}$. For a finite set $D$ of
constraints between types of~\cite{FCB08} and a set $\Delta$ of type variables,
$\textsf{Sol}_\Delta(D)$ denotes the finite set of substitutions that the algorithm
of~\cite{polyduce2} returns, whose domains avoid $\Delta$. It is sound and
complete~\cite{polyduce2}: every $\hat\sigma\in\textsf{Sol}_\Delta(D)$ satisfies $D$, and for
every $\hat\theta$ that satisfies $D$ and whose domain avoids $\Delta$ there are
$\hat\sigma\in\textsf{Sol}_\Delta(D)$ and $\hat\rho$ with
$\hat\theta(\alpha)\simeq\hat\sigma(\alpha)\hat\rho$ for every variable $\alpha$ of $D$.

The images of the substitutions in $\textsf{Sol}_\varnothing(\enc{C})$ need not be
encodings: a variable $\alpha$ used as the argument type of an arrow occurs in $\enc{C}$ as
$\alpha\lor\CPoint$, so its bounds may mention $\CPoint$, as in
$\alpha\mapsto\Tany\setminus\CPoint$. They are read back by the map $\dec{\cdot}$ of
Definition~\ref{def:readback}, defined on the class $\EncTypes$ of types
of~\cite{FCB08}, which contains the encodings (Appendix~\ref{app:enc-readback});
$\EncTypes$-substitutions, $X$-positions, and junk forms are as defined there.

\begin{definition}[Solutions]\label{def:solutions}
For a finite set $C$ of constraints between types,
$\textsf{Solutions}(C)=\Set{\dec{\hat\sigma}\mid\hat\sigma\in\textsf{Sol}_\varnothing(\enc{C})}$.
\end{definition}

That the elements of $\textsf{Sol}_\varnothing(\enc{C})$ can be read back, that is, that they
are $\EncTypes$-substitutions, depends on how the algorithm of~\cite{polyduce2} computes its
solutions, and not only on its soundness and completeness. We recall its three steps, as
its rules define them and without the simplifications that an implementation adds. The
\emph{atoms} of a type $t$ of~\cite{FCB08} are the type variables, the basic types (among
which the singleton types of the constants $\CPoint$, $\ell_{\Wch}$, $\ell_{\Wfun}$, and
$\Cnil$), the products, and the arrows that are reached from the root of $t$ by crossing
only connectives. By contractivity they are finitely many, and $t$ is a Boolean combination
of them. Below, an empty intersection is $\Tany$.
\begin{description}[leftmargin=*,itemsep=1pt,topsep=2pt]
\item[Normalization.] A constraint $s\issubt t$ is replaced by $s\land\neg t\issubt\Tempty$,
and $s\land\neg t$ by a disjunctive normal form, that is, a union of intersections of atoms
of $s\land\neg t$ and of negations of such atoms. Each of these intersections, the
\emph{summands}, must be empty. A summand is treated as follows.
\begin{itemize}[leftmargin=*,itemsep=1pt,topsep=1pt]
\item Some of its atoms are type variables. Let $\alpha$ be the smallest of them in a fixed
order in which the variables of $\Delta$ are the greatest, and let $u$ be the intersection
of the other literals of the summand, whatever their atoms are. If $\alpha\notin\Delta$ the
summand yields the constraint $\alpha\issubt\neg u$ when $\alpha$ occurs positively, and
$u\issubt\alpha$ when it occurs negatively. If $\alpha\in\Delta$, the literals on type
variables are removed and what remains of the summand is treated by the next case.
\item None of its atoms is a type variable. The atoms are of three kinds: basic types,
products, and arrows. If two atoms of different kinds occur positively, the summand is
empty and yields no constraint. If the atoms that occur positively are of a single kind, the
literals on the atoms of the other kinds are removed. If no atom occurs positively, the
summand is treated once for each kind, with the literals on the atoms of that kind. In all
cases what is left to treat is an intersection of literals on atoms of a single kind.
\begin{itemize}[leftmargin=*,itemsep=1pt,topsep=1pt]
\item Basic types: the intersection is tested for emptiness and yields no constraint.
\item Products, the intersection being
$\bigwedge_{p\in P}(s_p\times t_p)\land\bigwedge_{n\in N}\neg(s_n\times t_n)$: it yields,
for subsets $N'$ of $N$, constraints
$\bigwedge_{p\in P}s_p\land\bigwedge_{n\in N'}\neg s_n\issubt\Tempty$ and
$\bigwedge_{p\in P}t_p\land\bigwedge_{n\in N\setminus N'}\neg t_n\issubt\Tempty$.
\item Arrows, the intersection being
$\bigwedge_{p\in P}(s_p\to t_p)\land\bigwedge_{n\in N}\neg(s_n\to t_n)$: it yields, for
$n\in N$ and subsets $P'$ of $P$, constraints
$s_n\land\bigwedge_{p\in P'}\neg s_p\issubt\Tempty$ and, when $P'\neq P$,
$\bigwedge_{p\in P\setminus P'}t_p\land\neg t_n\issubt\Tempty$.
\end{itemize}
\end{itemize}
The decompositions of products and arrows offer alternatives (for products, one of the two
constraints for each $N'$; for arrows, one $n$ and, for each $P'$, one of the two
constraints): the algorithm explores them all, and only the form of the constraints matters
here. These constraints are normalized in turn, a memoization ensuring termination. The
result is a finite set of finite sets of constraints of the forms $\alpha\issubt u$ and
$u\issubt\alpha$, in which $u$ is a \emph{bound} of $\alpha$.
\item[Merging.] In each of these sets, the upper bounds of a variable are replaced by their
intersection and its lower bounds by their union; then, for every variable that has a lower
bound $t_1$ and an upper bound $t_2$, the constraint $t_1\issubt t_2$ is normalized and the
constraints obtained are added to the set. This is repeated until no new constraint
$t_1\issubt t_2$ arises.
\item[Solving.] Each set $E$ so obtained gives one substitution $\hat\sigma$. Every variable
$\alpha$ bounded in $E$, by $t_1$ from below and by $t_2$ from above ($\Tempty$ and $\Tany$
when a bound is missing), is given the equation $\alpha=(t_1\lor\alpha')\land t_2$, where
$\alpha'$ is a fresh variable. The substitution $\hat\sigma$ maps each of these variables to
its component of the solution of this system of equations in regular trees (which the
algorithm writes with a recursion operator), and is the identity on the other variables. So
$\hat\sigma(\alpha)=((t_1\lor\alpha')\land t_2)\hat\sigma$, an equality of trees, and these
trees are contractive~\cite{polyduce2}.
\end{description}

\begin{lemma}[Closure]\label{lem:tally-closure}
If $D$ is a finite set of constraints between types in $\EncTypes$ and $\Delta$ is a finite
set of type variables, then every $\hat\sigma\in\textsf{Sol}_\Delta(D)$ is an
$\EncTypes$-substitution.
\end{lemma}

\begin{proof}
Normalization replaces a constraint between products or between arrows by constraints
between their components, which need not be in $\EncTypes$: the second component of
$\Wch(X\to\CPoint)$ is an arrow. We therefore define a second set of types, $\mathcal{S}$,
which contains no type variable, and show that the algorithm produces only constraints
whose two sides are both in $\EncTypes$ or both in $\mathcal{S}$. A bound comes from a
summand with a type variable among its atoms, which is therefore in $\EncTypes$, and so is
the bound.

A \emph{tail} is a type $X_1\times(\cdots\times(X_n\times\Cnil))$ with $n\geq0$ and every
$X_k$ in $\EncTypes$ (it is $\Cnil$ when $n=0$). Let $\mathcal{S}$ be the set of the types
of~\cite{FCB08} each of whose atoms is $\Cnil$ or has one of the following forms:
\begin{gather*}
X\times L
\qquad \Tany\times T
\qquad \Wch(X\to\CPoint)\times(\CPoint\to X')\\
X\to\CPoint
\qquad \mathsf{In}(X_1,X_2,X_3)\to X_4
\qquad \CPoint\to X
\end{gather*}
where $X,X',X_1,\dots,X_4$ are any types in $\EncTypes$ and $L$ is any tail. The first line
lists products, the second arrows. The forms are chosen so that the first
component of a product that is an atom of a type in $\EncTypes$ or in $\mathcal{S}$ is in
$\EncTypes$ and its second component is in $\mathcal{S}$, the two components of an arrow in
$\mathcal{S}$ are in $\EncTypes$, and no type in $\EncTypes$ has an arrow among its atoms
(Steps~1 and~2). No atom of a type in $\mathcal{S}$ is a type variable. The atoms of a
Boolean combination of atoms $a_1,\dots,a_n$ are among $a_1,\dots,a_n$, since the only atom
of an atom is itself, its root not being a connective, and the atoms of a union or of a
negation are those of its operands. Hence a Boolean combination of atoms of types in
$\mathcal{S}$ is in $\mathcal{S}$. The tails and
$T=\Cnil\lor(\Tany\times T)$ are in $\mathcal{S}$, and $\Tempty$ and $\Tany$ are in
$\mathcal{S}$ and in $\EncTypes$.

\emph{Step 1 (atoms of the types in $\EncTypes$).} Every atom of a type $X$ in $\EncTypes$
is a type variable, a basic type, or a product of one of the forms
\begin{gather*}
X_1\times L
\qquad \Tany\times T
\qquad \ell_{\Wch}\times(X_1\to\CPoint)
\qquad \ell_{\Wfun}\times(\mathsf{In}(X_1,X_2,X_3)\to X_4)\\
(X_1\lor\CPoint)\times\bigl(\Wch(X_2\to\CPoint)\times(\CPoint\to X_3)\bigr)
\end{gather*}
with $X_1,\dots,X_4$ in $\EncTypes$ and $L$ a tail; in particular, no atom of $X$ is an
arrow. This is proved by induction on the number of connectives between the root of $X$ and
the atom, by cases on the form of $X$. If $X$ is $\Tempty$ it has no atom. If $X$ is a
union or a negation, the atom is an atom of an operand, which is in $\EncTypes$. If $X$ is
$T$, its atoms are $\Cnil$ and $\Tany\times T$. In every other case $X$ is its only atom:
a type variable, a basic type ($b$, $\Cnil$, $\CPoint$, $\ell_{\Wch}$, or $\ell_{\Wfun}$),
or a product of one of the five forms above, which are the tuple form, $\Tany\times T$,
the pid form, the function form, and the junk form $\mathsf{In}(X_1,X_2,X_3)$. Conversely
each of these atoms is in $\EncTypes$, which is closed under connectives. Hence a Boolean
combination of atoms of types in $\EncTypes$ is in $\EncTypes$.

\emph{Step 2 (normalization).} Say that a constraint is \emph{sorted} if its two sides are
both in $\EncTypes$ or both in $\mathcal{S}$. We show that the normalization of a sorted
constraint $s\issubt t$ yields only constraints $\alpha\issubt u$ and $u\issubt\alpha$ with
$u\in\EncTypes$. Since normalization terminates, we reason by induction on the number of
its steps: it suffices to show that the summands of the disjunctive normal form of
$s\land\neg t$ yield only such constraints and sorted constraints. Every summand is a
Boolean combination of atoms of $s$ and $t$. So, by Step~1 and by the definition of
$\mathcal{S}$, it is in $\EncTypes$ if $s$ and $t$ are, and in $\mathcal{S}$ if $s$ and $t$
are. This also holds for what remains of a summand once some of its literals are removed.

\textbf{Case some atoms are type variables.} The summand is in $\EncTypes$, since no atom of
a type in $\mathcal{S}$ is a type variable. The intersection $u$ of its literals other than
the one on $\alpha$ is in $\EncTypes$, and so is $\neg u$.

\textbf{Case basic types.} No constraint is produced.

\textbf{Case products, the summand being in $\EncTypes$.} By Step~1 the first
components of the atoms have the forms $X_1$, $\Tany$, $\ell_{\Wch}$, $\ell_{\Wfun}$, and
$X_1\lor\CPoint$, which are in $\EncTypes$. Their second components have the forms $L$,
$T$, $X_1\to\CPoint$, $\mathsf{In}(X_1,X_2,X_3)\to X_4$, and
$\Wch(X_2\to\CPoint)\times(\CPoint\to X_3)$, which are in $\mathcal{S}$. So the constraints
on the first components have their left-hand sides in $\EncTypes$, those on the second
components have them in $\mathcal{S}$, and all of them are sorted, their right-hand sides
being $\Tempty$.

\textbf{Case products, the summand being in $\mathcal{S}$.} The first
components of the atoms have the forms $X$, $\Tany$, and $\Wch(X\to\CPoint)$, which are in
$\EncTypes$. Their second components have the forms $L$, $T$, and $\CPoint\to X'$, which are
in $\mathcal{S}$. The constraints are sorted as in the previous case.

\textbf{Case arrows.} If the summand is in $\EncTypes$, then by Step~1 it has no arrow atom:
$N$ is empty and no constraint is produced. If it is in $\mathcal{S}$, the domains of its
atoms have the forms $X$, $\mathsf{In}(X_1,X_2,X_3)$, and $\CPoint$, and their codomains
the forms $\CPoint$, $X_4$, and $X$: all are in $\EncTypes$, so both constraints have their
left-hand sides in $\EncTypes$ and are sorted.

\emph{Step 3 (merging).} The constraints of $D$ are sorted. By Step~2, in the sets that the
normalization of $D$ returns every bound is in $\EncTypes$. Merging preserves this property:
the intersection of the upper bounds of a variable and the union of its lower bounds are in
$\EncTypes$, so each constraint $t_1\issubt t_2$ is sorted, and by Step~2 its normalization
adds only bounds in $\EncTypes$.

\emph{Step 4 (solving).} Let $\hat\sigma$ be the substitution obtained from a set $E$
returned by merging, and let $\EncTypes'=\Set{X\hat\sigma\mid X\in\EncTypes}$. The elements
of $\EncTypes'$ are contractive regular types, since the images of $\hat\sigma$ are. We
show that each of them has one of the forms that define $\EncTypes$, with components in
$\EncTypes'$. Since $\EncTypes$ is the greatest such set, this gives
$\EncTypes'\subseteq\EncTypes$, hence $\hat\sigma(\alpha)=\alpha\hat\sigma\in\EncTypes$ for
every variable $\alpha$. Let $X\in\EncTypes$. If $X$ is a variable that is not bounded in
$E$, then $X\hat\sigma=X$. If $X$ is a variable bounded in $E$ by $t_1$ and $t_2$, then
$X\hat\sigma=Y\hat\sigma$ for $Y=(t_1\lor\alpha')\land t_2$, which is in $\EncTypes$ by
Step~3 and is not a variable: we continue with $Y$ in place of $X$. If $X$ is not a
variable, then $X\hat\sigma$ has the form of $X$, in which every component $X_k$ is replaced
by $X_k\hat\sigma$, an element of $\EncTypes'$, the variables of $X$ occurring only in these
components.
\end{proof}

The junk forms are what makes Step~2 work. The first components of the pid and of the
function forms place the tags $\ell_{\Wch}$ and $\ell_{\Wfun}$ beside the first components
of the tuple forms; the second components of the same forms place the arrow $X\to\CPoint$
beside the arrow $\mathsf{In}(X_1,X_2,X_3)\to X_4$, hence $\mathsf{In}(X_1,X_2,X_3)$ beside
$X$ and $\CPoint$ beside $X_4$. A bound of a variable may therefore contain any of them.

The lemma depends on the order in which normalization proceeds: a type variable is singled
out in a summand before the kinds of the other atoms are examined, as in the normalization
rules of the extended version of~\cite{polyduce2}, where a summand is any intersection of
atoms and of negated atoms~\cite{CX11}.

\begin{lemma}[Tallying through the encoding]\label{lem:tally-readback}
Let $C$ be a finite set of constraints between types.
\begin{enumerate}[noitemsep,topsep=2pt]
\item \emph{(Soundness)} $\textsf{Solutions}(C)$ is a finite set of substitutions with regular
images, and each of its elements satisfies $C$.
\item \emph{(Completeness)} For every $\sigma'$ that satisfies $C$ there are
$\sigma\in\textsf{Solutions}(C)$ and a substitution $\rho$ such that
$\sigma'(\alpha)\simeq\sigma(\alpha)\rho$ for every variable $\alpha$ of $C$.
\end{enumerate}
\end{lemma}

\begin{proof}
Both items follow from Lemma~\ref{lem:readback}, the syntactic
facts~\eqref{eq:dec-syntactic}, and Lemma~\ref{lem:tally-closure}.

\emph{Item 1.} Let $\hat\sigma\in\textsf{Sol}_\varnothing(\enc{C})$ and $(s\issubt t)\in C$.
By the soundness of tallying, $\enc{s}\hat\sigma\subt_\textsf{FCB}\enc{t}\hat\sigma$. By
Lemma~\ref{lem:tally-closure}, $\hat\sigma$ is an $\EncTypes$-substitution, so both sides are in $\EncTypes$, and
Lemma~\ref{lem:readback} and~\eqref{eq:dec-syntactic} give
$s\,\dec{\hat\sigma}\subt t\,\dec{\hat\sigma}$. Finiteness is inherited from
$\textsf{Sol}_\varnothing(\enc{C})$, and regularity from $\dec{\cdot}$.

\emph{Item 2.} Let $\sigma'$ satisfy $C$.

\emph{Step 1 (renaming apart).} Let $\kappa$ rename the variables of the images
$\sigma'(\alpha)$, for $\alpha$ a variable of $C$, into variables that occur neither in $C$
nor in the images of the elements of $\textsf{Sol}_\varnothing(\enc{C})$, and let
$\sigma''(\alpha)=\sigma'(\alpha)\kappa$. Then $\sigma''$ satisfies $C$, subtyping being
preserved by substitutions.

\emph{Step 2 (first tally).} By Definition~\ref{def:subtyping} and
$\enc{t\sigma''}=\enc{t}\enc{\sigma''}$, the substitution $\enc{\sigma''}$ satisfies
$\enc{C}$. Completeness of tallying gives $\hat\sigma\in\textsf{Sol}_\varnothing(\enc{C})$ and
$\hat\rho$ with $\enc{\sigma''(\alpha)}\simeq\hat\sigma(\alpha)\hat\rho$ for every variable
$\alpha$ of $C$.

\emph{Step 3 (second tally).} The images of $\hat\rho$ need not be in $\EncTypes$ (an image
$\TInt\times\enc{\TTuple}$ does no harm under $\land(\TInt\times\Cnil)$), so $\hat\rho$ cannot
be read back: we replace it. Let $\Delta$ be the set of variables of the
$\enc{\sigma''(\alpha)}$, and
$D=\Set{\hat\sigma(\alpha)\issubt\enc{\sigma''(\alpha)},\ \enc{\sigma''(\alpha)}\issubt\hat\sigma(\alpha)\mid\alpha\text{ variable of }C}$.
The restriction of $\hat\rho$ to the variables of the $\hat\sigma(\alpha)$ has a domain
disjoint from $\Delta$ and satisfies $D$. So completeness of tallying gives some
$\hat\tau\in\textsf{Sol}_\Delta(D)$, and soundness gives
$\hat\sigma(\alpha)\hat\tau\simeq\enc{\sigma''(\alpha)}$, since $\hat\tau$ leaves $\Delta$
unchanged.

\emph{Step 4 (readback).} The constraints of $D$ are between types in $\EncTypes$, so
$\hat\tau$ is an $\EncTypes$-substitution by Lemma~\ref{lem:tally-closure}, and Lemma~\ref{lem:readback}
and~\eqref{eq:dec-syntactic} give
$\dec{\hat\sigma}(\alpha)\,\dec{\hat\tau}\simeq\sigma''(\alpha)$. Choose the fresh variables
of $\hat\tau$ outside the range of $\kappa$. Since $\kappa^{-1}$ fixes the variables of
$\dec{\hat\sigma}(\alpha)$ and preserves $\simeq$, the statement holds with
$\sigma=\dec{\hat\sigma}$ and $\rho$ the composition of $\dec{\hat\tau}$ followed by
$\kappa^{-1}$.
\end{proof}

\subsection{Soundness of inference}\label{app:elab-soundness}
Soundness is proved on elaborations (Lemma~\ref{lem:elab-sound}), using the soundness of
the constraint-generating pattern environments (Lemma~\ref{lem:pe-sound});
Theorem~\ref{thm:inf-sound} follows by Lemma~\ref{lem:decoration}.

\begin{lemma}[Pattern environments, soundness]\label{lem:pe-sound}
If $\ptrel{t}{p}{\Context_p}{D}$ and $\sigma$ satisfies $D$, then
$t\sigma\subt\accept{p}$ and $(\ptenv{t\sigma}{p})(x)\subt\Context_p\sigma(x)$ for every
capture variable $x$ of $p$.
\end{lemma}

\begin{proof}
By induction on $p$.

\textbf{Case $p=x$.} $\Context_p=\{x:t\}$ and $\accept{x}=\Tany$.

\textbf{Case $p=\tau$.} $D=\{t\issubt\tau\}$, and $p$ has no capture variable.

\textbf{Case $p=p_1\land p_2$.} The induction hypothesis on both premises.

\textbf{Case $p=\Tuple{p_1,\dots,p_n}$.} $D$ contains
$t\issubt\Tuple{\gamma_1,\dots,\gamma_n}$ and the constraints $D_k$ of the premises
$\ptrel{\gamma_k}{p_k}{\Context_k}{D_k}$. By the induction hypothesis
$\gamma_k\sigma\subt\accept{p_k}$, so
$t\sigma\subt\Tuple{\gamma_1\sigma,\dots,\gamma_n\sigma}\subt\accept{p}$. For $x$ in $p_k$,
$(\ptenv{t\sigma}{p})(x)=(\ptenv{\pi_k(t\sigma)}{p_k})(x)$, and
$\pi_k(t\sigma)\subt\gamma_k\sigma$ by Lemma~\ref{lem:penv-proj}. Then
Lemma~\ref{lem:penv-mono} and the induction hypothesis give
$(\ptenv{\pi_k(t\sigma)}{p_k})(x)\subt(\ptenv{\gamma_k\sigma}{p_k})(x)\subt\Context_k\sigma(x)$.
\end{proof}

\begin{lemma}[Soundness on elaborations]\label{lem:elab-sound}
If $\telab{\Context}{e}{t}[C]{f}$ and $\sigma$ satisfies $C$, then
$\wte[i\sigma][o\sigma]{\Context\sigma}{f\sigma}{t\sigma}$ in the system of
Section~\ref{sec:typing}, by a derivation that omits exactly the branches that the
deduction omits.
\end{lemma}

\begin{proof}
By induction on the deduction and by cases on the last rule applied. Refined types commute
with $\sigma$, test types containing no type variable.

\textbf{Case \rulename{i-const}, \rulename{i-pid}, \rulename{i-var}, \rulename{i-self}.}
The corresponding rule, with $(\Context\sigma)(x)=\Context(x)\sigma$ and
$\Self:\TChan{o\sigma}$.

\textbf{Case \rulename{i-abs}.} Let $f'$ be the elaborated body. For each $j$, the induction
hypothesis gives
$\wte[\alpha_j\sigma][\beta_j\sigma]{\Context\sigma,x:s_j\sigma}{f'\sigma}{t'_j\sigma}$, and
$t'_j\sigma\subt t_j\sigma$ since $\sigma$ satisfies $C$. We conclude by \rulename{t-sub} and
\rulename{t-abs}, the elaborated abstraction carrying
$\bigwedge_j s_j\sigma\to_{\beta_j\sigma}^{\alpha_j\sigma}t_j\sigma$.

\textbf{Case \rulename{i-app}.} The induction hypothesis types $f_1\sigma$ at $t_1\sigma$ and
$f_2\sigma$ at $t_2\sigma$, and
$t_1\sigma\subt t_2\sigma\to_{o\sigma}^{i\sigma}\gamma\sigma$ since $\sigma$ satisfies $C$.
We conclude by \rulename{t-sub} and \rulename{t-app}.

\textbf{Case \rulename{i-tuple}, \rulename{i-plus}, \rulename{i-send}.} The induction
hypothesis, \rulename{t-sub} with the emitted constraints, and the corresponding rule.

\textbf{Case \rulename{i-spawn}.} The induction hypothesis types the body at
$[\beta\sigma,\alpha\sigma]$. Since $\beta\sigma\subt t\sigma\subt\alpha\sigma$,
Theorem~\ref{thm:narrowing} moves the judgment to $[t\sigma,t\sigma]$, and we conclude by
\rulename{t-spawn}, the elaborated $\mkkeyword{spawn}$ carrying $t\sigma$.

\textbf{Case \rulename{i-case}.} The induction hypothesis types the scrutinee at $t\sigma$,
and $t\sigma\subt\bigvee_j\accept{p_j}$ since $\sigma$ satisfies $C$. The derivation omits
the branches $j\notin J$, which is allowed since $r_j\simeq\Tempty$ implies
$r_j\sigma\simeq\Tempty$, and types every branch $j\in J$, the omission being optional
(Section~\ref{sec:typing-rules}). For $j\in J$, Lemma~\ref{lem:pe-sound} gives
$(\ptenv{r_j\sigma}{p_j})(x)\subt\Context_j\sigma(x)$ for every capture variable $x$ of $p_j$;
the induction hypothesis types $f_j\sigma$ under $\Context\sigma,\Context_j\sigma$ at
$t_j\sigma$; Lemma~\ref{lem:env-sub} moves it under
$\Context\sigma,\ptenv{r_j\sigma}{p_j}$, and \rulename{t-sub} brings it to
$\bigvee_{k\in J}t_k\sigma$. We conclude by \rulename{t-case}.

\textbf{Case \rulename{i-receive}.} As for \rulename{i-case}, with the refined types computed
from $i\sigma$ and no exhaustiveness condition.

In every case the derivation omits exactly the branches that the deduction omits:
Lemma~\ref{lem:env-sub} only inserts \rulename{t-sub} below \rulename{t-var}, and the
derivation that Theorem~\ref{thm:narrowing} builds omits exactly the branches of the given
one.
\end{proof}

\noindent The case of \rulename{i-spawn} uses of the type $t$ only the constraints
$\beta\issubt t\issubt\alpha$. The lemma thus holds unchanged when a $\mkkeyword{spawn}$
carries an arbitrary type instead of an annotation
(Footnote~\ref{fn:spawn-variants}), as in Section~\ref{sec:reconstruction}; the elaborated
$\mkkeyword{spawn}$ then carries that type.

\InfSoundThm*

\begin{proof}
By Lemma~\ref{lem:elab-sound}, $\wte[i\sigma][o\sigma]{\Context\sigma}{f\sigma}{t\sigma}$ in
the system of Section~\ref{sec:typing}, for the elaboration $f$ of the deduction, by a
derivation that omits exactly the branches that the deduction omits. Since $f\sigma$ is a
decoration of $e$, this is, by Lemma~\ref{lem:decoration}, a derivation of
$\wte[i\sigma][o\sigma]{\Context\sigma}{e}{t\sigma}$ in the system of
Section~\ref{sec:annotations}.
\end{proof}

\subsection{Completeness of inference}\label{app:elab-completeness}
Completeness is proved on elaborations (Lemma~\ref{lem:elab-complete}), using the
completeness of the constraint-generating pattern environments
(Lemma~\ref{lem:pe-complete}); Theorem~\ref{thm:inf-complete} follows by
Lemmas~\ref{lem:decoration} and~\ref{lem:tally-readback}.

\begin{lemma}[Pattern environments, completeness]\label{lem:pe-complete}
If $\ptrel{r}{p}{\Context_p}{D}$, $\theta$ is a type substitution, and
$r\theta\subt\accept{p}$, then $\theta$ extends, on the variables minted by
\rulename{pe-tuple} in the deduction, to a substitution $\theta'$ that satisfies $D$ and
has $\Context_p\theta'(x)\simeq(\ptenv{r\theta}{p})(x)$ for every capture variable $x$
of $p$.
\end{lemma}

\begin{proof}
By induction on $p$.

\textbf{Case $p=x$.} $D=\varnothing$ and $\theta'=\theta$.

\textbf{Case $p=\tau$.} $D=\{r\issubt\tau\}$, which $\theta$ satisfies by hypothesis.

\textbf{Case $p=p_1\land p_2$.} The induction hypothesis on both premises, whose minted
variables are distinct.

\textbf{Case $p=\Tuple{p_1,\dots,p_n}$.} Let $\theta'(\gamma_k)=\pi_k(r\theta)$. Then
$r\theta\subt\Tuple{\pi_1(r\theta),\dots,\pi_n(r\theta)}$ by Lemma~\ref{lem:penv-proj},
and $\pi_k(r\theta)\subt\accept{p_k}$ since $r\theta\subt\accept{p}$. So the induction
hypothesis applies to $\ptrel{\gamma_k}{p_k}{\Context_k}{D_k}$ and extends $\theta'$ so that
it satisfies $D_k$ with $\Context_k\theta'(x)\simeq(\ptenv{\pi_k(r\theta)}{p_k})(x)
=(\ptenv{r\theta}{p})(x)$ for every capture variable $x$ of $p_k$.
\end{proof}

\begin{lemma}[Completeness on elaborations]\label{lem:elab-complete}
Let $\telab{\Context}{e}{t}[C]{f}$, let $\theta$ be a type substitution, and let
$\mathcal{D}$ be a derivation of $\wte[i\theta][o\theta]{\Context\theta}{f\theta}{t''}$ in
the system of Section~\ref{sec:typing} that is compatible with the deduction. Then some
substitution $\theta'$ satisfies $C$ and has $t\theta'\subt t''$, and it differs from $\theta$
only on the variables that \rulename{i-app}, \rulename{pe-tuple} and \rulename{i-spawn} mint
in the deduction.
\end{lemma}

\begin{proof}
By induction on the deduction, then on $\mathcal{D}$, and by cases on the last rule of the
deduction. The variables of \rulename{i-app} and \rulename{pe-tuple} are minted at each
application of their rule, so those of distinct premises are distinct; the variables
$\alpha,\beta$ of an \rulename{i-spawn} are minted once per occurrence, and the case below
maps them to a type that depends only on $\theta$ and $f$. None of them occurs in
$\Context$, $i$, $o$ or $f$. So the substitutions built for the premises of a rule agree
where they overlap, their union $\theta'$ differs from $\theta$ only on these variables, and
it leaves unchanged the elaborated subterms and the brackets of all the premises. If
$\mathcal{D}$ ends with \rulename{t-sub}, the statement for its premise, a smaller derivation
for the same deduction, implies the statement for $\mathcal{D}$; so we assume that it does
not, and $\mathcal{D}$ ends with the declarative counterpart of the last rule of the
deduction.

\textbf{Case \rulename{i-const}, \rulename{i-pid}, \rulename{i-var}, \rulename{i-self}.}
$C=\varnothing$, and $t\theta$ is the type in the conclusion of $\mathcal{D}$.

\textbf{Case \rulename{i-abs}.} The $j$-th premise of \rulename{t-abs} types the elaborated
body under $\Context\theta,x:s_j\theta$, at the bracket $[\beta_j\theta,\alpha_j\theta]$ and
at the type $t_j\theta$. The induction hypothesis gives $t'_j\theta'\subt t_j\theta$, that is,
the constraint that the rule emits, and $t\theta'=t\theta$ is the type in the conclusion of
\rulename{t-abs}.

\textbf{Case \rulename{i-app}.} The premises of \rulename{t-app} type $f_1\theta$ at
$s'\to_{o\theta}^{i\theta}t''$ and $f_2\theta$ at $s'$. The induction hypothesis gives
$t_1\theta'\subt s'\to_{o\theta}^{i\theta}t''$ and $t_2\theta'\subt s'$. Let
$\theta'(\gamma)=t''$: the constraint of the rule holds, since
$s'\to_{o\theta}^{i\theta}t''\subt t_2\theta'\to_{o\theta}^{i\theta}t''$ by contravariance of
the domain.

\textbf{Case \rulename{i-tuple}, \rulename{i-plus}.} The induction hypothesis on the
premises, those of \rulename{t-plus} being typed at $\TInt$.

\textbf{Case \rulename{i-spawn}.} The premise of \rulename{t-spawn} types the elaborated
body at $[t\theta,t\theta]$. Let $\theta'(\alpha)=\theta'(\beta)=t\theta$ for the variables
$\alpha,\beta$ of the rule, which satisfies $\beta\issubt t\issubt\alpha$; the induction
hypothesis applies to the premise, and $\TChan{t\theta}$ is the type in the conclusion of
\rulename{t-spawn}.

\textbf{Case \rulename{i-send}.} The premises of \rulename{t-send} type $f_1\theta$ at
$\TChan{t''}$ and $f_2\theta$ at $t''$. The induction hypothesis gives
$t\theta'\subt\TChan{t''}$ and $t'\theta'\subt t''$, hence
$t\theta'\subt\TChan{t'\theta'}$ by contravariance.

\textbf{Case \rulename{i-case}.} $\mathcal{D}$ types the scrutinee at some $s'$ with
$s'\subt\bigvee_j\accept{p_j}$, and the induction hypothesis gives $t\theta'\subt s'$. So the
exhaustiveness constraint holds, and $r_j\theta'\subt r'_j$ for every $j$, where $r'_j$ is
the refined type of the $j$-th branch in $\mathcal{D}$. A branch $j\notin J$ emits no
constraint and contributes no summand to the type of the deduction. Let $j\in J$. By
compatibility, $\mathcal{D}$ types the branch, under $\Context\theta,\ptenv{r'_j}{p_j}$ and at
$t''$. Lemma~\ref{lem:pe-complete} extends $\theta'$ so that it satisfies $D_j$ with
$\Context_j\theta'(x)\simeq(\ptenv{r_j\theta'}{p_j})(x)\subt(\ptenv{r'_j}{p_j})(x)$ for every
capture variable $x$ of $p_j$, by Lemma~\ref{lem:penv-mono}. Lemma~\ref{lem:env-sub}
re-types the branch under $\Context\theta,\Context_j\theta'$ by a derivation that omits the
same branches, since it only inserts \rulename{t-sub} below \rulename{t-var}; this
derivation is compatible with the deduction of the branch. The induction hypothesis gives
$t_j\theta'\subt t''$. Hence $\bigvee_{j\in J}t_j\theta'\subt t''$.

\textbf{Case \rulename{i-receive}.} As for \rulename{i-case}, the refined types being computed
from $i\theta$ on both sides.
\end{proof}

\InfCompleteThm*

\begin{proof}
Let $\telab[\alpha][\beta]{\Context}{e}{t'}[C]{f}$ be the deduction of $e$, which exists
since compatibility with inference is defined with respect to it
(Definition~\ref{def:compatible}). By Lemma~\ref{lem:decoration}, the given derivation is a derivation of
$\wte[i][o]{\Context}{f\theta}{t}$ in the system of Section~\ref{sec:typing}, for a
substitution $\theta$ on the variables that decorate the occurrences of $f$
(Definition~\ref{def:elaboration}); we extend it with $\theta(\alpha)=i$ and
$\theta(\beta)=o$. This derivation is compatible with the deduction
(Definition~\ref{def:compatible}). By Lemma~\ref{lem:elab-complete}, some $\theta'$ with
$\theta'(\alpha)=i$ and $\theta'(\beta)=o$ satisfies $C$ and has $t'\theta'\subt t$. By Lemma~\ref{lem:tally-readback}(2) there are
$\sigma\in\textsf{Solutions}(C)$ and $\rho$ such that $\rho\sigma$ and $\theta'$ map every
variable of $C$ to equivalent types. The variables of $t'$, $\alpha$ and $\beta$ that do not occur in $C$ are left unchanged by
$\sigma$ and do not occur in its images, the variables that tallying introduces being
fresh; let $\rho$ agree with $\theta'$ on them. Then $i\simeq\rho\sigma(\alpha)$,
$\rho\sigma(\beta)\simeq o$, and $t'\rho\sigma\simeq t'\theta'\subt t$.
\end{proof}

\subsection{Termination}\label{app:elab-termination}

\InferenceTerminatesProp*

\begin{proof}
\emph{Item 1.} By induction on $e$. Exactly one rule applies to each form of expression,
and its premises are on proper subterms; likewise $\rightsquigarrow$ is directed by the
pattern. The only test that the rules perform is $r_j\not\simeq\Tempty$ in \rulename{i-case}
and \rulename{i-receive}, which is decidable on types with type
variables~\cite[Section~3.3]{castagna2022programming}; the only choices are the names of
the variables that the rules mint and of the families of $\decorate\cdot$, whence uniqueness
up to renaming. Every rule mints finitely many variables and emits finitely many
constraints, so the derivation and $C$ are finite. The types that the rules build are
regular and contractive: $\decorate{\ta}$ by Lemma~\ref{lem:decorate-wd}, and the others
because the rules build them from types by type constructors and connectives.

\emph{Item 2.} By item~1, Lemma~\ref{lem:tally-readback} applies to $C$.

\emph{Item 3.} The item states two equivalences: the first for typability at some bracket,
the second for typability at a given ground bracket $[o',i']$. We prove them in turn. In
both, by Lemma~\ref{lem:tally-readback}, $\textsf{Solutions}(X)\neq\varnothing$ if and only
if $X$ is satisfiable, for every finite set $X$ of constraints.

\emph{Item 3, first equivalence (some bracket).} Let $\sigma$ satisfy $C$. The variables of
$\Context$ do not occur in $C$, so we may assume that $\sigma$ leaves them unchanged. Then
Theorem~\ref{thm:inf-sound} gives a derivation of $\wte[i\sigma][o\sigma]{\Context}{e}{t\sigma}$
that omits exactly the branches that the deduction omits (Lemma~\ref{lem:elab-sound}), hence
compatible with inference. The converse is Theorem~\ref{thm:inf-complete}.

\emph{Item 3, second equivalence (ground bracket $[o',i']$).} If $\sigma$ satisfies
$C\cup\{i'\issubt i,\;o\issubt o'\}$, then $\sigma$ satisfies $C$, the bracket $[o',i']$
narrows $[o\sigma,i\sigma]$, and Theorem~\ref{thm:narrowing} moves the derivation given by
the first equivalence to $[o',i']$, omitting the same branches. Conversely,
Theorem~\ref{thm:inf-complete} gives $\sigma\in\textsf{Solutions}(C)$ and $\rho$ with
$i'\subt\rho\sigma(i)$ and $\rho\sigma(o)\subt o'$; since $i'$ and $o'$ are ground, the
composition of $\sigma$ followed by $\rho$ satisfies the enlarged set.

\emph{Decidability.} Let $\Context$ contain no inference variable, and apply the rules of
Figure~\ref{fig:inference} to $e$ at fresh $i$ and $o$, by induction on $e$ as in item~1.
Every pattern has a rule of $\rightsquigarrow$, and every expression has a rule except a
variable outside the domain of the environment. If the rules reach such a variable, $e$ is
typable at no bracket by a derivation compatible with inference, by
Theorem~\ref{thm:inf-complete}. Otherwise they return $C$, and by item~3 $e$ is typable at
some bracket, respectively at the ground bracket $[o',i']$, if and only if
$\textsf{Solutions}(C)$, respectively $\textsf{Solutions}(C\cup\{i'\issubt i,\;o\issubt
o'\})$, is not empty, which tallying decides, both sets of constraints being finite
(Lemma~\ref{lem:tally-readback}).
\end{proof}

\noindent On the cost of generation: \rulename{i-abs} traverses the body once per arrow of
the declared intersection, so generation is linear in the size of $e$ when annotations are
single arrows, and in the worst case exponential in the nesting depth of the abstractions
that carry intersections.

\subsection{Uniform requirements}\label{app:elab-uniform}
Proposition~\ref{prop:requirement} follows from Definition~\ref{def:requirement}.
Lemma~\ref{lem:narrowing-closed} shows that the uniform requirements are those that have a
best decoration.

\RequirementProp*

\begin{proof}
\emph{Only if.} Let $\sigma$ satisfy $C$ with $\beta\sigma\subt\bigwedge_k s_k$ and
$\bigvee_k s_k\subt\alpha\sigma$. Composing $\sigma$ with a substitution that maps every
type variable to a ground type preserves these properties, subtyping being preserved by
type substitutions and the $s_k$ being ground; so we assume $\alpha\sigma$ and $\beta\sigma$
ground. By Definition~\ref{def:requirement}, the admitted bracket
$[\beta\sigma,\alpha\sigma]$ narrows $[o_0,i_0]$, hence
$o_0\subt\beta\sigma\subt s_k\subt\alpha\sigma\subt i_0$ for every $k$.

\emph{If.} The bracket $[o_0,i_0]$ narrows itself, so the requirement admits it: some
$\sigma$ satisfying $C$ has $\alpha\sigma\simeq i_0$ and $\beta\sigma\simeq o_0$, and then
$\beta\sigma\subt\bigwedge_k s_k$ and $\bigvee_k s_k\subt\alpha\sigma$.
\end{proof}

\begin{lemma}[Admitted brackets]\label{lem:narrowing-closed}
Let $\tinf[\alpha][\beta]{\Context}{e}{t}[C]$ with $\alpha,\beta$ fresh.
\begin{enumerate}[noitemsep,topsep=2pt]
\item If the requirement of $e$ admits a bracket, then it admits every ground bracket that
narrows it.
\item The requirement of $e$ is uniform if and only if it admits a bracket $[o_0,i_0]$ that
every bracket it admits narrows.
\end{enumerate}
\end{lemma}

\begin{proof}
\emph{Item 1.} Let $\sigma$ satisfy $C$ with $\alpha\sigma$ and $\beta\sigma$ ground, let the
ground bracket $[o',i']$ narrow $[\beta\sigma,\alpha\sigma]$, that is, $\beta\sigma\subt o'$
and $i'\subt\alpha\sigma$, and let $f$ be the elaboration of the deduction. By
Lemma~\ref{lem:elab-sound} and Theorem~\ref{thm:narrowing},
$\wte[i'][o']{\Context\sigma}{f\sigma}{t\sigma}$ by a derivation that omits exactly the
branches that the deduction omits. Let $\theta$ map $\alpha$ to $i'$, $\beta$ to $o'$, and
every other type variable as $\sigma$ does. Since $\alpha$ and $\beta$ occur neither in
$\Context$, being fresh, nor in $f$, whose type variables are minted for its occurrences, this
is a derivation of $\wte[\alpha\theta][\beta\theta]{\Context\theta}{f\theta}{t\sigma}$,
compatible with the deduction (Definition~\ref{def:compatible}). By
Lemma~\ref{lem:elab-complete}, some $\theta'$ that satisfies $C$ differs from $\theta$ only on
type variables minted by \rulename{i-app}, \rulename{pe-tuple} and \rulename{i-spawn}; so
$\alpha\theta'=i'$ and $\beta\theta'=o'$, and the requirement admits $[o',i']$.

\emph{Item 2.} If the requirement admits such a bracket, then by item~1 the brackets it
admits are exactly the ground narrowings of $[o_0,i_0]$. The converse holds since a bracket
narrows itself.
\end{proof}

\begin{remark}[Uniform requirements]\label{rem:uniform}
\emph{Runners.} When $\Context$ contains no inference variable and the requirement of $e$
is uniform, Proposition~\ref{prop:requirement} with $n=1$ and
Proposition~\ref{prop:inference-terminates}(3) give, for every ground type $s$: $e$ is
typable at the bracket $[s,s]$ by a derivation compatible with inference if and only if
$o_0\subt s\subt i_0$.

\emph{Best decorations.} By Lemma~\ref{lem:narrowing-closed}(2), a non-uniform requirement
has no best decoration. This is a property of $C$ as generated by $e$ alone: in a whole
program, the calls of a function bound its input decoration from below and its output
decoration from above.

\emph{Computing $i_0$ and $o_0$.} For ground types $q$, $m$ and $i$, we have
$i\land q\subt m$ if and only if $i\subt m\lor\neg q$. Hence, if eliminating the type
variables of $C$ other than $\alpha$ and $\beta$ leaves finitely many constraints
$\alpha\land q_j\issubt m_j$ and $u_h\issubt\beta$ with $q_j$, $m_j$, $u_h$ ground, in the
sense that these constraints have the same ground solutions for $\alpha$ and $\beta$ as
$C$, then the requirement is uniform with $i_0=\bigwedge_j(m_j\lor\neg q_j)$ and
$o_0=\bigvee_h u_h$. For \eelixir|halve|, eliminating the projection variables leaves the
single constraint~\eqref{eq:requirement}, and no constraint on $\beta$ since the body
neither sends nor uses $\Self$, so
$i_0=\Tuple{\Atom[{:}half],\TInt}\lor\neg\Tuple{\Atom[{:}half],\Tany}$ and $o_0=\Tempty$.

\emph{A greatest interface does not make a requirement uniform.} Let
$W=\Tuple{\Atom[{:}whoami],\Tany}$. In the branch of line~\ref{whoami}, $\Self$ has type
$\TChan\beta$, so \rulename{i-send} bounds the projection variable of \eelixir{from} by
$\TChan{\TChan\beta}$, and the requirement is
$\alpha\land W\issubt\Tuple{\Atom[{:}whoami],\TChan{\TChan\beta}}$. Its right-hand side is
monotone in $\beta$, so a ground interface $s$ can run the branch, that is, some $\sigma$
satisfying the requirement has $\beta\sigma\subt s\subt\alpha\sigma$, if and only if
$s\subt G(s)$, where $G(s)=\neg W\lor\Tuple{\Atom[{:}whoami],\TChan{\TChan s}}$. Since $G$ is
monotone and $s$ occurs in $G(s)$ only below type constructors, every such $s$ is below the
solution $s^\top$ of $s=G(s)$ (Lemma~\ref{lem:recon-guarded}(2)), which is therefore the
greatest interface that can run the branch. But
$s_1=s^\top\land W=\Tuple{\Atom[{:}whoami],\TChan{\TChan{s^\top}}}$, which is below $s^\top$,
cannot: that would require $\TChan{\TChan{s^\top}}\subt\TChan{\TChan{s_1}}$, that is,
$s^\top\subt s_1$, which fails since $\neg W\subt s^\top$ and $s_1\subt W$. Were the
requirement uniform, $o_0$ would be $\Tempty$, the bracket $[\Tempty,\Tempty]$ being
admitted, and Proposition~\ref{prop:requirement} would let every interface below $s^\top$
run the branch, $s_1$ included.
\end{remark}

\subsection{Type erasure and safety}\label{app:elab-erasure}
Theorem~\ref{thm:typerase} relates the reductions of a configuration and of its
erasure, and Corollary~\ref{cor:inferred-safety} transfers
Corollary~\ref{cor:soundness} to the erased program along it.

\begin{lemma}[Erasure and matching]\label{lem:erase-match}
For every value $v$, pattern $p$ and expression $e$:
\begin{enumerate}[noitemsep,topsep=2pt]
\item $\testof{\typeerase v}=\testof{v}$;
\item $\match{\typeerase v}{p}=\textsf{fail}$ if and only if $\match{v}{p}=\textsf{fail}$;
\item $\typeerase{e\subst{v}{p}}=\typeerase{e}\subst{\typeerase v}{p}$.
\end{enumerate}
\end{lemma}

\begin{proof}
The map $\testof\cdot$ discards exactly what $\typeerase\cdot$ erases, namely
$\testof{\PID^{t}}=\TChan\Tempty$ and $\testof{\Fun^{l}x.e}=\TFun$, and is homomorphic on
constants and tuples; $\match{v}{p}$ is defined from $\testof v$ at its leaves, and
$\typeerase\cdot$ changes no pattern.
\end{proof}

\TypeEraseThm*

\begin{proof}
Erasure maps evaluation contexts to evaluation contexts, commutes with structural
congruence, and is a bijection on each syntactic form except for the annotations it
erases. So it suffices to show, for each rule of Section~\ref{sec:opsem},
that it applies to a redex of $P$ if and only if its erased version applies to the erased
redex, with reducts that correspond.

\textbf{Case $\beta$-reduction, addition.} Immediate, by Lemma~\ref{lem:erase-match} at $p=x$.

\textbf{Case $\mkkeyword{case}$, $\mkkeyword{receive}$.} The side conditions of the rules
mention only $\match\cdot\cdot$. By Lemma~\ref{lem:erase-match}, the same branch is selected on
$\typeerase v$ as on $v$, and the reducts correspond.

\textbf{Case $\Self$, $\mkkeyword{spawn}$.} No side condition mentions an annotation: the
interface $s$ of $\Spawn{e}{s}$ only labels the new process, and the freshness of its pid is
unaffected.

\textbf{Case $\mkkeyword{send}$.} The two rules require the target to be $\PID[b]^{s}$ with
$s$ the interface of the recipient $\Process{\PID[b]}{s}{e_2}{\seqof w}$, whereas their
erased versions select the recipient by the name $\PID[b]$ alone. So item~1 is immediate. For
item~2, a redex $\Send{\PID[b]}{v'}$ of $\typeerase P$ is the erasure of a redex
$\Send{\PID[b]^{s}}{v}$ of $P$. By well-formedness (Definition~\ref{def:wf-config}), the process of pid $\PID[b]$ in $P$
has interface $s$, and, the pids of a configuration being pairwise distinct, it is the
process whose erasure the erased rule selects. So the annotated rule applies, and its reduct erases to the erased one.
\end{proof}

\noindent Well-formedness is needed only for item~2, and only for sends: without it,
$\typeerase P$ could deliver a message to a process whose interface differs from the
annotation of the pid, where $P$ is stuck.

\InferredSafetyCor*

\begin{proof}
Let $f$ be the elaboration of the deduction.

\emph{Step 1 (a well-typed configuration).} The composition of $\sigma$ followed by $\rho$
satisfies $C$, subtyping being preserved by type substitutions. By
Lemma~\ref{lem:elab-sound},
$\wte[\alpha\sigma\rho][\beta\sigma\rho]{\emptyset}{f\sigma\rho}{t\sigma\rho}$ in the system of
Section~\ref{sec:typing}. Since $\beta\sigma\rho\subt s\subt\alpha\sigma\rho$, $s$ being ground,
Theorem~\ref{thm:narrowing} moves the judgment to $[s,s]$. The term $f\sigma\rho$ is ground
and pid-free, so $P=\Process{\PID}{s}{f\sigma\rho}{\EmptyMailbox}$ is well typed by
\rulename{t-thread} and well-formed, and $\typeerase{P}=(\PID,\typeerase{e},\EmptyMailbox)$
since $\erase{f}=e$.

\emph{Step 2 (the erased run).} Let $Q$ be reachable from $\typeerase{P}$. Iterating
Theorem~\ref{thm:typerase}(2) along the reduction sequence, well-formedness being preserved
by reduction (Lemma~\ref{lem:config-inv}), gives $Q=\typeerase{Q'}$ for some $Q'$ reachable
from $P$, to which Corollary~\ref{cor:soundness} applies.

\emph{Item 1.} Every process of $Q'$ has terminated, waits, or can take a step. By
Theorem~\ref{thm:typerase}(1), a step of a process of $Q'$ is a step of the same process of
$Q$; termination and waiting are preserved by erasure, by Lemma~\ref{lem:erase-match}.

\emph{Item 2.} If the process of pid $\PID$ has terminated in $Q$ on $v$, then it has
terminated in $Q'$ on a value $v'$ with $\typeerase{v'}=v$, and
$\wtee{}{v'}{t\sigma\rho}$ by Corollary~\ref{cor:soundness}(3). When $v$ contains neither
abstractions nor pids, $v'=v$ and $\wte{}{v}{t\sigma\rho}$.
\end{proof}

\subsection{Elimination of bound variables}\label{app:elab-elimination}

Section~\ref{sec:elixir-tallying} describes how the prototype eliminates, before
tallying, the variables bounded in one direction by bare bounds only. We state the
condition under which a variable is eliminated and prove that elimination preserves
satisfiability.

We write $t\{u/\gamma\}$ for $t$ in which $u$ replaces $\gamma$, and
$\sigma[\gamma\mapsto\tau]$ for the substitution that maps $\gamma$ to $\tau$ and every
other variable as $\sigma$ does. A constraint $s\issubt t$ is \emph{antitone} in
$\gamma$ if $(s\wedge\neg t)\{\gamma\vee\delta/\gamma\}\subt s\wedge\neg t$ for a
variable $\delta$ that occurs neither in $s$ nor in $t$, that is, if $s\wedge\neg t$
decreases when $\gamma$ increases. It is \emph{monotone} in $\gamma$ if
$(s\wedge\neg t)\{\gamma\wedge\delta/\gamma\}\subt s\wedge\neg t$.

\begin{proposition}[Elimination]\label{prop:elimination}
Let $C$ be a constraint set and $\gamma$ a variable such that every constraint of $C$
in which $\gamma$ occurs either has the form $s\issubt u_k$ with $s\simeq\gamma$ and
$\gamma$ not occurring in $u_k$ ($k\in K$), or is antitone in $\gamma$. Let
$u=\bigwedge_{k\in K}u_k$, with $u=\Tany$ if $K$ is empty, and let $C'$ be obtained
from $C$ by removing the constraints $s\issubt u_k$ and replacing $\gamma$ by $u$ in
the others. Then a substitution $\sigma$ satisfies $C'$ if and only if
$\sigma[\gamma\mapsto\tau]$ satisfies $C$ for some type $\tau$, and $\tau=u\sigma$ is
one. The dual statement holds for lower bounds $l_k\issubt s$ with $s\simeq\gamma$,
every other constraint being monotone in $\gamma$, and $u=\bigvee_{k\in K}l_k$, with
$u=\Tempty$ if $K$ is empty.
\end{proposition}

\begin{proof}
The constraints in which $\gamma$ does not occur are the same in $C$ and $C'$, and
$\sigma$ satisfies them if and only if $\sigma[\gamma\mapsto\tau]$ does. We consider the
others, for upper bounds; the dual statement is symmetric.

\emph{If.} Let $\sigma[\gamma\mapsto\tau]$ satisfy $C$. Then $\tau\subt u\sigma$, by the
constraints $s\issubt u_k$. For every other constraint $s\issubt t$ of $C$ in which $\gamma$
occurs, instantiating its antitonicity by $\sigma[\gamma\mapsto\tau]$ and
$\delta\mapsto u\sigma$ gives
$(s\wedge\neg t)\{u/\gamma\}\sigma\subt(s\wedge\neg t)\sigma[\gamma\mapsto\tau]\simeq\Tempty$,
type equivalence being a congruence. So $\sigma$ satisfies the constraint
$s\{u/\gamma\}\issubt t\{u/\gamma\}$ of $C'$.

\emph{Only if.} Let $\sigma$ satisfy $C'$. Since $\gamma$ does not occur in $u$,
$t\{u/\gamma\}\sigma$ is $t\,\sigma[\gamma\mapsto u\sigma]$ for every type $t$. So
$\sigma[\gamma\mapsto u\sigma]$ satisfies every constraint $s\issubt t$ of $C$ whose image
$s\{u/\gamma\}\issubt t\{u/\gamma\}$ is in $C'$, and each $s\issubt u_k$, since
$s\,\sigma[\gamma\mapsto u\sigma]\simeq u\sigma\subt u_k\sigma$.
\end{proof}

\noindent Elimination is iterated while some variable qualifies, and a solution of
the final set extends to $C$ by assigning the eliminated variables in the reverse
order of their elimination. Only the eliminated variables lose their solutions, as
they receive an extreme one: $\{\alpha\issubt\TInt\}$ reduces to the empty set, whose
solution extends to $\alpha\mapsto\TInt$, of which $\alpha\mapsto\Tempty$ is not an
instance. This is why the variables of the types that the prototype reports are not
eliminated.

\section{Reconstruction: Proofs and Examples}
\label{app:reconstruction}
This appendix proves the statements of Section~\ref{sec:reconstruction} and works out the
examples it elides. Its first subsections follow the four steps of the reconstruction; the
proofs of the statements of Section~\ref{sec:reconstruction} come next
(Appendix~\ref{app:recon-proofs}), and the examples last (Appendix~\ref{app:recon-examples}).

The first subsections make formal what each step of Section~\ref{sec:reconstruction} computes.
The generation step (Appendix~\ref{app:recon-generation}): the form of the constraint set $C$
(Lemma~\ref{lem:recon-exptypes}) and the polarities of $\alpha$, $\beta$ and $\gamma$ in it
(Lemma~\ref{lem:recon-positive}). The derivation step (Appendices~\ref{app:recon-receives}
and~\ref{app:recon-demands}): the receives in $\rcv\alpha$ (Lemma~\ref{lem:recon-entail}), then
the type $m_j$ that a branch $j$ demands of its messages (Definition~\ref{def:recon-demand}); no rule emits $m_j$: it is read off the constraints on the types of capture variables, once
these constraints are decomposed (Definitions~\ref{def:recon-refinement}, \ref{def:recon-split}
and~\ref{def:recon-dec}). The solving step (Appendices~\ref{app:recon-overloaded}
and~\ref{app:recon-eliminating}): a call to an overloaded function that performs no
$\mkkeyword{receive}$ and evaluates no $\Self$ (a mailbox-pure function,
Definition~\ref{def:recon-pure}) has its constraint replaced by a bound on its argument
(Lemma~\ref{lem:recon-pure}); the constraints thus obtained, in which the $\mkkeyword{spawn}$s on
which this one depends are declared at their interfaces and the self-sends are left out
(Definition~\ref{def:recon-cstar}), give the extrema of the other type variables
(Definition~\ref{def:recon-elim}), and these turn the demands into the bound $F$ of
Definition~\ref{def:recon-bound}, whose properties are those of Lemma~\ref{lem:recon-elim}. The
checking step (Appendix~\ref{app:recon-checking}): Lemma~\ref{lem:recon-check} identifies its
constraints with those of $e[\vec{t}\,]$. The three propositions of
Section~\ref{sec:reconstruction} follow from these results.

The \emph{decoration variables} of a program $e$ are the type variables that its deduction
mints as decorations: those of $\decorate\cdot$, the $\alpha_j,\beta_j$ of \rulename{i-abs},
the $\alpha,\beta$ of \rulename{i-spawn} and \rulename{i-spawn-alt}, and the root bracket
$[\beta_0,\alpha_0]$; they include the input decorations of the uses of \rulename{i-receive}.
An \emph{expression type} is the type in the conclusion of a judgment of the deduction. An
\emph{annotation type} is a type $\decorate{\ta}$, or the type
$\bigwedge_j(s_j\to_{\beta_j}^{\alpha_j}t_j)$ that \rulename{i-abs} concludes: it is ground but
for decoration variables below its arrows.

\subsection{The generation step}\label{app:recon-generation}
The generation step yields the constraint set $C$ of the program
(Section~\ref{sec:recon-runs}). We describe its expression types and its constraints, and how
the readers and the writers of a mailbox occur in them.

\begin{definition}[Polarity]\label{def:recon-polarity}
An occurrence of a type variable in a type $t$, a path from the root of $t$ to a leaf labeled
by this variable, has \emph{polarity} $+$ if it crosses an even number of contravariant edges,
and $-$ otherwise; the contravariant edges lead from $\neg s$ to $s$, from $s\to_o^i s'$ to $s$
and to $i$, and from $\TChan{s}$ to $s$. A type variable occurs \emph{with polarity} $p$ in $t$
if one of its occurrences in $t$ has polarity $p$, and \emph{only with polarity} $p$ if all of
them have. An occurrence in a constraint $s\issubt t$ is \emph{positive} if it is an occurrence
in $s$ of polarity $-$ or an occurrence in $t$ of polarity $+$, and \emph{negative} otherwise.
We write $-p$ for the polarity opposite to $p$, and $pq$ for the product of two polarities.
\end{definition}

\begin{lemma}[Monotonicity]\label{lem:recon-monotone}
Let $t$ be a type and $\theta,\theta'$ type substitutions such that, for every type variable
$\xi$ of $t$: $\xi\theta\subt\xi\theta'$ if $\xi$ occurs in $t$ only with polarity $+$;
$\xi\theta'\subt\xi\theta$ if it occurs in $t$ only with polarity $-$; and
$\xi\theta\simeq\xi\theta'$ otherwise. Then $t\theta\subt t\theta'$.
\end{lemma}

\begin{proof}
We prove at once the statement and its dual, in which the hypotheses on the type variables of
polarity $+$ and of polarity $-$ are exchanged and the conclusion is $t\theta'\subt t\theta$, by
induction on the pair (element, type) that defines membership
(Appendix~\ref{app:enc-interpretation}).

\textbf{Case $t$ a type variable.} The hypothesis.

\textbf{Case $t$ a constant or a basic type.} Immediate.

\textbf{Case $t=\neg s$.} The dual for $s$, in which every occurrence has the opposite polarity.

\textbf{Case $t$ a union or an intersection.} The induction hypothesis on its operands.

\textbf{Case $t$ built by a type constructor.} The induction hypothesis on the components of
the element, and its dual at the contravariant arguments. By
Definition~\ref{def:interpretation} these are the domain of an arrow, its input decoration and
the argument of $\TChan\cdot$; the output decoration of an arrow is the argument of a
$\TChan\cdot$ in its domain, hence covariant, as is the codomain.
\end{proof}

\begin{lemma}[Expression types and constraints]\label{lem:recon-exptypes}
Let $C$ be the constraint set of a program $e$.
\begin{enumerate}[noitemsep,topsep=2pt]
\item Every expression type of the deduction of $e$ is generated by
\[
    \bar u\;::=\;g\mid\TChan{o}\mid\TChan{\gamma}\mid a\mid\zeta\mid\gamma'\mid
    (\bar u\land g)\setminus g'\mid(\delta\land g)\setminus g'\mid
    \Tuple{\bar u,\dots,\bar u}\mid\bar u\lor\bar u
\]
where $g,g'$ are ground types, $o$ is an output decoration, $\gamma$ the pid variable of an
unannotated $\mkkeyword{spawn}$, $a$ an annotation type, $\zeta$ a projection variable,
$\gamma'$ a result variable, and $\delta$ the input decoration of a use of \rulename{i-receive}.
\item Every constraint of $C$ is of one of the forms $\bar u\issubt g$, $\bar u\issubt a$,
$\bar u\issubt\Tuple{\zeta_1,\dots,\zeta_n}$, $\bar u\issubt\bar u'\to_o^i\gamma'$,
$\bar u\issubt\TChan{\bar u'}$, $\beta'\issubt a$ and $a\issubt\alpha'$, where $\bar u,\bar u'$ are
generated as in item~1, $\bar u$ is a projection variable or a type $(\bar u''\land g)\setminus g'$
or $(\delta\land g)\setminus g'$ in the third form, $[o,i]$ is the bracket of the call in the
fourth, and $a$ is an annotation type or a ground type and $\beta',\alpha'$ the decoration
variables of an annotated $\mkkeyword{spawn}$, or of the root bracket of a program whose interface
is declared, in the last two.
\item In a type $\bar u$ generated as in item~1:
\begin{enumerate}[label=(\alph*),ref=\alph*,noitemsep,topsep=1pt]
    \item\label{i:exptypes-atoms} every pid type and every arrow type is a $\TChan{o}$ or a
    $\TChan{\gamma}$, or occurs in a ground or an annotation type;
    \item\label{i:exptypes-top} the type variables that occur outside every type constructor are
    projection variables, result variables, and input decorations $\delta$ of uses of
    \rulename{i-receive}, the latter only in types $(\delta\land g)\setminus g'$, hence with
    polarity $+$;
    \item\label{i:exptypes-bracket} the decoration variables $\alpha,\beta$ at which
    \rulename{i-spawn-alt} types the body of an unannotated $\mkkeyword{spawn}$, its pid variable
    $\gamma$, and the root bracket $[\beta_0,\alpha_0]$ occur only in $\TChan{\beta}$, $\TChan{\gamma}$,
    $\TChan{\beta_0}$ and the types $(\alpha\land g)\setminus g'$ and $(\alpha_0\land g)\setminus g'$:
    $\beta$, $\gamma$ and $\beta_0$ only with polarity $-$, $\alpha$ and $\alpha_0$ only with polarity
    $+$.
\end{enumerate}
\end{enumerate}
\end{lemma}

\begin{proof}
\emph{Item 1.} By induction on the deduction and by cases on the last rule applied.

\textbf{Case \rulename{i-const}, \rulename{i-plus}.} A ground type.

\textbf{Case \rulename{i-self}.} The type $\TChan{o}$, with $o$ the output decoration of the
judgment.

\textbf{Case \rulename{i-spawn}, \rulename{i-spawn-alt}.} A pid type $\TChan{t}$, with $t$ an
annotation type or a ground type, which is an annotation type or a ground type, or
$\TChan{\gamma}$.

\textbf{Case \rulename{i-abs}.} An annotation type.

\textbf{Case \rulename{i-var}.} The type of the variable, which \rulename{i-abs} binds to an
annotation type and \rulename{pe-var} to its scrutinee: a projection variable
(\rulename{pe-tuple}) or the refined type $r_j$ of \rulename{i-case} or \rulename{i-receive},
of the form $(\bar u\land g)\setminus g'$ or $(\delta\land g)\setminus g'$. The program being
closed, no other binding occurs.

\textbf{Case \rulename{i-app}.} The fresh result variable.

\textbf{Case \rulename{i-tuple}, \rulename{i-case}, \rulename{i-receive}, \rulename{i-send}.} A
tuple or a union of expression types, or the type of the message.

\emph{Item 2.} By inspection of Figure~\ref{fig:inference}: \rulename{i-plus},
\rulename{i-case} and \rulename{pe-test} emit the first form, \rulename{i-abs} the second,
\rulename{pe-tuple} the third at its scrutinee, \rulename{i-app} the fourth, \rulename{i-send}
the fifth, and \rulename{i-spawn}, and the declaration of the interface of the program, the last
two; the other rules emit no constraint.

\emph{Item 3.} By induction on $\bar u$, all its formers being covariant on the operands that
are not ground. For (\ref{i:exptypes-bracket}), these variables are minted for a single
$\mkkeyword{spawn}$ or for the root, so they are not decoration variables of annotation types,
neither projection nor result variables, and no other decoration of a judgment.
\end{proof}

\begin{lemma}[Readers and writers]\label{lem:recon-positive}
Let $[\beta,\alpha]$ be the bracket at which \rulename{i-spawn-alt} types the body of an
unannotated $\mkkeyword{spawn}$ of a program of constraint set $C$, and $\gamma$ its pid
variable; or let $[\beta,\alpha]$ be the root bracket of a program whose interface is not
declared, and $\gamma$ a type variable not occurring in $C$.
\begin{enumerate}[noitemsep,topsep=2pt]
    \item Every occurrence of $\beta$ and of $\gamma$ in a constraint of $C$ is positive.
    \item Every occurrence of $\alpha$ in a constraint of $C$ is negative.
\end{enumerate}
\end{lemma}

\begin{proof}
By cases on the forms of Lemma~\ref{lem:recon-exptypes}(2). By
Lemma~\ref{lem:recon-exptypes}(3\ref{i:exptypes-bracket}), $\beta$ and $\gamma$ occur in an
expression type only with polarity $-$, and $\alpha$ only with polarity $+$; so their
occurrences in the left-hand side $\bar u$ of a constraint are positive for $\beta$ and $\gamma$
and negative for $\alpha$.

\textbf{Case $\bar u\issubt g$, $\bar u\issubt a$, $\bar u\issubt\Tuple{\zeta_1,\dots,\zeta_n}$,
$\beta'\issubt a$, $a\issubt\alpha'$.} The types $g$ and $a$, the projection variables and the
variables $\beta',\alpha'$ are none of $\alpha,\beta,\gamma$ and do not mention them.

\textbf{Case $\bar u\issubt\bar u'\to_o^i\gamma'$.} The occurrences in $\bar u'$, in the domain,
have the polarity opposite to that in $\bar u'$. The output decoration $o$, of polarity $+$, is
$\beta$ only when the call is typed at an output decoration $\beta$; the input decoration $i$, of
polarity $-$, is $\alpha$ only when it is typed at an input decoration $\alpha$; $\gamma'$ is fresh.

\textbf{Case $\bar u\issubt\TChan{\bar u'}$.} As in the previous case, $\bar u'$ being below
$\TChan\cdot$.
\end{proof}

\noindent The lemma concerns $C$ alone: the checking step adds $t_k\issubt\alpha_k$, and so does
the solving step for the processes that depend on the $k$-th (Definition~\ref{def:recon-cstar}).

\subsection{The derivation step: the receives that a process runs}\label{app:recon-receives}
The set $\rcv\alpha$ of Definition~\ref{def:recon-runs} contains uses of \rulename{i-receive},
not $\mkkeyword{receive}$s: the body of a function annotated with an intersection is typed once for
each arrow, so each of its $\mkkeyword{receive}$s is typed several times, and a process may call the
function at only some of these arrows. The set is computed by entailment, which the following
lemma reduces to finitely many subtyping tests.

\begin{lemma}[Entailment]\label{lem:recon-entail}
Let $C$ be a finite set of constraints and $\alpha,\delta$ type variables. Then
$C\models\alpha\issubt\delta$ if and only if $\alpha\sigma\subt\delta\sigma$ holds, as subtyping
between types with type variables, for every $\sigma\in\textsf{Solutions}(C)$. Hence
entailment is decidable.
\end{lemma}

\begin{proof}
\emph{Only if.} Every $\sigma\in\textsf{Solutions}(C)$ satisfies $C$
(Lemma~\ref{lem:tally-readback}(1)).

\emph{If.} Let $\theta$ satisfy $C$. By Lemma~\ref{lem:tally-readback}(2) there are
$\sigma\in\textsf{Solutions}(C)$ and $\rho$ with $\theta(\xi)\simeq\sigma(\xi)\rho$ for every type
variable $\xi$ of $C$. A type variable that does not occur in $C$ is left unchanged by $\sigma$
and does not occur in its images, whose type variables are those of $C$ or fresh: we extend
$\rho$ by $\rho(\xi)=\theta(\xi)$ on these variables, and the equivalence holds for $\alpha$ and
$\delta$ too. Hence $\alpha\theta\simeq\alpha\sigma\rho\subt\delta\sigma\rho\simeq\delta\theta$, since
subtyping between types with type variables is preserved by type substitution
(Appendix~\ref{app:enc-variables}).

\emph{Decidability.} $\textsf{Solutions}(C)$ is finite and computable, and subtyping between
types with type variables is decidable~\cite{CX11,polyduce1}.
\end{proof}

\noindent The entailment holds as well at every decoration variable $\delta$ such that $C$
entails $\alpha\issubt g$ and $g\issubt\delta$ for a ground type $g$, whoever runs the
$\mkkeyword{receive}$s typed at $\delta$. For $g=\Tany$, the $\mkkeyword{receive}$s of a function
called by a process declared at \eelixir{term()} are in every $\rcv\alpha$.

An implementation can avoid computing $\textsf{Solutions}(C)$: it suffices to follow, from
$\alpha$, the constraints between type variables of the set $\mathsf{dec}_{\models}(C)$ of
Definition~\ref{def:recon-dec}, each of which $C$ entails (Lemma~\ref{lem:recon-dec}). This
sufficient condition yields a subset $R$ of $\rcv\alpha$, possibly strict.
Proposition~\ref{prop:recon-canonical} holds, in its three items, for the type $F$ built from
$R$ in place of $\rcv\alpha$ (the coverage, the unknowns, \ref{h:polarity}, the extrema and
the servable types being those of $R$): its proof uses $C\models\alpha\issubt\delta$ only for the
$\mkkeyword{receive}$s from which $F$ is built. Proposition~\ref{prop:recon-sound} does not depend
on how the interfaces are computed.

\subsection{The derivation step: demands}\label{app:recon-demands}
The demand of a branch is not emitted by a rule of its own: it is the composition of
constraints that the constraint-generating pattern environment of Section~\ref{sec:cg-penv} and
the branch body already produce. Composing them requires splitting some constraints on tuples.
A $\mkkeyword{case}$ gives the variables it binds the type of its scrutinee filtered by ground
types, the patterns of its branches; we name these filtered types, the \emph{refinements}, so that a
bound on one yields a bound on the type it filters.

\begin{definition}[Refinement]\label{def:recon-refinement}
The \emph{refinements} of a type $s$, each with its \emph{ground part}, are defined
inductively: $s$ is a refinement of $s$ of ground part $\Tany$; if $r$ is a refinement of
$s$ of ground part $g$ and $g_1,g_2$ are ground types, then $(r\land g_1)\setminus g_2$ is
a refinement of $s$ of ground part $(g\land g_1)\setminus g_2$.
\end{definition}

\noindent A type is a refinement of $s$ by its form, not up to equivalence. By induction, a
refinement $r$ of $s$ of ground part $g$ satisfies $r\sigma\simeq s\sigma\land g$ for every
type substitution $\sigma$.

\noindent A constraint on a tuple bounds one of its components only if the others are nonempty in
every solution: we say syntactically when this is so (the \emph{inhabited types}) and which bounds the
components then receive (the \emph{component bounds}).

\begin{definition}[Splittable constraint and inhabited types]\label{def:recon-split}
The \emph{inhabited} types form the least set of types that contains the ground nonempty
types in which no arrow and no pid type occur, the pid types and their refinements whose ground
part $g$ satisfies $\TChan{\Tempty}\subt g$, the intersections of arrow types and their
refinements whose ground part $g$ satisfies $\TFun\subt g$, every tuple of its elements, and
every union of which an operand is one of its elements.
Let $c$ be a constraint $\Tuple{t_1,\dots,t_n}\issubt u$, let $\bar t_h$ be $t_h$ if $t_h$
is ground and no arrow and no pid type occur in it, and $\Tany$ otherwise, and let
$u_h=\pi_h(u\land\Tuple{\bar t_1,\dots,\bar t_n})$ for $1\leq h\leq n$. The types
$u_1,\dots,u_n$ are the \emph{component bounds} of $c$, and $c$ is \emph{splittable at} $i$
if $t_h$ is inhabited for every $h\neq i$ and
$u\land\Tuple{\bar t_1,\dots,\bar t_n}\simeq\Tuple{u_1,\dots,u_n}$.
\end{definition}

\noindent When $c$ is splittable, its components $\bar t_h\neq\Tany$, such as tags, select a single
tuple type of $u$. A ground type in which an arrow or a pid type occurs neither selects nor is
inhabited for being ground: so substituting ground types for the pid variables and for the
decoration variables of the mailbox-pure abstractions, as the solving step does (Definition~\ref{def:recon-cstar}), changes neither the
types $\bar t_h$ nor which components of its tuple constraints are inhabited.

\begin{lemma}[Inhabited types]\label{lem:recon-inhabited}
If $t$ is inhabited, then $t\sigma\not\simeq\Tempty$ for every type substitution $\sigma$.
\end{lemma}

\begin{proof}
By induction on the definition of inhabited types, in the direct interpretation, which induces
the same subtyping (Corollary~\ref{cor:same-subtyping}).

\textbf{Case $t$ ground and nonempty.} Then $t\sigma=t$.

\textbf{Case $t$ a pid type.} For any element $d$ of the domain, the approximation
$\{(d,\mho)\}$, which rejects no message, belongs to every pid type
(Definition~\ref{def:interpretation}).

\textbf{Case $t$ an intersection of arrow types.} For any constant $c_0$, the approximation
$\{((v,c_0,r),\Omega)\}$ belongs to every arrow type, since $c_0$ is not a write capability and
the triple lies outside every domain.

\textbf{Case $t$ a refinement of $s$, of ground part $g$.} Then $t\sigma\simeq s\sigma\land g\simeq
s\sigma$, every pid type being below $\TChan{\Tempty}$ and every arrow below $\TFun$; then the two
previous cases.

\textbf{Case $t$ a tuple or a union.} The induction hypothesis on its components, or on its
inhabited operand.
\end{proof}

\begin{definition}[Decompositions]\label{def:recon-dec}
For a set $X$ of constraints, $\mathsf{dec}(X)$ and $\mathsf{dec}_{\models}(X)$ are the least
supersets $Y$ of $X$ such that:
\begin{enumerate}[noitemsep,topsep=2pt]
    \item if $(s_1\lor s_2\issubt u)\in Y$, then $s_1\issubt u$ and $s_2\issubt u$ are in $Y$;
    \item if $(r\issubt s'\to_{o'}^{i'}t')\in Y$, with $r$ a refinement of $s\to_o^i t$ whose ground
    part $g$ satisfies $\TFun\subt g$, then $s'\issubt s$, $i'\issubt i$, $t\issubt t'$ and
    $o\issubt o'$ are in $Y$;
    \item if $(r\issubt\TChan{s})\in Y$, with $r$ a refinement of $\TChan{t}$ whose ground part $g$
    satisfies $\TChan{\Tempty}\subt g$, then $(s\issubt t)\in Y$;
    \item if $(\Tuple{t_1,\dots,t_n}\issubt u)\in Y$ has component bounds $u_1,\dots,u_n$ and
    is splittable at $i$, then $t_h\issubt u_h$ is in $Y$ for every $h$, for
    $\mathsf{dec}(X)$, and $t_i\issubt u_i$ is in $Y$, for $\mathsf{dec}_{\models}(X)$.
\end{enumerate}
\end{definition}

\noindent The refinements in clauses 2 and 3 are the types of the variables that a
$\mkkeyword{case}$ binds: \eelixir|me = self(); e| stands for \eelixir|case self() do me -> e end|,
which gives \eelixir{me} the type $(\TChan{o}\land\Tany)\setminus\Tempty$, $o$ being the output
decoration of the judgment. The set $\mathsf{dec}(X)$ also bounds the inhabited
components of a tuple whose remaining component may be empty, which $X$ need not entail.

An implementation need not apply these clauses by form. It can extract the bounds of a
constraint from the normal form that tallying computes for it
(Section~\ref{sec:elixir-tallying}), which is equivalent to the constraint, so that every bound
extracted is entailed, as Lemma~\ref{lem:recon-dec} states for $\mathsf{dec}_{\models}$. To
decompose a bound $\zeta\issubt u$ further, for the type variables that occur in $u$ outside
every type constructor, $\zeta$ must first be renamed to a fresh type variable: the type
variables that tallying keeps fixed are universally quantified, so with $\zeta$ fixed the
normal form is that of ``$\zeta\issubt u$ for every $\zeta$'', which the constraint does not
entail (from $\zeta\issubt(\eta\land\Tuple{\xi})\lor\neg\Tuple{\Tany}$ it yields
$\Tany\issubt\xi$). This reading is finer than the clauses: it finds a bound of $\zeta$ in
every constraint $(\zeta\land g_1)\setminus g_2\issubt t$, whereas
Definition~\ref{def:recon-demand} does so only when $\zeta$ is the type of a capture variable
(see the examples in Appendix~\ref{app:recon-examples}).

\begin{lemma}[Decompositions]\label{lem:recon-dec}
For every set $X$ of constraints, $X\models c$ for every $c\in\mathsf{dec}_{\models}(X)$.
\end{lemma}

\begin{proof}
By induction on the derivation of $c\in\mathsf{dec}_{\models}(X)$ and by cases on the last
clause of Definition~\ref{def:recon-dec} applied.

\textbf{Case $c\in X$.} Immediate.

\textbf{Case clause 1.} A union is below $u$ if and only if both its operands are.

\textbf{Case clause 2.} For every type substitution $\sigma$,
$r\sigma\simeq(s\to_o^i t)\sigma\land g\simeq(s\to_o^i t)\sigma$, since every arrow is below
$\TFun\subt g$; then Proposition~\ref{prop:arrow-factor}.

\textbf{Case clause 3.} As in the previous case, $r\sigma\simeq\TChan{t\sigma}$, every pid type
being below $\TChan{\Tempty}$; then Proposition~\ref{prop:pid-laws}(1).

\textbf{Case clause 4.} Let $\sigma$ satisfy $\Tuple{t_1,\dots,t_n}\issubt u$, of component
bounds $u_1,\dots,u_n$ and splittable at $i$. Since $\sigma$ leaves the ground components
$\bar t_h$ unchanged and preserves equivalence,
\[
    \Tuple{t_1,\dots,t_n}\sigma\;\subt\;u\sigma\land\Tuple{\bar t_1,\dots,\bar t_n}
    \;\simeq\;\Tuple{u_1\sigma,\dots,u_n\sigma},
\]
and $t_h\sigma\not\simeq\Tempty$ for every $h\neq i$ by Lemma~\ref{lem:recon-inhabited}. If
$t_i\sigma\simeq\Tempty$, then $t_i\sigma\subt u_i\sigma$. Otherwise $\Tuple{t_1,\dots,t_n}\sigma$
is a nonempty tuple type below a tuple type, hence below it componentwise, and
$t_i\sigma\subt u_i\sigma$.
\end{proof}

\begin{definition}[Self-send]\label{def:recon-selfsend}
Let $C$ be the constraint set of a program $e$ and $\beta$ an output decoration of its
deduction. An output decoration $\delta$ of the deduction is \emph{linked to} $\beta$ if
$\mathsf{dec}(C)$ contains the constraints $\delta_{h-1}\issubt\delta_h$, $1\leq h\leq n$, for some
$n\geq0$ and type variables $\delta_0,\dots,\delta_n$ with $\delta_0=\delta$ and $\delta_n=\beta$.
A \emph{self-send of} $\beta$ is a use of \rulename{i-send} in the deduction of $e$ whose
constraint $t\issubt\TChan{t'}$ satisfies $\TChan{\delta}\subt t$ for some $\delta$ linked to
$\beta$. We write $S_\beta$ for the set of the constraints of the self-sends of $\beta$.
\end{definition}

\noindent A self-send of $\beta$ need not write to the mailbox of the process of output
decoration $\beta$: a closure that another runner passes to a function it shares with that
process is linked too. Leaving out its constraint is harmless, since only the derivation and
solving steps leave it out. A write whose target is a pid of the process extracted from a tuple
or returned by a function is not a self-send, since its target has a type variable as its
type: its constraint stays.

\begin{definition}[Demand]\label{def:recon-demand}
Let $C$ be the constraint set of a program $e$, $\beta$ an output decoration of its
deduction, and $j$ a branch of a use of \rulename{i-receive} at input decoration $\delta$ in
the deduction of $e$, with pattern $p_j$, such that
$r_j=(\delta\land\accept{p_j})\setminus\bigvee_{h<j}\accept{p_h}$ is not empty, and let
$\ptrel{r_j}{p_j}{\Gamma_j}{D_j}$ be the premise of the rule for $j$.
For a finite set $X$ of constraints and every capture variable $x$ of $p_j$, let
\[
    u^X_j(x)\;=\;\textstyle\bigwedge\{\,t\lor\neg g\mid (r\issubt t)\in
    \mathsf{dec}_{\models}(X\setminus S_\beta),\ r\text{ a refinement of }\Gamma_j(x)\text{ of
    ground part }g\,\}.
\]
Let $m_j^\beta[X]=p_j[u^X_j]$, where, for an assignment $u$ of types to capture variables,
$x[u]=u(x)$, $\tau[u]=\tau$,
$\Tuple{p_1,\dots,p_n}[u]=\Tuple{p_1[u],\dots,p_n[u]}$ and
$(p_1\land p_2)[u]=p_1[u]\land p_2[u]$, and let $m_j^\beta=m_j^\beta[C]$. The \emph{demand of $j$
at} a type $i$ is the constraint $i\land q_j\issubt m_j^\beta$, where
$q_j=\accept{p_j}\setminus\bigvee_{h<j}\accept{p_h}$. The \emph{demands of the process} of
bracket $[\beta,\alpha]$ are the demands at $\alpha$ of the branches $j$ of the uses of \rulename{i-receive} in $\mathsf{rcv}_e(\alpha)$.
\end{definition}

\noindent We write $u_j$ for $u_j^{C}$ and $m_j$ for $m_j^\beta$ when $\beta$ is clear, and say
that a type variable \emph{occurs in} a demand, and that a demand is \emph{ground}, according to
$m_j$ alone; an empty meet is $\Tany$. Since $r_j\simeq\delta\land q_j$, the demand of $j$ at
$\delta$ is equivalent to $r_j\issubt m_j$. A branch with $r_j\simeq\Tempty$ is skipped by
\rulename{i-receive}: it has no premise, hence no demand, and from now on the branches of a use
of \rulename{i-receive} are those that the rule types, also in~\eqref{eq:demands}
and~\eqref{eq:coverage}. Since every input decoration of the deduction is a type variable, a
skipped branch has $q_j\simeq\Tempty$, and leaving it out changes neither $\dem\alpha$ nor
$\cov\alpha$. The type $\Gamma_j(x)$ is $r_j$ when $x$ occurs in
$p_j$ outside every tuple pattern, and otherwise one of the \emph{projection variables} of $j$,
the type variables that \rulename{pe-tuple} mints in the premise. The type $m_j$ is computed from
the whole of $C$, so that it also collects the bounds that the code after the
$\mkkeyword{receive}$ places on the value that $j$ returns. The test on $\Gamma_j(x)$ is syntactic
because the bounds $\delta\issubt\alpha_f$, for the input decoration $\alpha_f$ of a callee, that
the decomposition of the constraint of \rulename{i-app} yields at a call are not constraints on a
refinement of
$\Gamma_j(x)$, even when $\Gamma_j(x)\simeq\delta$, as for a bare-variable first branch.
Refinements let $m_j$ take into account the bounds that a $\mkkeyword{case}$ on a capture
variable places in its branches (see the examples in Appendix~\ref{app:recon-examples}).

\begin{lemma}[Demands]\label{lem:recon-demand}
With the notation of Definition~\ref{def:recon-demand}, $D_j\cup X\models r_j\issubt m_j^\beta[X]$
for every finite set $X$ of constraints.
\end{lemma}

\begin{proof}
By induction on the sub-patterns $p'$ of $p_j$ we show $D_j\cup X\models t_{p'}\issubt p'[u^X_j]$,
where $t_{p'}$ is the scrutinee of $p'$ in the derivation of $\ptrel{r_j}{p_j}{\Gamma_j}{D_j}$:
$r_j$ at the root, the fresh $\zeta_h$ of \rulename{pe-tuple} at the $h$-th component of a tuple
pattern. At the root this is the claim.

\textbf{Case $p'=x$.} By \rulename{pe-var}, $t_x=\Gamma_j(x)$. Let
$(r\issubt t)\in\mathsf{dec}_{\models}(X\setminus S_\beta)$, with $r$ a refinement of
$\Gamma_j(x)$ of ground part $g$. By Lemma~\ref{lem:recon-dec}, $X\setminus S_\beta\models r\issubt t$; since
$r\sigma\simeq\Gamma_j(x)\sigma\land g$ for every $\sigma$, $X\models\Gamma_j(x)\issubt t\lor\neg g$.
Hence $X\models t_x\issubt u^X_j(x)$.

\textbf{Case $p'=\tau$.} \rulename{pe-test} puts $t_{p'}\issubt\tau$ in $D_j$, and
$\tau[u^X_j]=\tau$.

\textbf{Case $p'=\Tuple{p_1,\dots,p_n}$.} \rulename{pe-tuple} puts
$t_{p'}\issubt\Tuple{\zeta_1,\dots,\zeta_n}$ in $D_j$, and the induction hypothesis gives
$\zeta_h\issubt p_h[u^X_j]$ for every $h$.

\textbf{Case $p'=p_1\land p_2$.} By \rulename{pe-and} both sub-patterns have the scrutinee
$t_{p'}$, and the induction hypothesis gives both bounds.
\end{proof}

\subsection{The solving step: overloaded callees}\label{app:recon-overloaded}
A call to a function of type $\bigwedge_{j\in J}(s_j\to_{\beta_j}^{\alpha_j}t_j)$ with
$|J|\geq2$, at a bracket $[o,i]$, emits $t_f\issubt t_2\to_{o}^{i}\gamma'$, which does not
split componentwise: the caller selects the arrows it uses, and the selection depends on its
bracket (Lemma~\ref{lem:folded-decomp}). It is trivial when $i\subt\alpha_j$ and
$\beta_j\subt o$ for every $j$ (Lemma~\ref{lem:recon-select}), as for the mailbox-pure functions
once their arrows are decorated at $[\Tempty,\Tany]$, which preserves satisfiability; the
constraint then entails a ground upper bound on the argument (Lemma~\ref{lem:recon-pure}).

\begin{definition}[Mailbox-pure abstraction]\label{def:recon-pure}
The \emph{mailbox-pure} abstractions of a program $e$ form the least set $M$ of abstractions of
$e$ that contains every abstraction $\lambda$ such that:
\begin{enumerate}[noitemsep,topsep=2pt]
    \item $\decorate{\sa_j}$ and $\decorate{\ta_j}$ are ground for every arrow $\sa_j\to\ta_j$ of
    the annotation of $\lambda$;
    \item the body of $\lambda$ contains neither $\mkkeyword{receive}$ nor $\Self$;
    \item the head of every application in the body of $\lambda$, that is, the expression that is
    applied, is an $M$-head;
\end{enumerate}
where the \emph{$M$-heads}, each with its \emph{abstraction}, are defined inductively: an
abstraction of $M$ is an $M$-head, whose abstraction is itself; and a variable that a
$\mkkeyword{case}$ binds, through a variable pattern, to a scrutinee that is an $M$-head is an
$M$-head, whose abstraction is that of the scrutinee. We write $\theta_M$ for the type
substitution that maps the
decoration variables $\alpha_j$ and $\beta_j$ that \rulename{i-abs} mints for the abstractions
of $M$ to $\Tany$ and $\Tempty$, respectively.
\end{definition}

\noindent A mailbox-pure function is first order, performs no $\mkkeyword{receive}$, evaluates
no $\Self$, and applies only abstractions of the program: never a parameter, in particular
never the variable that the fixpoint combinator binds. Recall that \eelixir|p = e1; e2| stands for
\eelixir|case e1 do p -> e2 end|. The second case of $M$-heads is that of an abstraction that is
named before it is applied, as in

\centerline{\eelixir|fn y -> g = fn x -> x + 1 end; g(y) end|}\par
\noindent where the $\mkkeyword{case}$ binds \eelixir{g}, through a variable pattern, to the inner
abstraction. If the inner abstraction is in $M$, the head \eelixir{g} of the application
\eelixir|g(y)| is an $M$-head, of that abstraction, and the outer abstraction satisfies
condition~(3). By contrast, in

\centerline{\eelixir|fn p -> {f, x} = p; f(x) end|}\par
\noindent the $\mkkeyword{case}$ binds \eelixir{f}, through a tuple pattern, to a component of the
scrutinee: the head \eelixir{f} is no $M$-head, since we do not know which abstraction it names,
and the outer abstraction is not in $M$.

\begin{lemma}[Arrow selection]\label{lem:recon-select}
Let $J$ be finite and nonempty, and $[o,i]$ a bracket such that $i\subt i_j$ and $o_j\subt o$
for every $j\in J$. Then $\bigwedge_{j\in J}(s_j\to_{o_j}^{i_j}t_j)\subt s\to_o^i t$ if and only
if, for every $P\subseteq J$, either $P\neq\varnothing$ and $s\subt\bigvee_{j\in P}s_j$, or
$P\neq J$ and $\bigwedge_{j\in J\setminus P}t_j\subt t$.
\end{lemma}

\begin{proof}
By Lemma~\ref{lem:folded-decomp}, the subtyping holds if and only if for every $P\subseteq J$
either (a) for every $\kappa:P\to\{1,2,3\}$ one of $\kappa^{-1}(1)\neq\varnothing$ and
$s\subt\bigvee_{\kappa^{-1}(1)}s_j$, or $o_j\subt o$ for some $j\in\kappa^{-1}(2)$, or
$i\subt i_j$ for some $j\in\kappa^{-1}(3)$; or (b) $P\neq J$ and
$\bigwedge_{J\setminus P}t_j\subt t$. Since $o_j\subt o$ and $i\subt i_j$ for every $j$, the
second or the third alternative of (a) holds for every $\kappa$ that sends some $j$ to $2$ or
$3$. So (a) holds if and only if the first alternative holds for the $\kappa$ that sends every
$j\in P$ to $1$, that is, if and only if $P\neq\varnothing$ and $s\subt\bigvee_{j\in P}s_j$.
\end{proof}

\begin{definition}[Inverse image]\label{def:recon-inverse}
Let $f$ be a mailbox-pure abstraction with annotation $\bigwedge_{j\in J}(\sa_j\to\ta_j)$, and
let $s_j=\decorate{\sa_j}$ and $t_j=\decorate{\ta_j}$. For a ground type $g$,
\[
    f^{-1}(g)\;=\;\textstyle\bigwedge\Bigl\{\,\bigvee_{j\in P}s_j\;\Bigm|\;\varnothing\neq
    P\subseteq J,\ P=J\text{ or }\bigwedge_{j\in J\setminus P}t_j\not\subt g\,\Bigr\}.
\]
\end{definition}

\noindent When $\bigwedge_{j\in J}t_j\subt g$, it is the largest argument type on which every
arrow of $f$ that may apply returns within $g$. Otherwise no argument type satisfies the
constraint of a call whose result is below $g$ (the case $P=\varnothing$ of
Lemma~\ref{lem:recon-select}), and $f^{-1}(g)$ is not that type: for
$(\TInt\to\TInt)\land(\TInt\lor\TAtom\to\TInt\lor\TAtom)$, $f^{-1}(\TAtom)=\TInt$.

\begin{lemma}[Mailbox-pure functions]\label{lem:recon-pure}
Let $C$ be the constraint set of a program $e$.
\begin{enumerate}[noitemsep,topsep=2pt]
    \item Let $c=(t_f\issubt t_2\to_o^i\gamma')$ be the constraint of a use of \rulename{i-app}
    whose head is an $M$-head of abstraction $f$, $g$ a ground type, and $\theta'$ a type
    substitution that is $\theta_M$ on the decoration variables of the abstractions of $M$ and
    does not change $\gamma'$. Then $\{c\theta',\ \gamma'\issubt g\}\models t_2\theta'\issubt f^{-1}(g)$.
    \item Let $B$ be a set of constraints between the variables $\alpha_0,\beta_0$ of the root
    bracket of $e$ and ground types. If $C\cup B$ is satisfiable, so is $\theta_M(C)\cup B$.
\end{enumerate}
\end{lemma}

\begin{proof}
For an abstraction $f\in M$ with annotation $\bigwedge_{j\in J}(\sa_j\to\ta_j)$, let
$s_j=\decorate{\sa_j}$ and $t_j=\decorate{\ta_j}$, and let
$a_f=\bigwedge_{j\in J}(s_j\to_{\beta_j}^{\alpha_j}t_j)$ be the type that \rulename{i-abs} gives
to $f$.

\emph{The type of an $M$-head.} Let $t_f$ be the type of an $M$-head of abstraction $f$, and
$\theta'$ a type substitution that is $\theta_M$ on the decoration variables of the abstractions
of $M$. If the head is $f$, $t_f$ is $a_f$; otherwise it is the type of a variable that a
$\mkkeyword{case}$ binds through a variable pattern, the refined type
$(t\land\accept{p})\setminus\bigvee_h\accept{p_h}$ of \rulename{i-case}, where $t$ is the type of an
$M$-head of abstraction $f$. By induction on the head, $t_f$ is a refinement of $a_f$
(Definition~\ref{def:recon-refinement}) whose ground part $G$ is a Boolean combination of
accepted types of patterns, hence a test type. Since $a_f\subt\TFun$ and $\TFun\land G$ is
$\Tempty$ or $\TFun$, by induction on the test type $G$ (Section~\ref{sec:syntax}),
$t_f\sigma\simeq a_f\sigma\land G$ is $a_f\sigma$ or empty for every $\sigma$; it is not empty,
since a branch whose refined type is empty is omitted. So
$t_f\theta'\simeq\bigwedge_{j\in J}(s_j\to_{\Tempty}^{\Tany}t_j)$, a ground type.

\emph{Item 1.} Let $\sigma$ satisfy $c\theta'$ and $\gamma'\issubt g$. By the type of an
$M$-head,
$\bigwedge_{j\in J}(s_j\to_{\Tempty}^{\Tany}t_j)\subt t_2\theta'\sigma\to_{o\theta'\sigma}^{i\theta'\sigma}\gamma'\sigma$.
Since $i\theta'\sigma\subt\Tany$ and $\Tempty\subt o\theta'\sigma$, Lemma~\ref{lem:recon-select}
applies: for every $P\subseteq J$, either $P\neq\varnothing$ and
$t_2\theta'\sigma\subt\bigvee_{j\in P}s_j$, or $P\neq J$ and
$\bigwedge_{j\in J\setminus P}t_j\subt\gamma'\sigma\subt g$. Hence
$t_2\theta'\sigma\subt\bigvee_{j\in P}s_j$ for every $P$ of the meet that defines $f^{-1}(g)$, and
$t_2\theta'\sigma\subt f^{-1}(g)$.

\emph{Item 2.} Let $\sigma$ satisfy $C\cup B$, and let $\sigma'$ be $\sigma$ but on the decoration
variables $\alpha_j,\beta_j$ that \rulename{i-abs} mints for the abstractions of $M$, which it maps
to $\Tany$ and $\Tempty$, respectively. For every constraint $c$, $c\sigma'$ is $c\theta_M\sigma$,
and $B$ mentions none of these variables; so it suffices that $\sigma'$ satisfies $C$.

\emph{Where the decorations of $M$ occur.} Let $f\in M$. The rules that read the bracket of their judgment
are \rulename{i-self}, \rulename{i-receive} and \rulename{i-app}, and the first two do not occur in
the body of $f$ (Definition~\ref{def:recon-pure}). So $\alpha_j,\beta_j$ occur in $C$ only in $a_f$,
within expression types, and as the bracket of the constraints of the uses of \rulename{i-app} in
the body of $f$ that no nested abstraction or $\mkkeyword{spawn}$ encloses; the annotation types on
the right of the constraints of \rulename{i-abs} and \rulename{i-spawn} are types $\decorate{\ta}$.
In an expression type $a_f$ occurs only below covariant formers
(Lemma~\ref{lem:recon-exptypes}(1)), so $\alpha_j$ occurs only with polarity $-$ and $\beta_j$ only
with polarity $+$; in the constraints of Lemma~\ref{lem:recon-exptypes}(2), expression types occur
on the left, or on the right in the domain of an arrow or below a pid type.

\textbf{Case a constraint $s\issubt t$ in which the decorations of $M$ occur only in expression
types.} By Lemma~\ref{lem:recon-monotone}, $s\sigma'\subt s\sigma$ and $t\sigma\subt t\sigma'$, so
$\sigma'$ satisfies it.

\textbf{Case the constraint $t_1\issubt t_2\to_{\beta_j}^{\alpha_j}\gamma'$ of a use of
\rulename{i-app} in the body of $f$.} As in the previous case, $t_1\sigma'\subt t_1\sigma$ and
$t_2\sigma'\subt t_2\sigma$, and $\gamma'\sigma'=\gamma'\sigma$. The head of the application is an
$M$-head, of an abstraction $f'$ with annotation $\bigwedge_{l\in L}(\sa'_l\to\ta'_l)$; by the type
of an $M$-head, $t_1\sigma'=t_1\theta_M\sigma\simeq\bigwedge_{l\in L}(s'_l\to_{\Tempty}^{\Tany}t'_l)$,
with $s'_l=\decorate{\sa'_l}$ and $t'_l=\decorate{\ta'_l}$. Then
$t_1\sigma'\subt t_1\sigma\subt t_2\sigma\to_{\beta_j\sigma}^{\alpha_j\sigma}\gamma'\sigma$. All the
arrows of $t_1\sigma'$ are decorated at $[\Tempty,\Tany]$, so Lemma~\ref{lem:recon-select} applies
at every bracket, with a condition that does not depend on the bracket: it holds at
$[\beta_j\sigma,\alpha_j\sigma]$, hence at $[\Tempty,\Tany]$, and
$t_1\sigma'\subt t_2\sigma\to_{\Tempty}^{\Tany}\gamma'\sigma\subt t_2\sigma'\to_{\Tempty}^{\Tany}\gamma'\sigma'$.
\end{proof}

\noindent In item 1, when $t_2$ is the type of a capture variable, or a tuple from which
$\mathsf{dec}_{\models}$ extracts such a type, the bound is a bound of that type, which
Definition~\ref{def:recon-demand} collects.

\subsection{The solving step: eliminating the other variables}\label{app:recon-eliminating}
Throughout this subsection we fix a program $e$, of constraint set $C$, and the index $k$ of an
unannotated $\mkkeyword{spawn}$ of $e$, or $k=0$ for $e$ itself, and we write $\alpha,\beta$ for
$\alpha_k,\beta_k$; the index $j$ ranges over the unannotated $\mkkeyword{spawn}$s of $e$ and
over $0$, and there is no $\gamma_0$.

\begin{definition}[Dependency]\label{def:recon-depends}
Let $C^k=\mathsf{dec}_{\models}(\theta_M(C\setminus S_\beta))$. Let $G_0$ be the set of the types
$\Gamma_\ell(x)$ of the capture variables $x$ of the branches $\ell$ of the uses of
\rulename{i-receive} in $\mathsf{rcv}_e(\alpha)$, and $G$ the union of $G_0$ and of the set of the
result variables $\gamma'$ of the constraints $t_f\issubt t_2\to_o^i\gamma'$ of the uses of
\rulename{i-app} whose head is an $M$-head whose abstraction has an annotation with at least two
arrows, and in whose $t_2$ a type of $G_0$ occurs. Let $Q$ be the least set of pairs $(\zeta,p)$ of
a type variable $\zeta\notin\{\alpha,\beta\}$ and a polarity $p$ such that:
\begin{enumerate}[noitemsep,topsep=2pt]
    \item $(\zeta,p)\in Q$ if $\zeta$ occurs with polarity $p$ in $u$, for a constraint
    $(s\issubt u)\in C^k$ such that a type of $G$ occurs in $s$;
    \item $(\eta,p)\in Q$ if $(\zeta,+)\in Q$, $(\zeta\issubt u)\in C^k$ and $\eta$ occurs with
    polarity $p$ in $u$;
    \item $(\eta,-p)\in Q$ if $(\zeta,-)\in Q$, $(u\issubt\zeta)\in C^k$ and $\eta$ occurs with
    polarity $p$ in $u$.
\end{enumerate}
The $k$-th $\mkkeyword{spawn}$ \emph{depends on} the $j$-th, written $j\prec k$, if $j\neq k$
and $\alpha_j$, $\beta_j$ or $\gamma_j$ is the first component of a pair of $Q$.
\end{definition}

\noindent The relation $\prec$ is read before any $\gamma_j$ is replaced and any bracket is
added, since it decides which are. The solving step takes the $\mkkeyword{spawn}$s in an order
that extends $\prec$, and fails if $\prec$ has a cycle; from now on, the interfaces $t_j$ of
the $j\prec k$ are computed.

\begin{definition}[Reduced constraint set]\label{def:recon-cstar}
Let $\theta$ be $\theta_M$ extended by $\gamma_j\mapsto t_j$ for every $j\prec k$, let
$B=\{\beta_j\issubt t_j,\ t_j\issubt\alpha_j\mid j\prec k\}$, and let
$C_1=\mathsf{dec}_{\models}(\theta(C\setminus S_\beta)\cup B)$. Let $R$ be the set of the
constraints $c=(t_f\issubt t_2\to_o^i\gamma')$ of the uses of \rulename{i-app} in the deduction
of $e$ such that:
\begin{enumerate}[label=(\arabic*),ref=(\arabic*),noitemsep,topsep=2pt]
    \item\label{i:cstar-pure} the head of the application is an $M$-head whose abstraction $f$
    has an annotation with at least two arrows;
    \item a type $\Gamma_\ell(x)$ of a capture variable $x$ of a branch $\ell$ of a use of
    \rulename{i-receive} occurs in $t_2$;
    \item $u$ is ground for every $(\gamma'\issubt u)\in C_1$.
\end{enumerate}
For $c\in R$, let $\hat g_c=\textstyle\bigwedge\{u\mid(\gamma'\issubt u)\in C_1\}$, and let
\[
    C^\circ\;=\;\theta\bigl(C\setminus(S_\beta\cup R)\bigr)\;\cup\;
    \{\,t_2\theta\issubt f^{-1}(\hat g_c)\mid c\in R\,\},
\]
where $t_2$ and $f$ are those of $c$. The \emph{reduced constraint set} is
$C^\ast=\mathsf{dec}_{\models}(C^\circ\cup B)$, and the \emph{reduced demand type} of a branch
$\ell$ of a use of \rulename{i-receive} is $m^\circ_\ell=m_\ell^\beta[C^\circ]$.
\end{definition}

\noindent The substitution $\theta$ instantiates the mailbox-pure functions as
Lemma~\ref{lem:recon-pure}(2) allows, and treats the $\mkkeyword{spawn}$s on which $k$ depends as
declared at their interfaces. A constraint $c$ of $R$ is replaced by a bound on its argument,
which $c\theta$ entails with the bounds of $\gamma'$ in $C_1$ (Lemma~\ref{lem:recon-pure}(1)); the
constraints of the other calls to overloaded functions stay, bound no type variable, and the
extrema miss them. The upper bounds of $\gamma'$ are read in $C_1$, before any replacement.

\begin{definition}[Unknowns]\label{def:recon-unknowns}
Let $P$ be the least set of pairs $(\zeta,p)$ of a type variable $\zeta\notin\{\alpha,\beta\}$
and a polarity $p$ such that:
\begin{enumerate}[noitemsep,topsep=2pt]
    \item $(\zeta,p)\in P$ if $\zeta$ occurs with polarity $p$ in $m^\circ_\ell$, for a branch
    $\ell$ of a use of \rulename{i-receive} in $\mathsf{rcv}_e(\alpha)$;
    \item $(\eta,p)\in P$ if $(\zeta,+)\in P$, $(\zeta\issubt u)\in C^\ast$ and $\eta$ occurs
    with polarity $p$ in $u$;
    \item $(\eta,-p)\in P$ if $(\zeta,-)\in P$, $(u\issubt\zeta)\in C^\ast$ and $\eta$ occurs
    with polarity $p$ in $u$.
\end{enumerate}
An \emph{unknown} is a type variable $\zeta$ such that $(\zeta,p)\in P$ for some $p$; it is
\emph{covariant} if $(\zeta,+)\in P$ and \emph{contravariant} if $(\zeta,-)\in P$.
\end{definition}

\begin{definition}[Polarity hypothesis]\label{def:recon-hyp}
The polarity hypothesis \ref{h:polarity} holds at $k$ if no unknown is both covariant and
contravariant.
\end{definition}

\begin{definition}[Extrema]\label{def:recon-elim}
Assume \ref{h:polarity} at $k$. The \emph{extrema} $\zeta^\ast$ of the unknowns $\zeta$ form
the solution, greatest on the covariant unknowns and least on the contravariant ones, of
the system
\[
    \zeta^\ast \;=\; \textstyle\bigwedge\{\,u^\ast\mid (\zeta\issubt u)\in C^\ast\,\}
    \ \text{ ($\zeta$ covariant)},\qquad
    \zeta^\ast \;=\; \textstyle\bigvee\{\,u^\ast\mid (u\issubt \zeta)\in C^\ast\,\}
    \ \text{ ($\zeta$ contravariant)},
\]
where $u^\ast$ is $u$ with every unknown $\eta$ replaced by $\eta^\ast$.
\end{definition}

\noindent An empty meet is $\Tany$ and an empty join $\Tempty$; Lemma~\ref{lem:recon-elim}(1)
shows that the solution exists. Under \ref{h:polarity}, we write $p_\zeta$ for the polarity of an
unknown $\zeta$, $m_j^\ast$ for $m^\circ_j$ in which every unknown is replaced by its extremum,
and $m_j(a,b)$ for $m_j^\ast[\alpha:=a,\beta:=b]$. In~\eqref{eq:recon-req}, the types $m_j$ of
$\mathsf{dem}^\ast_e(\alpha)$ are the $m^\circ_j$: since the $q_j$ are ground,
$F(a,b)=\cov\alpha\land\bigwedge_j(m_j(a,b)\lor\neg q_j)$, with $j$ ranging over the branches of
the uses of \rulename{i-receive} in $\rcv\alpha$.

\begin{lemma}[Bounds]\label{lem:recon-bounds}
In the following, an \emph{annotation part} is a type built by Boolean connectives from ground
types and from the subtrees of annotation types that are not type variables; its type variables are
decoration variables, below arrows. A \emph{union operand} of a type $s'$ is $s'$ itself or, if $s'=s_1\lor s_2$, a union
operand of $s_1$ or of $s_2$. Expression types and annotation types are taken under $\theta$, which maps the atoms
$\TChan{\gamma_j}$, $j\prec k$, to ground types and the annotation types of the abstractions of
$M$ to ground types.
\begin{enumerate}[noitemsep,topsep=2pt]
\item\label{i:bounds-upper} If $(s\issubt u)\in C^\ast$ and $s$ is a refinement of a type variable
$\zeta$, then $u$ is of one of the forms:
\begin{enumerate}[label=(u\arabic*),ref=u\arabic*,noitemsep,topsep=1pt,itemindent=1em]
    \item\label{i:bounds-u1} a decoration variable, $s$ being a decoration variable;
    \item\label{i:bounds-u2} a pid variable or an output decoration $u'$, with
    $r\issubt\TChan{s'}$ in $C^\ast$ for a refinement $r$ of $\TChan{u'}$ and a type $s'$ of which $s$
    is a union operand;
    \item\label{i:bounds-u3} an annotation part;
    \item\label{i:bounds-u4} a tuple of projection variables;
    \item\label{i:bounds-u5} an arrow $\bar u\to_o^i\gamma''$, with $\bar u$ an expression type,
    $o,i$ decoration variables or ground types, and $\gamma''$ a result variable;
    \item\label{i:bounds-u6} a pid type $\TChan{\bar u}$, with $\bar u$ an expression type.
\end{enumerate}
\item\label{i:bounds-lower} If $(u\issubt\zeta)\in C^\ast$ and $\zeta$ is a type variable, then
$u$ is of one of the forms:
\begin{enumerate}[label=(l\arabic*),ref=l\arabic*,noitemsep,topsep=1pt,itemindent=1em]
    \item\label{i:bounds-l1} a decoration variable, $\zeta$ being a decoration variable;
    \item\label{i:bounds-l2} an annotation part;
    \item\label{i:bounds-l3} an expression type $\bar u$, $\zeta$ being a pid variable or an output
    decoration, with $r\issubt\TChan{s'}$ in $C^\ast$ for a refinement $r$ of $\TChan{\zeta}$ and a type
    $s'$ of which $\bar u$ is a union operand.
\end{enumerate}
\end{enumerate}
\end{lemma}

\begin{proof}
The constraints of $C^\circ\cup B$ are those of Lemma~\ref{lem:recon-exptypes}(2) under $\theta$,
whose expression types are still generated as in Lemma~\ref{lem:recon-exptypes}(1), those of
$B$, between decoration variables and ground types, and the bounds $t_2\theta\issubt f^{-1}(\hat g_c)$,
of ground right-hand side. By Lemma~\ref{lem:recon-exptypes}(3\ref{i:exptypes-top}), an
expression type that is a refinement of a type variable is a refinement of a projection variable,
of a result variable, or of an input decoration. We prove item~2, item~1 for every constraint of
$C^\ast$ whose left-hand side is not an annotation part (the refinements of type variables are
not), and that the right-hand side of a constraint whose left-hand side is an annotation part is a
type variable or of one of the forms~\eqref{i:bounds-u3}--\eqref{i:bounds-u6}, by induction on the
derivation of the constraint in $C^\ast$ and by cases on the last clause of
Definition~\ref{def:recon-dec}.

\textbf{Case a constraint of $C^\circ\cup B$.} By the forms above: $\bar u\issubt g$ and
$\bar u\issubt a$ give \eqref{i:bounds-u3}, $\bar u\issubt\Tuple{\zeta_1,\dots,\zeta_n}$
gives~\eqref{i:bounds-u4}, $\bar u\issubt\bar u'\to_o^i\gamma'$ gives~\eqref{i:bounds-u5},
$\bar u\issubt\TChan{\bar u'}$ gives~\eqref{i:bounds-u6}, and $\beta'\issubt a$, $a\issubt\alpha'$
and $B$ give~\eqref{i:bounds-u3} and~\eqref{i:bounds-l2}.

\textbf{Case clause 1.} The right-hand side $u$ is that of the decomposed constraint
$s_1\lor s_2\issubt u$, to which form~\eqref{i:bounds-u1} does not apply, a union being no
decoration variable. If $s_i$ is not an annotation part, neither is $s_1\lor s_2$; if $s_i$ is one and
$s_1\lor s_2$ is not, $u$ is of a form~\eqref{i:bounds-u2}--\eqref{i:bounds-u6}, hence a type
variable or of a form~\eqref{i:bounds-u3}--\eqref{i:bounds-u6}. A union operand of $s_i$ is one of
$s_1\lor s_2$, so the side condition of~\eqref{i:bounds-u2} carries over. If $u$ is a type variable,
$s_1\lor s_2$ is of the form~\eqref{i:bounds-l2} or~\eqref{i:bounds-l3}, and so is $s_i$.

\textbf{Case clause 2.} By Lemma~\ref{lem:recon-exptypes}(3\ref{i:exptypes-atoms}), the left-hand
side of the decomposed constraint is a refinement of an arrow in an annotation part, and the arrow
on its right is
in an annotation part or is the arrow $\bar u\to_o^i\gamma'$ of \rulename{i-app}. The constraints
produced relate two decoration variables, which gives~\eqref{i:bounds-u1}
and~\eqref{i:bounds-l1}; a decoration variable and a ground type, which
gives~\eqref{i:bounds-u3} and~\eqref{i:bounds-l2}; an expression type and a subtree of an
annotation type, which gives~\eqref{i:bounds-u3} and~\eqref{i:bounds-l2}; or two subtrees of
annotation types, which gives~\eqref{i:bounds-u3} and~\eqref{i:bounds-l2}.

\textbf{Case clause 3.} By Lemma~\ref{lem:recon-exptypes}(3\ref{i:exptypes-atoms}), the left-hand
side is a refinement of a $\TChan{o}$, of a $\TChan{\gamma}$, or of a pid type in an annotation
part, and the pid type on
the right is the $\TChan{\bar u}$ of \rulename{i-send} or in an annotation part. The constraint
produced is $\bar u\issubt o$ or $\bar u\issubt\gamma$, which gives~\eqref{i:bounds-u2}
and~\eqref{i:bounds-l3}, or a constraint with an annotation part on one side,
which gives~\eqref{i:bounds-u3} and~\eqref{i:bounds-l2}.

\textbf{Case clause 4.} The component bounds $\pi_h(u\land\Tuple{\bar t_1,\dots,\bar t_n})$ are
computed on the disjunctive normal form, discarding the literals on type variables and the
empty tuple types (proof of Lemma~\ref{lem:penv-proj}). By the induction hypothesis $u$ is a type
variable or of one of the forms~\eqref{i:bounds-u3}--\eqref{i:bounds-u6}, but not~\eqref{i:bounds-u4},
which only the constraints of \rulename{pe-tuple} have, their left-hand side being no tuple. So the
component bounds are annotation parts: form~\eqref{i:bounds-u3}.
\end{proof}

\begin{lemma}[Guarded systems]\label{lem:recon-guarded}
Let $U$ be a finite set of type variables, each with a polarity $p_\zeta$, and $(b_\zeta)_{\zeta\in U}$
regular types whose type variables are in $U\cup\{\alpha,\beta\}$, in which every occurrence of
a variable of $U$ is below a type constructor.
\begin{enumerate}[noitemsep,topsep=2pt]
\item The system $\zeta=b_\zeta$, $\zeta\in U$, has a unique solution $(\zeta^\ast)_{\zeta\in U}$,
formed of regular types.
\item Assume that every $\eta\in U$ occurs in every $b_\zeta$ only with polarity $p_\zeta p_\eta$,
and $\beta$ only with polarity $p_\zeta$. Let $\sigma$ be a type substitution with $\alpha\sigma=a$
and $\beta\sigma\subt b$, such that $\zeta\sigma\subt b_\zeta\sigma$ for every $\zeta$ with
$p_\zeta=+$ and $b_\zeta\sigma\subt\zeta\sigma$ for every $\zeta$ with $p_\zeta=-$. Then
$\zeta\sigma\subt\zeta^\ast[\alpha:=a,\beta:=b]$ if $p_\zeta=+$, and
$\zeta^\ast[\alpha:=a,\beta:=b]\subt\zeta\sigma$ if $p_\zeta=-$.
\item Under the assumption of item~2, if moreover $\alpha$ occurs in every $b_\zeta$ only with
polarity $-p_\zeta$, then, in every $\zeta^\ast$, $\beta$ occurs only with polarity $p_\zeta$ and
$\alpha$ only with polarity $-p_\zeta$.
\end{enumerate}
\end{lemma}

\begin{proof}
\emph{Item 1.} A system of regular equations in which every unknown is below a type constructor
has a unique solution among trees, and this solution is regular.

\emph{Item 2.} Write $t(a,b)$ for $t[\alpha:=a,\beta:=b]$. Let $\zeta^\ast_0$ be $\Tany$ if
$p_\zeta=+$ and $\Tempty$ if $p_\zeta=-$, and $\zeta^\ast_{n+1}$ be $b_\zeta$ with every $\eta$
replaced by $\eta^\ast_n$.

\emph{Claim A.} For every $n$, $\zeta\sigma\subt\zeta^\ast_n(a,b)$ if $p_\zeta=+$ and
$\zeta^\ast_n(a,b)\subt\zeta\sigma$ if $p_\zeta=-$. By induction on $n$; the case $n=0$ is
immediate. Let $p_\zeta=+$, and $\theta'$ map every $\eta\in U$ to $\eta^\ast_n(a,b)$, $\alpha$ to $a$
and $\beta$ to $b$. In $b_\zeta$, every $\eta$ occurs only with polarity $p_\eta$, and $\beta$ only
with polarity $+$; by the induction hypothesis and Lemma~\ref{lem:recon-monotone},
$\zeta\sigma\subt b_\zeta\sigma\subt b_\zeta\theta'=\zeta^\ast_{n+1}(a,b)$. The case $p_\zeta=-$ is
dual.

\emph{Claim B.} Call \emph{height} of an element of the domain of
Definition~\ref{def:interpretation} its height as a finite term, tags ignored. For every $n$,
$\zeta^\ast_n(a,b)$ and $\zeta^\ast(a,b)$ contain the same elements of height smaller than $n$. By
induction on $n$: whether an element $d$ belongs to $b_\zeta$ with every $\eta$ replaced by a type
$s_\eta$ depends only on the elements of height smaller than that of $d$ that belong to the
$s_\eta$, by induction on the pair (element, type) that defines membership, since every $\eta$ is
below a type constructor.

By Claim A, an element of $\zeta\sigma$, if $p_\zeta=+$, is in $\zeta^\ast_n(a,b)$ for $n$ above its
height, hence in $\zeta^\ast(a,b)$ by Claim B; so $\zeta\sigma\subt\zeta^\ast(a,b)$. The case
$p_\zeta=-$ is dual.

\emph{Item 3.} By induction on the length of the path from the root of $\zeta^\ast$ to an
occurrence, since $\zeta^\ast$ is $b_\zeta$ with every $\eta$ replaced by $\eta^\ast$, at occurrences
of polarity $p_\zeta p_\eta$.
\end{proof}

\begin{lemma}[Elimination]\label{lem:recon-elim}
Assume \ref{h:polarity}.
\begin{enumerate}[noitemsep,topsep=2pt]
    \item The system of Definition~\ref{def:recon-elim} has a solution, greatest on the
    covariant unknowns and least on the contravariant ones, formed of regular types, and $\alpha$
    and $\beta$ occur in every $m_j(\alpha,\beta)$ only below a type constructor.
    \item For every type substitution $\sigma$ that satisfies $C^\ast$ with $\alpha\sigma=a$ and
    $\beta\sigma\subt b$, and every branch $j$ of a use of \rulename{i-receive} in $\rcv\alpha$,
    $m^\circ_j\sigma\subt m_j(a,b)$.
    \item In every $m_j(\alpha,\beta)$, $\alpha$ occurs only with polarity $-$ and $\beta$ only
    with polarity $+$.
    \item If $\sigma$ satisfies $C^\ast$ and $\alpha\sigma\subt\delta\sigma$ for the input decoration
    $\delta$ of every use of \rulename{i-receive} in $\rcv\alpha$, then
    $\alpha\sigma\land q_j\subt m_j(\alpha\sigma,\beta\sigma)$ for every branch $j$ of these uses.
\end{enumerate}
\end{lemma}

\begin{proof}
We replace the system of Definition~\ref{def:recon-elim} by a guarded system with the same
extremal solution (Step~1), apply Lemma~\ref{lem:recon-guarded} to it (Step~2), and derive the
items (Step~3). Bounds are classified by Lemma~\ref{lem:recon-bounds}. By
Lemma~\ref{lem:recon-positive}, $\alpha$ occurs in $C^\ast$ only negatively, and $\beta$ and
the pid variables $\gamma_j$, $j\not\prec k$, only positively, since $\theta$, $B$, the replaced
bounds and the clauses of Definition~\ref{def:recon-dec} preserve the polarity of their
occurrences in a constraint: $\alpha$ is no upper bound, $\beta$ no lower bound, and these
$\gamma_j$ have no upper bound.

\emph{Step 1 (the treated system).} For an unknown $\zeta$, let $V(\zeta)$ be the set of the
unknowns $\eta$ for which $C^\ast$ contains a chain $\zeta=\eta_0\issubt\eta_1$, \dots,
$\eta_{m-1}\issubt\eta_m=\eta$ ($m\geq0$) if $\zeta$ is covariant, and the reversed chain if it is
contravariant; by Definition~\ref{def:recon-unknowns}, the members of $V(\zeta)$ have the polarity
of $\zeta$. Let $B_0(\zeta)$ be the set of the upper bounds, if $\zeta$ is covariant, and of the lower
bounds, if it is contravariant, of the members of $V(\zeta)$, except those that are unknowns, and
$B(\zeta)$ the set obtained
from $B_0(\zeta)$ by replacing, in each bound of form~\eqref{i:bounds-l3}, every unknown $\eta'$
that occurs outside every type constructor by $\bigvee B_0(\eta')$. Let $b_\zeta=\bigwedge B(\zeta)$
if $\zeta$ is covariant and $b_\zeta=\bigvee B(\zeta)$ if it is contravariant.
\begin{enumerate}[label=(\alph*),ref=\alph*,noitemsep,topsep=2pt]
    \item\label{i:elim-guarded} \emph{Every unknown occurs in every $b_\zeta$ only below a type
    constructor.} In the forms~\eqref{i:bounds-u3}--\eqref{i:bounds-u6} and~\eqref{i:bounds-l2}
    every type variable is below a type constructor, and a bound of form~\eqref{i:bounds-u1},
    \eqref{i:bounds-u2} or~\eqref{i:bounds-l1} that is an unknown is not in $B_0(\zeta)$. In a
    bound of form~\eqref{i:bounds-l3}, an unknown outside every type constructor is a projection
    variable, a result variable or an input decoration
    (Lemma~\ref{lem:recon-exptypes}(3\ref{i:exptypes-top})), and the bounds in $B_0$ of these are,
    respectively, none (no constraint of $C^\ast$ has a projection variable as its right-hand side),
    annotation parts, and $\alpha$ and annotation parts, every other type variable of a bound of an
    unknown being an unknown.
    \item\label{i:elim-polarity} \emph{Every unknown $\eta$ occurs in every $b_\zeta$ only with
    polarity $p_\zeta p_\eta$, $\beta$ only with polarity $p_\zeta$, and $\alpha$ only with polarity
    $-p_\zeta$.} Let $\eta$ occur with polarity $q$ in $u\in B_0(\zeta)$. The bound $u$ is a bound of a
    member of $V(\zeta)$, of polarity $p_\zeta$, so $(\eta,p_\zeta q)$ is in the set $P$ of
    Definition~\ref{def:recon-unknowns}, and $q=p_\zeta p_\eta$ by \ref{h:polarity}. A replaced
    $\eta'$ has polarity $+$ in its bound, hence $p_{\eta'}=p_\zeta$, and every unknown $\eta$ of
    $\bigvee B_0(\eta')$ occurs in it with polarity $p_{\eta'}p_\eta=p_\zeta p_\eta$. Since $\beta$
    occurs in $C^\ast$ only positively and $\alpha$ only negatively, $\beta$ occurs in an upper
    bound only with polarity $+$ and in a lower bound only with polarity $-$, and $\alpha$
    conversely; a replacement $\bigvee B_0(\eta')$, a join of lower bounds at an occurrence of
    polarity $+$ in a lower bound, keeps these polarities. So $\beta$ occurs in $b_\zeta$ only with
    polarity $p_\zeta$, and $\alpha$ only with polarity $-p_\zeta$.
    \item\label{i:elim-ineq} \emph{If $\sigma$ satisfies $C^\ast$, then $\zeta\sigma\subt b_\zeta\sigma$
    if $\zeta$ is covariant and $b_\zeta\sigma\subt\zeta\sigma$ if it is contravariant.} The inequalities
    of a chain compose, and a replacement preserves the inequality by
    Lemma~\ref{lem:recon-monotone}, applied to the bound with a fresh type variable at the replaced
    occurrences, which have polarity $+$, since $\bigvee B_0(\eta')\sigma\subt\eta'\sigma$.
    \item\label{i:elim-same} \emph{A solution of the treated system solves the system of
    Definition~\ref{def:recon-elim}, and every solution $(s_\zeta)$ of the latter satisfies the
    inequalities of~\eqref{i:elim-ineq} with $\zeta\mapsto s_\zeta$, $\alpha\mapsto\alpha$ and
    $\beta\mapsto\beta$ in place of $\sigma$.} The bounds of $\zeta$ that are unknowns $\eta$ contribute
    the members of $B(\eta)$, which are in $B(\zeta)$, and a replaced $\eta'$ equals $\bigvee B_0(\eta')$
    at a solution, $\eta'$ having no bound of form~\eqref{i:bounds-l3}. The proof
    of~\eqref{i:elim-ineq} uses only the bounds of the unknowns, which every solution of the system
    of Definition~\ref{def:recon-elim} satisfies.
\end{enumerate}

\emph{Step 2.} By~\eqref{i:elim-guarded} and Lemma~\ref{lem:recon-guarded}(1), the treated system
has a unique solution $(\zeta^\ast)$, formed of regular types; by~\eqref{i:elim-polarity},
Lemma~\ref{lem:recon-guarded}(2) and~(3) apply to it.

\emph{Step 3 (the items).}

\emph{Item 1.} By~\eqref{i:elim-same}, $(\zeta^\ast)$ solves the system of
Definition~\ref{def:recon-elim}, and, by Lemma~\ref{lem:recon-guarded}(2) with $a=\alpha$ and
$b=\beta$, it is greatest on the covariant unknowns and least on the contravariant ones. The type
$m^\circ_j$ is $p_j[u^{C^\circ}_j]$, with each $u^{C^\circ}_j(x)$ a meet of types $t\lor\neg g$, $t$
the right-hand side of a constraint of $\mathsf{dec}_{\models}(C^\circ\setminus S_\beta)\subseteq C^\ast$
on a refinement of $\Gamma_j(x)$, hence of a
form~\eqref{i:bounds-u2}--\eqref{i:bounds-u6}, $\Gamma_j(x)$ being no decoration variable; outside
every type constructor, $m^\circ_j$ contains only the type variables $t$ of
form~\eqref{i:bounds-u2}, which are not
$\alpha$, which is no upper bound, nor $\beta$, since the self-sends are removed. Such a $t$ comes
from a send to a pid whose type is a refinement of $\TChan{t}$ (clause~3 of
Definition~\ref{def:recon-dec}), and is
a pid variable $\gamma_j$, $j\not\prec k$, or an output decoration not linked to $\beta$, a covariant
unknown. Since $\gamma_j$ has no upper bound, $\gamma_j^\ast=\Tany$. The constraints of $C^\ast$ between two type variables are among
those of $\mathsf{dec}(C)$. Indeed, $B$ and the replaced bounds add none, and in $C$ the type variables
that $\theta$ maps to ground types occur only below arrows and pid types, and in expression types only
in the atoms $\TChan{\gamma_j}$ and in the annotation types of the abstractions of $M$
(Lemma~\ref{lem:recon-exptypes}(1) and~(3\ref{i:exptypes-bracket}), proof of
Lemma~\ref{lem:recon-pure}), which $\theta$ maps to pid types and to intersections of arrow types. So
$\theta$ changes neither which clauses of Definition~\ref{def:recon-dec} apply, nor the types $\bar t_h$
of Definition~\ref{def:recon-split}, nor which types are inhabited: a type in which $\theta$ changes an
atom contains an arrow or a pid type, so it is inhabited, before and after $\theta$, only by one of the
other forms of Definition~\ref{def:recon-split}, which $\theta$ preserves in both directions. And
$\theta$ makes a tuple constraint splittable only by making its right-hand side ground, after which
every constraint derived from it has a ground side. Hence $B_0$
of an output decoration not linked to $\beta$ contains neither $\beta$ nor $\alpha$; by~\eqref{i:elim-guarded}, $\alpha$ and $\beta$ occur in
its extremum only below a type constructor.

\emph{Item 2.} Let $\theta'$ map every unknown $\eta$ to $\eta^\ast(a,b)$, $\alpha$ to $a$ and $\beta$
to $b$. By~\eqref{i:elim-ineq} and Lemma~\ref{lem:recon-guarded}(2), $\eta\sigma\subt\eta\theta'$ for
every covariant $\eta$ and $\eta\theta'\subt\eta\sigma$ for every contravariant one. In $m^\circ_j$,
every type variable other than $\alpha,\beta$ is an unknown, which occurs only with its polarity
(Definition~\ref{def:recon-unknowns}), and $\beta$ occurs only with polarity $+$, as in the upper
bounds~\eqref{i:elim-polarity}. By Lemma~\ref{lem:recon-monotone},
$m^\circ_j\sigma\subt m^\circ_j\theta'=m_j(a,b)$.

\emph{Item 3.} In $m^\circ_j$, built from upper bounds by covariant formers, $\beta$ occurs only
with polarity $+$ and $\alpha$ only with polarity $-$, and every unknown $\zeta$ only with
polarity $p_\zeta$; by~\eqref{i:elim-polarity} and Lemma~\ref{lem:recon-guarded}(3), $\beta$ occurs
in $\zeta^\ast$ only with polarity $p_\zeta$ and $\alpha$ only with polarity $-p_\zeta$. So, in
$m_j(\alpha,\beta)$, $\beta$ occurs only with polarity $+$ and $\alpha$ only with polarity $-$.

\emph{Item 4.} By Lemma~\ref{lem:recon-demand} with $X=C^\circ$,
$D_j\cup C^\circ\models r_j\issubt m^\circ_j$, and $D_j\subseteq C^\circ\subseteq C^\ast$, since $\theta$
leaves $D_j$ unchanged and $D_j$ meets neither $S_\beta$ nor $R$. So
$\alpha\sigma\land q_j\subt\delta\sigma\land q_j\simeq r_j\sigma\subt m^\circ_j\sigma$, and
$m^\circ_j\sigma\subt m_j(\alpha\sigma,\beta\sigma)$ by item~2.
\end{proof}

\begin{lemma}[Contractivity of the reconstructed interface]\label{lem:recon-contractive}
Let $\alpha$ be the input decoration of an unannotated $\mkkeyword{spawn}$, and assume
\ref{h:polarity}. Then $\alpha$ and $\beta$ occur in $F(\alpha,\beta)$ only below a type constructor,
so the equation $x=F(x,x)$ has a unique solution, which is regular and contractive.
\end{lemma}

\begin{proof}
The types $\cov\alpha$ and $q_j$ are ground, so $\alpha$ and $\beta$ occur in $F(\alpha,\beta)$ only in the
$m_j(\alpha,\beta)$, below a type constructor by Lemma~\ref{lem:recon-elim}(1). So $x$ occurs in
$F(x,x)$ only below a type constructor, and Lemma~\ref{lem:recon-guarded}(1) gives a unique
regular solution, contractive since every path from an occurrence of $x$ to its unfolding
crosses a type constructor.
\end{proof}

\subsection{The checking step}\label{app:recon-checking}
The checking step runs tallying on a constraint set that Lemma~\ref{lem:recon-check}
identifies with that of $e[\vec{t}\,]$, up to the constraints on the root bracket.

\begin{lemma}[Constraints of the checking step]\label{lem:recon-check}
Let $e$ be a program with constraint set $C$, let $\alpha_k,\beta_k,\gamma_k$ be the
variables of its $k$-th unannotated $\mkkeyword{spawn}$ ($k\geq1$), and let $\vec{t}$
be ground types $t_k$ for the $k$ of a set $S$ of such indices. The deduction of
$e[\vec{t}\,]$, in which only these $\mkkeyword{spawn}$s are replaced, is the deduction of $e$ in which the \rulename{i-spawn-alt} node of
each $k\in S$ is replaced by an \rulename{i-spawn} node with $t=t_k$, and its
constraint set is, up to the renaming of fresh variables,
\[
    C[\gamma_k:=t_k]_{k\in S}\;\cup\;\{\,\beta_k\issubt t_k,\ t_k\issubt\alpha_k\;\mid\;k\in S\,\}.
\]
\end{lemma}

\begin{proof}
By induction on the deduction of $e$, which is unique up to renaming
(Proposition~\ref{prop:inference-terminates}(1)), and by cases on the last rule applied.

\textbf{Case \rulename{i-spawn-alt} at the $k$-th $\mkkeyword{spawn}$, $k\in S$.} It concludes
with $\TChan{\gamma_k}$ and emits no constraint. \rulename{i-spawn} with $t=t_k$ has the same
premise, at the same bracket $[\beta_k,\alpha_k]$, minted once per occurrence
(Section~\ref{sec:inference-rules}), and concludes with $\TChan{t_k}$ and the constraints
$\beta_k\issubt t_k$ and $t_k\issubt\alpha_k$.

\textbf{Case \rulename{i-case}, \rulename{i-receive}.} A branch is omitted when its refined type
is empty. When this type mentions $\gamma_k$, through a scrutinee $\TChan{\gamma_k}$, the decision
is the same for $\TChan{t_k}$: test types observe a pid only at $\TChan\Tempty$
(Section~\ref{sec:syntax}), above every pid type, so $\TChan{u}\land\tau$ is empty or not
independently of $u$.

\textbf{Other cases.} No other rule inspects the shape of a type in its premises:
\rulename{i-app} posits a fresh result, \rulename{pe-tuple} defers the projections, and
\rulename{i-case} emits exhaustiveness as a constraint.

So the same rules apply at the same nodes, with the same fresh variables. The type variable
$\gamma_k$ is fresh, so it occurs in $C$ only through the conclusion type of its node, which the
enclosing rules copy into their constraints and conclusion types: replacing $\TChan{\gamma_k}$ by
$\TChan{t_k}$ replaces every occurrence of $\gamma_k$ by $t_k$, and nothing else.
\end{proof}

\subsection{Proofs of the statements of Section~\ref{sec:reconstruction}}\label{app:recon-proofs}

\ReconComputableProp*

\begin{proof}
Every test of the reconstruction is an emptiness, subtyping or equivalence test between types
with type variables, decidable~\cite[Section~3.3]{castagna2022programming}, or an entailment,
decidable by Lemma~\ref{lem:recon-entail}; every object it computes is a finite set of
constraints between regular types, or a regular type. We follow
Definition~\ref{def:reconstruction}.

\textbf{Generation.} The deduction of $e$ and its constraint set $C$ are finite and computable
(Proposition~\ref{prop:inference-terminates}(1)).

\textbf{Receives and coverage.} The uses of \rulename{i-receive} and their patterns are recorded
in the deduction, and membership in $\rcv{\alpha_k}$ is an entailment; so $\rcv{\alpha_k}$ and
$\cov{\alpha_k}$ are computable.

\textbf{Decompositions.} Each clause of Definition~\ref{def:recon-dec} yields constraints between
subtrees of the types it decomposes, or between such subtrees and projections, which are
Boolean combinations of subtrees computed on the disjunctive normal form (proof of
Lemma~\ref{lem:penv-proj}). Splittability is decidable: inhabitedness is syntactic, but for
emptiness tests on ground types and subtyping tests on ground parts, and the last condition is an
equivalence test; the
condition of clauses 2 and 3 on refinements is syntactic, but for a subtyping test on the ground
part. A
regular type has finitely many subtrees, and finitely many Boolean combinations of them up to
equivalence, so $\mathsf{dec}(X)$ and $\mathsf{dec}_{\models}(X)$ are finite, up to equivalence,
and computable for every finite $X$.

\textbf{Self-sends.} The output decorations linked to $\beta$ are found by reachability over the
finitely many constraints between type variables of $\mathsf{dec}(C)$, and a self-send by
finitely many subtyping tests $\TChan\delta\subt t$.

\textbf{Mailbox-pure abstractions and dependencies.} $M$ is the least fixpoint of a monotone
operator on the finite set of the abstractions of $e$, whose conditions are syntactic. $Q$ is
computed by reachability over $C^k$, so $\prec$, and whether it has a cycle, are decidable.

\textbf{Reduced constraint set.} For each $k$, in an order that extends $\prec$, the $t_j$ of the
$j\prec k$ are computed by induction. Then $\theta$, $B$ and $C_1$ are computable, the conditions
of Definition~\ref{def:recon-cstar} are syntactic, $\hat g_c$ is a meet of finitely many ground
types, and $f^{-1}(\hat g_c)$ a meet over the finitely many $P\subseteq J$ whose side condition is a
subtyping test; so $C^\ast$ and the types $m^\circ_j$ are computable.

\textbf{Polarity hypothesis and extrema.} The unknowns and their polarities are computed by
reachability over $C^\ast$, which decides \ref{h:polarity}. The extrema are the solution of the
finite guarded system of the proof of Lemma~\ref{lem:recon-elim}, the regular type that this
system describes.

\textbf{Interfaces and checking step.} Each interface is the regular solution of the guarded
equation $x=F(x,x)$ (Lemma~\ref{lem:recon-contractive}). Tallying terminates on every finite set
of constraints between regular types and returns a finite, sound and complete set of
solutions (Lemma~\ref{lem:tally-readback}), so whether
the set of Definition~\ref{def:reconstruction} has a solution is decidable.
\end{proof}

\ReconSoundnessProp*

\begin{proof}
Let $e$ have $n$ unannotated $\mkkeyword{spawn}$s, let $C$ be its constraint set and
$[\beta_0,\alpha_0]$ its root bracket, let $t_0,\dots,t_n$ be the interfaces that the solving step
computes ($t_0$ being the declared one, if there is one), and let
$C'=C[\gamma_k:=t_k]_{k\geq1}\cup\{\beta_k\issubt t_k,\ t_k\issubt\alpha_k\mid 0\leq k\leq n\}$, which
has a solution since $e$ is accepted.

\emph{Step 1 (the deduction of $e[\vec{t}\,]$).}
\begin{itemize}[noitemsep,topsep=2pt]
    \item $e[\vec{t}\,]$ is a program of Section~\ref{sec:ann-inference} whose reconstructed
    $\mkkeyword{spawn}$s carry a type, as \rulename{i-spawn} allows
    (Footnote~\ref{fn:spawn-variants} in Section~\ref{sec:inference-rules}).
    \item By Lemma~\ref{lem:recon-check}, its constraint set, with the constraints on the root
    bracket, is $C'$.
    \item Its elaboration (Appendix~\ref{app:elab-elaboration}) is a term $f$ that differs from
    $e[\vec{t}\,]$ only in the annotations and types it carries; its $k$-th reconstructed
    $\mkkeyword{spawn}$ is $\Spawn{f_k}{t_k}$.
    \item Since $C'$ has a solution, Corollary~\ref{cor:inferred-safety} applies to
    $e[\vec{t}\,]$ at the interface $t_0$: it rests on Lemma~\ref{lem:elab-sound}, which
    covers a $\mkkeyword{spawn}$ that carries a type.
    \item Extending $\typeerase\cdot$ to a $\mkkeyword{spawn}$ that carries a type, or none, as
    Section~\ref{sec:typerase} defines it on one that carries an annotation,
    $\typeerase{e[\vec{t}\,]}=\typeerase{e}$.
\end{itemize}
This proves the first claim. The corollary does not mention the processes that $e[\vec{t}\,]$
spawns: we derive the second claim from its proof.

\emph{Step 2 (the annotated run).} Let $\sigma\in\textsf{Solutions}(C')$ and let $\rho$ map every
type variable to a ground type. By Step~1 of the proof of Corollary~\ref{cor:inferred-safety},
$P_0=\Process{\PID}{t_0}{f\sigma\rho}{\EmptyMailbox}$ is well typed and well-formed, and, by
Corollary~\ref{cor:soundness}, in every configuration reachable from $P_0$ every message $w$ in
the mailbox of a process $\Process{\PID[b]}{s}{e'}{\seqof v}$ has $\wtee{}{w}{s}$. In the
derivation of $f\sigma\rho$ that this proof builds, the $k$-th reconstructed $\mkkeyword{spawn}$
is typed by \rulename{t-spawn} at the interface $t_k$: the case of \rulename{i-spawn} in the
proof of Lemma~\ref{lem:elab-sound} moves its body from $[\beta_k\sigma\rho,\alpha_k\sigma\rho]$ to
$[t_k,t_k]$ by Theorem~\ref{thm:narrowing}, since $\beta_k\sigma\rho\subt t_k\subt\alpha_k\sigma\rho$, and $t_k$
is ground. Erasing the type from this instance of \rulename{t-spawn} gives an instance of
\rulename{t-spawn-alt} with $t=t_k$.

\emph{Step 3 (the reconstructed processes).} Only the $\mkkeyword{spawn}$ rule of
Section~\ref{sec:opsem} creates processes: for a redex $\Spawn{e'}{s}$ it creates
$\Process{\PID[b]}{s}{e'}{\EmptyMailbox}$ with $\PID[b]$ fresh, and no rule changes the interface
of a process. The redexes that come from the $k$-th reconstructed $\mkkeyword{spawn}$ are of the
form $\Spawn{e'}{t_k}$, since reduction substitutes values for variables in $f_k\sigma\rho$ but
leaves its ground type $t_k$ in place. So every process that this $\mkkeyword{spawn}$ creates has
interface $t_k$, and, by Step~2, every message in its mailbox has type $t_k$ in the enriched
system, and in the system of Section~\ref{sec:typing} when it contains neither abstractions nor
pids (Lemma~\ref{lem:vee-principal}).

\emph{Step 4 (the erased run).} Since $f\sigma\rho$ differs from $e[\vec{t}\,]$ only in the
annotations and types it carries, $P_0$ erases to $(\PID,\typeerase{e},\EmptyMailbox)$. By
Theorem~\ref{thm:typerase}, iterated along a reduction sequence as in the proof of
Corollary~\ref{cor:inferred-safety}, every configuration reachable from the erased $P_0$ is the
erasure of a configuration reachable from $P_0$, process by process
(Section~\ref{sec:typerase}). So a process created by the $k$-th $\mkkeyword{spawn}$ in the erased
run is the erasure of a process of interface $t_k$ whose mailbox holds messages of type $t_k$.
\end{proof}

\ReconCanonicalProp*

\begin{proof}
Let $k$ be the index of the $\mkkeyword{spawn}$, so that $\alpha,\beta,\gamma$ are
$\alpha_k,\beta_k,\gamma_k$. By Lemma~\ref{lem:recon-elim}(3) and Lemma~\ref{lem:recon-monotone},
every $m_j(a,b)$, and so $F(a,b)$, is antitone in $a$ and monotone in $b$.

\emph{Item 1.} $t\simeq F(t,t)$, so $t$ is servable. Let $s$ be servable with $t\subt s$. Then
$s\subt F(s,s)\subt F(t,s)$. The type variable $x$ occurs in $F(t,x)$ only below a type
constructor (Lemma~\ref{lem:recon-contractive}) and only with polarity $+$
(Lemma~\ref{lem:recon-elim}(3)), and $t$ is the solution of $x=F(t,x)$. By
Lemma~\ref{lem:recon-guarded}(2), with $U=\{x\}$, $p_x=+$ and $x\sigma=s$, we have $s\subt t$, hence
$s\simeq t$.

\emph{Item 2.} Let $s\subt\cov\alpha$ be such that inference accepts $e[\vec{t}\,]$ with $s$ as the
interface of the $k$-th $\mkkeyword{spawn}$. Let $e'$ be $e$ in which the $j$-th unannotated
$\mkkeyword{spawn}$ carries $t_j$ for every $j\prec k$ and the $k$-th carries $s$, run at the
interface $t_0$ if $0\prec k$ and at $s$ if $k=0$.
\begin{itemize}[noitemsep,topsep=2pt]
    \item By Lemma~\ref{lem:recon-check}, the constraint set of $e'$ is
    $C'=C[\gamma_j:=t_j]_{j\prec k}[\gamma:=s]\cup B\cup\{\beta\issubt s,\ s\issubt\alpha\}$, with $B$ as in
    Definition~\ref{def:recon-cstar}. The constraint set of $e[\vec{t}\,]$ with $s$ contains $C'$
    with the other $\gamma_j$ replaced by $t_j$, so $C'$ is satisfiable.
    \item By Lemma~\ref{lem:recon-pure}(2), applied to $e'$ and to the constraints of $C'$ on the root
    bracket, $\theta_M(C')$ has a solution $\sigma$.
    \item Extended by $\gamma\mapsto s$, $\sigma$ satisfies $\theta(C)$, hence $\theta(C\setminus(S_\beta\cup R))$,
    $B$, and $C_1$ by Lemma~\ref{lem:recon-dec}; so it satisfies the bounds
    $t_2\theta\issubt f^{-1}(\hat g_c)$ (Lemma~\ref{lem:recon-pure}(1)), and $C^\ast$ by
    Lemma~\ref{lem:recon-dec}.
    \item The substitution that applies $\theta$ and then $\sigma$ satisfies $C$, and agrees with $\sigma$
    on $\alpha$ and on the input decorations of the uses of \rulename{i-receive}. Since
    $C\models\alpha\issubt\delta$ for the input decoration $\delta$ of every use of
    \rulename{i-receive} in $\rcv\alpha$, $\alpha\sigma\subt\delta\sigma$.
\end{itemize}
By Lemma~\ref{lem:recon-elim}(4), $\alpha\sigma\land q_j\subt m_j(\alpha\sigma,\beta\sigma)$ for every branch
$j$ of these uses. Since $s\subt\alpha\sigma$ and $\beta\sigma\subt s$,
$s\land q_j\subt\alpha\sigma\land q_j\subt m_j(\alpha\sigma,\beta\sigma)\subt m_j(s,s)$. With
$s\subt\cov\alpha$, $s\subt\cov\alpha\land\bigwedge_j(m_j(s,s)\lor\neg q_j)=F(s,s)$.

\emph{Item 3.} If $F(\alpha,\beta)$ does not depend on $\alpha$, then $F(s,s)=F(t,s)$ for every
$s$, and the argument of item~1 applies to every servable $s$ without assuming $t\subt s$: $t$ is
the greatest servable type. If $F(\alpha,\beta)$ depends on neither $\alpha$ nor $\beta$, it is
ground, the equation is $x=F(\alpha,\beta)$, and $t$ is $F(\alpha,\beta)$.
\end{proof}

\begin{remark}[Written decorations]\label{rem:recon-written}
With the written decorations of Section~\ref{sec:elixir-decorations}, an input decoration of
the deduction can be a ground type, and $C$ can bound $\alpha$ by a ground type that is not a
demand. Let $\mathsf{ub}_e(\alpha)$ be the meet of the ground types $g$ such that
$(\delta\issubt g)\in\mathsf{dec}_{\models}(C)$ for a type variable $\delta$ that is $\alpha$ or
is reached from it by a chain $\alpha\issubt\delta_1\issubt\dots\issubt\delta$ of constraints of
$\mathsf{dec}_{\models}(C)$ between type variables; it is $\Tany$ when no decoration is
written. Since $C\models\alpha\issubt\mathsf{ub}_e(\alpha)$ (Lemma~\ref{lem:recon-dec}), every
interface that the checking step accepts is below $\mathsf{ub}_e(\alpha)$, whereas the solution of
$x=F(x,x)$ need not be: for a process that calls a function of written input decoration
\eelixir|{:add, integer()}| whose only branch, of pattern \eelixir|{:add, k}|, does not use
\eelixir{k}, this solution is \eelixir|{:add, term()}|, which the checking step rejects, since
it places the interface below $\alpha$. An implementation therefore solves
$x=F(x,x)\land\mathsf{ub}_e(\alpha)$. Since $\mathsf{ub}_e(\alpha)$ is ground, this equation is
guarded and its right-hand side has the polarities of $F$: its solution $t'$ is unique, it is
servable, since $t'\subt F(t',t')$, and it is maximal among the servable types below
$\mathsf{ub}_e(\alpha)$, by the proof of item~1 of Proposition~\ref{prop:recon-canonical}
applied to $F\land\mathsf{ub}_e(\alpha)$. At a ground input decoration, \rulename{i-receive} may
also skip a branch that selects some messages: this branch is left out of the demands and of the
coverage, as Definition~\ref{def:recon-demand} prescribes.
\end{remark}

\subsection{Examples}\label{app:recon-examples}

\paragraph{The maximal interfaces of the extended loop}
Write $u_0$ for the union of the first three summands of
equation~\eqref{eq:recon-run}, and for $u\subt u_0$ let
$s_u=u\lor\Tuple{\Atom[{:}run],\Tuple{}\to_{s_u}^{s_u}\Tany}$, a
contractive regular type. It is servable: $u$ satisfies the ground demands and
the coverage, and $s_u\land q_{\mathtt{run}}$ is
$\Tuple{\Atom[{:}run],\Tuple{}\to_{s_u}^{s_u}\Tany}$, which is
$m_{\mathtt{run}}(s_u,s_u)$. It is maximal: let $s$ with $s_u\subt s$ be
servable; then
$\Tuple{\Atom[{:}run],\Tuple{}\to_{s_u}^{s_u}\Tany}\subt s\land q_{\mathtt{run}}\subt\Tuple{\Atom[{:}run],\Tuple{}\to_{s}^{s}\Tany}$,
the tuple is nonempty, and by Proposition~\ref{prop:arrow-factor} the
contravariance of the input decoration gives $s\subt s_u$. Since
$s_u\land\neg q_{\mathtt{run}}=u$, distinct $u$ give distinct and, by
maximality, incomparable types: for instance
$u=\Atom[{:}bump]\lor\Tuple{\Atom[{:}add],n}$ for each integer $n$. The join
of two of them is strictly above one of them, hence not servable.

\paragraph{The reply server}
Consider a process whose $\mkkeyword{receive}$ has the branch\par
\centerline{\eelixir|{:run, from, job} -> send(from, job.())|}\par
\noindent typed at $[\beta,\alpha]$ (after
the elimination of the chains). \rulename{pe-tuple} gives
$\alpha\land q\issubt\Tuple{\zeta_{\mathtt{tag}},\zeta_{\mathtt{from}},\zeta_{\mathtt{job}}}$ with
\eelixir{from}$\,:\zeta_{\mathtt{from}}$ and \eelixir{job}$\,:\zeta_{\mathtt{job}}$; \rulename{i-app} gives
$\zeta_{\mathtt{job}}\issubt\Tuple{}\to_{\beta}^{\alpha}\gamma'$, at the bracket of the branch;
\rulename{i-send} gives $\zeta_{\mathtt{from}}\issubt\TChan{\gamma'}$. In the
composed bound $\Tuple{\Atom[{:}run],\TChan{\gamma'},\Tuple{}\to_{\beta}^{\alpha}\gamma'}$
the variable $\gamma'$ occurs contravariantly under $\TChan\cdot$ and covariantly as
the codomain, so \ref{h:polarity} fails and no extremum weakens both occurrences.
No most permissive interface exists. For a ground $t$, the type
$s_t=\Tuple{\Atom[{:}run],\TChan t,\Tuple{}\to_{s_t}^{s_t}t}$ lets the branch
typecheck at interface $s_t$ (take $\gamma'=t$ and $\alpha=\beta=s_t$: the arrow
demanded of the job is then decorated with $s_t$ itself). Let $s$ be an interface above both $s_{t_1}$ and $s_{t_2}$ at which
the branch typechecks, by some $\sigma$. Then
$s_{t_i}\subt s\land q\subt\Tuple{\zeta_{\mathtt{tag}},\zeta_{\mathtt{from}},\zeta_{\mathtt{job}}}\sigma$, so
$\TChan{t_i}\subt\zeta_{\mathtt{from}}\sigma\subt\TChan{\gamma'\sigma}$ gives
$\gamma'\sigma\subt t_1\land t_2$ (Proposition~\ref{prop:pid-laws}(1)), while
$\Tuple{}\to_{s_{t_i}}^{s_{t_i}}t_i\subt\zeta_{\mathtt{job}}\sigma\subt\Tuple{}\to_{\beta\sigma}^{\alpha\sigma}\gamma'\sigma$
gives $t_1\lor t_2\subt\gamma'\sigma$ (Proposition~\ref{prop:arrow-factor}). For
$t_1=\TInt$ and $t_2=\TBool$ the two are contradictory: no interface lies above both
$s_{\TInt}$ and $s_{\TBool}$, and the servable interfaces of the reply server have
no greatest element and no canonical maximal one. The correlation between
\eelixir{from} and \eelixir{job} that no single type expresses is what the pattern
environments of Section~\ref{sec:typing} give up when they project a tuple componentwise
(Section~\ref{sec:typing-rules} indicates how it could be recovered at a declared interface).

\paragraph{Refinements}
With patterns $p'_1,\dots,p'_n$, a $\mkkeyword{case}$ on a capture variable $x$ of $p_j$ emits
the bound $\Gamma_j(x)\issubt\bigvee_h\accept{p'_h}$ (\rulename{i-case}), and its pattern
environments, and the uses of the capture variables they bind, place on the left of constraints
the refinements $(\Gamma_j(x)\land\accept{p'_h})\setminus\bigvee_{l<h}\accept{p'_l}$ of $\Gamma_j(x)$.
In the branch\par
\centerline{\eelixir|{:m, x} -> case x do {:k, z} -> z + 1; _ -> 0 end|}\par
\noindent for instance, \rulename{pe-tuple} emits
$(\Gamma_j(x)\land\Tuple{\Atom[{:}k],\Tany})\setminus\Tempty\issubt\Tuple{\zeta_1,\zeta_z}$,
with $\zeta_1,\zeta_z$ fresh, which contributes
$\Tuple{\zeta_1,\zeta_z}\lor\neg\Tuple{\Atom[{:}k],\Tany}$ to $u_j(x)$. The extrema of
$\zeta_1$ and $\zeta_z$ are their upper bounds $\Atom[{:}k]$ (\rulename{pe-test}) and $\TInt$
(\rulename{i-plus}), at which $m_j$ is equivalent to
$\Tuple{\Atom[{:}m],\Tuple{\Atom[{:}k],\TInt}\lor\neg\Tuple{\Atom[{:}k],\Tany}}$; without
refinements it would be $\Tuple{\Atom[{:}m],\Tany}$.

\paragraph{Self-sends}
Without Definition~\ref{def:recon-selfsend}, some sends would place $\beta$, or an output
decoration whose extremum is a meet with $\beta$, at the top of an upper bound of the type of a
capture variable, and make the equation $x=F(x,x)$ unguarded. A target type with a summand
$\TChan\delta$, where $\delta$ is linked to $\beta$, as in
\eelixir|send(case other do :a -> self(); _ -> w end, other)|, decomposes into
$\Gamma_j(\mathtt{other})\issubt\delta$ (clauses 1 and 3 of Definition~\ref{def:recon-dec}). So does a send,
in a closure that the process runs, to a $\Self$ captured outside it, as in
\eelixir|me = self(); receive do m -> (fn -> send(me, m) end).() end|, whose constraint would
reduce the equation to $x=x$. Since expression types are built from $\TChan\delta$ by covariant
formers only (Lemma~\ref{lem:recon-exptypes}(1)), a target type with the summand $\TChan\delta$ is
above $\TChan\delta$, which is the criterion of Definition~\ref{def:recon-selfsend}.

\paragraph{A closure sent to a received pid}
The decorations $\alpha_c,\beta_c$ of a closure created in a branch and sent to a pid that the
branch received, as in the third case of Section~\ref{sec:recon-limits}, are unknowns of
Definition~\ref{def:recon-unknowns}, $\alpha_c$ covariant and $\beta_c$ contravariant. The
$\mkkeyword{receive}$s of the closure never bound $\alpha_c$: they place on it the demands
$\alpha_c\land q\issubt m$, which are not bounds of $\alpha_c$. So $\alpha_c$ is taken, in general,
at $\Tany$, an over-approximation that Lemma~\ref{lem:recon-elim}(2) allows and that the checking
step rejects when these $\mkkeyword{receive}$s constrain what they read. The sends of the closure to
its own $\Self$, and those of the functions it calls, bound $\beta_c$ from below, unless the
process runs the closure too: they are then self-sends of the process and are removed, and
$\beta_c$ is taken at $\Tempty$. An implementation that extracts the bounds from normal forms
(see the comments on Definition~\ref{def:recon-dec}) is finer here: for it the demand
$\alpha_c\land q\issubt m$ is the bound $\alpha_c\issubt m\lor\neg q$, and for the branch of
Section~\ref{sec:recon-limits} it reconstructs the interface
$\Tuple{\Atom[{:}job],\TChan{\Tuple{}\to_{\Tempty}^{i_c}\TInt}}$, with
$i_c=\neg\Tuple{\Atom[{:}int],\neg\TInt}$, which the checking step accepts.

\paragraph{An effect-overloaded function}
Condition~\ref{i:cstar-pure} of Definition~\ref{def:recon-cstar}, under which the constraint
of a call is replaced by a bound, is needed: if the function were not mailbox-pure, the
selection would depend on the interface itself. Let \eelixir{f} be
\begin{elixircode}
fn x -> case x do
  n when is_integer(n) -> 0
  a when is_atom(a)    -> (receive do {:m, y} -> y end) + 1
end end
\end{elixircode}
\noindent
annotated at $(\TInt\to\TInt)\land(\TAtom\to\TInt)$. Under the first arrow the
second branch is skipped (its refined type $(\TInt\land\TAtom)\setminus\TInt$ is empty), so nothing
constrains that arrow's input decoration, which may be $\Tany$; under the second,
the $\mkkeyword{receive}$ requires $\alpha_2\land\Tuple{\Atom[{:}m],\Tany}\subt\Tuple{\Atom[{:}m],\TInt}$,
whose greatest solution is
$i_2=\Tuple{\Atom[{:}m],\TInt}\lor\neg\Tuple{\Atom[{:}m],\Tany}$. So
\eelixir{f} has type
$(\TInt\to_{\Tempty}^{\Tany}\TInt)\land(\TAtom\to_{\Tempty}^{i_2}\TInt)$.
A process running \eelixir|receive do {:v, x} -> f(x); {:m, _} -> :ok end| in a
loop, at interface $s$, selects the second arrow if and only if $s\subt i_2$, that
is, if and only if its \eelixir{:m} payloads are integers, and by
Lemma~\ref{lem:folded-decomp} the type of its \eelixir{x} must then be below $\TInt$ or below
$\TInt\lor\TAtom$ accordingly. Inference accepts the program with either
$s_1=\Tuple{\Atom[{:}v],\TInt}\lor\Tuple{\Atom[{:}m],\Tany}$ or
$s_2=\Tuple{\Atom[{:}v],\TInt\lor\TAtom}\lor\Tuple{\Atom[{:}m],\TInt}$ as the interface,
each is maximal among such interfaces (enlarging the \eelixir{:m} payload of $s_2$ loses the
second arrow, and $s_1$ already has the largest payloads its selection allows),
and it rejects $s_1\lor s_2$, which selects the first arrow alone and asks it to
accept atoms. No greatest interface exists, although $\alpha$ occurs in no demand: the
upper bound of the type of \eelixir{x} depends on the interface, so no ground bound can replace the constraint
of this call, which is why Definition~\ref{def:recon-cstar} replaces only the constraints of calls
to mailbox-pure functions. The constraint of this call stays, the extrema miss it, and the
checking step rejects the reconstructed interface
$\Tuple{\Atom[{:}v],\Tany}\lor\Tuple{\Atom[{:}m],\Tany}$. (Read in normal form, the constraint
of the call bounds the type of \eelixir{x} by $\TInt\lor\TAtom$, the union of the domains, which
it entails whatever the selection; the interface is then $s_1\lor s_2$, rejected as well.) The same
example shows that $\rcv\alpha$, hence the coverage, is sensitive to
the selection: whether the $\mkkeyword{receive}$ of \eelixir{f} is in $\rcv\alpha$
depends on whether $C\models\alpha\issubt\alpha_2$, which fails as soon as some
solution selects the first arrow only.

\section{Prototype Measurements}
\label{app:measurements}
This appendix gives the measurements that Section~\ref{sec:elixir-tallying} summarizes. They
are produced by the script \texttt{bench/bench.sh} of the prototype, which reproduces them on
any machine. The programs are those of the paper and are small (at most fifty constraints):
the figures give orders of magnitude for programs of this size, nothing more.
The figures were measured on a machine with the processor 11th Gen Intel(R) Core(TM) i7-1185G7 @ 3.00GHz
(8 logical cores, 31~GB of memory), under
Fedora Linux 43 (Workstation Edition), with Elixir~1.20.4 on Erlang/OTP~29
(commit 5d68891 of the prototype, 2026-11-11). Each line is the
fastest of 3 runs, each in a fresh virtual machine. Times are in
milliseconds and do not count the start of the virtual machine (385~ms
on this machine).

In the tables, \emph{constr.} is the number of constraints given to tallying and \emph{calls}
the number of calls to it: one for each set of constraints that share type variables, and more
when errors must be located. \emph{Generation} is the time of parsing and of constraint
generation, and \emph{total} the time of the whole check. All times are in milliseconds.

\begin{table}[ht]
\caption{The server of Section~\ref{sec:example}, without and with its jobs: decorations
written or inferred, interfaces declared or reconstructed.}
\label{tab:meas-configs}
\footnotesize\setlength{\tabcolsep}{4.5pt}
\begin{tabular}{@{}lllrrrrrr@{}}
\toprule
  program & decorations & interfaces & constr. & calls & generation & reconstr. & tallying & total \\
\midrule
  server & written & declared & 15 & 3 & 5.7 & -- & 3.4 & 10.1 \\
  server & written & reconstructed & 16 & 4 & 5.6 & 3.4 & 3.5 & 14.8 \\
  server & inferred & declared & 34 & 2 & 6.5 & -- & 14.1 & 22.5 \\
  server & inferred & reconstructed & 35 & 3 & 6.9 & 8.9 & 13.5 & 34.6 \\
  server with jobs & written & declared & 26 & 4 & 7.8 & -- & 110 & 119 \\
  server with jobs & written & reconstructed & 26 & 4 & 7.6 & 6.1 & 53.4 & 71.5 \\
  server with jobs & inferred & declared & 50 & 1 & 13.9 & -- & 591 & 609 \\
  server with jobs & inferred & reconstructed & 50 & 1 & 9.9 & 25.8 & 525 & 571 \\
\bottomrule
\end{tabular}

\end{table}

Table~\ref{tab:meas-configs} checks the same program in four ways: with its decorations
written or inferred, and with its interfaces declared, which is the inference of
Section~\ref{sec:ann-inference}, or reconstructed. The column \emph{reconstr.} is the time of
the steps of the reconstruction that precede the checking step. For the server with jobs, only
the interface of the worker is reconstructed (when no decoration is written, the prototype does
not reconstruct both interfaces). The time of a check is essentially the time of tallying. It
grows with the decorations that are left to infer, whereas omitting the interfaces changes it
by a few tens of milliseconds at most, in one direction or the other.

\begin{table}[ht]
\caption{The steps of the reconstruction.}
\label{tab:meas-steps}
\footnotesize\setlength{\tabcolsep}{3.5pt}
\begin{tabular}{@{}lrrrrrrrrr@{}}
\toprule
  example & constr. & interf. & generation & bounds & derivation & depend. & solving & checking & total \\
\midrule
  bank account & 33 & 2 & 4.9 & 3.0 & 0.4 & 0.5 & 1.7 & 8.9 & 20.4 \\
  server & 35 & 2 & 8.0 & 7.8 & 0.6 & 0.5 & 0.6 & 18.3 & 37.7 \\
  server with jobs & 50 & 1 & 10.1 & 22.8 & 1.0 & 0.9 & 1.3 & 548 & 589 \\
  forwarding & 25 & 2 & 4.3 & 2.4 & 0.4 & 0.2 & 1.5 & 5.0 & 14.4 \\
  appendix examples & 32 & 3 & 5.9 & 5.0 & 0.5 & 0.5 & 0.7 & 11.6 & 25.2 \\
  closure sent to a pid & 19 & 1 & 2.9 & 1.0 & 0.2 & 0.4 & 0.4 & 4.6 & 9.9 \\
  interface too large (rejected) & 50 & 3 & 5.8 & 7.1 & 0.7 & 1.2 & 1.9 & 67.2 & 86.0 \\
\bottomrule
\end{tabular}

\end{table}

Table~\ref{tab:meas-steps} details the reconstruction on examples of
Section~\ref{sec:reconstruction} and of Appendix~\ref{app:recon-examples}. \emph{Bounds} is the
extraction of the bounds from the normal forms of the constraints, with the search for the
mailbox-pure functions; \emph{derivation} is the search for the $\mkkeyword{receive}$s that
each process runs, with their demands and their coverage; \emph{depend.} is the dependency
order; \emph{solving} is the computation of the polarities, of the extrema and of the
interfaces; \emph{checking} is the checking step. Only the last one calls tallying. The four
others together stay below thirty milliseconds, most of which is spent on the bounds.

\begin{table}[ht]
\caption{The server with jobs: locating errors, and inference without a change of tallying.}
\label{tab:meas-tallying}
\footnotesize\setlength{\tabcolsep}{4.5pt}
\begin{tabular}{@{}llllrrr@{}}
\toprule
  program & decorations & tallying & outcome & calls & time in it & total \\
\midrule
  ill-typed sends & written & as described & rejected (4) & 22 & 206 & 218 \\
  ill-typed sends & written, except for the jobs & as described & rejected (4) & 20 & 10\,302 & 10\,330 \\
  server with jobs & inferred & as described & accepted & 1 & 605 & 624 \\
  server with jobs & inferred & no elimination & accepted & 1 & 7\,299 & 7\,351 \\
  server with jobs & inferred & all the solutions & accepted & 1 & 614 & 633 \\
\bottomrule
\end{tabular}

\end{table}

Table~\ref{tab:meas-tallying} concerns the server with jobs. Its first two lines are the
program with the two ill-typed messages and the two jobs that overstep: locating the four
errors solves about twenty prefixes of the constraints, which is cheap when every decoration
is written and takes seconds when those of the jobs are inferred. Its last three lines are
inference on the program without decorations: as Section~\ref{sec:elixir-tallying} describes
it, without the elimination of bound variables, and with all the solutions computed instead
of the first one. On this program the elimination gains a factor of more than ten, and stopping at
the first solution gains nothing. The first change of Section~\ref{sec:elixir-tallying}, the
single system of equations, cannot be switched off in the prototype.

\behavior{%
\input{app-behavioral-types}
\input{app-behavioral-domain}
}

\end{document}